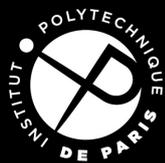

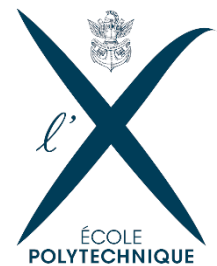

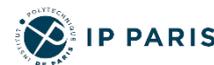

Thèse de doctorat

# Statistical properties of partonic configurations and diffractive dissociation in high-energy electron-nucleus scattering

Thèse de doctorat de l'Institut Polytechnique de Paris
préparée à l'École polytechnique

École doctorale n°626 de l'Institut Polytechnique de Paris (EDIPP)
Spécialité de doctorat: Physique

Thèse présentée et soutenue à Palaiseau, France, le 19 Novembre 2021, par

## Anh Dung Le

Composition du Jury :

**Grégory Schehr**
Directeur de recherche - LPTHE, Sorbonne Université,
CNRS (France)                                               Président

**Yuri Kovchegov**
Professeur - Department of Physics, The Ohio State
University (USA)                                            Rapporteur

**Tuomas Lappi**
Professeur - Department of Physics, Jyväskylä University
(Finland)                                                   Rapporteur

**Nestor Armesto**
Professeur - IGFAE, Universidade de Santiago de
Compostela (Spain)                                          Examinateur

**Cédric Lorcé**
Professeur - CPHT, École polytechnique (France)             Examinateur

**Stéphane Munier**
Directeur de recherche - CPHT, CNRS, École polytechnique
(France)                                                    Directeur de thèse

*Cette thèse est dédiée à ma femme et ma jolie fille LE Minh-Anh!*

# Acknowledgements

*"He who is unable to live in society, or who has no need because he is sufficient for himself, must be either a beast or a god."*

– Aristotle –

A three-year period has elapsed since the date I entered CPHT as a doctoral student, neither so long nor so short. And it is now the time to finish the PhD journey. There is no doubt that I would be not able to complete my Ph.D. study without the help and support from many people around. Therefore, before continuing with the scientific discussions, let me express my appreciation to them.

My sincere gratitude should go first to my thesis supervisor, Prof. Stéphane Munier, for driving me to the physics of small-$x$ dynamics and of branching-diffusion processes, and for his thorough support, navigation and valuable discussions through the whole of my doctoral study. It was not just once, but several times, that he spent part of his days off for my problems. Actually I have learnt a lot from him, in both physics and the way he works, over the past years.

I am grateful to Prof. Alfred H. Mueller. To be honest, his handwritten working notes in the form of letters really impressed me, and helped me appreciate the considered problems better. I also enjoyed a few conversations with him before the pandemic.

My gratitude is also dedicated to my PhD committee, Prof. Christoph Kopper and Dr. Stéphane Peigne, for their advices and suggestions, as well as for their valuable questions and comments in the thesis mid-term defense.

It would be a great gap not to mention CPHT. I am really fortunate to be a part of CPHT with very friendly and kind people. I would like to thank them for providing an enjoyable working environment and for being ready for any help. A special thanks is to Cédric Lorcé, for giving me some insights into the light-cone formalism at the beginning of my PhD. I would like to show my gratitude also to the informatic and technical team for providing and maintaining the computational resources for numerical calculations.

To all of my friends, thank you for their help in daily life and for numerous relaxed conversations. Many thanks also goes to my neighbors with lovely childrens.

Especially, I am indebted to my family for their invaluable support. My wife and my adorable daughter Minh-Anh, they are always beside me. They are the greatest asset of mine after all.

Finally, I would like to forward my thanks to Fondation de l'X for their financial support for my Master+PhD track during last five years in École polytechnique.

# Abstract

**Title**: *Statistical properties of partonic configurations and diffractive dissociation in high-energy electron-nucleus scattering.*

In the high-energy scattering of a quark-antiquark color dipole off a hadron, the quantum states of the former are represented by a stochastic set of dipoles generated by a binary branching process, in the so-called color dipole model of quantum chromodynamics (QCD). It was found that there is a profound connection between this QCD description and the branching-diffusion processes studied in statistical physics from which different properties of the scattering in the high-energy regime are revealed. Our work in this thesis is aimed to exploit the cross-fertilization between QCD and statistical physics to study the detailed partonic content of the Fock states of a color dipole subject to high-energy evolution in the scattering off a large nucleus. We also produce predictions for diffractive dissociation in electron-ion collisions, based on the QCD dipole picture.

In the first place, the scattering events of a color dipole, when parameters are set in such a way that the total cross section is small, are triggered by configurations containing large-transverse-size dipoles. The latter are due to rare partonic fluctuations, which look different as seen from different reference frames, from the rest frame of the nucleus to frames in which the rapidity is shared between the projectile dipole and the target nucleus. It turns out that the freedom to select a frame allows to deduce an asymptotic analytic expression for the rapidity distribution of the first branching of the slowest parent dipole of the set of those which scatter, which provides an estimator for the correlations of the latter. In another aspect, the study implies the importance of the characterization of particle distribution near the extremal particles, referred to as the "tip", in the states generated by the QCD dipole branching, and more generally, by any one-dimensional branching random walk model. To this aim, we develop a Monte Carlo algorithm to generate the tip of a binary branching random walk on a real line evolving to a predefined time, which allows to study both rare and typical configurations.

The above statistical description proves advantageous for calculating diffractive cross section demanding a minimal rapidity gap $Y_0$ and the distribution of rapidity gaps $Y_{\text{gap}}$ in the diffractive dissociation of a small dipole off a large nucleus, in a well-defined parametric region. They are the asymptotic solutions to the Kovchegov-Levin equation, which was established more than 20 years ago to describe the diffractive dipole dissociation at high energy. Additionally, we present predictions for the distribution of rapidity gaps in realistic kinematics of future electron-ion machines, based on the numerical solutions to the original Kovchegov-Levin equation and of its next-to-leading extension taking into account the running of the strong coupling. The outcomes for the former reflect in a qualitative way our asymptotic analytical result already at rapidities accessible at future electron-ion colliders.

# Résumé

**Titre**: *Propriétés statistiques des configurations partoniques et dissociation diffractive dans la diffusion électron-noyau à haute énergie.*

Dans le cadre de la chromodynamique quantique (QCD), la théorie microscopique de l'interaction forte, on montre que les états quantiques d'un quarkonium pertinents dans les collisions hadroniques à très haute énergie, dans la limite paramétrique théorique d'un grand nombre de couleurs, peuvent être représentés par un ensemble stochastique de dipôles de couleur générés par un processus de branchement binaire particulier. Cette image des états quantiques hadroniques est appelée "modèle des dipôles de couleur". Ce modèle peut être analysé à l'aide d'outils généraux développés pour l'étude de processus de branchement diffusif en physique statistique et en mathématiques. On sait par exemple que l'équation de Balitsky-Kovchegov établie dans le cadre du modèle des dipôles de couleur et qui régit l'évolution avec l'énergie d'amplitudes de diffusion d'un quarkonium sur un noyau atomique lourd, appartient à la classe d'universalité de l'équation de Fisher-Kolmogorov-Petrosky-Piscounov (F-KPP) qui régit, entre autre, l'évolution temporelle de la distribution de la position des particules extrêmes dans le mouvement brownien branchant. Dans cette thèse, nous exploitons ce lien entre physique des particules et physique statistique pour étudier le contenu partonique détaillé des états de Fock d'un dipôle dans la diffusion à haute énergie sur un ion lourd, dont nous déduisons le comportement asymptotique des sections efficaces de dissociation diffractive d'un quarkonium. Nous présentons également des prédictions pour les sections efficaces de collision électron-ion.

En premier lieu, les événements de diffusion nucléaire d'un petit dipôle de couleur, lorsque les paramètres sont réglés de sorte que la section efficace totale soit petite, sont induits par des configurations contenant des dipôles de grande taille transverse. Ces dernières sont dues à des fluctuations partoniques rares, distribuées différemment selon le référentiel choisi, du référentiel de repos du noyau aux référentiels dans lesquels la rapidité est partagée entre le dipôle projectile et le noyau cible. Il s'avère que la liberté de sélectionner un référentiel permet de déduire une expression analytique asymptotique de la distribution de la rapidité du premier branchement du dipôle parent de l'ensemble des dipôles qui interagissent, ce qui fournit un estimateur des corrélations de ces derniers. Dans un autre aspect, notre étude montre l'importance de la caractérisation de la distribution des particules au voisinage des particules extrémales dans les états générés par le processus de branchement de dipôles en QCD, et plus généralement, par tout modèle de marche aléatoire branchante unidimensionnelle. Dans le but d'étudier quantitativement cette distribution, nous développons un algorithme de Monte Carlo pour générer la région frontalière d'une marche aléatoire unidimensionnelle avec branchements binaires évoluée à grand temps, qui permet d'étudier à la fois des configurations typiques et les configurations rares conditionnées de sorte que la particule extrême au temps final ait une position très différente de la position typique ou moyenne.

Un autre résultat de notre travail est l'observation que la diffusion d'un petit dipôle de couleur sur un noyau lourd possède une interprétation probabiliste pour les sections efficaces de diffusion :

la section efficace totale de diffusion est le double de la probabilité d'avoir au moins un dipôle en interaction dans l'état du dipôle initial à la rapidité de diffusion, et la section efficace de diffusion diffractive est le double de la probabilité d'avoir un nombre pair de dipôles en interaction. Cette interprétation probabiliste ainsi que la description statistique ci-dessus permettent de dériver les expressions analytiques asymptotiques de la section efficace diffractive conditionnée à un "gap" de rapidité minimal Y0 ou, de manière équivalente, la distribution des "gaps" de rapidité Ygap dans la dissociation diffractive d'un petit dipôle sur un grand noyau, dans une région paramétrique bien définie. Nous obtenons ainsi les solutions asymptotiques de l'équation de Kovchegov-Levin, qui a été établie il y a plus de 20 ans pour décrire la dissociation diffractive d'un dipôle sur un noyau dans des collisions à haute énergie. De plus, nous présentons des prédictions pour la distribution des "gaps" de rapidité dans la cinématique des futurs collisionneurs électrons-ions, sur la base des solutions numériques de l'équation originale de Kovchegov-Levin et de son extension à une constante de couplage forte courante. Les résultats sont en accord qualitatif avec nos formules analytiques asymptotiques déjà à des rapidités accessibles aux futurs collisionneurs électron-ion.

# Table of Contents

# General introduction

Strong interactions of quarks and gluons are described by quantum chromodynamics (QCD), a Yang-Mills gauge theory whose gauge field is characterized by the color quantum number. Due to the color confinement, quarks and gluons do not stay isolated, but are trapped together to form composite bound states known as hadrons. Among puzzles of QCD, the dynamics of hadronic matter in the regime of high energy involves intriguing issues, and has been queried for a long time. Theoretical studies on such topic are also supported by a massive amount of high-energy collision data, which have been collected at various colliders around the world. To understand the behaviors of hadronic matter in high energy collisions is also a main physical goal of the research programmes at many proposed future colliders, such as the Large Hadron Electron Collider (LHeC) [1] and the Future Circular Collider (FCC) [2] at CERN, or the Electron-Ion Collider (EIC) [3] at Brookhaven.

Many high-energy collision machines are motivated by deep-inelastic scattering of a lepton off a hadron, which is an outstanding process to probe a variety of properties of hadronic matter, and has been closely associated with the development of QCD from the beginning. As an example, the observations in the MIT-SLAC experiment on electron-proton collisions during the late 1960s and early 1970s provided the experimental evidences to support the existence of quarks and the parton model (for a review, see Ref. [4]). In this scattering process, the interaction between the lepton with the hadron is mediated by a virtual photon with a high-enough virtuality in order to be able to resolve the partonic level. In an appropriate frame, the photon could be replaced by a quark-antiquark dipole, which therefore gives rise to the study of the dipole-hadron scattering.

The scope of the thesis is limited to discussions of deep-inelastic scattering off a large nucleus, and hence, the dipole-nucleus scattering. As a matter of fact, the latter is a remarkable process to understand theoretically, not only by the fact that it can be factorized from the deep-inelastic scattering of a virtual photon at high energy. Actually, it is the simplest dilute-dense interaction process. A dipole may be a good starting point to model dilute systems, such as heavy mesons, or maybe even specific states of proton, in order to understand some of their properties. On another aspect, in proton-nucleus collisions, it turns out that an appropriate Fourier transform of the dipole-nucleus total cross section is mathematically identical to the differential cross section for producing a semi-hard jet of a given transverse momentum [5], at least at next-to-leading logarithmic accuracy [6], which is usually referred to as *transverse momentum broadening*.

For the dipole-nucleus scattering, if the dipole is subject to a high-energy boost, it does not appear as a bare quark-antiquark state when traversing the nucleus, but as a complex state dominated by soft gluons characterized by small longitudinal momentum fractions $x$, as a result of quantum evolution. At low density, the growth of this gluonic system with the rapidity is linear, and the behavior of its mean density is controlled by the Balitsky-Fadin-Kuraev-Lipatov evolution equation [7, 8], which resums the leading logarithmic series of the parameter $\bar{\alpha} \ln(1/x)$ . Such linear evolution is tamed when the parton density becomes sufficiently high by nonlinear effects. A prominent example of equations encoding these nonlinear effects is the Jalilian–Marian–Iancu–McLerran–Weigert–Leonidov–Kovner equation [9–16]. When the number of colors is taken to be large, it boils down to the Balitsky-Kovchegov evolution equation [17, 18], which lies at the basis of our studies presented in this thesis.

Apart from the mean-field evolution, there could be fluctuations in the quantum states of both the projectile dipole and the target nucleus, which generate rare gluonic scattering configurations. In many cases, those fluctuations can play an important role [19–23]. When fluctuations enter the game, the foremost problem is to construct a model to describe them properly. Such description should capture the main features of the physics we are considering, in this case, the QCD evolution. It will then provide us with a picture of the scattering, and enable us to address certain observables and/or to draw some consequences.

The large part of this thesis will be dedicated to discuss the nuclear scattering of a small dipole. We shall assume that the target nucleus follows the deterministic evolution, consequently fluctuations in the target are neglected. The scattering is then triggered by fluctuations in the content of the Fock state of the dipole. In fact, by the analogy between the QCD dipole evolution and a branching-diffusion process, Mueller and Munier [20] adapted a description of fluctuations in the latter process [24] to the former, and yielded some properties of QCD scattering amplitudes. This stochastic picture also enabled them [22, 23] to deduce an (incomplete) estimation for the rapidity gap distribution in the diffractive onium-nucleus scattering. In this thesis, we shall improve that description by developing a model of dipole distribution, which allows us to study the configurations of onia in the scattering off a nucleus and a related genealogical problem.

As another remark, we will construct a formulation for diffraction of a small dipole. This formulation, together with the description of rare fluctuations, will enable us to address important observables of interests in diffractive dissociation.

In additional to the dipole-nucleus scattering, we shall also investigate diffractive virtual photon-nucleus scattering base on the numerical solutions to the QCD evolution equations. The aim of this investigation is to produce predictions for future electron-ion colliders.

The main content of this thesis consists in four chapters, which are organized as follows:

- **Chapter 1 - QCD evolution of hadronic matter toward high energy**: this chapter is to review some backgrounds for the discussions in the thesis: light-cone formalism, deep-inelastic scattering (DIS) at high energy and QCD color dipole model.

- **Chapter 2 - QCD evolution in analogy with branching-diffusion processes**: this chapter is aimed to introduce the QCD nonlinear evolution for the onium-nucleus scattering at high energy in connection to branching-diffusion processes in statistical physics, and to present a Monte Carlo algorithm [25] to generate particles close to an extreme particle in a one-dimensional branching random walk.

- **Chapter 3 - Nuclear scattering of small onia**: in this chapter, we shall present our investigation [26] on the nuclear scattering configuration of a small onia subject to high-energy evolution and a related genealogical structure.

– **Chapter 4 - Diffractive dissociation**: this chapter is aimed at presenting our studies [27, 28] on diffractive dissociation of a small onium and a virtual photon. For the former, we shall introduce a theoretical formulation of diffraction from which one can derive the observables of interest. For the diffraction of a virtual photon, we shall present a numerical study in the framework of the color dipole formulation and produce some predictions for future electron-ion colliders.

We shall then conclude the discussions in the thesis by summarizing the main results together with some possible future developments. Three appendices gather some technical details for the calculations presented in the main chapters.

# QCD evolution of hadronic matter toward high energy

## Contents

This chapter is aimed to review the theoretical description of high energy evolution in the framework of color dipole formalism [29–32] in QCD. We shall start with a brief introduction of the light-cone perturbation theory (LCPT) and the deep-inelastic scattering in the dipole picture. We shall then present the Balitsky-Fadin-Kuraev-Lipatov (BFKL) [7, 8] equation, which governs the linear evolution of the gluonic content at high-energy.

## 1.1 QCD Lagrangian

Quarks and gluons, which constitute hadrons, are fundamental degrees of freedom of QCD. Their strong interaction is associated to the color charge, which is an analog to electric charge in the electromagnetic interaction. For a general number of colors $N_c$, the gauge group of QCD is given by the special unitary group $SU(N_c)$. A quark of flavor $f$ and color index $i$ is represented by a four-component spinor $\psi_i^f(x)$ ($1 \le i \le N_c$), which is the component $i$ of a vector of size $N_c$ in the fundamental representation of $SU(N_c)$. Meanwhile, a gluon, which carries the strong interaction (i.e, gauge boson), is described by a massless vector field $A_\mu^a(x)$ (gauge field) with color index $a$ in the adjoint representation of the $SU(N_c)$ (hence, $a$ runs from 1 to $N_c^2 - 1$). The Lagrangian which describes the dynamics of those fields and their mutual couplings reads

$$\mathcal{L}_{QCD} = \underbrace{-\frac{1}{4} F_{\mu\nu}^a(x) F_a^{\mu\nu}(x)}_{\mathcal{L}_{YM}} + \underbrace{\bar{\psi}_i^f(x) \left[ i\gamma^\mu D_\mu^{ij} - m_f \delta^{ij} \right] \psi_j^f(x)}_{\mathcal{L}_{Dirac}}, \tag{1.1}$$

where $m_f$ is the mass of a quark flavor $f$, and the sums over color, flavor and Lorentz indices are understood. The covariant derivative $D$ reads $D_\mu^{ij} = \delta^{ij}\partial_\mu - ig_s A_\mu^a(x) t_a^{ij}$, where $g_s$ is the strong coupling constant and $t^a$ are generators of $SU(N_c)$ in the fundamental representation. The gluon field strength tensor $F_{\mu\nu}^a$ is given by

$$F_{\mu\nu}^a = \partial_\mu A_\nu^a - \partial_\nu A_\mu^a + g_s f^{abc} A_\mu^b A_\nu^c. \tag{1.2}$$

The real numbers $f^{abc}$ in the above expression of the field strength are the structure constants of $SU(N_c)$, which are coefficients of the linear extension of the Lie brackets of pairs of generators, $[t^a, t^b] = if^{abc}t^c$. The first term in Eq. (1.1), $\mathcal{L}_{YM}$, is the Yang-Mills term concerning the dynamics of gluons and their interactions. Unlike QED, there is an additional term in the field strength tensor (1.2), the third term, due to the non-abelian nature of the strong interaction. This term induces gluon self-coupling, making QCD a theory with an intriguingly rich coupling structure. The second term in Eq. (1.1), $\mathcal{L}_{Dirac}$, is the Dirac Langrangian, which encodes the dynamics of quarks and their coupling to gluons. The QCD Lagrangian (1.1) is invariant under a local gauge transformation with respect to the $SU(N_c)$ group which acts on the elementary fields as

$$\psi_i^f(x) \to [\Omega(x)]_{ij} \, \psi_j^f(x), \qquad \bar{\psi}_i^f(x) \to \bar{\psi}_j^f(x) \left[ \Omega^\dagger(x) \right]_{ji}, \tag{1.3a}$$

$$A_\mu(x) \to \Omega(x) A_\mu(x) \Omega^\dagger(x) - \frac{i}{g_s} \left[ \partial_\mu \Omega(x) \right] \Omega^\dagger(x), \tag{1.3b}$$

in such a way that the covariant derivative and the field strength tensor transform in the adjoint representation:

$$D_\mu \to \Omega(x) D_\mu \Omega^\dagger(x), \tag{1.4a}$$

$$F_{\mu\nu}(x) \to \Omega(x) F_{\mu\nu}(x) \Omega^\dagger(x), \tag{1.4b}$$

where $A_\mu = A_\mu^a t^a$, $F_{\mu\nu} = F_{\mu\nu}^a t^a$, and $\Omega(x) = e^{i\theta^a(x)t^a} \in \mathrm{SU}(N_c)$, with $\theta^a(x)$ being real-valued functions. Under the gauge transformation, each gauge field configuration develops into a class of gauge-equivalent configurations, or a gauge orbit, in the configuration space. Gauge symmetry implies that the physics is invariant along each such orbit. Therefore, it is, "at heart, a redundancy in our description of the world" [33]. To avoid unphysical degrees of freedom due to the gauge redundancy in the quantization, we select from each orbit a particular configuration by imposing a condition on the gauge fields, which procedure is referred to as *gauge fixing* [1]. In this thesis, we fix the gauge according to the so-called light-cone gauge condition,

$$\eta \cdot A^a = 0, \quad \eta^2 = 0. \tag{1.5}$$

An advantage of this gauge is that the gluons have only physical transverse degrees of freedom. Consequently, the theory is free of unphysical ghost fields.

For the sake of simplicity, we will hereafter omit the flavor and color indices in the notation of the quark's spinor. Therefore, the Dirac term in the QCD Lagrangian can be rewritten as

$$\mathcal{L}_{Dirac} = \bar{\psi}(x) \left[ i\gamma^\mu D_\mu - m \right] \psi(x), \tag{1.6}$$

where the sums over flavor and color indices are implicitly understood.

## 1.2 Light-cone formulation

Field theory is usually quantized in a Lorentz frame,

$$\begin{aligned}
\tilde{x}^\mu &= (x^0, x^1, x^2, x^3) \equiv (t, x, y, z), \\
g_{\mu\nu} &= \eta_{\mu\nu} \equiv \mathrm{diag}(1, -1, -1, -1).
\end{aligned} \tag{1.7}$$

This parametrization is usually referred to as the instant form of Hamiltonian dynamics in which the Hamiltonian of a physical system drives the evolution of the system along the ordinary time $t$. It turns out that Eq. (1.7) is not the only choice: there are various possibilities to cast the ordinary spacetime coordinates into another representation. Dirac [35] pointed out that, there are three inequivalent spacetime parameterizations, including the instant form, in the sense that they cannot be mapped to each other by a finite Lorentz transform. In this thesis, we shall deal with one of them which is known as the lightcone parameterization [2].

---

[1] This procedure however does not fixed the gauge completely, due to Gribov copies [34].

[2] In fact, under an infinite Lorentz boost ($\beta = 1$) along the $x^3$ direction, the instant form and the lightcone form are mathematically equivalent. Therefore, Kogut and Soper [36] interpreted the infinite-momentum limit as the lightcone reparametrization of the spacetime coordinates to avoid the limiting procedure.

### 1.2.1 Light-cone kinematics

In the light-cone notation, the spacetime coordinates are given by $x^\mu = (x^+, x^-, x^\perp)$, which are related to the components in the instant form (1.7) as

$$x^+ = \frac{1}{\sqrt{2}}(x^0 + x^3), \quad x^- = \frac{1}{\sqrt{2}}(x^0 - x^3), \quad x^\perp = (x^1, x^2). \tag{1.8}$$

The component $x^+$ is conventionally chosen to be the light-cone time. The light-cone time derivative is denoted as $\partial_+ \equiv \partial/\partial x^+$, and the longitudinal derivative is $\partial_- \equiv \partial/\partial x^-$. Note that $\partial_+ = \partial^-$, and $\partial_- = \partial^+$. In general, for a four-vector $u$, the $u^+$ and $u^-$ are referred to as the "time-like" and the longitudinal components, respectively, while $u^\perp$ are the transverse components. The metric tensor $g_{\mu\nu}$ in this notation reads

$$g_{\mu\nu} = \begin{pmatrix} 0 & 1 & 0 & 0 \\ 1 & 0 & 0 & 0 \\ 0 & 0 & -1 & 0 \\ 0 & 0 & 0 & -1 \end{pmatrix}, \tag{1.9}$$

and the scalar product of two four-vectors $u$ and $v$ is given by

$$u \cdot v = u^+ v^- + u^- v^+ - u^\perp \cdot v^\perp \tag{1.10}$$

From Eq. (1.10), the light-cone energy of a free particle on the mass shell with four-momentum $P$ is given by

$$P^- = \frac{(P^\perp)^2 + m^2}{2P^+}, \tag{1.11}$$

where $m$ is the mass of the particle. Comparing to the expression of the energy in the instant form, $P^0 = \sqrt{m^2 + |\vec{P}|^2}$, one can see that the light-cone energy (1.11) are free of square root, and hence, the issue of negative energies can be avoided, as pointed out by Dirac. This square-root-free feature simplifies the perturbative calculations when the light-cone coordinates are employed. In the end, physical results should be unchanged, since this formulation is just a spacetime reparametrization in its nature.

### 1.2.2 QCD Hamiltonian on the light cone

With the light-cone parameterization, one can rewrite the QCD Lagrangian (1.1) as

$$\mathcal{L}_{QCD} = \frac{1}{2}F_a^{+-}F_a^{+-} + F_a^{+m}F_a^{-m} - \frac{1}{4}(F_a^{mn})^2 + \bar{\psi}\left(i\gamma^+ D^- + i\gamma^- D^+ - i\gamma^\perp D^\perp - m\right)\psi, \tag{1.12}$$

with indices $m, n \in \{1, 2\}$. The elementary fields can be decomposed as follows:

$$A_a^\mu = \left(0, A_a^-, A_a^\perp\right), \tag{1.13a}$$

$$\psi = \psi_+ + \psi_-, \quad \psi_\pm = \Lambda_\pm \psi \equiv \frac{\gamma^0 \gamma^\pm}{\sqrt{2}} \psi, \tag{1.13b}$$

where $\Lambda_\pm$ are projection operators:

$$\Lambda_+ + \Lambda_- = \mathbb{1}, \quad \Lambda_\pm \Lambda_\mp = 0, \quad (\Lambda_\pm)^2 = \Lambda_\pm, \tag{1.14}$$

and $\gamma^\pm = (\gamma^0 \pm \gamma^3)/\sqrt{2}$. In Eq. (1.13), we have employed the light-cone gauge condition (1.5). In particular, we choose $\eta = (0, 1, 0^\perp)$, and hence, get rid of the plus component of the gauge field, $A_a^+ = 0$.

We are going to review the structure of the QCD Hamiltonian $H_{QCD}$, which is related to the Lagrangian (1.12) through a Legendre transform,

$$H_{QCD} = \int dx^- d^2 x^\perp \left[ \sum_\phi \Pi_\phi \partial_+ \phi - \mathcal{L}_{QCD} \right], \tag{1.15}$$

where $\phi$ are the field components appearing in Eq. (1.13), and $\Pi_\phi = \frac{\delta \mathcal{L}_{QCD}}{\delta(\partial_+ \phi)}$ are their corresponding conjugate momenta. From the Lagrangian (1.12), the latter reads

$$\Pi_{A_a^m} = \partial^+ A_a^m, \quad \Pi_{A_a^-} = 0, \tag{1.16a}$$

$$\Pi_{\psi_+} = i\sqrt{2} \psi_+^\dagger, \quad \Pi_{\psi_-} = 0. \tag{1.16b}$$

The field components $A_a^-$ and $\psi_-$ have zero conjugate momenta: they are not dynamical fields. Consequently, the usual canonical quantization procedure cannot be applied on such fields. However, they can be expressed in terms of the dynamical fields $A_a^m$ and $\psi_+$ by the virtue of the equations of motion.

### Quark fields

The Dirac equation in the light-cone notation reads

$$\left(i\gamma^+ D^- + i\gamma^- D^+ - i\gamma^\perp D^\perp - m\right) \psi(x) = 0. \tag{1.17}$$

Acting the "plus" projector on the Dirac equations (1.17) from the left gives

$$i\sqrt{2} \gamma^0 \partial_- \psi_-(x) = \left(i\gamma^\perp \cdot D^\perp + m\right) \psi_+(x). \tag{1.18}$$

Hence,

$$\psi_-(x) = \frac{1}{\sqrt{2}} \frac{\gamma^0}{i\partial_-} \left(i\gamma^\perp \cdot D^\perp + m\right) \psi_+(x) = \frac{1}{2} \frac{\gamma^+}{i\partial_-} \left(i\gamma^\perp \cdot D^\perp + m\right) \psi_+(x), \tag{1.19}$$

where $1/\partial_-$ is the antiderivative operator, i.e. an integral with respect to $x^-$. Eq. (1.19) includes the coupling with the color gauge field encoded in the covariant derivative. If one sets this coupling to zero, we obtain the so-called free "minus" components,

$$\widetilde{\psi}_-(x) = \frac{1}{2} \frac{\gamma^+}{i\partial_-} \left( i\gamma^\perp \cdot \partial^\perp + m \right) \psi_+(x), \tag{1.20}$$

and the free Dirac spinors reads $\widetilde{\psi} = \psi_+ + \widetilde{\psi}_-$.

**Gluon fields**

The Euler-Lagrange equations for the gluon fields are Yang-Mills equations:

$$\partial_\mu F_a^{\mu\nu} = J_a^\nu, \tag{1.21}$$

where the current is given by

$$J_a^\nu = -g_s f^{abc} F_b^{\nu\sigma} A_\sigma^c - g_s \bar{\psi} \gamma^\nu t^a \psi. \tag{1.22}$$

The equation for $(\nu = +)$ reads

$$-\partial_-^2 A_a^- - \partial_-(\nabla^\perp \cdot A_a^\perp) = -g_s f^{abc}(\partial_- A_b^i) \cdot A_i^c - g_s \bar{\psi} \gamma^+ t^a \psi. \tag{1.23}$$

The component $A^-$ can then be expressed in terms of $A^\perp$ and $\psi_+$ as

$$A_a^- = -\frac{1}{\partial_-}(\nabla^\perp \cdot A_a^\perp) + \frac{g_s}{\partial_-^2} \left[ f^{abc}(\partial_- A_b^i) A_i^c + \bar{\psi} \gamma^+ t^a \psi \right]. \tag{1.24}$$

The first term in Eq. (1.24) is free of the coupling constant $g_s$, and is referred to as the "free component":

$$\widetilde{A}_a^- \equiv -\frac{1}{\partial_-}(\nabla^\perp A_a^\perp). \tag{1.25}$$

The free gauge vector field is then

$$\widetilde{A}_a^\mu = (0, \widetilde{A}_a^-, A_a^\perp). \tag{1.26}$$

**Free and interaction Hamiltonians**

With the help of the free fields, we can write the QCD Hamiltonian on the light cone as $H_{QCD} = H_0 + H_{int}$, where $H_0$ and $H_{int}$ are free and interaction parts, respectively, and are given by

$$H_0 = \frac{1}{2} \int_\Sigma dx^- d^2 x^\perp \left\{ \overline{\widetilde{\psi}} \gamma^+ \frac{m^2 - \nabla_\perp^2}{i\partial^+} \widetilde{\psi} - \widetilde{A}_a^\mu \nabla_\perp^2 \widetilde{A}_\mu^a \right\}, \tag{1.27}$$

$$H_{int} = \int_\Sigma dx^- d^2x^\perp \left\{ 2g_s Tr\left(i\partial^\mu \widetilde{A}^\nu \left[\widetilde{A}_\mu, \widetilde{A}_\nu\right]\right) - g_s \overline{\widetilde{\psi}}\gamma^\mu \widetilde{A}_\mu \widetilde{\psi} + \frac{g_s^2}{2} Tr\left(\left[\widetilde{A}^\mu, \widetilde{A}^\nu\right]\left[\widetilde{A}_\mu, \widetilde{A}_\nu\right]\right) \right.$$

$$+ g_s^2 \overline{\widetilde{\psi}}(\gamma^\mu \widetilde{A}_\mu) \frac{\gamma^+}{2i\partial_-}(\gamma^\nu \widetilde{A}_\nu)\widetilde{\psi} + g_s^2 Tr\left(\left[i\partial_- \widetilde{A}^\mu, \widetilde{A}_\mu\right] \frac{1}{(i\partial_-)^2}\left[i\partial_- \widetilde{A}^\nu, \widetilde{A}_\nu\right]\right) \tag{1.28}$$

$$\left. - g_s^2 \overline{\widetilde{\psi}} \frac{\gamma^+}{(i\partial_-)^2}\left[i\partial_- \widetilde{A}^\mu, \widetilde{A}_\mu\right]\widetilde{\psi} + \frac{g_s^2}{2}\overline{\widetilde{\psi}}\gamma^+ t^a \widetilde{\psi} \frac{1}{(i\partial_-)^2}\overline{\widetilde{\psi}}\gamma^+ t^a \widetilde{\psi}\right\}.$$

### Quantization

To quantize the theory, we first decompose the dynamical fields $\psi_+$ and $A^\perp$ into Fourier modes. Since the subspace image of the projection $\Lambda_+$ is two-dimensional, the spinor $\psi_+$ can be expressed in a basis $\{w_r, r = \pm 1/2\}$ of that subspace as

$$\psi_{+,i}(x) = \sum_r \int \frac{d^2p^\perp dp^+}{(2\pi)^3 2p^+}\left\{2^{1/4} w_r(p)b_{r,i}(p)e^{-ip\cdot x} + 2^{1/4}w_{-r}(p)d^\dagger_{r,i}(p)e^{ip\cdot x}\right\}\Theta(p^+), \tag{1.29}$$

where we have recovered the quark color index $i$. The basis spinors are chosen to obey the following completeness and orthogonality relations:

$$\sum_r w_r(p)w_r^\dagger(p) = p^+\Lambda_+; \qquad w_r^\dagger(p)w_s(p) = p^+\delta_{rs}. \tag{1.30}$$

In the same manner, the transverse gauge field can be expanded in modes as follows:

$$A_a^\perp(x) = \int \frac{d^2k^\perp dk^+}{(2\pi)^3 2k^+}\sum_\lambda \left\{\epsilon_\lambda^\perp a_{\lambda,a}(k)e^{-ik\cdot x} + \epsilon_\lambda^{\perp *}a_{\lambda,a}^\dagger(k)e^{ik\cdot x}\right\}\Theta(k^+), \tag{1.31}$$

where the two transverse polarization vectors $\epsilon_\lambda^\perp$ are chosen to be normalised as

$$\sum_\lambda \epsilon_\lambda^{\perp j}(\epsilon_\lambda^{\perp j'})^* = \delta^{jj'}; \qquad \epsilon_\lambda^\perp \cdot \epsilon_{\lambda'}^{\perp *} = \delta_{\lambda\lambda'}. \tag{1.32}$$

We then treat the fermionic and bosonic coefficients in the mode expansions (1.29) and (1.31), respectively, as operators. Their (anti)commutation relations read

$$\left\{b_{r,i}(p), b_{s,j}^\dagger(q)\right\} = (2\pi)^3 2p^+\delta(p^+ - q^+)\delta^2(p^\perp - q^\perp)\delta_{rs}\delta_{ij}, \tag{1.33a}$$

$$\left\{d_{r,i}(p), d_{s,j}^\dagger(q)\right\} = (2\pi)^3 2p^+\delta(p^+ - q^+)\delta^2(p^\perp - q^\perp)\delta_{rs}\delta_{ij}, \tag{1.33b}$$

$$\left[a_{a,\lambda}(k), a_{b,\lambda'}^\dagger(k')\right] = (2\pi)^3 2k^+\delta(k^+ - k'^+)\delta^2(k^\perp - k'^\perp)\delta_{ab}\delta_{\lambda\lambda'}. \tag{1.33c}$$

All other possible commutation (resp. anticommutation) relations between bosonic (resp. fermionic) are identically zero.

Since the Hamiltonians (1.27) and (1.28) are expressed in terms of the free fields $\widetilde{\psi}$ and $\widetilde{A}$ The spinor $\tilde{\psi}$ can then be decomposed as

$$\widetilde{\psi}^i(x) = \psi_+^i + \widetilde{\psi}_-^i = \sum_{r=\pm 1/2} \int \frac{d^2 p^\perp dp^+}{(2\pi)^3 2p^+} \left\{ b_r^i(p) u_r(p) e^{-ip\cdot x} + d_r^{\dagger i}(p) v_r(p) e^{ip\cdot x} \right\} \Theta(p^+), \tag{1.34}$$

where the basis spinors $u$ and $v$ are given by

$$u_r(p) = w_r + \frac{\gamma^+}{2p^+}(\gamma^\perp \cdot p^\perp + m) w_r, \tag{1.35a}$$

$$v_r(p) = w_{-r} + \frac{\gamma^+}{2p^+}(\gamma^\perp \cdot p^\perp - m) w_{-r}. \tag{1.35b}$$

In a similar manner, we can also decompose the gauge field as

$$\widetilde{A}^\mu(x) = \int \frac{d^2 k^\perp dk^+}{(2\pi)^3 2k^+} \sum_\lambda \left\{ a_a^\lambda(k) \epsilon_\lambda^\mu(k) e^{-ik\cdot x} + a_a^{\lambda\dagger}(k) \epsilon_\lambda^{\mu*}(k) e^{ik\cdot x} \right\} \Theta(k^+), \tag{1.36}$$

where $\epsilon_\lambda^\mu(k)$ ($\lambda = 1, 2$) are polarization vectors. Bosonic creation and annihilation operators $a_a^{\lambda\dagger}$ and $a_a^\lambda$ satisfy following commutation relations:

$$\begin{aligned} \left[ a_a^\lambda(k), a_b^{\lambda'}(k') \right] &= \left[ a_a^{\lambda\dagger}(k), a_b^{\lambda'\dagger}(k') \right] = 0, \\ \left[ a_a^\lambda(k), a_b^{\lambda'\dagger}(k') \right] &= 2k^+ (2\pi)^3 \delta(k^+ - k'^+) \delta^2(k^\perp - k'^\perp) \delta_{ab} \delta_{\lambda\lambda'}. \end{aligned} \tag{1.37}$$

In Eqs. (1.29), (1.34) and (1.36), we use the following Lorentz-invariant integral measure:

$$\int \frac{d^4 k}{(2\pi)^4} (2\pi) \delta(k^2 - m^2) = \int \frac{d^2 k^\perp dk^+}{(2\pi)^3 2k^+} \Theta(k^+). \tag{1.38}$$

Using the Fourier decompositions (1.34) and (1.36) together with underlying (anti-)commutation relations, we can construct the light-cone pertubation theory (LCPT) based on the light-cone QCD Hamiltonian. The LCPT rules can be found, for example, in Ref. [37]. In the following, we shall introduce the interaction vertices in the LCPT, which are written in terms of the quantized fields.

**Interaction vertices**

Now, with the notion of quantized fields, we can interpret the terms in the interaction Hamiltonians (1.28).

The first three terms in Eq. (1.28) correspond to the usual QCD vertices:

Three-gluon vertex: $\quad 2g_s \int_\Sigma dx^- d^2x^\perp \left\{ Tr \left( i\partial^\mu \widetilde{A}^\nu \left[ \widetilde{A}_\mu, \widetilde{A}_\nu \right] \right) \right\} \quad = $ 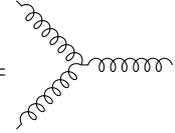 (1.39)

Quark-gluon coupling vertex: $\int_\Sigma dx^- d^2x^\perp \left\{ -g_s \overline{\widetilde{\psi}} \gamma^\mu \widetilde{A}_\mu \widetilde{\psi} \right\} \qquad = $ 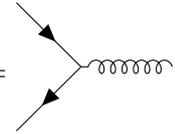 (1.40)

Four-gluon vertex: $\qquad \int_\Sigma dx^- d^2x^\perp \left\{ \frac{g_s^2}{2} Tr \left( \left[ \widetilde{A}^\mu, \widetilde{A}^\nu \right] \left[ \widetilde{A}_\mu, \widetilde{A}_\nu \right] \right) \right\} = $ 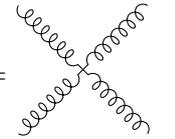 (1.41)

The remaining terms in Eq. (1.28) are referred to as instantaneous effective vertices, and can be represented diagrammatically as

$$\int_\Sigma dx^- d^2x^\perp \left\{ g_s^2 \overline{\widetilde{\psi}} (\gamma^\mu \widetilde{A}_\mu) \frac{\gamma^+}{2i\partial_-} (\gamma^\nu \widetilde{A}_\nu) \widetilde{\psi} \right\} \qquad = $$ 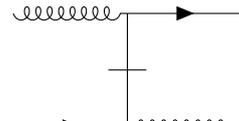 (1.42)

$$\int_\Sigma dx^- d^2x^\perp \left\{ g_s^2 Tr \left( \left[ i\partial_- \widetilde{A}^\mu, \widetilde{A}_\mu \right] \frac{1}{(i\partial_-)^2} \left[ i\partial_- \widetilde{A}^\nu, \widetilde{A}_\nu \right] \right) \right\} = $$ 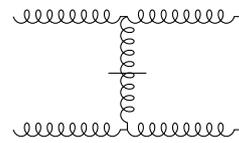 (1.43)

$$\int_\Sigma dx^- d^2x^\perp \left\{ -g_s^2 \overline{\widetilde{\psi}} \frac{\gamma^+}{(i\partial_-)^2} \left[ i\partial_- \widetilde{A}^\mu, \widetilde{A}_\mu \right] \widetilde{\psi} \right\} \qquad = $$ 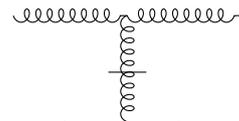 (1.44)

$$\int_\Sigma dx^- d^2x^\perp \left\{ \frac{g_s^2}{2} \overline{\widetilde{\psi}} \gamma^+ t^a \widetilde{\psi} \frac{1}{(i\partial_-)^2} \overline{\widetilde{\psi}} \gamma^+ t^a \widetilde{\psi} \right\} \qquad = $$ 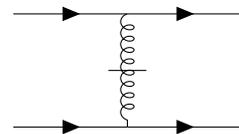 (1.45)

where instantaneous quark and gluon lines are depicted by regular quark and gluon lines with a short line segment cross.

## 1.2.3 Perturbative expansion on the light-cone

Due to quantum effects, the initial state of a system at the asymptotic light-cone time $x^+ = -\infty$ may fluctuate into another quantum state at the considered light-cone time. The quantum evolution of the system from $x^+ = -\infty$ to $x^+ = 0$ is governed by the so-called evolution operator $U(0, -\infty)$, which is a solution of the Schroedinger equation, as

$$|\Psi(x^+ = 0)\rangle = U(0, -\infty)|\Psi(x^+ = -\infty)\rangle \equiv \mathcal{T} \exp\left\{-i \int_{-\infty}^{0} dx'^+ \mathcal{H}_{int}\right\} |\Psi(x^+ = -\infty)\rangle, \quad (1.46)$$

where $\mathcal{T}$ is the light-cone time order product, and $\mathcal{H}_{int}$ is the interaction Hamiltonian in the interaction representation, which is related to the interaction Hamiltonian in the Schroedinger picture $H_{int}$ as

$$\mathcal{H}_{int} = e^{iH_0 x^+} H_{int} \, e^{-iH_0 x^+}. \quad (1.47)$$

Expanding the evolution operator $U$, one gets

$$|\Psi(x^+ = 0)\rangle = \sum_{n=0}^{+\infty} \frac{(-i)^n}{n!} \int dx_1^+ \cdots dx_n^+ \mathcal{T}\left\{\mathcal{H}_{int}(x_1^+) \cdots \mathcal{H}_{int}(x_n^+)\right\} |\Psi(x^+ = -\infty)\rangle. \quad (1.48)$$

Let us denote $\{|\omega\rangle\}$ as the complete set of the eigenstates of the free Hamiltonian $H_0$ corresponding to the light-cone energy $p_\omega^-$, $H_0|\omega\rangle = p_\omega^-|\omega\rangle$. The asymptotic state $|\omega_0\rangle \equiv |\Psi(x^+ = -\infty)\rangle$ also belongs to this set, corresponding to the energy $p_{\omega_0}^-$. They are chosen to be normalised as

$$\langle \omega|\omega'\rangle = \prod_i 2k_{(\omega)i}^+(2\pi)^3 \delta(k_{(\omega)i}^+ - k_{(\omega')_i}^+)\delta^{(2)}(k_{(\omega)i}^\perp - k_{(\omega')i}^\perp)\delta_{\{\lambda_{(\omega)i}\}\{\lambda_{(\omega')i}\}}, \quad (1.49)$$

where $k_{(\omega)i}$ and $\{\lambda_{(\omega)i}\}$ ($k_{(\omega')i}$ and $\{\lambda_{(\omega')i}\}$) are momenta and quantum indices of $n_\omega$ ($n_{\omega'}$) particles in the state $|\omega\rangle$ ($|\omega'\rangle$). The state $|\Psi(x^+ = 0)\rangle$ can then be expanded in this basis as

$$\begin{aligned}
|\Psi(x^+ = 0)\rangle &= \sqrt{Z}\left(|\omega_0\rangle + \sum_{\omega \neq \omega_0} \phi_\omega|\omega\rangle\right), \\
&= \sqrt{Z}\left(|\omega_0\rangle + \sum_{\omega \neq \omega_0} \int \left[\prod_{i=1}^{n_\omega} \frac{dk_{(\omega)i}^+ d^2 k_{(\omega)i}^\perp}{(2\pi)^3 2k_{(\omega)i}^+}\right] \phi_\omega|\{k_{(\omega)j}^+\}, \{k_{(\omega)j}^\perp\}, \{\lambda_{(\omega)j}\}\rangle_{j=\overline{1,n_\omega}}\right),
\end{aligned} \quad (1.50)$$

where $\sqrt{Z}$ is the renormalization factor for the onium wave function. The wave function $\phi_\omega$ of a particular quantum fluctuation $q\bar{q} \to \omega$ is defined by $\phi_\omega = \langle\omega|\Psi(x^+ = 0)\rangle$. From the expansion (1.48), we obtain:

$$\begin{aligned}
\phi_{\omega \neq \omega_0} &= \sum_{n=0}^{+\infty} \frac{(-i)^n}{n!} \int dx_1^+ \cdots dx_n^+ \langle\omega|\mathcal{T}\left\{\mathcal{H}_{int}(x_1^+) \cdots \mathcal{H}_{int}(x_n^+)\right\} |\omega_0\rangle \\
&= \underbrace{\langle\omega|\omega_0\rangle}_{=0} + \frac{\langle\omega|H_{int}|\omega_0\rangle}{p_{\omega_0}^- - p_\omega^- + i\epsilon} + \sum_{\omega'} \frac{\langle\omega|H_{int}|\omega'\rangle\langle\omega'|H_{int}|\omega_0\rangle}{(p_{\omega_0}^- - p_\omega^- + i\epsilon)(p_{\omega_0}^- - p_{\omega'}^- + i\epsilon)} + \cdots.
\end{aligned} \quad (1.51)$$

Energy denominators in Eq. (1.51) contain light-cone energy difference the asymptotic state and intermediate states. The states $|\omega\rangle$ and $|\omega'\rangle$ are not identical to $|\omega_0\rangle$ (up to constant factor). The terms with $|\omega\rangle$ and $|\omega'\rangle$ indistinguishable from $|\omega_0\rangle$ are absorbed into the renormalization factor $\sqrt{Z}$, as suggested by Ref. [38] (see also Ref. [39]).The Fock state expansion (1.50) together with Eq. (1.51) prove to be useful in constructing light-cone wave function of a system subject to quantum evolution from the lowest perturbative order.

## 1.3 Deep-inelastic scattering in the dipole picture

Deep-inelastic scattering (DIS) is a scattering process to resolve the internal structure of a hadron using a leptonic particle like electron. Typically in DIS, the hadron is probed by a virtual photon, which usually shatters hadron, resulting in the production of a set of hadrons $X$ in the final state. In this section, we are going to discuss the DIS on a nucleus $A$ at high energy, which is conveniently described by the so-called dipole picture. This constitutes the main framework of the discussions in the dissertation. We shall begin with a short introduction of kinematic variables in DIS.

### 1.3.1 DIS kinematics

An illustration of the deep-inelastic electron-nucleus collision is sketched in Fig. 1.1. We denote $p = (p^+, p^-, p^\perp)$ and $p' = (p'^+, p'^-, p'^\perp)$ for the four-momenta of ingoing and outgoing electrons, respectively, $P = (P^+, P^-, P^\perp)$ for the four-momentum of the nucleus, and $q = (q^+, q^-, q^\perp)$ for the four-momentum of the virtual photon. In addition, the nucleus is supposed to move in the $x^-$ direction, i.e. $P^\perp = 0$. The DIS can be described by following Lorentz-invariant quantities:

$$\hat{s} = (P+q)^2, \quad Q^2 = -q^2,$$

$$x_{Bj} = \frac{Q^2}{2P \cdot q}, \quad \eta = \frac{P \cdot q}{P \cdot p}. \tag{1.52}$$

The quantity $\hat{s}$ is the squared center-of-mass energy of the $\gamma^*A$ scattering process. $Q^2$ is called the virtuality of the virtual photon. For the process to be deep inelastic, the photon should be highly virtual, or $Q^2 \gg \Lambda_{QCD}^2$. Otherwise, when $Q^2$ is negligibly small, i.e. $Q^2 \approx 0$, the process is referred to as photoproduction. Therefore, the photon's virtuality $Q^2$ provides a natural hard scale in the DIS.

To see the physical interpretations of the quantity $\eta$, let us consider the process in the rest frame of the nucleus in which the four-momentum of the nucleus reads $P^\mu = (\frac{M}{\sqrt{2}}, \frac{M}{\sqrt{2}}, 0^\perp)$, where $M$ is the mass of the nucleus. We have:

$$\eta = \frac{P \cdot q}{P \cdot p} = \frac{q^+ + q^-}{p^+ + p^-} = \frac{q^0}{p^0} = \frac{E - E'}{E}, \tag{1.53}$$

where $q^0$ and $p^0$ are the energy components of the four-momenta of the photon and the ingoing electron written in the instant form, and $E$ and $E'$ are the energies of the electron before and after the scattering, respectively. Therefore, in the nucleus's rest frame, $\eta$ is the fraction of the electron's energy transferred to the nucleus.

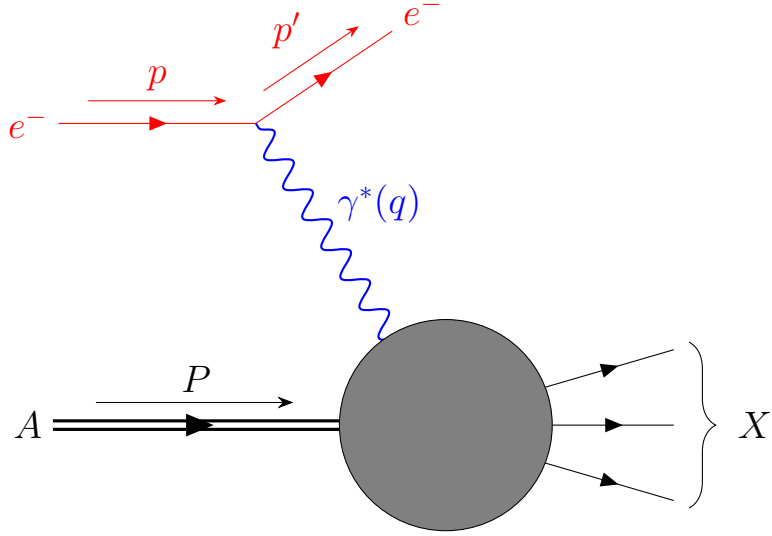

Figure 1.1: Schematic diagram of the DIS of an electron ($e^-$) on a nucleus ($A$) by exchanging a virtual photon ($\gamma^*$). The nucleus is disintegrated into an inclusive set $X$ of hadrons in the final state.

Let us now interpret the Bjorken $x$ variable, $x_{Bj}$. For this purpose, it is convenient to work in the so-called Breit frame in which the nucleus moves very fast in the $x^-$ direction ($P^+ \gg M$),

$$P = (P^+, \frac{M^2}{2P^+}, 0^\perp), \tag{1.54}$$

and the photon's momentum reads

$$q = \frac{1}{\sqrt{2}}(-Q, Q, 0^\perp). \tag{1.55}$$

From Eqs. (1.52), (1.54) and (1.55), we have the following relation:

$$\frac{Q}{x_{Bj}} = \sqrt{2}P^+ - \frac{M^2}{\sqrt{2}P^+}. \tag{1.56}$$

Solving this equation for $P^+$, one get

$$P^+ = \frac{Q}{x_{Bj}\sqrt{2}} \frac{1 + \sqrt{1 + \frac{4x_{Bj}^2 M^2}{Q^2}}}{2} \tag{1.57}$$

Since the lepton current can be factorized out, from now onwards we will consider the DIS as the deep-inelastic virtual photon-nucleus scattering. The typical time scale for the photon-nucleus interaction is then

$$\tau_{scatter} \sim \frac{1}{Q}. \tag{1.58}$$

Meanwhile, the partons inside the nucleus can interact mutually. For the nucleus at rest, the typical time scale for such mutual interactions is of order of the size of the nucleus $R \sim \frac{1}{\Lambda_{QCD}}$. In the Breit frame, this time is dilated by the Lorentz factor $\gamma = P^+/M$. Therefore, the time scale for partons' mutual interactions in the Breit frame is given by

$$\tau_{m.i.} \sim \frac{P^+}{M\Lambda_{QCD}}. \tag{1.59}$$

As $P^+ Q \gg M\Lambda_{QCD}$, we deduce that $\tau_{scatter} \ll \tau_{m.i.}$. In other words, the partons are effectively independent during the scattering. This is the basis idea of the parton model.

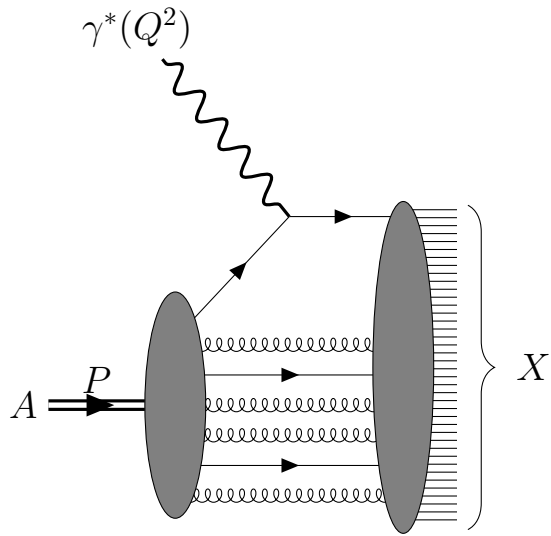

Figure 1.2: DIS in the Breit frame. The photon kicks out a parton (a quark in this figure), which is effectively independent of remaining partons during the interaction. The partons after the interaction are hadronized to form an inclusive set of hadrons $X$ in the final state.

In the parton model, the virtual photon does not kick the hadron as a whole, but a single parton of momentum

$$k^\mu = (k^+, \frac{(k^\perp)^2 + k^2}{2k^+}, k^\perp), \tag{1.60}$$

see Fig. 1.2. This parton carries a light-cone longitudinal momentum fraction $\xi = \frac{k^+}{P^+}$ of the nucleus. After the scattering, the scattered parton carries the momentum $k'^\mu$. Since in the Breit frame the partons can be treated as free during the interaction, there is an energy-momentum conservation across the electromagnetic vertex. Therefore, the parton before and after scattering is on-shell. And since it is assumed to be massless, $k^2 = k'^2 = 0$. In addition, as the nucleus is moving fast along the $x^-$ axis, we can assume that the parton is collinear, $k^\perp = 0$. We can then approximate the four-momentum of the struck quark as

$$k^\mu \approx (k^+, 0, 0^\perp). \tag{1.61}$$

Furthermore, the conservation of four-momenta reads $k' = q + k$. Therefore,

$$
\begin{aligned}
0 = q^2 + 2q \cdot k &= -Q^2 + \sqrt{2}\xi P^+ Q \\
&= -Q^2 + \frac{\xi Q^2}{x_{Bj}} \frac{1 + \sqrt{1 + \frac{4x_{Bj}^2 M^2}{Q^2}}}{2}.
\end{aligned}
\tag{1.62}
$$

We end up with following relation between $x_{Bj}$ and $\xi$:

$$
\xi = \frac{2x_{Bj}}{1 + \sqrt{1 + \frac{4x_{Bj}^2 M^2}{Q^2}}} \underset{M^2 \ll Q^2}{\simeq} x_{Bj}.
\tag{1.63}
$$

If $M^2 \ll Q^2$, $\xi$ and $x_{Bj}$ are approximately identical. In other words, in this limit, $x_{Bj}$ can be interpreted as the light-cone longitudinal momentum fraction of the nucleus carried by the struck parton.

## 1.3.2 Dipole picture for DIS

Let us return to the restframe of the nucleus, where $P^\mu = (M/\sqrt{2}, M/\sqrt{2}, 0^\perp)$, and choose the axis such that the transverse components of the photon are zero, $q^\mu = (q^+, q^-, 0^\perp)$. The components $q^+$ and $q^-$ obey following expressions

$$
\begin{aligned}
2q^+ q^- &= -Q^2, \\
\frac{Q^2}{2x_{Bj}} &= P \cdot q = \frac{M}{\sqrt{2}}(q^+ + q^-).
\end{aligned}
\tag{1.64}
$$

Solving this system of equations in terms of $Q$, $x$, and $M$, one gets

$$
q^- = -\frac{\sqrt{2}Mx_{Bj}}{1 + \sqrt{1 + \frac{4M^2 x_{Bj}^2}{Q^2}}} \sim -Mx_{Bj}.
\tag{1.65}
$$

The coherent length, which is defined as the typical light-cone longitudinal distance of the interaction, is given by

$$
\Delta\tau_{coh} \sim \frac{1}{|q^-|} \sim \frac{1}{Mx_{Bj}}.
\tag{1.66}
$$

It is also called the Ioffe time [40]. When $x_{Bj}$ decreases, the coherent length increases. For small $x_{Bj}$, $\Delta\tau_{coh}$ becomes much larger than the size of the nucleus. Therefore, at high energy, the virtual photon does not interact directly with the nucleus. Instead, it will fluctuate into a quark-antiquark dipole, which is hereafter referred to as an onium, before the scattering. This onium, possibly equiped with quantum corrections, then interact with the gluonic field inside the nucleus (see Fig. 1.3). Moreover, the positions of the quark and the antiquark are frozen in the transverse plane: the onium does not change size during the interaction!

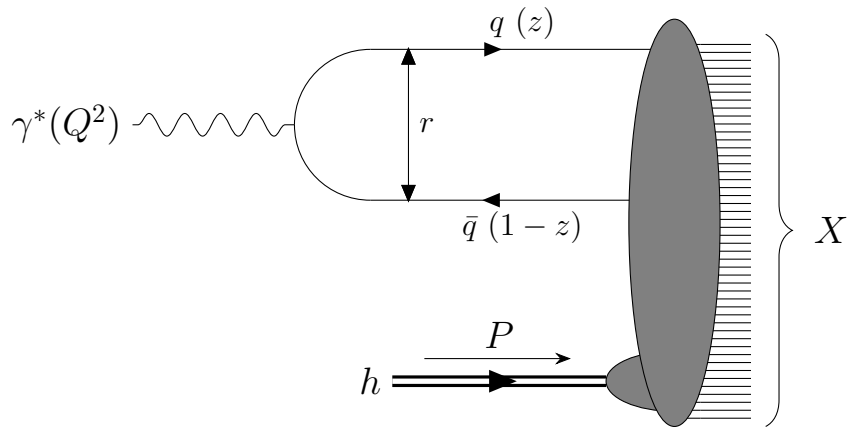

Figure 1.3: DIS in the dipole picture. The virtual photon splits into an onium (a $q\bar{q}$ pair) of tranverse size $r$ before the interaction with the nucleus. The quark carries a momentum fraction $z$ of the photon, while the antiquark carries the remainder $(1-z)$.

The small-$x_{Bj}$ limit, which is mentioned in the previous paragraph, is equivalent to the limit of high energy. Indeed, for large $\hat{s}$ such that $\hat{s} \gg Q^2, M^2$, we have

$$\frac{\hat{s} + Q^2 - M^2}{Q^2} = \frac{2P \cdot q}{Q^2} = \frac{1}{x_{Bj}}. \tag{1.67}$$

Therefore,

$$\frac{\hat{s}}{Q^2} \approx \frac{1}{x_{Bj}}. \tag{1.68}$$

For the sake of convenience, we introduce a representation of the energy, which is called as the rapidity $Y$ defined by

$$Y \equiv \ln \frac{\hat{s} + Q^2}{Q^2} \approx \ln \frac{1}{x_{Bj}}. \tag{1.69}$$

Since the virtual photon interacts with the nucleus via the onium, we can write down the following dipole factorization for the total cross section $\sigma^{\gamma^* A}$:

$$\sigma_{tot}^{\gamma^* A}(Y, Q^2) = \int_0^1 dz \int d^2 r^\perp \left| \Psi^{\gamma^* \to q\bar{q}}(r, z; Q^2) \right|^2 \sigma_{tot}^{onium}(r^\perp, Y), \tag{1.70}$$

$|\Psi^{\gamma^* \to q\bar{q}}(r, z; Q^2)|^2$ is the probability density for the photon with the virtuality $Q^2$ to dissociate into an onium of transverse size $r \equiv |r^\perp|$ and a fraction $z$ of the photon's longitudinal momentum (their expressions can be found in Refs. [31, 37, 38]; see also Chapter 4). The quantity $\sigma_{tot}^{onium}(r, Y)$ is the total cross section of the scattering of an onium of transverse size $r$ at the total relative rapidity $Y$. By the optical theorem, it is related to the forward elastic scattering amplitude $T_1$ by

$$\sigma_{tot}^{onium}(r^\perp, Y) = 2 \int d^2 b^\perp T_1(r^\perp, b^\perp, Y). \tag{1.71}$$

In this thesis, we shall thoroughly assume the impact parameter independence in such a manner that the $b$-integration results in an overall constant $\sigma_0$, $\sigma_{tot}^{onium}(r, Y) = \sigma_0 2T_1(r, Y)$. This is basically a good approximation for centered scatterings off a large, homogeneous target. From Eq. (1.70), it is important to understand the scattering process from the onium level at high energy. The studies of the onium-nucleus scattering are the main discussions of the work presented in this thesis.

## 1.4 Dipole evolution and BFKL equation

As mentioned in the previous section, the onium-nucleus scattering is the backbone of the high-energy DIS in the dipole picture. The onium may interact with the gluonic state of the nucleus by its bare state or its evolved state by the virtue of quantum corrections, depending on the setting. In this thesis, we are interested in the frame where the onium is highly evolved. Therefore, it is essential to understand the wave function of the onium subject to the high energy evolution.

### 1.4.1 High-energy evolution of the onium

At $x^+ = -\infty$, the onium is a bare color-singlet quark-antiquark dipole. Therefore, its asymptotic state can be written as

$$|\Psi\rangle_{-\infty} = \int \frac{\overline{d\hat{k}}}{2(p^+ - \hat{k}^+)} \phi_{ij}^{(0)}{}_{rs}(\hat{k}^\perp, \zeta) |q(\hat{k}; r, i)\bar{q}(p - \hat{k}; s, j)\rangle, \qquad (1.72)$$

where $|q(\hat{k}; r, i), \bar{q}(p - \hat{k}; s, j)\rangle$ is the $q\bar{q}$ state in which the quark of color index $i$ carries a momentum $\hat{k}$ and has the helicity $r$, while $p - \hat{k}$, $s$ and $j$ are of the antiquark, and $\zeta \equiv \hat{k}^+/p^+$ is the longitudinal momentum fraction carried by the quark. For the sake of convenience, in Eq. (1.72), we use the following notation for the integral measure:

$$\overline{dk} \equiv \frac{dk^+ d^2 k^\perp}{(2\pi)^3 2k^+} \Theta(k^+). \qquad (1.73)$$

At the observation time $x^+ = 0$, the wave function of the onium can be dressed by gluons by the virtue of quantum radiation. Since we consider small-$x_{Bj}$ limit, in which only soft gluon emissions are taken into account, the quark contribution is negligible. Using Eq. (1.50), we can write the state of the onium at $x^+ = 0$ as

$$|\Psi\rangle_0 = |\Psi\rangle_{-\infty} + \underbrace{\sum_{\lambda, a} \int \frac{\overline{dk dl}}{2(p^+ - k^+ - l^+)} \phi_{\lambda r s}^{(1)}{}_{aij}(k^\perp, l^\perp, \xi, \xi') |q(k; r, i)\bar{q}(p - k - l; s, j)g(l; \lambda, a)\rangle}_{\text{one gluon}} + \cdots,$$

$$(1.74)$$

where the second term is the lowest-order correction taking into account one-gluon emission. The momenta $l$, $k$, and $p - k - l$ are correspondingly of the emitted gluon, quark and the antiquark. The indices $\lambda$ and $a$ are the polarization and color index of the gluon. $\xi = k^+/p^+$ and $\xi' = l^+/p^+$ are the momentum fractions carried by the quark and the gluon, respectively. Higher-order terms are corresponding to further gluon emissions in the Fock state of the onium. In the above formula, the quantum numbers of quark and antiquark are suppressed. We are going to explore the quantum fluctuations in the onium's state from the lowest order.

**One-gluon emission**

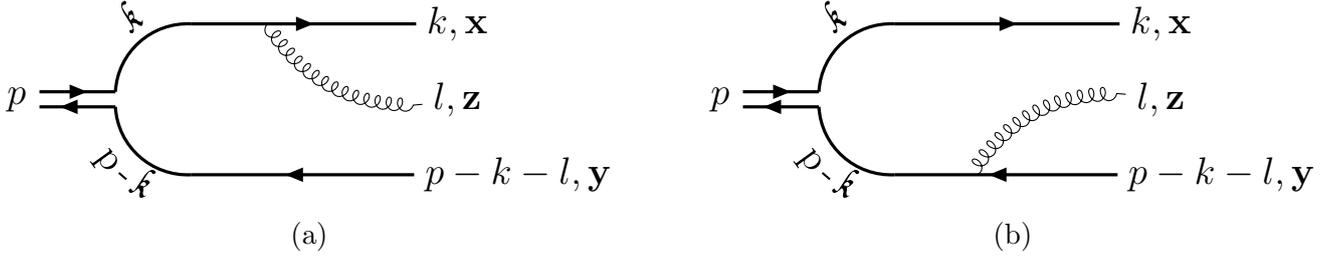

(a)                     (b)

Figure 1.4: Real one-gluon emission either from (a) quark or (b) antiquark of the onium.

Using Eq. (1.51) and keeping only the lowest-order term, the wave function of the onium with a single emitted gluon is given by

$$
\phi^{(1)}_{\substack{\lambda r s \\ a i j}}(k^\perp, l^\perp, \xi, \xi') = \sum_{\substack{r_0, s_0 \\ i_0, j_0}} \int \frac{d\bar{\hat{k}}}{2(p^+ - \hat{k}^+)}
$$
$$
\times \frac{\langle q(k; r, i)\bar{q}(p - k - l; s, j)g(l; \lambda, a)|H^{qqg}_{int}|q(\hat{k}; r_0, i_0)\bar{q}(p - \hat{k}; s_0, j_0)\rangle}{(p - \hat{k})^- + (\hat{k})^- - k^- - l^- - (p - k - l)^-} \ \phi^{(0)}_{\substack{r_0 s_0 \\ i_0 j_0}}(\hat{k}^\perp, \zeta),
$$

(1.75)

where $\hat{k}$ is the momentum of the quark before the gluon emission. The Hamiltonian $H^{qqg}_{int}$ is the quark-gluon coupling term (1.40),

$$
H^{qqg}_{int} = \int_\Sigma dx^+ d^2x^\perp : g_s \overline{\widetilde{\psi}}\gamma^\mu \widetilde{A}^a_\mu t^a \widetilde{\psi} :,
$$

(1.76)

where as usual, the operators are subject to the normal ordering ($: \mathcal{O} :$). Now notice that

$$
|q(k; r, i)\bar{q}(p - k - l; s, j)g(l; \lambda, a)\rangle = a^{\lambda\dagger}_a(l)b^{i\dagger}_r(k)d^{j\dagger}_s(p - k - l)|0\rangle,
$$
$$
|q(\hat{k}; r_0, i_0)\bar{q}(p - \hat{k}; s_0, j_0)\rangle = b^{i_0\dagger}_{r_0}(\hat{k})d^{j_0\dagger}_{s_0}(p - \hat{k})|0\rangle,
$$
$$
\langle 0|0\rangle = 1,
$$

(1.77)

where $|0\rangle$ is the normalized vacuum state. When expressing the field operators in the Hamiltonian (1.76) in Fourier modes using Eqs. (1.34) and (1.36), only two terms survive in the bra-ket sandwich, which can be written explicitly as follows

$$
\begin{aligned}
\langle q\bar{q}g|H_{int}^{qqg}|q\bar{q}\rangle = g_s &\int \overline{dk'dl'dh'} \langle 0|a_a^\lambda(l)b_r^i(k)d_s^j(p-k-l)b_{r'}^{i'\dagger}(k')a_{a'}^{\lambda'\dagger}(l')b_{s'}^{j'}(h')b_{r_0}^{i_0\dagger}(\hat{k})d_{s_0}^{j_0\dagger}(p-\hat{k})|0\rangle \\
&\times \bar{u}_{r'}(k')\not{\epsilon}_{\lambda'}(l')t^a u_{s'}(h')\int_\Sigma dx^+d^2x^\perp e^{i(\vec{k}'+\vec{l}'-\vec{h}')\cdot\vec{x}} \\
- g_s &\int \overline{dk'dl'dh'}\langle 0|a_a^\lambda(l)b_r^i(k)d_s^j(p-k-l)d_{s'}^{j'\dagger}(h')a_{a'}^{\lambda'\dagger}(l')d_{r'}^{j'}(k')b_{r_0}^{i_0\dagger}(\hat{k})d_{s_0}^{j_0\dagger}(p-\hat{k})|0\rangle \\
&\times \bar{v}_{r'}(k')\not{\epsilon}_{\lambda'}(l')t^a v_{s'}(h')\int_\Sigma dx^+d^2x^\perp e^{i(\vec{h}'+\vec{l}'-\vec{k}')\cdot\vec{x}}.
\end{aligned}
\tag{1.78}
$$

Using associated (anti-)commutation relations for the creation and annihilation operators, after some manipulations, we end up with following expression:

$$
\begin{aligned}
\phi_{\substack{\lambda_r rs\\aij}}^{(1)}(k^\perp,l^\perp,\xi,\xi') = g_s(2\pi)^3 \sum_{\substack{r_0,s_0\\i_0,j_0}} \Bigg\{ &\frac{1}{2(k^++l^+)}\frac{\left[\bar{u}_r(k)\gamma^\mu\epsilon_\mu^{\lambda*}(l)t^a u_{r_0}(k+l)\right]\phi_{\substack{r_0 s\\i_0 j_0}}^{(0)}(k^\perp+l^\perp,\xi+\xi')\delta_{ii_0}}{(k+l)^--k^--l^-} \\
&-\frac{1}{2(p^+-k^+)}\frac{\left[\bar{v}_{s_0}(p-k)\gamma^\mu\epsilon_\mu^{\lambda*}(l)t^a v_s(p-k-l)\right]\phi_{\substack{rs_0\\i_0 j_0}}^{(0)}(k^\perp,\xi)\delta_{jj_0}}{(p-k)^--(p-k-l)^--l^-}\Bigg\}.
\end{aligned}
\tag{1.79}
$$

The first term is corresponding to the case in which the gluon is emitted from the quark, while the second one is from the antiquark (Fig. 1.4). The Kronecker deltas of quark color indices represent the color rotation of (anti-)quark after emitting the gluon. Assuming the soft-gluon emission, i.e. the gluon emission is eikonal, from Eqs. (1.30) and (1.35) together with some algebras, we get

$$
\begin{aligned}
\bar{u}_r(k)\gamma^\mu\epsilon_\mu^{\lambda*}(l)t^a u_{r_0}(k+l) &\approx \bar{u}_r(k)\gamma^\mu\epsilon_\mu^{\lambda*}(l)t^a u_{r_0}(k) = 2k^\mu\epsilon_\mu^{\lambda*}(l)\delta_{rr_0}, \\
\bar{v}_{s_0}(p-k)\gamma^\mu\epsilon_\mu^{\lambda*}(l)t^a v_s(p-k-l) &\approx \bar{v}_{s_0}(p-k)\gamma^\mu\epsilon_\mu^{\lambda*}(l)t^a v_s(p-k) = 2(p-k)^\mu\epsilon_\mu^{\lambda*}(l)\delta_{ss_0}.
\end{aligned}
\tag{1.80}
$$

In the light-cone gauge, the polarization vectors read

$$
\epsilon_\lambda^\mu(l) = (0,\frac{l^\perp\cdot\epsilon_\lambda^\perp}{l^+},\epsilon_\lambda^\perp),
\tag{1.81}
$$

and we will choose $\epsilon_\lambda^\perp$ to be real. The denominators containing light-cone energies can be evaluated as

$$
\begin{aligned}
(k+l)^--k^--l^- &= \frac{(k^\perp+l^\perp)^2}{2(k^++l^+)}-\frac{(k^\perp)^2}{2k^+}-\frac{(l^\perp)^2}{2l^+}\approx-\frac{(l^\perp)^2}{2l^+}, \\
(p-k)^--(p-k-l)^--l^- &= \frac{(p^\perp-k^\perp)^2}{2(p^+-k^+)}-\frac{(p^\perp-k^\perp-l^\perp)^2}{2(p^+-k^+-l^+)}-\frac{(l^\perp)^2}{2l^+}\approx-\frac{(l^\perp)^2}{2l^+}.
\end{aligned}
\tag{1.82}
$$

In the end, the light-cone wave function of the onium taking into account one-gluon emission in the momentum space reads

$$\phi^{(1)}_{\substack{\lambda rs\\aij}}(k^\perp, l^\perp, \xi, \xi') \approx -2(2\pi)^3 g_s t^a \frac{l^\perp \cdot \epsilon_\lambda^\perp}{(l^\perp)^2} \left[ \phi^{(0)}_{\substack{rs\\ij}}(k^\perp + l^\perp, \xi + \xi') - \phi^{(0)}_{\substack{rs\\ij}}(k^\perp, \xi) \right]. \tag{1.83}$$

Now we employ the mix representation by transforming Eq. (1.83) into the transverse coordinate space while keeping the longitudinal component intact. One gets

$$\begin{aligned}
\bar{\phi}^{(1)}_{\substack{\lambda rs\\aij}}(x^\perp, y^\perp, z^\perp, \xi, \xi') &= \int \frac{d^2 l^\perp}{(2\pi)^2} \frac{d^2 k^\perp}{(2\pi)^2} \phi^{(1)}_{\substack{\lambda rs\\aij}}(k^\perp, l^\perp, \xi, \xi') e^{ik^\perp \cdot (x^\perp - y^\perp) + il^\perp \cdot (x^\perp - z^\perp)} \\
&= 2ig_s t^a \bar{\phi}^{(0)}_{rs}(x^\perp - y^\perp, \xi) \left[ \frac{(x^\perp - z^\perp) \cdot \epsilon_\lambda^\perp}{(x^\perp - z^\perp)^2} - \frac{(y^\perp - z^\perp) \cdot \epsilon_\lambda^\perp}{(y^\perp - z^\perp)^2} \right],
\end{aligned} \tag{1.84}$$

where $\bar{\phi}$ denotes for the transverse Fourier image of $\phi$, and $x^\perp, y^\perp$ and $z^\perp$ are the relative transverse positions of the quark, the antiquark and the emitted gluon, respectively (see Fig. 1.4). To arrive at the second line of Eq. (1.84), we employ the formula Eq. (B.22) in Appendix B and also suppress the quark color indices in the wave function of the onium before emitting the gluon. Squaring the wave function $\bar{\phi}^{(1)}_{\substack{\lambda rs\\aij}}$, summing over all possible quantum numbers and integrating over the longitudinal momentum and over the transverse position of the gluon, the leading-$\alpha_s$ order probability to find a soft gluon in the wave function of the onium is then given by

$$\int_{\xi_0}^{min(\xi, 1-\xi)} \frac{d\xi'}{\xi'} \int d^2 r'^\perp \frac{\alpha_s C_F}{\pi^2} \frac{r^{\perp 2}}{r'^{\perp 2}(r^\perp - r'^\perp)^2} \sum_{r,s} |\bar{\phi}^{(0)}_{rs}(r^\perp, \xi)|^2, \tag{1.85}$$

where $\alpha_s = g_s^2/(4\pi)$, $r^\perp \equiv x^\perp - y^\perp$, $r'^\perp \equiv x^\perp - z^\perp$. The factor $C_F = (N_c^2 - 1)/(2N_c)$ is the fundamental Casimir. We see that, in the coordinate space, the wave function of the bare onium totally factorizes. Furthermore, there are two types of logarithmic singularity occuring in Eq. (1.85). The first type is the soft divergence corresponding to the limit $\xi' \to 0$. Therefore, we introduce an IR cutoff $\xi_0$ in the integration. The second one is realized when $z^\perp$ approaches either $x^\perp$ or $y^\perp$. In such cases we have collinear divergence.

**Large-$N_c$ limit**

Let us now consider the onium wave function in the limit of large $N_c$, which was introduced by 't Hooft [41, 42]. As we shall shortly see, this limit will eliminate a class of diagrams which matter at higher-order gluon emissions, which simplifies the construction of the onium wave function.

We start with the Fierz identity for the $SU(N_c)$ generator $t^a$, which reads

$$
t^a_{ij} t^a_{kl} = \frac{1}{2} \delta_{il} \delta_{kj} - \frac{1}{2N_c} \delta_{ij} \delta_{kl}
$$

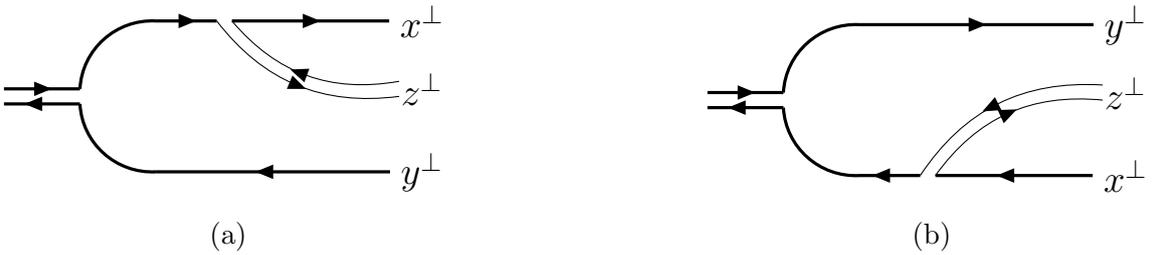

(1.86)

When the number of colors $N_c$ is taken to be large, the second term in Eq. (1.86) is negligible. The gluon is then equivalent to a zero-size quark-antiquark pair. In this limit, the emission of one soft gluon of momentum fraction $x$ at position $z^\perp$ from the initial onium is essentially a dipole branching process: the initial onium of size $r^\perp \equiv x^\perp - y^\perp$ splits into two daughter dipoles of size $r'^\perp \equiv x^\perp - z^\perp$ and $r^\perp - r'^\perp$ (see Fig. 1.5). From Eq. (1.85), the probability of this process, up to $dx$ and $d^2 z^\perp$, given by

$$
\bar{\alpha} \frac{dx}{x} \underbrace{\frac{d^2 r'^\perp}{2\pi} \frac{r^{\perp 2}}{r'^{\perp 2}(r^\perp - r'^\perp)^2}}_{dp_{1\to2}(r^\perp, r'^\perp)},
$$

(1.87)

with $C_F \simeq N_c/2$ at large $N_c$ and $\bar{\alpha} \equiv (\alpha_s N_c)/\pi$. The dipole splitting rate $dp_{1\to2}$ in Eq. (1.87) can be decomposed as

$$
dp_{1\to2}(r^\perp, r'^\perp) \equiv \frac{1}{2\pi} \frac{r^{\perp 2}}{r'^{\perp 2}(r^\perp - r'^\perp)^2} d^2 r'^\perp = \frac{1}{2\pi} \left( \frac{1}{r'^{\perp 2}} + \frac{1}{(r^\perp - r'^\perp)^2} + \frac{2 r'^\perp \cdot (r^\perp - r'^\perp)}{r'^{\perp 2}(r^\perp - r'^\perp)^2} \right) d^2 r'^\perp.
$$

(1.88)

(a)          (b)

Figure 1.5: The large $N_c$ limit of the Fig. 1.4

The first two terms in Eq. (1.88) is corresponding to the first and the second diagrams on the right of Fig. 1.6, while the last term is from the last two inteference diagrams. The full dipole kernel is represented by the diagram on the left of Fig. 1.6.

The emission of one soft gluon, or a single dipole branching, given by the probability (1.87) is one step of the evolution of the dipole when boosting the bare onium to a higher rapidity. Instead of the momentum fraction $x$, we can write the splitting probability in term of the evolution rapidity,

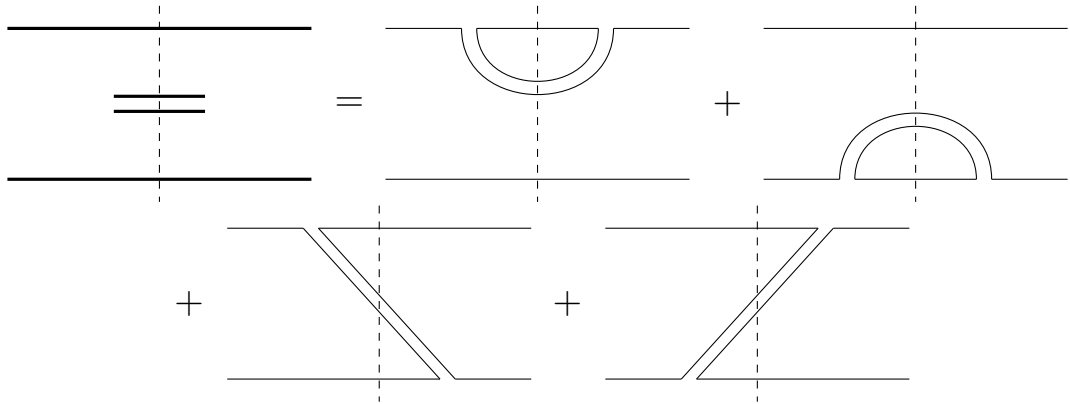

Figure 1.6: Squared onium wave function with a single soft gluon at large $N_c$. The dashed line separates the wave function (to the left) from its complex conjugate (to the right).

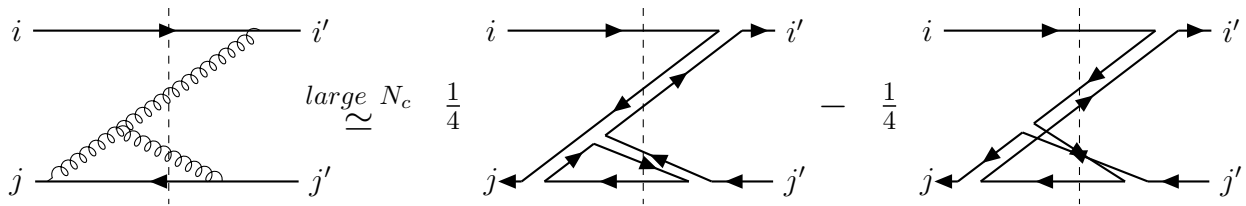

Figure 1.7: Triple gluon vertex in the large $N_c$ limit.

which is given by the logarithm of $x$, $Y \equiv \ln(1/x)$. The probability (1.87) can be rewritten as

$$\bar{\alpha} dY dp_{1\to 2}(r^\perp, r'^\perp), \tag{1.89}$$

which is the probability of dipole branching into a pair of dipoles of sizes $r'^\perp$ and $r^\perp - r'^\perp$ up to $d^2 r'^\perp$ when advancing the rapidity by a step $dY$. This one-gluon emission is the first-order modification to the wave function of the onium in the leading-$\alpha_s \ln(1/x)$ approximation (LLA).

Moving on to the higher-order corrections requires the triple-gluon vertex. In the eikonal limit, both the quark-gluon and the triple-gluon vertices are the same, in particular of the form $2g_s k \cdot \epsilon$, where $k^\mu$ and $\epsilon^\mu$ are the momentum and the polarization of the emitted soft gluon, respectively. Furthermore, in the large-$N_c$ limit, the triple gluon vertex, as shown in Fig. 1.7, has two configurations: the planar color flow (the first diagram on the right of Fig. 1.7), and the non-planar one (the second diagram on the right of Fig. 1.7). When taking the trace of color matrices, which is tantamount to connect $i$ and $j$, and $i'$ and $j'$ in Fig. 1.7, the former consists of three color loops ($N_c^3$) and hence, its contribution is of order $(\alpha_s N_c)^2$, after averaging over all colors. Meanwhile, the contribution from the non-planar diagram is of order $(\alpha_s N_c)^2/N_c^2$, which is suppressed by the square of $N_c$. The triple gluon vertex can then be represented by the planar configuration at large $N_c$, which is similar to the quark-gluon coupling. In general, in diagrams of the same order of $\alpha_s N_c$, non-planar diagrams are suppressed by powers of $N_c$. Therefore, we can neglect all non-planar diagrams at large $N_c$. This is an important consequence of the large $N_c$ limit, which greatly simplifies the analyses of the evolution of the onium Fock state.

**Higher-order corrections**

When adding additional gluons to the Fock state of the onium, to obtain contributions at the LLA, a gluon emitted later in the light-cone time must be softer than another gluon emitted earlier. In other words, there should be a strong ordering in the light-cone longitudinal momenta of gluons, $l_1^+ \gg l_2^+ \gg l_3^+ \gg \cdots$, in accordance to their emission times (see, for e.g., Fig. 1.8). Consequently, the diagrams containing instantaneous interaction vertices are subleading.

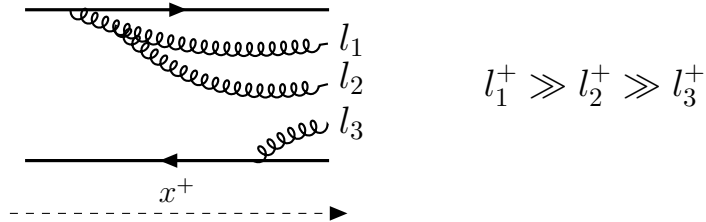

Figure 1.8: An example of a LLA diagram. The minimal value for the longitudinal momenta of the emitted soft gluons is fixed by the total rapidity of the evolution.

The treatment of the higher-order soft-gluon emissions can be simplified due to the following facts. First, as in Eq. (1.85), in the transverse coordinate space each step of evolution well factorises from the previous step. Second, the eikonal emissions of gluons from a quark and from a gluon can be treated identically. Finally, since the non-planar diagrams are suppressed, each subsequent dipole in the Fock state of the onium at each evolution step evolves independently. Therefore, in the large-$N_c$ and eikonal limits, the evolution of the onium toward high energy (or high rapidity) is essentially the iteration of dipole branching (Fig. 1.9). Eventually, the Fock state of the onium is a stochastic set of dipoles with different transverse sizes.

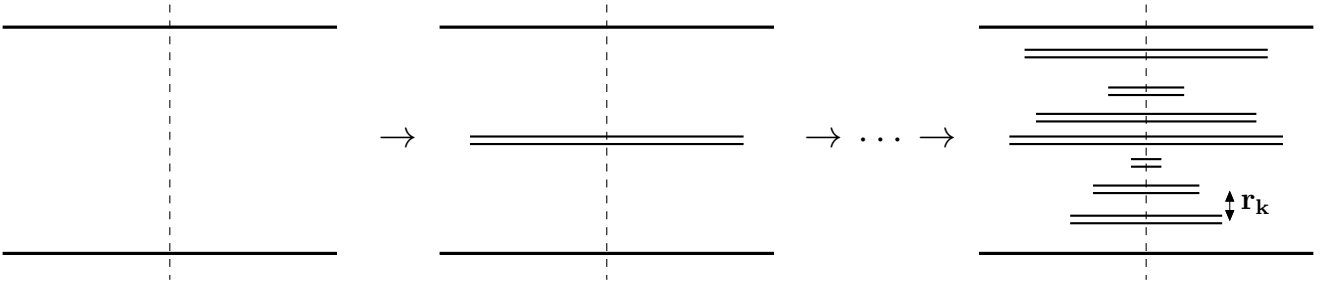

Figure 1.9: A realization of the squared wave function of the evolved onium at large $N_c$.

The formalism presented in this section is referred to as Mueller's color dipole model [32]. Knowing the structure of the wave function of the onium subject to high energy evolution at LLA, it is important to resum all the LLA terms. It was done by the virtue of evolution equations.

### 1.4.2 Dipole number and BFKL evolution

Consider an onium of size $r$ evolved to a rapidity $Y > 0$. Let us denote by $n(r, Y; R)$ the mean number of dipoles of (scalar) transverse size $R$ in the Fock state of the onium $r \equiv |r^\perp|$. We are going to derive an equation to control the rapidity evolution $n(r, Y; R)$ at LLA.

Let us start with a bare onium at $Y = 0$, and evolve it to $Y + dY$. After an evolution step $dY$, there are two possibilities. In the first place, the bare onium can split into two dipoles of sizes $r' \equiv |r'^\perp|$ and $|r^\perp - r'^\perp|$, with the probability given by Eq. (1.89). These two daughter dipoles evolves in the rapidity interval of width $Y$, from $dY$ to $Y + dY$. The total mean number of dipoles then comes from the contributions of both offsprings. On the other hand, the initial onium may not split and hence, the mean number of dipoles is unchanged after the boost $dY$. In this case, there may be totally no soft-gluon emission in the wave function of the onium during the rapidity step $dY$, or virtual emissions: the soft gluon is emitted and reabsorbed before $dY$. Therefore, the mean number of dipoles at the rapidity $Y + dY$ is given by

$$
\begin{aligned}
n(r, Y + dY; R) = \bar{\alpha} dY \int dp_{1\to2}(r^\perp, r'^\perp) \left[ n(r', Y; R) + n(|r^\perp - r'^\perp|, Y; R) \right] \\
+ \left[ 1 - \bar{\alpha} dY \int dp_{1\to2}(r^\perp, r'^\perp) \right] n(r, Y; R).
\end{aligned}
\tag{1.90}
$$

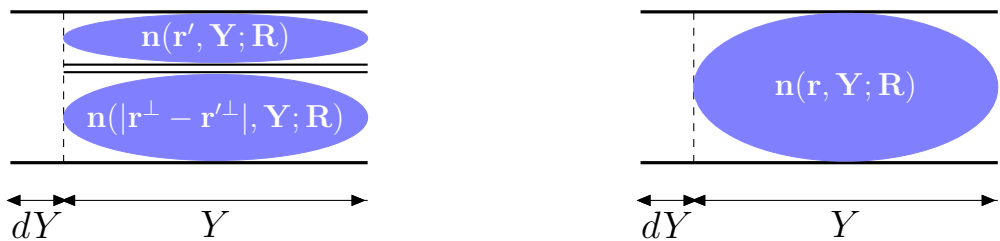

Figure 1.10: Illustration of two terms in Eq. (1.90). The first one is due to the real emission, while the second one represents the virtual correction.

Taking the limit $dY \to 0$, the equation above turns into the following integro-differential evolution equation:

$$
\frac{\partial}{\partial Y} n(r, Y; R) = \bar{\alpha} \int dp_{1\to2}(r^\perp, r'^\perp) \left[ n(r', Y; R) + n(|r^\perp - r'^\perp|, Y; R) - n(r, Y; R) \right].
\tag{1.91}
$$

Since at $Y = 0$, the onium is in the bare state, the initial condition for $n$ reads $n(r, Y = 0; R) = \delta\left(\frac{r}{R}\right)$.

The equation (1.91) is known as the BFKL equation [7, 8] (see also Refs.[32, 43–45]) written in the transverse coordinate space. Its more general form includes the impact parameter dependence. However, as mentioned previously, we shall neglect such dependency in this thesis.

### 1.4.3 Solution to the BFKL equation

The equation (1.91) can be rewritten as

$$
\partial_Y n(r, Y; R) = K_{BFKL} \otimes n(r, Y; R),
\tag{1.92}
$$

where $K_{BFKL}$ is the integral kernel in the BFKL equation, which acts on the mean dipole number $n$. We observe that,

$$K_{BFKL} \otimes \left(\frac{r^2}{R^2}\right)^\gamma = \left\{ \bar{\alpha} \int \frac{d^2 r'^\perp}{2\pi} \frac{r^2}{r'^2 |r^\perp - r'^\perp|^2} \left[ \left(\frac{r'^2}{r^2}\right)^\gamma + \left(\frac{|r^\perp - r'^\perp|^2}{r^2}\right)^\gamma - 1 \right] \right\} \left(\frac{r^2}{R^2}\right)^\gamma. \quad (1.93)$$

We can rewrite the integral in the curly bracket as

$$\int \frac{d^2 r'^\perp}{2\pi} \left\{ \frac{r^2}{r'^2 |r^\perp - r'^\perp|^2} \left[ \left(\frac{r'^2}{r^2}\right)^\gamma + \left(\frac{|r^\perp - r'^\perp|^2}{r^2}\right)^\gamma \right] - \frac{r^\perp \cdot r'^\perp + r^\perp \cdot (r^\perp - r'^\perp)}{r'^2 |r^\perp - r'^\perp|^2} \right\}$$
$$= \int \frac{dr' d\theta}{2\pi} \frac{r}{r^2 + r'^2 - 2rr'\cos\theta} \left[ \left(\frac{r'}{r}\right)^{2\gamma - 1} - \cos\theta \right]. \quad (1.94)$$

Performing the angular integration and putting $u = r'/r$, we obtain

$$\int_0^\infty du \frac{2u^{2\gamma} - (u^2 + 1) + |1 - u^2|}{2u|1 - u^2|} = \int_0^1 du \frac{u^{2\gamma - 1} + u^{1 - 2\gamma} - 2u}{1 - u^2}. \quad (1.95)$$

Using the integral representation for the digamma function $\psi(z)$ (the Appendix B), we end up with the following final expression for the integral in Eq. (1.93), which we shall hereafter denote by $\chi(\gamma)$:

$$\chi(\gamma) = 2\psi(1) - \psi(\gamma) - \psi(1 - \gamma). \quad (1.96)$$

Therefore, the functions $(r^2/R^2)^\gamma$ are the eigenfunctions of the BKFL kernel, corresponding to the eigenvalues $\bar{\alpha}\chi(\gamma)$. The function $\chi(\gamma)$ has simple poles at integers, $\gamma = k$, $k \in \mathbb{Z}$. The principal branch of $\chi(\gamma)$ lies on the domain $0 < \text{Re}(\gamma) < 1$. A graphical illustration for real arguments showing its principal branch and two other branches on both sides of the principal one is plotted in Fig. 1.11.

With the help of the eigenfunctions, the general solution of the BFKL equation (1.91) can be written as

$$n(r, Y; R) = \int_{c - i\infty}^{c + i\infty} \frac{d\gamma}{2\pi i} \left(\frac{r^2}{R^2}\right)^\gamma \widetilde{n}(\gamma, Y), \quad (1.97)$$

with $c$ a real constant. The coefficient function $\widetilde{n}(\gamma, Y)$ satisfies the following differential equation:

$$\partial_Y \widetilde{n}(\gamma, Y) = \bar{\alpha}\chi(\gamma)\widetilde{n}(\gamma, Y). \quad (1.98)$$

with the initial condition $\widetilde{n}(\gamma, Y = 0) = 1$. Consequently, $\widetilde{n}(\gamma, Y)$ is given by

$$\widetilde{n}(\gamma, Y) = e^{\bar{\alpha}\chi(\gamma)Y}; \quad (1.99)$$

and the solution (1.97) becomes

$$n(r, Y; R) = \int_{c - i\infty}^{c + i\infty} \frac{d\gamma}{2\pi i} \left(\frac{r^2}{R^2}\right)^\gamma e^{\bar{\alpha}\chi(\gamma)Y}. \quad (1.100)$$

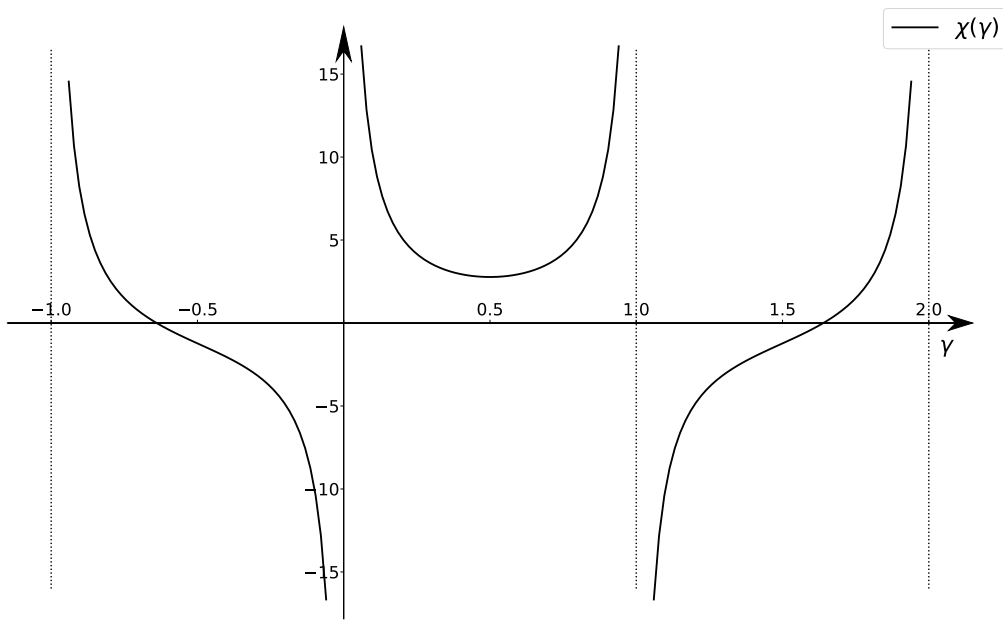

Figure 1.11: The function $\chi(\gamma)$ on the real domain. The principal branch is on the range $(0, 1)$.

Now let us evaluate the mean number of dipoles for a special case in which $r/R \ll 1$. In this limit the integral in the solution (1.100) is dominated by the region $\gamma \sim 1$, and then the eigenfunction $\chi(\gamma)$ can be approximated as $\chi(\gamma) \simeq 1/(1 - \gamma)$. Eq. (1.100) can be rewritten as

$$n_{r \ll R}(r, Y; R) \simeq \int_{c-i\infty}^{c+i\infty} \frac{d\gamma}{2\pi i} \exp\left[\gamma \ln \frac{r^2}{R^2} + \frac{\bar{\alpha} Y}{1 - \gamma}\right]. \tag{1.101}$$

The saddle point is located at

$$\gamma_{SD} = 1 - \sqrt{\frac{\bar{\alpha} Y}{\ln(R^2/r^2)}}, \tag{1.102}$$

provided that $\ln(R^2/r^2) \gg \bar{\alpha} Y$. Setting $c = \gamma_{SD}$ and performing the integration in Eq. (1.101) in the saddle point approximation, one yields

$$n_{DLA}(r, Y; R) \simeq \frac{r^2}{2R^2} \frac{(\bar{\alpha} Y)^{1/4}}{[\ln(R^2/r^2)]^{3/4}} \exp\left[2\sqrt{\bar{\alpha} Y \ln \frac{R^2}{r^2}}\right] \tag{1.103}$$

This solution is referred to as the double logarithmic approximation (DLA), as it resums two large logarithms per each power of $\bar{\alpha}$: $\bar{\alpha} \ln(1/x) \ln(R^2/r^2)$.

When $\ln(R^2/r^2)$ is not large, and hence is not important to resum, but $\bar{\alpha} \ln(1/x)$ is still large, the integration in Eq. (1.100) is dominated by the saddle point at $\chi'(\gamma_c) = 0$, or $\gamma_c = 1/2$. Using the saddle point approximation with the contour being the line parallel to the imaginary axis and passing through the saddle point, we get the following result:

$$n_{LLA}(r, Y; R) \simeq \frac{r}{R} e^{\bar{\alpha} \chi \left( \frac{1}{2} \right) Y} \frac{\exp \left[ - \frac{\ln^2 \frac{r^2}{R^2}}{2\chi''(1/2)\bar{\alpha}Y} \right]}{\sqrt{2\pi \chi''(1/2)\bar{\alpha}Y}} \tag{1.104}$$

with $\chi(1/2) = 4\ln 2$ and $\chi''(1/2) = 28\zeta(3)$ ($\zeta(z)$ is the Riemann zeta function). We see that the density exhibits an exponential growth in the rapidity with the slope $\alpha_{\mathbb{P}} - 1 \equiv \bar{\alpha}\chi(1/2)$. This slope is usually referred to as the intercept of the BFKL pomeron, in reference to the Regge phenomenology (for a review, see [46]).

## 1.5 Summary

We close this chapter by summarizing some remarks. The process of deep-inelastic scattering at high energy can be conveniently described using the dipole formulation in which the virtual photon interacts with the target via its onium state. This formulation enables us to turn the discussion of virtual photon-hadron scattering into that of the onium-nucleus interaction, which ,to a certain extent, requires the understanding of the onium wave function at high energy. The latter is dominated by soft gluons and turns out, in the large $N_c$ limit, to be a set of color dipoles of various sizes. The resummation of small-$x$ gluon emissions at the leading logarithmic accuracy can be done by the virtue of the linear BFKL evolution equation. The discussions in this chapter, especially on the onium wave function at high energy, are the basis for further discussions in the thesis, which are mainly on the onium-nucleus scattering.

# QCD evolution in analogy with branching-diffusion processes

## Contents

In the previous chapter, we dealt with the wave function of an onium subject to a high-energy evolution. In the present chapter, we shall introduce the Balitsky-Kovchegov (BK) equation [17, 18], which governs the high-energy onium-nucleus interaction, and demonstrate its relationship to the Fisher-Kolmogorov-Petrovsky-Piscounov (F-KPP) equation [47, 48], which describes the reaction-diffusion processes in statistical physics. We shall then report on one of our original contributions [25]. In particular, we will present a Monte-Carlo algorithm to generate particles of a branching random walk (BRW) in the vicinity of a leading particle, or the "tip" of the BRW, which allows to investigate the particle density in the tip as well as the structure of the evolution.

## 2.1 Nuclear scattering of onia

### 2.1.1 Balitsky-Kovchegov evolution equation

Consider the scattering of an onium of size $r$ off an nucleus of mass number $A$ at total relative rapidity $Y$. We denote by $S(r, Y)$ the S-matrix elements for that scattering process. At high energy, cross sections are purely absorptive, so the S-matrix elements are real. To remind, we assumed that $S$ depend neither on the impact parameter nor on the orientation of the dipole in the transverse plane.

The evolution equation for $S$ can be established using the same technique as for the mean number of dipoles $n$. Let us stay in the rest frame of the nucleus, and boost the onium to the rapidity $Y + dY$. In this frame, $S(r, Y)$ can be intepreted as the probability that an onium of size $r$ evolving to the rapidity $Y$ does not interact with the nucleus at rest. After an evolution step $dY$ at the beginning of the dipole evolution, the initial onium $r$ may split into two dipoles of sizes $r'$ and $|r^\perp - r'^\perp|$, with the probability given by Eq. (1.89); or it may stay unchanged. Hence, the S-matrix elements at $Y + dY$, $S(r, Y + dY)$, reads

$$S(r, Y+dY) = \bar\alpha dY \int dp_{1\to2}(r^\perp, r'^\perp) S(r', Y) S(|r^\perp - r'^\perp|, Y) + \left[1 - \bar\alpha dY \int dp_{1\to2}(r^\perp, r'^\perp)\right] S(r, Y). \tag{2.1}$$

Taking the limit $dY \to 0$, $S(r, Y)$ solves the following evolution equation:

$$\partial_Y S(r, Y) = \bar\alpha \int dp_{1\to2}(r^\perp, r'^\perp) \left[S(r', Y) S(|r^\perp - r'^\perp|, Y) - S(r, Y)\right]. \tag{2.2}$$

Equivalently, one can write the evolution equation for the forward elastic scattering amplitude $T_1(r, Y) = 1 - S(r, Y)$, which reads

$$\partial_Y T_1(r, Y) = \bar\alpha \int dp_{1\to2}(r^\perp, r'^\perp) \left[T_1(r', Y) + T_1(|r^\perp - r'^\perp|, Y) - T_1(r, Y)\right. \\ \left. - T_1(r', Y) T_1(|r^\perp - r'^\perp|, Y)\right]. \tag{2.3}$$

The equations (2.2) and (2.3) are two equivalent forms of the Balitsky-Kovchegov (BK) nonlinear evolution equation [17, 18] written for the S-matrix elements and the forward elastic scattering amplitude, respectively. From now on, we shall refer to the former as the S-type BK equation, and to the latter as the T-type BK equation. If one neglects the nonlinear term, the T-type BK equation (2.3) becomes the BFKL equation for the forward elastic scattering amplitude. The initial conditions for $S$ and $T$ are scattering profiles determined at a particular rapidity $Y_0$. Normally, they are set at zero rapidity, which can be chosen, for example, to be the McLerran-Venugopalan (MV) [49, 50] amplitude

$$T_1(r, Y=0) = 1 - S(r, Y=0) = 1 - \exp\left[-\frac{r^2 Q_A^2}{4} \ln\left(e + \frac{1}{r^2 \Lambda_{QCD}^2}\right)\right], \tag{2.4}$$

or the Golec-Biernat-Wusthoff (GBW) [51, 52] amplitude

$$T_1(r, Y = 0) = 1 - S(r, Y = 0) = 1 - \exp\left[-\frac{r^2 Q_A^2}{4}\right],\tag{2.5}$$

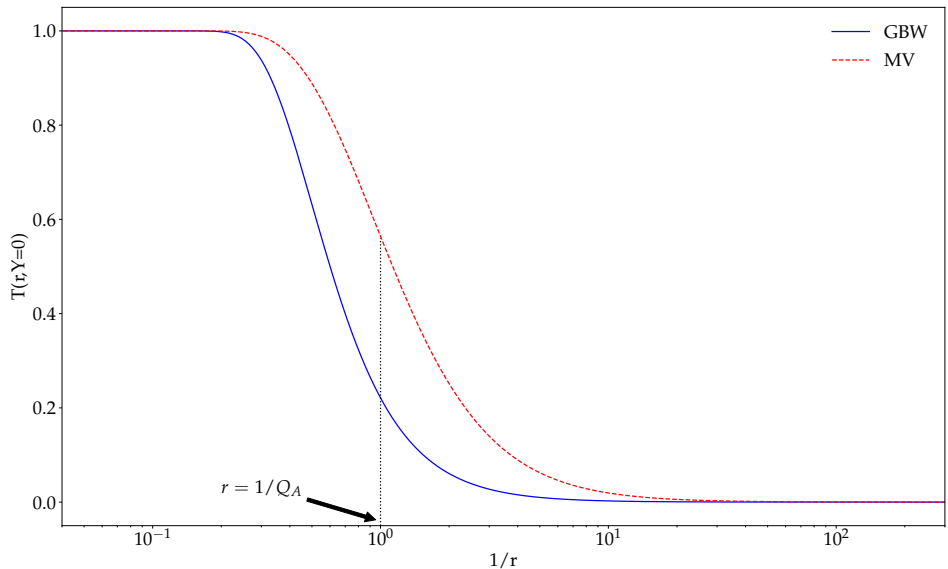

Figure 2.1: Two initial conditions for the BK equation: the MV model (dashed curve) and the GBW model (solid curve). The inverse saturation scale $1/Q_A$ is shown, which separates the $T_1 \sim 1$ regime from the $T_1 \sim 0$ regime. Parameters are set as $Q_A = 1$ GeV and $\Lambda_{QCD} = 0.2$ GeV.

where the A-dependent momentum $Q_A$ is called the saturation momentum characteristics of the nucleus. In such models, the scattering amplitude monotonically decreases as the size $r$ becomes smaller, and asymptotically reaches 1 when $r \to \infty$ (black-disk limit) and 0 when $r \to 0$ (color transparency limit). The transition between the two regimes $T_1 \sim 1$ and $T_1 \sim 0$ occurs around $r \sim 1/Q_A$ (see Fig. 2.1). Due to the evolution, the saturation scale will acquire a rapidity dependence, as we shall shortly see.

Before finishing this paragraph, let us write the T-type BK equation (2.3) in the transverse momentum space by the following Fourier transformation

$$\widetilde{T}_1(k^\perp, Y) = \int \frac{d^2 r^\perp}{2\pi r^2} e^{-ik^\perp \cdot r^\perp} T_1(r, Y) = \int_0^\infty \frac{dr}{r} J_0(kr) T_1(r, Y),\tag{2.6}$$

which shows that $\widetilde{T}_1$ depends only on the magnitude of $k^\perp$, $k \equiv |k^\perp|$. We first deal with the linear BFKL sector of Eq. (2.3). Expressing $T_1(r, Y)$ by the virtue of the inverse Mellin transformation, one obtained:

$$\begin{aligned}
\widetilde{T}_1(k^\perp, Y) &= \int_{\frac{1}{2}-i\infty}^{\frac{1}{2}+i\infty} \frac{d\gamma}{2\pi i} \mathcal{T}_\gamma(Y) \int_0^\infty d(rQ_A)(rQ_A)^{2\gamma-1} J_0(kr) \\
&= \int_{\frac{1}{2}-i\infty}^{\frac{1}{2}+i\infty} \frac{d\gamma}{2\pi i} \mathcal{T}_\gamma(Y) 2^{2\gamma-1} \left(\frac{k}{Q_A}\right)^{-2\gamma} \frac{\Gamma(\gamma)}{\Gamma(1-\gamma)}.
\end{aligned}\tag{2.7}$$

Reminding that $(rQ_A)^{2\gamma}$ are eigenfunctions of the BFKL kernel with the eigenvalues $\bar{\alpha}\chi(\gamma)$, we have

$$
\begin{aligned}
\int_0^\infty \frac{dr}{r} J_0(kr) K_{BFKL} \otimes T_1(r, Y) &= \bar{\alpha} \int_0^{+\infty} \frac{dr}{r} J_0(kr) \int_{\frac{1}{2}-i\infty}^{\frac{1}{2}+i\infty} \frac{d\gamma}{2\pi i} \chi(\gamma)(rQ_A)^{2\gamma} \mathcal{T}_\gamma(Y) \\
&= \bar{\alpha} \int_{\frac{1}{2}-i\infty}^{\frac{1}{2}+i\infty} \frac{d\gamma}{2\pi i} \mathcal{T}_\gamma(Y) 2^{2\gamma-1} \chi(\gamma) \left(\frac{k}{Q_A}\right)^{-2\gamma} \frac{\Gamma(\gamma)}{\Gamma(1-\gamma)} = \bar{\alpha}\chi(-\partial_L)\widetilde{T}_1(k^\perp, Y),
\end{aligned}
\tag{2.8}
$$

where $L \equiv \ln(k^2/Q_A^2)$.

For the nonlinear term, its Fourier transform reads

$$
\begin{aligned}
\bar{\alpha} \int \frac{d^2 r^\perp}{2\pi r^2} e^{-ik^\perp \cdot r^\perp} &\int \frac{d^2 r'^\perp}{2\pi} \frac{r^2}{r'^2 |r^\perp - r'^\perp|^2} T_1(r', Y) T_1(|r^\perp - r'^\perp|, Y) \\
&= \bar{\alpha} \int \frac{d^2 r'^\perp}{2\pi r'^2} e^{-ik^\perp \cdot r'^\perp} T_1(r', Y) \int \frac{d^2(r^\perp - r'^\perp)}{2\pi |r^\perp - r'^\perp|^2} e^{-ik^\perp \cdot (r^\perp - r'^\perp)} T_1(r_2, Y) = \bar{\alpha} \left[\widetilde{T}_1(k, Y)\right]^2.
\end{aligned}
\tag{2.9}
$$

To the end, the BK equation in the transverse momentum space is given by

$$
\partial_Y \widetilde{T}_1(k, Y) = \bar{\alpha}\chi(-\partial_L)\widetilde{T}_1(k, Y) - \bar{\alpha}\widetilde{T}_1^2(k, Y).
\tag{2.10}
$$

One can easily observe that, the linear parts of the BK equations in the coordinate and momentum spaces are similar. In particular, the coordinate-space BK equation in the linear regime reads

$$
\partial_Y T_1(\rho, Y) = \bar{\alpha}\chi(-\partial_\rho) T_1(\rho, Y),
\tag{2.11}
$$

with $\rho = \ln\left[1/(r^2 Q_A^2)\right]$.

### 2.1.2 Solution to the BK equation

#### a. General properties of the solution

The BK equation (2.3) has two fixed points: the stable $T_1 = 1$ and the unstable $T_1 = 0$. In the vicinity of the latter, the amplitude is small, $T_1 \ll 1$, and one can neglect the nonlinear term in Eqs. (2.3) and (2.11). The BK equation then becomes the BFKL equation, and the amplitude grows exponentially as in Eq. (1.104), $T_1 \sim e^{\bar{\alpha}\chi(1/2)Y}$. When $T_1$ approaches the stable fixed point, the nonlinear term becomes important. The effect of this term is to compensate the growth of the amplitude, and to cause it to saturate at the value $T_1 = 1$ when $Y \to \infty$. In other words, the nonlinear term in the BK equation comes from the requirement of the unitarity of the scattering amplitude.

Therefore, the scattering amplitude $T_1(r, Y)$ solving the BK equation is a smooth curve connecting two states $T_1 = 1$ and $T_1 = 0$. The transition between two limits occurs at some scale $r_s = 1/Q_s(Y)$, where $Q_s(Y)$ is the saturation momentum at the rapidity $Y$. The value of the latter at $Y = 0$ is the momentum $Q_A$ in the initial condition.

### b. Asymptotic solution outside the saturation region

Let us now derive a solution for the BK equation at an asymptotic large rapidity for the case $T_1 \ll 1$, taking into account the saturation correction near $r \sim 1/Q_s(Y)$. We begin with the linear equation (2.11) whose general solution reads

$$T_1(\rho, Y) = \int_{1/2-i\infty}^{1/2+i\infty} \frac{d\gamma}{2\pi i} \mathcal{T}_\gamma \exp\left[-\gamma\left(\rho - v(\gamma)\bar{\alpha}Y\right)\right], \tag{2.12}$$

where $v(\gamma) = \chi(\gamma)/\gamma$. The solution Eq. (2.12) can be interpreted as a linear superposition of elementary travelling waves of the form $e^{-\gamma(\rho-v(\gamma)\bar{\alpha}Y)}$. They propagate at different velocities given by $v(\gamma)$. The minimum velocity of such waves is $v(\gamma_0)$, with $0 < \gamma_0 < 1$ solving the equation

$$v'(\gamma_0) = 0 \Leftrightarrow \chi'(\gamma_0) = \frac{\chi(\gamma_0)}{\gamma_0}. \tag{2.13}$$

In numerical values, $\gamma_0 \approx 0.6275$, $\quad \chi(\gamma_0) \approx 3.0645$, $\quad \chi''(\gamma_0) \approx 48.5176$.

Now let us expand the kernel $\chi(-\partial_\rho)$ around $\gamma_0$ and truncate the series at the second-order term. This truncation would limit the applicability of the approach: in particular, it does not work in the DLA limit. The equation (2.11) then becomes

$$\partial_Y T_1(\rho, Y) = \bar{\alpha}\chi(\gamma_0)\partial_Y T_1(\rho, Y) - \bar{\alpha}\chi'(\gamma_0)(\partial_\rho + \gamma_0)T_1(\rho, Y) + \frac{1}{2}\bar{\alpha}\chi''(\gamma_0)(\partial_\rho + \gamma_0)^2 T_1(\rho, Y). \tag{2.14}$$

The equation (2.14) is the so-called diffusive approximation of the BFKL equation. It is equivalent to saddle point method, with the saddle point located at $\gamma_0$. We find the solution to Eq. (2.14) in the form

$$T_1(\rho, Y) = e^{-\gamma_0 \Delta} G(\Delta, Y), \tag{2.15}$$

where $\Delta \equiv \rho - v(\gamma_0)\bar{\alpha}Y$. From Eq. (2.14), the function $G$ solves the following diffusion equation:

$$\partial_Y G(\Delta, Y) = \frac{1}{2}\bar{\alpha}\chi''(\gamma_0)\partial_\Delta^2 G(\Delta, Y). \tag{2.16}$$

The presence of the diffusion equation is natural: the dipole evolution is essentially a dipole branching process in which the subsequent dipoles diffuse in size. For the initial condition, we can approximate the MV or the GBW conditions by a step function at $\rho = 0$. Then the function $G$ at $Y = 0$ can be approximated by a Dirac delta function,

$$G(\Delta, Y = 0) = \delta(\Delta). \tag{2.17}$$

The diffusion equation (2.16) with the initial condition (2.17) has the following solution:

$$\frac{1}{\sqrt{2\pi\chi''(\gamma_0)\bar{\alpha}Y}} \exp\left(-\frac{\Delta^2}{2\chi''(\gamma_0).\bar{\alpha}Y}\right). \tag{2.18}$$

For the nonlinear term, it is challenging to treat it in a direct way. Instead, we notice that its effect is to tame the exponential growth of the amplitude predicted by the linear BFKL evolution. The evolution is then driven by the linear kernel. We can treat the nonlinear effect on the solution by putting an absorptive boundary which moves as the rapidity increases. In particular, we require that

$$G(\Delta = -C, Y) = 0, \tag{2.19}$$

where $C$ is a constant. To satisfy this boundary condition, one should substract from the solution Eq. (2.18) a similar gaussian term centered at $\Delta = -2C$. This is the basic idea of the method of image. Consequently, the solution for $G$ in the presence of the absorptive boundary (2.19) is given by

$$G(\Delta, Y) = \frac{1}{\sqrt{2\pi\chi''(\gamma_0)\bar{\alpha}Y}} \left[ \exp\left(-\frac{\Delta^2}{2\chi''(\gamma_0)\bar{\alpha}Y}\right) - \exp\left(-\frac{(\Delta+2C)^2}{2\chi''(\gamma_0)\bar{\alpha}Y}\right) \right]. \tag{2.20}$$

For $\Delta$ and $C$ small compared to $\sqrt{\bar{\alpha}Y}$, the amplitude $T_1$ reads

$$
\begin{aligned}
T_1(\rho, Y) &\simeq c_T \frac{\Delta + C}{(\bar{\alpha}Y)^{3/2}} e^{-\gamma_0\Delta} \exp\left(-\frac{\Delta^2}{2\chi''(\gamma_0)\bar{\alpha}Y}\right) \\
&= c_T\left(\rho - v(\gamma_0)\bar{\alpha}Y + C\right) e^{-\gamma_0\left(\rho - v(\gamma_0)\bar{\alpha}Y + \frac{3}{2\gamma_0}\ln(\bar{\alpha}Y)\right)} \exp\left(-\frac{(\rho - v(\gamma_0)\bar{\alpha}Y)^2}{2\chi''(\gamma_0)\bar{\alpha}Y}\right).
\end{aligned}
\tag{2.21}
$$

We require the amplitude is a constant of order unity along the saturation line $\rho = \rho_s \equiv \ln(Q_s^2(Y)/Q_A^2)$. To this aim, we pull back the boundary by $3/(2\gamma_0)\ln(\bar{\alpha}Y)$. Eq. (2.21) then becomes

$$T(\rho, Y) \simeq c_T(\rho - \rho_s + \text{const})e^{-\gamma_0(\rho-\rho_s)} \exp\left[-\frac{(\rho - \rho_s)^2}{2\chi''(\gamma_0)\bar{\alpha}Y}\right], \tag{2.22}$$

where

$$\rho_s = v(\gamma_0)\bar{\alpha}Y - \frac{3}{2\gamma_0}\ln(\bar{\alpha}Y). \tag{2.23}$$

Returning to the physical variables, the forward elastic scattering amplitude $T_1(r, Y)$ solving the BK equation at large rapidity is given by

$$T_1(r, Y) \simeq c_T\left[\ln\frac{1}{r^2 Q_s^2(Y)} + \text{const}\right](r^2 Q_s^2(Y))^{\gamma_0} \exp\left[-\frac{\ln^2(r^2 Q_s^2(Y))}{2\chi''(\gamma_0)\bar{\alpha}Y}\right], \tag{2.24}$$

and the saturation momentum reads

$$Q_s^2(Y) = Q_A^2 \exp\left(\bar{\alpha}\chi'(\gamma_0)Y - \frac{3}{2\gamma_0}\ln(\bar{\alpha}Y)\right). \tag{2.25}$$

The solution (2.24) is valid for $1 < \ln(1/r^2 Q_s^2(Y)) < \sqrt{\chi''(\gamma_0)\bar{\alpha}Y}$, i.e. within the diffusion radius. When $\ln(1/r^2 Q_s^2(Y)) \ll \sqrt{\chi''(\gamma_0)\bar{\alpha}Y}$, we can neglect the gaussian term, and the dipole scattering amplitude $T_1$ effectively becomes a function of a single variable $\ln(1/r^2 Q_s^2(Y))$, which is referred to as the scaling variable. This properties is known as the "geometric scaling". It was manifested in the analysis of the HERA data on the electron-proton collision [53]. The geometric scaling of the solution to the BK equation outside the saturation region was first presented in [54], but the subleading term at large Y was incorrect therein. The derivation presented here is based on the method in Ref. [55] by replacing the nonlinearity by an absorptive barrier. The results (2.24) and (2.25) can also be obtained by exploiting the fact that the BK equation is in a universal class of the reaction-diffusion equations (see below), and at large rapidity its solution converges to traveling waves [56]. The geometric scaling is then corresponding to the traveling wave solution.

We have derived the solution to the BK equation for small onia outside the saturation region. The solution deep inside the saturation domain, $r \gg 1/Q_s(Y)$, can also be obtained by realizing that, for such large onium's sizes, $T_1 \approx 1$. Equivalently, the S-matrix is small, and one can neglect the nonlinear term of in the BK equation (2.2) for the S-matrix. We can eventually obtain the following expression [57]:

$$T_1^{sat}(r, Y) \simeq 1 - S_0 \exp\left[-\frac{\ln^2(r^2 Q_s^2(Y))}{2\chi'(\gamma_0)}\right]. \tag{2.26}$$

Several comments are in order. Firstly, the geometric scaling also manifests inside the saturation regime. At very large onium's size, the amplitude (2.26) approaches the black-disk limit value $T_1 = 1$. The unitarity is then preserved, as discussed previously. The amplitude (2.26) also grows with the rapidity. However, this growth gets slower at larger rapidity or at larger onium's size. At asymptotic rapidity, it eventually terminates at the black-disk limit.

### c. Numerical solutions

We now present numerical solutions to the BK equation, with the initial condition given by the MV model (2.4) at the rapidity $Y = 0$. The initial saturation scale is chosen to be $Q_A = 0.25$ GeV. Other parameters are set as in Appendix C.

The dipole scattering amplitude $T_1(r, Y)$ is plotted in Fig. 2.2. As the rapidity $Y$ increases, the perturbation around the unstable state $T_1 = 0$ moves toward the saturation value $T_1 = 1$. It then remains unchanged at this value. As a result, the solution is pushed forward to small onium sizes.

The asymptotic solution (2.24) can be visualized by plotting the function $T_1(r, Y)e^{\gamma_0 \ln(1/r^2 Q_s^2(Y))}$ as the function of the scaling variable $\ln(1/r^2 Q_s^2(Y))$ (see Fig. 2.3). We see that for a positive-value domain of the scaling variable close to 0, the rescaled amplitude is roughly linear, which is more evidently at higher rapidities. At small onium sizes far from the saturation line, the gaussian suppression becomes significant.

We conclude this paragraph by reminding that, the BK equation in Eq. (2.3) (or equivalently Eq. (2.2)) is written at leading order (LO). Its next-to-leading order (NLO) extension was already known [58–65]. In addition, it was shown [66–70] that the BK equations at both LO and NLO could describe HERA data on electron-proton collisions.

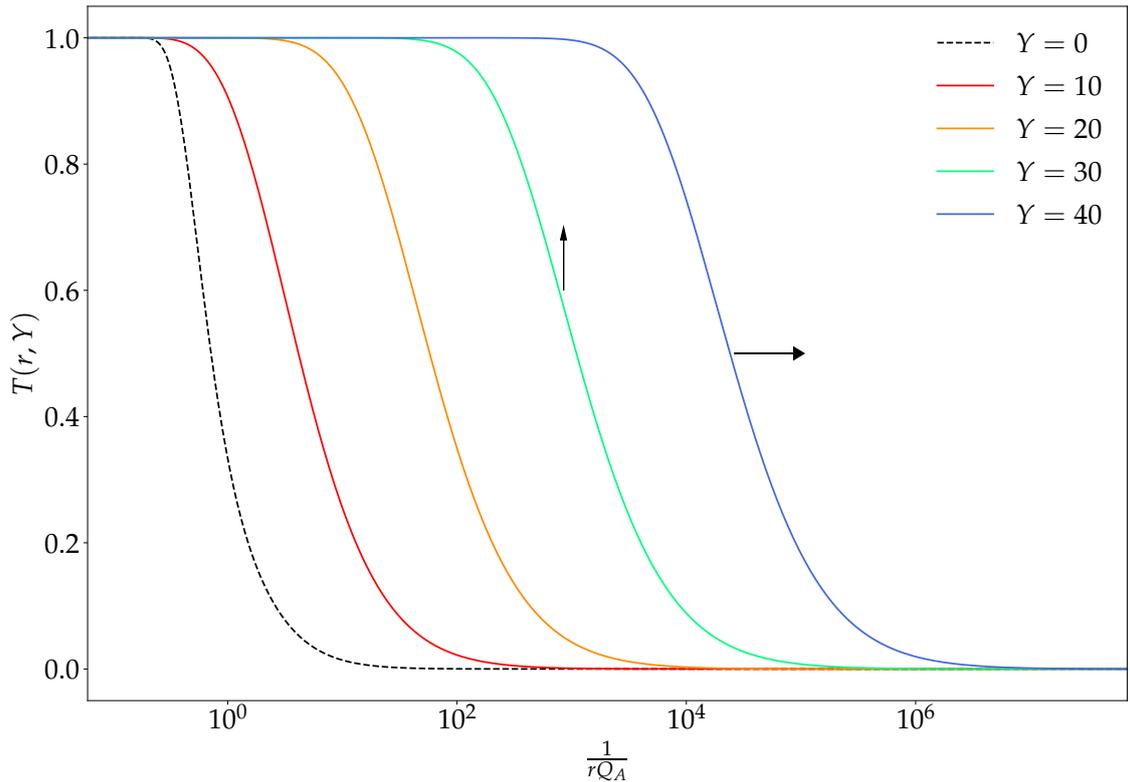

Figure 2.2: Dipole scattering amplitude $T_1(r, Y)$ at various rapidities. Due to the nonlinear evolution, the amplitude $T_1$ is driven toward small values of $r$ (horizontal arrow). Meanwhile, at a fixed size $r$, the amplitude approaches the unitary limit as $Y$ increases (vertical arrow).

### 2.1.3 Dual intepretation of the BK equation

From the above discussion, the BK equation (2.3) appears as the equation governing the nonlinear evolution of the forward elastic scattering amplitude $T_1(r, Y)$ of the nuclear scattering of an onium with size $r$ off a large nucleus at the total relative rapidity $Y$. In the restframe of the onium, this is equivalent to the deterministic evolution of the set of gluons in the nucleus in rapidity characterized by the nuclear saturation scale $Q_s(Y)$. The scattering then just measures the opacity of this gluonic system.

The BK equation also accepts another probabilistic interpretation in a frame where the onium is evolved to, for e.g, a rapidity $\widetilde{Y} \leq Y$. At the rapidity $\widetilde{Y}$, the Fock state of the onium in the large-$N_c$ limit is essentially a stochastic set of color dipoles of various sizes generated by the dipole branching with probability given by Eq. (1.87). We define $P(r, \widetilde{Y}; R)$ as the probability of having at least one dipole larger than $R$ in the onium Fock state. It is then straightforward to show that, the probability $P(r, \widetilde{Y}; R)$ solves the BK equation (2.3). The initial condition for $P$ is given by

$$P(r, 0; R) = \theta \left[ \ln \frac{r^2}{R^2} \right].$$
(2.27)

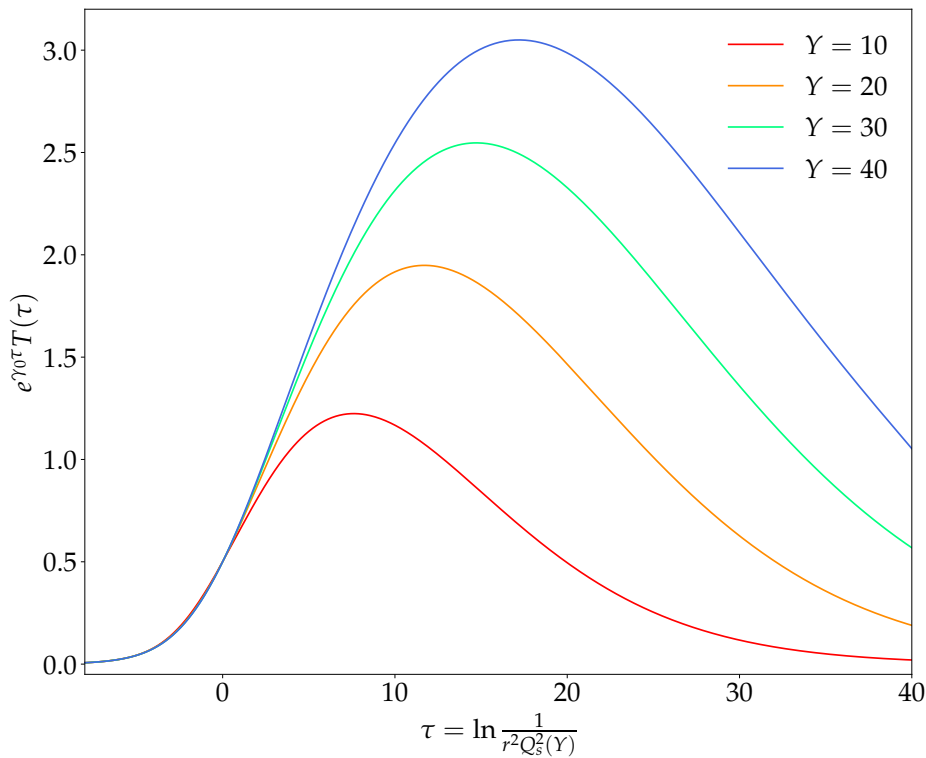

Figure 2.3: The dipole scattering amplitude $T_1(r, Y)$ rescaled by $e^{-\gamma_0 \tau}$ as a function of the scaling variable $\tau = \ln \frac{1}{r^2 Q_s^2(Y)}$ at different values of rapidity.

Therefore, the BK equation is also the evolution equation for a measure of the stochastic evolution of the onium Fock state. As we shall shortly see, although the initial conditions for $T$ and $P$ are different, the asymptotic solutions at large rapidity for them fall into the same universality class. One then can identify the QCD scattering amplitude $T(r, Y)$ with the probability $P(r, Y; R = 1/Q_A)$ of having at least one dipole larger than the inverse saturation momentum of the nucleus at rest in the onium Fock state at $Y$, starting from a bare dipole $r$ at $Y = 0$,

$$T_1(r, Y) \overset{\text{large } Y}{\simeq} P(r, Y; 1/Q_A). \tag{2.28}$$

In the spirit of the probabilistic interpretation, the BK equation is an equation controlling the rapidity evolution of an observable on a branching-diffusion process. It is therefore natural to relate the QCD dipole evolution to branching random walks described in the statistical physics.

## 2.2 Dipole evolution and branching random walk

### 2.2.1 Mapping the BK equation to the F-KPP equation

We are going to show that, under the aforementioned diffusive approximation, the BK equation can be mapped into a nonlinear partial differential equation which is well-known in the context of the statistical physics.

Let us return to the BK equation in the momentum space (2.10). In the diffusive approximation, it reads

$$\partial_y \widetilde{T}_1 = -\chi'(\gamma_0)\partial_L \widetilde{T}_1 + \frac{1}{2}\chi''(\gamma_0)(\partial_L + \gamma_0)^2 \widetilde{T}_1 - \widetilde{T}_1^2$$

$$= \frac{1}{2}\chi''(\gamma_0)\partial_L^2 \widetilde{T}_1 + [\gamma_0\chi''(\gamma_0) - \chi'(\gamma_0)]\,\partial_L \widetilde{T}_1 + \frac{1}{2}\gamma_0^2\chi''(\gamma_0)\widetilde{T}_1 - \widetilde{T}_1^2 \tag{2.29}$$

where $y \equiv \bar{\alpha}Y$, and the relation $\gamma_0\chi'(\gamma_0) = \chi(\gamma_0)$ is used. We perform the following change of variables $(y, L) \to (t, x)$ [71]:

$$y = \frac{2}{\gamma_0^2\chi''(\gamma_0)}t,$$

$$L = \frac{x}{\gamma_0} - \frac{2\,[\gamma_0\chi''(\gamma_0) - \chi'(\gamma_0)]}{\gamma_0^2\chi''(\gamma_0)}t. \tag{2.30}$$

The equation 2.29 then becomes

$$\frac{\gamma_0^2\chi''(\gamma_0)}{2}\partial_t \widetilde{T}_1 = \frac{\gamma_0^2\chi''(\gamma_0)}{2}\partial_x^2 \widetilde{T}_1 + \frac{\gamma_0^2\chi''(\gamma_0)}{2}\widetilde{T}_1 - \widetilde{T}_1^2. \tag{2.31}$$

Defining a new function $U = [2/(\gamma_0^2\chi''(\gamma_0))]\,\widetilde{T}_1$, we end up with following equation for $U$:

$$\partial_t U(t, x) = \partial_x^2 U(t, x) + U(t, x) - U^2(t, x). \tag{2.32}$$

This equation shows the manifestation of the Galilean non-relativistic symmetry on the transverse plane within the light-cone description (with the "time" variable $t \sim y$ and the "space" variable $x \sim |k_\perp|$). It belongs to a family given by [48]

$$\partial_t U(t, x) = D\partial_x^2 U(t, x) + F(U), \tag{2.33}$$

where $D$ is the diffusion coefficient, and $F(U)$ is a smooth function satisfying following conditions:

$$F(0) = F(1) = 0,$$

$$F'(0) = r > 0,$$

$$F(U) > 0 \ (0 < U < 1), \ F'(U) < r \ (0 < U \le 1). \tag{2.34}$$

The general equation (2.33) is known as the F-KPP equation [47, 48] describing reaction-diffusion processes. For a comprehensive review on the F-KPP equation, see Ref. [72].

We have presented a rigorous mapping between the QCD BK equation in the diffusive approximation and the F-KPP equation in the context of the statistical physics. However, as the dipole evolution is similar to branching-diffusion processes, it would be expected to have a more profound relation between them. It is indeed the case: the BK equation is in the same universality class of the F-KPP equation. We shall address this universality shortly, after discussing the emergence of the F-KPP equation in a stochastic process of interest: one-dimensional branching random walk

(BRW).

## 2.2.2 F-KPP equation and branching random walks

### a. General features of the solution to the F-KPP equation

We now consider the following particular form of the F-KPP equation:

$$\partial_t U(t,x) = D\partial_x^2 U(t,x) + U(t,x) - U^2(t,x). \tag{2.35}$$

The F-KPP equation of the form (2.35) has two fixed points at $U = 0$ and $U = 1$. If we start by a small perturbation in the vicinity of the latter, it will move back to that fixed point. Therefore, $U = 1$ is called the stable fixed point. On the other hand, if we start with a small perturbation around $U = 0$, it will grow towards $U = 1$. $U = 0$ is then the unstable fixed point of the equation. At a sufficiently large time, $U$ becomes close to unity and the nonlinear term is essential. In such case, it tames the evolution and makes the solutions saturate at the stable fixed point $U = 1$. This property of the solutions to the F-KPP equation is similar to those of the BK equation.

Let us now come into the so-called travelling wave solution of the F-KPP equation. First, we notice that the eigenvalue of the linear kernel $\chi_F(\partial_x) \equiv D\partial_x^2 + 1$ of the F-KPP equation corresponding to the eigenfunction $e^{-\gamma x}$ is given by

$$\chi_F(\gamma) = D\gamma^2 + 1. \tag{2.36}$$

We denote by $\gamma_0$ the solution of the equation $\chi_F(\gamma) = \gamma\chi_F'(\gamma)$, or $\gamma_0 = 1/\sqrt{D}$.

We choose an initial condition such that it falls monotonically and smoothly from 1 to 0 as $x$ goes from $-\infty$ to $\infty$. In addition, at large positive $x$, it behave as

$$U(0,x) \sim e^{-\beta x}, \quad (\beta > \gamma_0). \tag{2.37}$$

For a certain point of $x$ such that $U(0,x) \ll 1$, when the time elapses, the solution grows, and then stops at $U = 1$, as discussed above. However, for larger values of $x$, the growth continues, and a wave front establishes and moves toward large positive $x$ as a traveling wave. The traveling wave is characterised by its position $X(t)$, which can be defined by, for e.g., the requirement $U(t, x = X(t)) = 1/2$. At asymptotically large values of $t$, $U$ is effectively a function of a single variable $x - X(t)$,

$$U(t,x) \simeq \mathcal{U}(x - X(t)), \tag{2.38}$$

and the position of the front is given by:

$$X(t) = \chi_F'(\gamma_0)t - \frac{3}{2\gamma_0}\ln t + O(1). \tag{2.39}$$

This result constitutes a part of the Bramson's theorem [73] for the traveling wave solution to the F-KPP equation. We see that the characteristics of the traveling wave is determined by the linear kernel. Interestingly enough, the position of the front $X(t)$ is similar to the logarithm of the saturation scale (2.25) in the case of the BK equation, up to some appropriate substitutions.

### b. A simple BRW

Consider a one-dimensional lattice labelled by $x$, with lattice spacing $\delta x$. Let us start with a single particle located at $x = 0$ at time $t = 0$, and evolve the system forward in time. After a time step $\delta t$, a particle at a site $x$ can

  (i) jump left or right with the same probability $\mu$, or

  (ii) duplicate with probability $\lambda$, or

  (iii) remain unchanged with probability $1 - 2\mu - \lambda$,

conditioned that $2\mu + \lambda < 1$.

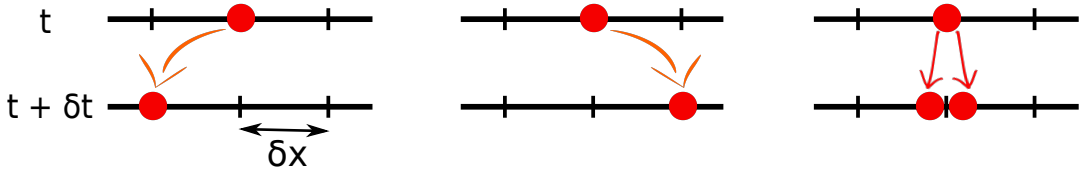

Figure 2.4: Elementary processes for a simple BRW: diffusion to adjacent sites and branching. The case in which particle remains unchanged is not shown in this illustration.

We define by $u(t, x)$ the probability to have at least one particle located to the right of position $x$ (including $x$) at time $t$ (or after the evolution in the time period $t$). The equation for $u(t + \delta t, x)$ can be obtained by tracking the initial particle at $(t, x) = (0, 0)$ as follows:

**Proba. to have particle(s) to the left of $x$ at time $t + \delta t$) =**

  (*Proba. that the initial particle at $t = 0$ jumps left after $\delta t$*)

    $\times$ (Proba. to have particle(s) to the right of $x + \delta x$ after the time period $t$ from $\delta t$ to $t + \delta t$)

 + (*Proba. that the initial particle at $t = 0$ jumps right after $\delta t$*)

    $\times$ (Proba. to have particle(s) to the right of $x - \delta x$ after the time period $t$ from $\delta t$ to $t + \delta t$)

 + (*Proba. that the initial particle at $t = 0$ duplicates after $\delta t$*)

    $\times$ (Proba. that at least one offspring generates particle(s) to the right of $x$

                                after the time period $t$ from $\delta t$ to $t + \delta t$)

 + (*Proba. that the initial particle at $t = 0$ does nothing*)

    $\times$ (Proba. to have particle(s) to the right of $x$ after the time period $t$ from $\delta t$ to $t + \delta t$).

The above rule can be written in terms of $u$ and the probabilities $\lambda$ and $\mu$ as follows:

$$
\begin{aligned}
u(t+\delta t, x) &= \mu\left[u(t, x+\delta x)+u(t, x-\delta x)\right]+\lambda\left[2u(t, x)\left(1-u(t, x)\right)+u^2(t, x)\right] \\
&\qquad\qquad\qquad\qquad\qquad\qquad\qquad\qquad + (1-2\mu-\lambda)u(t, x) \quad (2.40) \\
&= u(t, x)+\mu\left[u(t, x+\delta x)-2u(t, x)+u(t, x-\delta x)\right]+\lambda\left[u(t, x)-u^2(t, x)\right].
\end{aligned}
$$

We set $\lambda = \delta t$ and $\mu\delta x^2 = \delta t$, and go to the continuous limit $\delta t, \delta x \to 0$ (the limit of branching Brownian motion (BBM)). The equation (2.40) then becomes

$$
\partial_t u(t, x) = \partial_x^2 u(t, x) + u(t, x) - u^2(t, x), \tag{2.41}
$$

which is exactly the F-KPP equation. As, we start with a single dipole at the origin, the initial condition for $u(t, x)$ is simply a step function,

$$
u(0, x) = 1 - \Theta(x). \tag{2.42}
$$

In reference to the QCD dipole evolution, one can see that both the BK and the F-KPP equations control the evolution of the probabilities of the same type: $u(t, x)$ is analogous to $P(r, Y; R)$ defined previously.

Before moving to the next case, let us derive an equation for the mean particle number $\langle n(t, x)\rangle$ on a site $x$ at time $t$, starting from a single particle at $(t, x) = (0, 0)$. The contribution to the mean particle number $\langle n(t+\delta t, x)\rangle$ at $t+\delta t$ comes from following contributions. First, a portion $\mu\langle n(t, x)\rangle$ is extracted from the total mean number $\langle n(t, x)\rangle$ by the diffusion either to the left or to the right of the site $x$. A component $\lambda\langle n(t, x)\rangle$ is added to $\langle n(t, x)\rangle$ due to the duplication. Furthermore, $\mu\langle n(t, x+\delta x)\rangle$ and $\mu\langle n(t, x-\delta x)\rangle$ particles from the sites $x+\delta x$ and $x-\delta x$, respectively, diffuse to the site $x$ after $\delta t$. Combining all such contributions, the equation for the mean number $\langle n(t+\delta t, x)\rangle$ reads

$$
\langle n(t+\delta t, x)\rangle = \langle n(t, x)\rangle - 2\mu\langle n(t, x)\rangle + \lambda\langle n(t, x)\rangle + \mu\left[\langle n(t, x-\delta x)\rangle + \langle n(t, x+\delta x)\rangle\right]. \tag{2.43}
$$

Similar to the above discussion, let us set $\lambda = \delta t$ and $\mu\delta x^2 = \delta t$, and take the limit $\delta t, \delta x \to 0$. The time evolution of the mean particle number $\langle n(t, x)\rangle$ is given by

$$
\partial_t\langle n(t, x)\rangle = \partial_x^2\langle n(t, x)\rangle + \langle n(t, x)\rangle. \tag{2.44}
$$

Eq. (2.44) differs from the F-KPP equation by a nonlinear term. In the following example, we will see that, by including a nonlinear recombination mechanism, this term can be recovered.

### c. A BRW with recombination

We now introduce to the simple one-dimensional BRW defined aboved an additional recombination process: two arbitrary particles located at the site $x$ can recombine to become a single particle at the same site with probability $\lambda/\mathcal{N}$, where $\lambda$ is the duplication rate as before and $\mathcal{N}$ is a new parameter. Then the equation for the mean particle number at $t + \delta t$ reads

$$
\begin{aligned}
\langle n(t+\delta t,x)\rangle = \langle n(t,x)\rangle - 2\mu\langle n(t,x)\rangle + \lambda\langle n(t,x)\rangle & + \mu\left[\langle n(t,x-\delta x)\rangle + \langle n(t,x+\delta x)\rangle\right] \\
& - \frac{\lambda}{\mathcal{N}}\langle n(t,x)(n(t,x)-1)\rangle,
\end{aligned} \tag{2.45}
$$

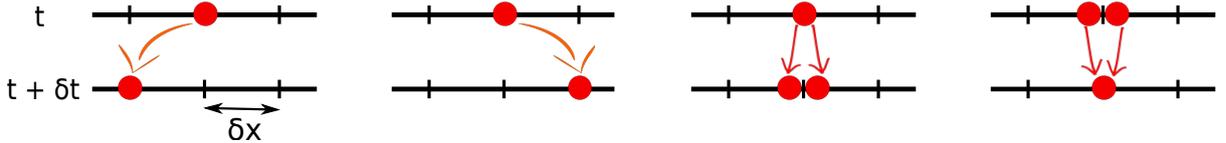

Figure 2.5: Elementary processes for a BRW with recombination: diffusion to adjacent sites, branching, and recombination. The case in which particle remains unchanged is not shown in this illustration.

where the first four terms are the same to Eq. (2.43), and the last term encodes the recombination mechanism. Due to the presence of the latter, this is not a closed equation. Now assuming that the number of particles is large, we can employ the mean-field approximation and get

$$
\begin{aligned}
\langle n(t+\delta t,x)\rangle = \langle n(t,x)\rangle - 2\mu\langle n(t,x)\rangle + \lambda\langle n(t,x)\rangle & + \mu\left[\langle n(t,x-\delta x)\rangle + \langle n(t,x+\delta x)\rangle\right] \\
& - \frac{\lambda}{\mathcal{N}}\langle n(t,x)\rangle^2.
\end{aligned} \tag{2.46}
$$

Taking the continuous limit as before, we arrive at the following equation:

$$
\partial_t\langle n(t,x)\rangle = \partial_x^2\langle n(t,x)\rangle + \langle n(t,x)\rangle - \frac{1}{\mathcal{N}}\langle n(t,x)\rangle^2. \tag{2.47}
$$

We rescale the mean particle number by $\mathcal{N}$ by introducing a novel function $w(t,x) = \langle n(t,x)\rangle/\mathcal{N}$. From Eq. (2.47), this rescaled mean particle number obeys the F-KPP equation,

$$
\partial_t w(t,x) = \partial_x^2 w(t,x) + w(t,x) - w^2(t,x). \tag{2.48}
$$

If we start with a single particle at the origin and take $\mathcal{N}$ to be a large number, the initial condition for $w$ is a small perturbation around the unstable fixed point. Following the previous discussions, when the time is not sufficiently large such that the mean particle number is small compared to $\mathcal{N}$, the linear part of Eq. (2.48) dominates the evolution, and hence the recombination is negligible. However, when the particle number becomes comparable to $\mathcal{N}$, the contribution of the nonlinear recombination effect is essential, causing $w$ to saturate at the stable fixed point $w = 1$. Therefore, $\mathcal{N}$ can be interpreted as the saturated value of the mean particle number, and the saturation is due to the recombination.

The above-mentioned situation is in analogy to the QCD in the regime of high parton density. For the latter, the QCD dynamics is dominated by saturation effects (including the gluon recombination) which kill certain partons resulted from the evolution (or, maybe more realistically, slow down their splittings) and, hence, tame the rapid rise of the linear evolution. It is described by nonlinear equations, for example the BK equation, which is analogous to the F-KPP mentioned above. In the context of two BRW models hitherto, the BK and the F-KPP equations are similar in their interpretations.

### 2.2.3 The BK equation in the Fisher-KPP universality class

Let us return to the QCD high-energy evolution. From previous discussions, there is very close correspondance between it and the branching-reaction models introduced above. In the first place, the dipole evolution is a branching-diffusion process: dipoles diffuse basically in the $\ln(1/r^2)$ space during the rapidity evolution. The dipole elastic scattering amplitude $T(r, Y)$, which is essentially equal to the number of dipoles of size $r$ in the onium Fock state multiplied by $\alpha_s^2$, is corresponding to the rescaled particle density in a BRW with recombination. This saturation dynamics is also observed in the model of BRW with recombination.

The BK equation and the F-KPP equation are also very similar. The F-KPP equation has two interpretations as shown in two above BRW models, which are corresponding to two interpretations of the BK equation discussed previously. More importantly, two equations have a lot in common in their structure. They can be divided into linear and nonlinear parts: the linear parts are differential kernels, and the nonlinear parts are quadratic. In addition, they have two fixed points at 0 and 1, with the former is unstable while the latter is stable. When the solutions are in the vicinity of the former, the linear kernels dominate the evolution. The nonlinear corrections become important when approaching the stable fixed point: the solutions hence saturate. In both QCD evolution and a BRW model with recombination, saturation is shown to happen when the system becomes dense.

From these considerations, the BK equation is in the same universality class of the F-KPP equation. One can make a correspondance between the two equations, as summarized in Table 2.1.

Table 2.1: The correspondance of the FKPP equation and the BK equation

|  | Reaction-diffusion FKPP equation | QCD dipole evolution BK equation |
|---|---|---|
| Linear kernel | $\chi_F(-\partial_x)$ | $\chi(-\partial_\rho)$, $\rho \equiv \ln\left[1/(r^2 Q_A^2)\right]$ |
| Evolution variable | time $t$ | rescaled rapidity $\bar\alpha Y$ |
| Diffusion space | spatial axis $x$ | logarithmic transverse size $\rho$ |
| Characteristics | Wave front position $X(t)$ | Logarithmic saturation scale $\ln\left[Q_s^2(Y)/Q_A^2\right]$ |

One can check this correspondance by looking into the asymptotic solution (2.22) of the BK equation. In appropriate limits, it is actually in agreement with the travelling wave solution (2.39) in the F-KPP case. The logarithmic saturation scale Eq. (2.23) is similar to the F-KPP front position (2.39). In addition, the relevant initial conditions (either MV or GBW) for the BK equation, at small $r$ ($r \ll 1/Q_A$), can be rewritten as

$$T_1(r, Y = 0) \sim 1 - \exp\left[-\frac{1}{4} e^{-\ln \frac{1}{r^2 Q_A^2}}\right] \overset{r \ll 1/Q_A}{\simeq} \frac{1}{4} e^{-\ln \frac{1}{r^2 Q_A^2}}, \tag{2.49}$$

which is steeper than $e^{-\gamma_0 \ln \frac{1}{r^2 Q_A^2}}$, with $\gamma_0$ defined in Eq. (2.13). Therefore, the asymptotic solution of the BK equation outside the saturation region obeys the Bramson's theorem for the travelling wave solution of the F-KPP equation.

## 2.3 Particles in the tip of BRW: a Monte-Carlo algorithm

### 2.3.1 Motivation

BRW and its continuous limit, BBM, [74] are important stochastic processes which appear in many contexts in different fields including physics, biology, chemistry, computer science and economical science [75–80]. Particularly, from the discussions in the previous section, the QCD dipole evolution is in analogy to one-dimensional BRW; and to some extent, one can replace the highly-evolved onium state in the high-energy onium-nucleus scattering by a state created by a BRW.

As we shall see in the next chapters, the nuclear scattering of a small onium is dominatedly triggered by largest dipoles in the onium Fock state. Such largest dipoles are corresonding to particles located in the vicinity of the rightmost (or leftmost) particle of BRW (BBM), which is referred to as the "tip" region. In many applications of BBM and BRW, it is important to understand the distribution of particles residing on the tip [81–84]. From the first model of BRW (BBM) presented in the previous section, the F-KPP equation can be used to characterize the tip; therefore one method is to explore the solutions to the F-KPP equation. However, such method is not omnipotent: for example, the genealogical structure of the rightmost particles [85] cannot be obtained in this way.

One available method to study the tip of a BRW is to generate it and, then, measure observables of interest from resulting realizations. Nevertheless, direct Monte Carlo simulations are impractical for large time, due to the exponential increase of the number of particles with time. In addition, it is almost impossible to use it to study the events in which the rightmost (or leftmost) particle is far from its expected position, since they are particularly rare.

In the following, we shall report our work [25] on establishing a Monte Carlo algorithm to follow only a tree of selected particles in the tip to cure the above issue. In particular, that algorithm is designed to generate all particles which are close to the rightmost particle, when the latter is conditioned to arrive to the right or exactly at some given position. This algorithm can be used to study the tip in both typical (with the rightmost particle located in the vicinity of its expected position) or rare events evolved to a very large time.

### 2.3.2 Generating particles in the tip of a BRW

**Model definition**

We consider a one-dimensional BRW on the $x$ axis with lattice spacing $\delta x$, and in discrete time with step $\delta t$. The system starts with one single particle located at $x = 0$ at time $t = 0$. After each time step, a particle at $x$ can

$$
\begin{cases}
\text{⌊⚫⌝} & \text{jump from } x \text{ to } x + \delta x \quad \text{probability } p_r, \\
\text{⌐⚫⌋} & \text{jump from } x \text{ to } x - \delta x \quad \text{probability } p_l, \\
\text{⚫⚫} & \text{branch without moving} \quad \text{probability } r,
\end{cases}
\tag{2.50}
$$

with $p_r + p_l + r = 1$. Let us denote by $u(t, x)$ the probability to have at least one particle located to the right of $x$ at time $t$. Following a similar discussion as in Section 2.2.2, it satisfies the following equation:

$$
\begin{aligned}
u(t + \delta t, x) &= p_l u(t, x + \delta x) + p_r u(t, x - \delta x) + r u(t, x)\left[2 - u(t, x)\right] \\
&\equiv \mathcal{L} u(x, t) - r u^2(x, t),
\end{aligned}
\tag{2.51}
$$

with the initial condition $u(0, x) = 1 - \Theta(x)$. $\mathcal{L}$ is a linear operator acting on $u$ defined as follows:

$$
\mathcal{L} u(x, t) \equiv p_l u(t, x + \delta x) + p_r u(t, x - \delta x) + 2 r u(t, x).
\tag{2.52}
$$

Eq. (2.51) is in the universality class of the F-KPP equation. The mean position of the rightmost particle is at $m_t$, which reads

$$
m_t = \chi_b'(\gamma_0) t - \frac{3}{2\gamma_0} \ln t + \mathcal{O}(1),
\tag{2.53}
$$

where $\chi_b(\gamma)$ is the eigenvalue of the linear operator $\mathcal{L}$ corresponding to the eigenfunction $e^{-\gamma x}$, in analogy to Eq. (2.36), and $\gamma_0$ solves $\chi_b(\gamma) = \gamma \chi_b'(\gamma)$. The expression of $\chi_b(\gamma)$ is

$$
\chi_b(\gamma) = \frac{1}{\delta t} \ln\left[p_r e^{\gamma \delta x} + p_l e^{-\gamma \delta x} + 2r\right].
\tag{2.54}
$$

We take a large time $T$, a position $X$, and a distance $\Delta$. By the time evolution, a particle at an intermediate time $t < T$ will bring about descendant(s) at $T$, and among the latter, the rightmost one is either to the right of $X$, to the left of $X$ or exactly at $X$. For the sake of convenience, let us define following particular sets of particles.

**Definition 1.** *A particle is* **red** *if its rightmost offspring at $T$ resides in $[X, +\infty)$.*

**Definition 2.** *A particle is* **orange** *if its rightmost offspring at $T$ resides in $[X - \Delta, X)$.*

**Definition 3.** *A particle is* **blue** *if its rightmost offspring at $T$ resides in $(-\infty, X - \Delta)$.*

(See Fig. 2.6). In addition, we introduce following notations:

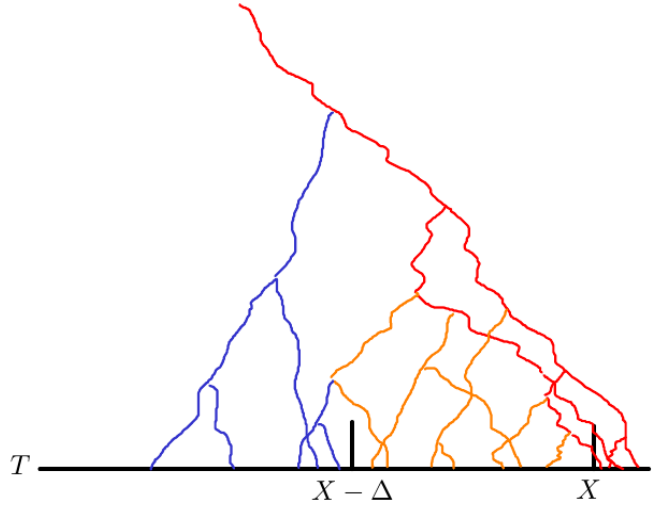

Figure 2.6: An illustration of the **red**, **orange** and **blue** particles.

- $\mathbb{P}(A)$: the probability for $A$ to happen,

- $\mathbb{P}(A; B)$: the joint probability for $A$ and $B$ to happen together,

- $\mathbb{P}(A|B) = \mathbb{P}(A; B)/\mathbb{P}(B)$: the conditional probability for $A$ to happen provided that $B$ is realized.

**Ultimate goal**

We now aim to *generate all particles in the interval $[X - \Delta, +\infty)$, provided that the rightmost particle is located in $[X, +\infty)$*. In other words, it is to track all the **red** and **orange** particles, starting from an initial **red** particle. In a variant, the rightmost particle will be fixed exactly at $X$. The algorithm is constructed as follows.

**Generator**

**a. First goal: Tracing particles arriving in $[X, +\infty)$**

The first target of the algorithm is to keep track of the paths of all **red** particles in the BRW, provided that the initial particle is **red**.

Let us define $\mathcal{U}(t, x)$ to be the probability that a particle at $(t, x)$ is **red**. By definition, $\mathcal{U}(t, x)$ is related to $u(t, x)$ by

$$\mathcal{U}(t, x) := \mathbb{P}(\bullet \text{ is } \textbf{red} \ ) = u(T - t, X - x). \tag{2.55}$$

The probability that a particle at $(t, x)$ is **red** and jumps right is

$$\mathbb{P}(\sqcup\bullet\widehat{\bullet}\sqcup \ ; \bullet \text{ is } \textbf{red}) = p_r \mathcal{U}(t + \delta t, x + \delta x). \tag{2.56}$$

Then the probablity that a particle at $(t, x)$ jumps right given that it is **red** reads

$$\mathbb{P}(\llcorner\bullet\widehat{\bullet}\lrcorner \mid \bullet \text{ is } \textbf{red}) = p_r \frac{\mathcal{U}(t + \delta t, x + \delta x)}{\mathcal{U}(t, x)}. \tag{2.57}$$

Similarly, the probablity that a particle at $(t, x)$ jumps left given that it is **red** is given by

$$\mathbb{P}(\widehat{\bullet}\bullet\lrcorner \mid \bullet \text{ is } \textbf{red}) = p_l \frac{\mathcal{U}(t + \delta t, x - \delta x)}{\mathcal{U}(t, x)}. \tag{2.58}$$

In the case of a particle at $(t, x)$ branching into two offsprings at $(t + \delta t, x)$, there are two relevant situations. If those two offsprings are both **red**, then

$$\mathbb{P}\left(\bullet\!\!\prec_{\bullet\textbf{red}}^{\bullet\textbf{red}} \mid \bullet \text{ is } \textbf{red}\right) = r \frac{\mathcal{U}^2(t + \delta t, x)}{\mathcal{U}(t, x)}. \tag{2.59}$$

Otherwise, only one of them is **red**. In the latter case, the conditional probability for a **red** particle to branch into a **red** and a non-**red** (non-**red** = **orange** + **blue**) is

$$\mathbb{P}\left(\bullet\!\!\prec_{\bullet\text{non-}\textbf{red}}^{\bullet\textbf{red}} \mid \bullet \text{ is } \textbf{red}\right) = r \frac{2\mathcal{U}(t + \delta t, x) - \mathcal{U}^2(t + \delta t, x)}{\mathcal{U}(t, x)}. \tag{2.60}$$

With Eq. (2.51), one can check that the sum of the conditional probabilities in Eqs. (2.57) to (2.60) is unity. Those probabilities allow to generate realizations of the trajectories of all the **red** particles given that the initial particle are **red**. In the last case given by the probability (2.60), we can ignore the non-**red** child in the next evolution steps.

### b. Second goal: Tracing particles arriving in $[X - \Delta, X)$

We can now extend the above algorithm to furthermore track the paths of all the particles arriving in $[X - \Delta, X)$ (the **orange** particles), in addition to the red ones.

Introduce $\mathcal{V}_\Delta(t, x)$ as the probability that a particle at $(t, x)$ is **orange**. $\mathcal{V}_\Delta$ is related to $\mathcal{U}$ by

$$\mathcal{V}_\Delta(t, x) \coloneqq \mathbb{P}(\bullet \text{ is } \textbf{orange}) = \mathcal{U}(t, x + \Delta) - \mathcal{U}(t, x) \tag{2.61}$$

An **orange** particle can be created by the branching of a **red** particle. The conditional probability for a **red** particle to branch into a **red** and a **orange** is

$$\mathbb{P}\left(\bullet\!\!\prec_{\bullet\textbf{orange}}^{\bullet\textbf{red}} \mid \bullet \text{ is } \textbf{red}\right) = r \frac{2\mathcal{U}(t + \delta t, x)\mathcal{V}_\Delta(t + \delta t, x)}{\mathcal{U}(t, x)}. \tag{2.62}$$

Similarly,

$$\mathbb{P}\left(\bullet\!\!\prec_{\bullet\textbf{blue}}^{\bullet\textbf{red}} \mid \bullet \text{ is } \textbf{red}\right) = r \frac{2\mathcal{U}(t + \delta t, x)\left[1 - \mathcal{U}(t + \delta t, x + \Delta)\right]}{\mathcal{U}(t, x)}. \tag{2.63}$$

Once the **orange** particles are created, we should follow their trajectories. Given that a particle is **orange**, it can jump right and left with following corresponding conditional probabilities:

$$\mathbb{P}(\llcorner \bullet \widehat{\curvearrowright} \lrcorner \mid \bullet \text{ is } \textbf{orange}) = p_r \frac{\mathcal{V}_\Delta(t + \delta t, x + \delta x)}{\mathcal{V}_\Delta(t, x)}, \tag{2.64}$$

$$\mathbb{P}(\llcorner \widehat{\curvearrowleft} \bullet \lrcorner \mid \bullet \text{ is } \textbf{orange}) = p_l \frac{\mathcal{V}_\Delta(t + \delta t, x - \delta x)}{\mathcal{V}_\Delta(t, x)}. \tag{2.65}$$

Additionally, it can branch into either two **orange** or one **orange** and one **blue**. Their corresponding conditional probabilities read

$$\mathbb{P}\left(\bullet \mathrel{\substack{\bullet \textbf{orange} \\ \bullet \textbf{orange}}} \mid \bullet \text{ is } \textbf{orange}\right) = r \frac{\mathcal{V}_\Delta^2(t + \delta t, x)}{\mathcal{V}_\Delta(t, x)}, \tag{2.66}$$

$$\mathbb{P}\left(\bullet \mathrel{\substack{\bullet \textbf{orange} \\ \bullet \textbf{blue}}} \mid \bullet \text{ is } \textbf{orange}\right) = r \frac{2\mathcal{V}_\Delta(t + \delta t, x)\left[1 - \mathcal{U}(t + \delta t, x + \Delta)\right]}{\mathcal{V}_\Delta(t, x)}. \tag{2.67}$$

In the cases given by probabilities (2.67) and (2.63), we ignore the further evolution of **blue** particles. Combining the two goals, it is possible to generate realizations in which all the trajectories of the particles arriving in $[X - \Delta, +\infty)$ at time $T$ are tracked, provided that there is at least one particle to the right of $X$.

To implement the algorithm, we present the state of the system at a given time $t$ by two arrays indexed by bins in the $x$ axis containing the numbers of **red** and **orange** particles. To advance the system from $t$ to $t + \delta t$, one observes that on each bin in each array, the numbers of particles undergoing the different possible events obey multinomial laws with parameters that we can compute from $u(t, x)$. This requires to integrate numerically Eq. (2.51) before the event generation begins.

We set the probabilities of the elementary processes to $p_r = p_l = \frac{1}{2}(1 - \delta t)$, $r = \delta t$, and lattice spacing to $\delta t = 0.01$ and $\delta x = 0.1$. Relevant front parameters for this model are $\gamma_0 = 1.43195 \cdots$ and $\chi_b(\gamma_0) = 1.99666 \cdots$. A realization of this conditioned BRW is shown in Fig. 2.7.

In order to validate the algorithm and its implementation, we measured the expected number of particles at distance $a$ from the rightmost particle. Based on the formalism developed in Ref. [82], this observable can be computed using the formula

$$\langle n(a) \rangle = \partial_a^2 \int dx R_a(t, x), \tag{2.68}$$

where the function $R_a(t, x)$ obeys the following equation:

$$R(t + \delta t, x) = p_l R(t, x + \delta x) + p_r R(t, x - \delta x) + 2r\left[1 - u(t, x)\right] R(t, x), \tag{2.69}$$

with $u(t, x)$ defined in Eq. (2.51). The initial condition for $R(t, x)$ reads

$$R(t = 0, x) = \begin{cases} 1 & \text{for } 0 \leq x < a, \\ 0 & \text{otherwise.} \end{cases} \tag{2.70}$$

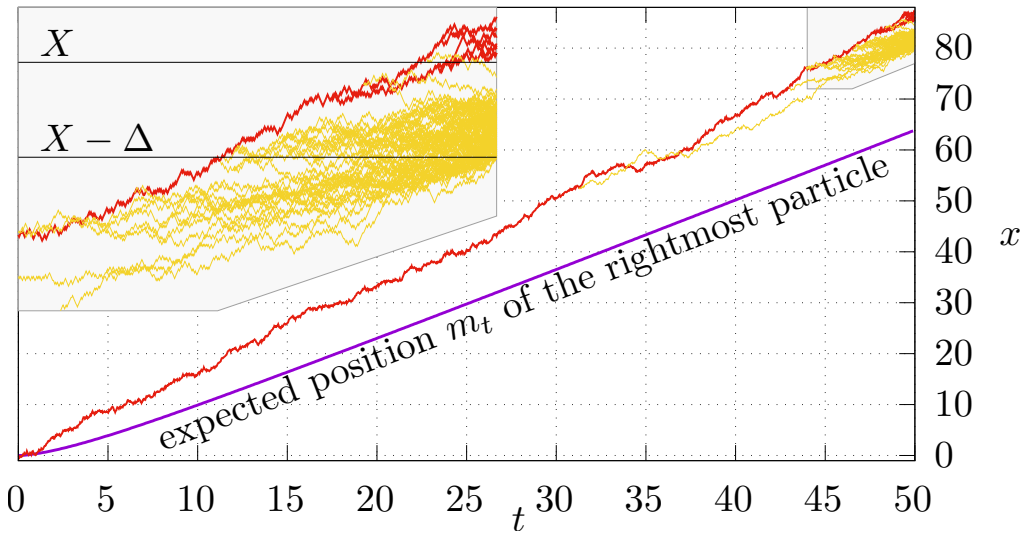

Figure 2.7: A realization of the desired conditioned BRW up to $T = 50$ with $X = 85.1 \simeq m_t + 3\sqrt{T}$ and $\Delta = 5$, where $m_t$ is the expected position of the rightmost particle at time $t$. The inset is a zoom of the final times showing the **red** and **orange** particles. Figure is adapted from Ref. [25]

We solved Eqs. (2.51), (2.68) and (2.69) numerically and compared to the measured values from the Monte Carlo simulation. The results shown in Fig. 2.8 exhibit a perfect agreement between both methods within statistical uncertainties.

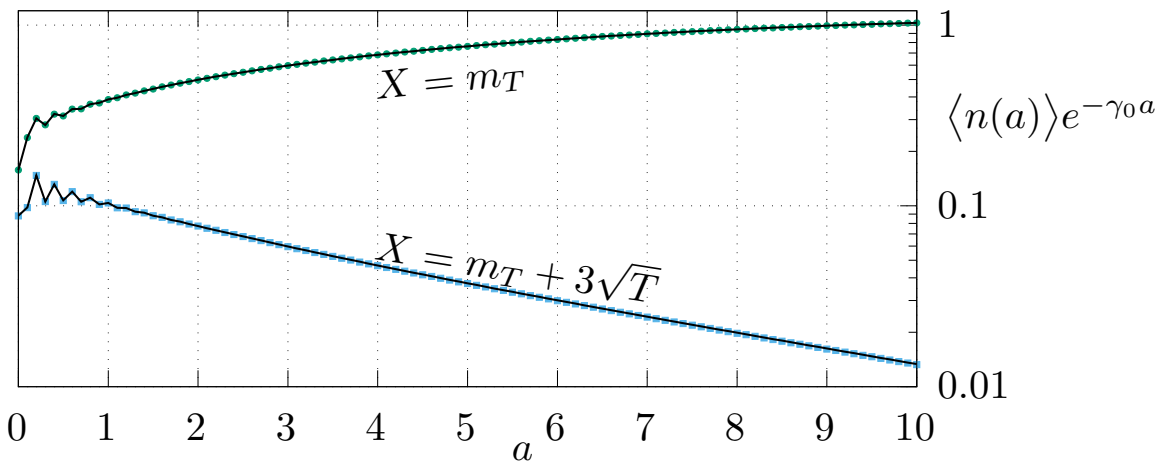

Figure 2.8: The expected number of particles $\langle n(a) \rangle$ at distance $a$ from the tip particle, rescaled by $e^{-\gamma_0 a}$, as a function of $a$, for $T = 400$ and two values of $X$. The dotted are obtained from $3 \times 10^6$ realizations of the BRW generated by the established algorithm. The lines are from numerical solutions obtained from Eqs. (2.51), (2.68) and (2.69). For $a = 0$, we removed 1 from the count of particles. Figure is adapted from Ref. [25].

The above-presented algorithm enables the study of tip observables of a BRW for which no other method is available to date. For example, we measured the distribution of the number of particles at distance $a$ to the left of the rightmost particle, in typical and rare realizations, for which a heuristic calculation have been published recently [86]. The result is shown in Fig. 2.9.

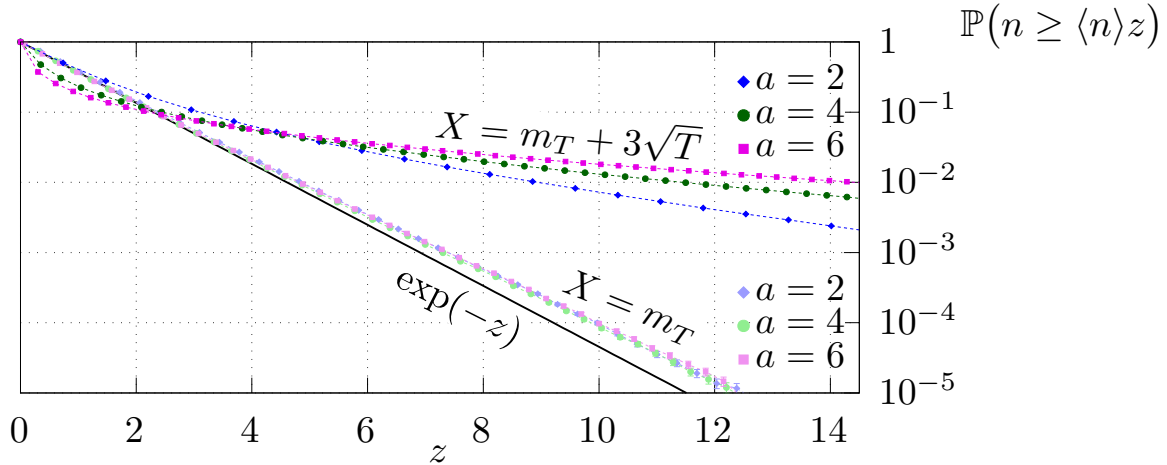

Figure 2.9: Rescaled tail distribution $\mathbb{P}(n \geq \langle n \rangle z)$ as a function of the rescaled factor $z$ for the number of particles at distance $a$ to the left of the rightmost particle with $T = 400$, for two values of $X$ and three values of $a$. The distribution of $n$ is roughly exponential in the typical case ($X = m_t$), while it exhibits a much fatter tail for the rare configuration ($X = m_t + 3\sqrt{T}$). Figure is adapted from Ref. [25].

#### c. Variant: Fixing the rightmost particle

Instead of letting the rightmost particle to be located in $[X, +\infty)$, we can vary the algorithm in such a way that the rightmost offspring of the **red** particles is fixed *exactly* at position $X$. The probability for a particle currently at $(t, x)$ to reach $(T, X)$ is

$$\widetilde{\mathcal{U}}(t, x) := u(T - t, X - x) - u(T - t, X - x + \delta x). \tag{2.71}$$

Then following the same argument as above, the evolution probablities for these "new **red**" particles are given by Eqs. (2.57) to (2.59) and (2.62) with $\mathcal{U}$ replaced by $\widetilde{\mathcal{U}}$, and by Eq. (2.63) with the two $\mathcal{U}$ outside the square brackets replaced by $\widetilde{\mathcal{U}}$. With these new probabilities, one can follow "new **red**" and **orange** particles (the definition of **orange** particle is kept the same).

### 2.3.3  Continuous limit: Conditioning the BBM

BBM is the continuous version of BRW. Passing to the continuous limit, each of the particles in the above BRW model follows an independent Brownian motion, and branches with rate 1 (during each infinitesimal time $dt$, each particle splits with probablility $dt$). The algorithm developed above can be adapted to the BBM. We are going to discuss about it, with the goal being to offer a starting point to analytical studies of the tip of the BBM, not to generate realizations.

Introduce as before the probability $u(t, x)$ that the rightmost particle at $t$ is on the right of $x$. It satisfies the F-KPP equation

$$\partial_t u = \frac{1}{2} \partial_x^2 u + u - u^2, \tag{2.72}$$

with the initial condition $u(0, x) = 1 - \Theta(x)$. Eq. (2.72). The only difference compared to the previous version of the F-KPP equation (as in Eq. (2.41)) is in the diffusion coefficient, which would not change the discussion. This version of the F-KPP equation is in fact the continuous limit of the discrete equation (2.51) when the parameters $p_r$, $p_l$, $r$ and $(\delta t, \delta x)$ are set properly, for instant, as in arriving at results shown in Figs. 2.8 and 2.9.

Given $X$ and $T$, we define **red** particles as above. The probability of a particle at $(t, x)$ to be **red** is still $\mathcal{U}(t, x) = u(T - t, X - x)$. The conditional probability for a **red** particle to branch into two **red** between $t$ and $t + \delta t$ can be obtained from Eq. (2.59) when $\delta t$ is infinitesimally small,

$$\mathbb{P}\left( \bullet \!\begin{smallmatrix} \bullet \text{\textbf{red}} \\ \bullet \text{\textbf{red}} \end{smallmatrix} \mid \bullet \text{ is \textbf{red}} \right) = \delta t \, \mathcal{U}(t, x) + \mathcal{O}(\delta t^2), \tag{2.73}$$

with $r = \delta t$. Similarly, the conditional probability for a **red** particle to branch into a **red** and a non-**red** is

$$\mathbb{P}\left( \bullet \!\begin{smallmatrix} \bullet \text{\textbf{red}} \\ \bullet \text{non-\textbf{red}} \end{smallmatrix} \mid \bullet \text{ is \textbf{red}} \right) = \delta t \, 2 \left[ 1 - \mathcal{U}(t, x) \right] + \mathcal{O}(\delta t^2). \tag{2.74}$$

The conditional probability that a particle at $(t, x)$ moves during $\delta t$ by $\Delta x \in [\epsilon, \epsilon + d\epsilon]$ (for $a \in [b, b + db]$ we will hereafter write $a \in db$ for short) provided that it is **red** is

$$\mathbb{P}\left( \Delta x \in d\epsilon \mid \bullet \text{ is \textbf{red}} \right) = \frac{e^{-\frac{\epsilon^2}{2\delta t}}}{\sqrt{2\pi \delta t}} d\epsilon \times \frac{\mathcal{U}(t + \delta t, x + \epsilon)}{\mathcal{U}(t, x)}. \tag{2.75}$$

Multiplying the probability (2.75) by $\epsilon$ and integrating over $\epsilon$, we obtain after expanding for small $\delta t$ the average drift distance of a **red** particle:

$$\langle \Delta x \mid \bullet \text{ is \textbf{red}} \rangle = \delta t \, \partial_x \ln \mathcal{U}(t, x) + \mathcal{O}(\delta t^2). \tag{2.76}$$

With Eqs. (2.73) and (2.76), we thus obtain the following result:

*"The trajectories of the particles in a BBM ending on the right of $X$ ($X$ included) at time $T$, conditioned on the event that there is at least one of them, is a BBM with a space- and time-dependent drift $\partial_x \ln \mathcal{U}(t, x)$ and a space- and time-dependent branching rate $\mathcal{U}(t, x)$."*

If **orange** particles are needed, one checks that a **red** particle branches out an **orange** particle at rate $2\mathcal{V}_\Delta(t, x)$, that an **orange** particle branches into two **orange** at rate $\mathcal{V}_\Delta(t, x)$, and that **orange** particles have a drift $\partial_x \ln \mathcal{V}_\Delta(t, x)$.

Similar to the BRW, one can modify the algorithm to fix the rightmost particle at $X$ in the BBM. In that case, the probability for a particle to end in the range $[X, X + dX]$ is $\partial_x \mathcal{U}(t, x) dX$. The "new **red**" particles follows a Brownian motion with drift $\partial_x \ln[\partial_x \mathcal{U}(t, x)]$.

There is another way to construct the tree of **red** particles in the BBM. Consider a particle at $(t, x)$ and call $(\tau_1, \xi_1)$ the time and position of the next branching event ($\tau_1 > t$). For $x_1 \in \mathbb{R}$ and $t_1 > t$, one has

$$\mathbb{P}(\tau \in dt_1 \; ; \; \xi_1 \in dx_1) = e^{-(t_1 - t)} dt_1 \times \frac{e^{-\frac{(x_1 - x)^2}{2(t_1 - t)}}}{2\pi(t_1 - t)} dx_1. \tag{2.77}$$

For $t_1 < T$, the conditional probability of a **red** particle to branch at $(\tau_1, \xi_1)$ is

$$\mathbb{P}(\tau \in dt_1 \; ; \; \xi_1 \in dx_1 \mid \bullet \text{ is } \mathbf{red}) = e^{-(t_1-t)}dt_1 \times \frac{e^{-\frac{(x_1-x)^2}{2(t_1-t)}}}{2\pi(t_1-t)}dx_1 \times \frac{\mathcal{U}(t_1, x_1)\left[2 - \mathcal{U}(t_1, x_1)\right]}{\mathcal{U}(t, x)}. \quad (2.78)$$

This probability is not normalized: the integration over $x_1 \in \mathbb{R}$ and $t_1 \in [t, T]$ is smaller than 1. Its unitary complement comes from the event that the next branching occurs after the time $T$. In such case, the trajectory up to $T$ of the **red** particle is simply a Brownian motion conditioned to arrive to the right of $X$.

With the probability (2.78) one can draw the coordinates $(\tau_1, \xi_1)$ of the next branching event. The trajectory between $t$ and $t_1$ is then a Brownian motion conditioned to be at position $(\tau_1, \xi_1)$. With no conditioning, the probability to branch into two **red** is $U^2(\tau_1, \xi_1)$, and the probability to branch into one **red** and one non-**red** is $2U(\tau_1, \xi_1)\left[1 - U(\tau_1, \xi_1)\right]$. Then, given that the branching particle is **red**, the probability that it branches into two **red** is $U(\tau_1, \xi_1)/\left[2 - U(\tau_1, \xi_1)\right]$. Its complement is the probability that only one offspring is **red**. The algorithm is then repeated at each branching point.

## 2.4 Summary

In this chapter, we have reviewed the BK nonlinear evolution equation at leading order, which controls the rapidity evolution of the forward elastic scattering amplitude (or, equivalently, the elastic S-matrix element) of the onium-nucleus scattering. It was shown to be in the same universality class of the F-KPP equation, which describes branching-diffusion processes in statistical physics. This universality enables to treat the dipole evolution as a peculiar BRW, and hence, the highly-evolved onium Fock state as a state generated by a BRW. Consequently, many asymptotic features between two sectors can be linked together, creating a cross-fertilization with many potential applications.

With the need to characterize the frontier region of the BRW, which is important for both QCD and statistical physics, we developed a Monte Carlo algorithm to follow only a tree of rightmost particles at a time $T$, with the rightmost particle is conditioned to located to the right of $X$, or to be exactly at $X$ in a variant. It enables the study of observables of the tip as well as the structure of the evolution. When $X$ is larger than the expected position $m_T$ of the rightmost particle at $T$, our algorithm allows to study rare realizations, while with $X$ close to $m_T$, it allows to generate more typical realizations.

One potential application of the algorithm is to investigate the distribution of the genealogical tree of the particles in the tip [85, 87]. The algorithm may also be used for a numerical analysis of the order statistics [81, 82, 88–90] of particles near one tip, or the statistics of the spatial span [91], i.e. the distance between the leftmost and the rightmost particles, of a BRW.

Furthermore, the algorithm is believed, by the virtue of the aforementioned universality, to be able to generate the tip of realizations the Fock state of an onium subject to high-energy evolution. Therefore, one can use this algorithm to check certain results in QCD. For another possible develop-ment, we can also extend the algorithm for a real QCD dipole evolution: it just amounts to replace the rule (2.50) in the considered BRW by the dipole branching with the probability (1.89), and the discrete equation (2.51) by the BK equation. As a matter of fact, there have been Monte Carlo studies on the high-energy QCD dipole evolution (see, for e.g., OEDIPUS [92–95]), but without conditioning. An extension to the case of conditioned events may, therefore, be worthy.

While our main focus is BRW, we have also extended the algorithm to provide a theoretical description of the conditioned BBM. This may be useful to construct a mathematical description of the tip as in Ref. [84], in order to compute tip observables in a systematic way.

# Nuclear scattering of small onia

## Contents

In this chapter, we shall report on our recent study [26] on the configuration of the Fock state of a small onium in the scattering off a large nucleus. The analysis is based on a phenomenological model of dipole distribution, which is constructed from the understanding on the asymptotic solutions of the BK equation, and on the statistical properties of the QCD dipole evolution, which were presented in the previous chapter. To initiate, let us figure out the motivation of the problem.

# 3.1 Why scattering configuration matters?

As being discussed in the previous chapters, the QCD evolution towards high energies of an onium is a gluon branching process which, in the limit of large number of colors, boils down to the iteration of independent one-to-two dipole splittings. This process results in realizations of a specific branching random walk. In the nuclear scattering of a highly-evolved onium, the scattering amplitude is a collective quantity, which depends on how each single dipole in the Fock state interacts with the nucleus. This consequently leads to the question of dipole distribution. In addition, for a particular scattering set up, there should be a class of realizations of dipole distribution which produce the dominant contribution to the scattering. Therefore, it is of interest to look into the detail of the Fock state of the onium in the interaction with the nucleus, and the relevant evolution configurations. The latter naturally prompts the investigation of the correlation among participating dipoles, or their genealogical tree, which is analogous to a similar problem for a specific set of particles generated by a one-dimensional BRW (or BBM) [85, 87].

In this chapter, we shall investigate the scattering amplitudes for onium-nucleus collisions at high energies in the geometric scaling region in which, as discussed in the previous chapter, the forward elastic scattering amplitude is small and its asymptotic expression is known to effectively depend only on the scaling variable of the rapidity and the size of the onium. In particular, we shall study the amplitude in the framework of the color dipole model, in terms of fluctuations of the partonic content of the onium, in different reference frames related to each other through longitudinal boosts. The use of boost invariance is natural. First, it is the fundamental symmetry of scattering amplitudes. An interesting issue arising here is to understand theoretically how this symmetry is manifested at the microscopic level in the regime in which the onium can be considered as a set of independent dipoles generated by a branching process. In addition, it is already well-known that using boost invariance helps formulate the calculation of observables. In our case, we will take advantage of boost invariance to select a specific class of frames which enables us to derive a particular distribution, which is a priori very challeging to calculate.

The main outcome of the investigation is a partonic picture of the scattering in different frames. Such picture allows to extract the asymptotic expression of the probability distribution of the rapidity at which the latest ancestor of the dipoles in the Fock state of the onium effectively interacting with the nucleus has branched. The latter characterizes the rapidity correlation of the interacting dipoles. And, as mentioned previously, this is in analogy to the genealogical problem to figure out the splitting time of the last common ancestor of a specific set of particles, which is of particular interest in the study of BRWs.

We will start by introducing formulations for scattering amplitudes. An approximation scheme for the dipole distribution for the scattering is then constructed. Finally we shall use this scheme to investigate the scattering configuration of the onium in different reference frames, and to compute observables of interests.

## 3.2 Formulations for scattering amplitudes

We are interested in the following quantities:

(i) the forward elastic scattering amplitude $T_1$,

(ii) the probability to have at least two dipoles in the onium Fock state to scatter off the nucleus $T_2$, and

(iii) the distribution of the branching rapidity of the last common ancestor of the dipoles which scatter $G$, which is a particular derivative of $T_2$ (see below).

All the above quantities are defined at a fixed impact parameter. We are going to construct two different formulations for these quantities. The first formulation is the evolution equations. We already knew the equation for the amplitude $T_1$, which is the BK equation. We shall hence derive equations for the two remaining quantities. Despite the fact that it is difficult to find the exact analytical solutions to these equations, they can be solved numerically, and hence, can provide a cross check for the results obtained from other possible approaches. The second formulation is a frame-dependent representation of the solutions to those QCD evolution equations. This representation is the starting point to construct the above-mentioned approximation scheme for further calculations.

### 3.2.1 Exact evolution equations in the color dipole model

In the previous chapter, we introduced the BK equation for the forward elastic amplitude $T_1(r, Y)$. As a matter of fact, $T_1(r, Y)$ can be interpreted as the probability to have at least one dipole in the Fock state of the onium of size $r$ to scatter with the nucleus, when the total relative rapidity is $Y$.

Let us now consider also the case of multiple scatterings. We define $T_2(r, Y; Y_0)$ as the probability that *at least two dipoles* in the Fock state of the onium in the reference frame where the nucleus is boosted to a rapidity $Y_0 < Y$ and the onium evolves to the remaining rapidity $\widetilde{Y}_0 = Y - Y_0$ are involved in the scattering. We can derive the evolution equation for $T_2$ using the similar technique to derive the BK and BFKL equations. If one increases the total rapidity by $dY$ while keeping the rapidity of the nucleus fixed at $Y_0$, the former is then an infinitesimal boost of the onium. Furthermore, we can place that infinitesimal boost at the beginning of the evolution of the onium. In such set up, after the interval $dY$ from zero rapidity, the initial onium can branch into two dipoles with the probability $\bar{\alpha} dY \, dp_{1 \to 2}(r^\perp, r'^\perp)$, or stay a single dipole with the probability $1 - \bar{\alpha} dY \int dp_{1 \to 2}(r^\perp, r'^\perp)$.

For the ensemble with no branching, $T_2(r, Y + dY; Y_0)$ is just $T_2(r, Y; Y_0)$. Instead, for events in which the initial onium $r$ branches into two daughter dipoles $r'$ and $|r^\perp - r'^\perp|$, there are two possible cases. If one of two daughter dipoles (either $r'$ or $|r^\perp - r'^\perp|$) results in no offspring scattering with the nucleus, the set of offspring of the other should contain at least two interacting dipoles. Otherwise, each daughter dipole should give at least one interacting offspring. Therefore the equation for

$T_2(r, Y + dY; Y_0)$ reads

$$
\begin{aligned}
T_2(r, Y + dY; Y_0) = {} & \left(1 - \bar\alpha dY \int dp_{1\to2}(r^\perp, r'^\perp)\right) T_2(r, Y; Y_0) \\
& + \bar\alpha dY \int dp_{1\to2}(r^\perp, r'^\perp) \big[\, T_1(r', Y) T_1(|r^\perp - r'^\perp|, Y) \\
& + T_2(r', Y; Y_0) S(|r^\perp - r'^\perp|, Y) + T_2(|r^\perp - r'^\perp|, Y; Y_0) S(r', Y) \,\big] \,.
\end{aligned}
\tag{3.1}
$$

The coupling constant $\bar\alpha$ always enters as a scaling factor of the rapidity. Therefore, for convenience, we will absorb it into the rapidity variable by defining the rescaled rapidity $y \equiv \bar\alpha Y$. From now on, we will use this rescaled rapidity and keep calling it as the "rapidity", when there is no further notice. Enforcing the continuous limit $dy \equiv \bar\alpha dY \to 0$ and writing $S = 1 - T_1$, we obtain the evolution equation for $T_2$ in the form of the following integrodifferential equation:

$$
\begin{aligned}
\partial_y T_2(r, y; y_0) = \int dp_{1\to2}(r^\perp, r'^\perp) \big[ & T_2(r', y; y_0) + T_2(|r^\perp - r'^\perp|, y; y_0) - T_2(r, y; y_0) \\
& - T_2(r', y; y_0) T_1(r^\perp - r'^\perp|, y) - T_2(|r^\perp - r'^\perp|, y; y_0) T_1(r', y) + T_1(r', y) T_1(|r^\perp - r'^\perp|, y) \big] \,.
\end{aligned}
\tag{3.2}
$$

In the frame where the nucleus is boosted to the total rapidity $y_0 = y$, the onium appears just as a bare dipole. Therefore, there is no possibility to pick at least two dipoles in the Fock state of the onium, $T_2(r, y_0; y_0) = 0$. This identity is used as the initial condition for Eq. (3.2).

When tracking backward the evolution, since we start with a single dipole (the initial onium), and since the evolution in rapidity is driven by elementary $1 \to 2$ dipole splitting processes, the set of dipoles which are involved in the interaction with the nucleus necessarily stem from the branching of a single dipole at a certain rapidity. This dipole is called as the "last common ancestor" of that particular set. We are going to address the calculation of the distribution of the branching rapidity $y_1$ with respect to the nucleus of this ancestor.

To quantify this problem, let us introduce $G(r, y; y_1)$, the joint probability distribution that there are at least two dipoles in the Fock state of the onium of size $r$ involved in the interaction with the nucleus at the total rapidity $y$, and that their last common ancestor has splitted at the rapidity $y_1$. Using the same method as for $S$ and $T_2$, we could derive the evolution equation for $G$. However, the latter can be obtained by using the following simple relation between $G$ and $T_2$:

$$
T_2(r, y; y_0) = \int_{y_0}^{y} dy_1 G(r, y; y_1),
\tag{3.3}
$$

or,

$$
G(r, y; y_1) = - \left.\frac{\partial T_2(r, y; y_0)}{\partial y_0}\right|_{y_0 = y_1} \,.
\tag{3.4}
$$

Taking the derivative with respect to $y_0$ of Eq. (3.2) and using the relation (3.4), we end up with the following evolution equation for $G$:

$$\partial_y G(r, y; y_1) = \int dp_{1\to2}(r^\perp, r'^\perp) \left[ G(r', y; y_1) S(|r^\perp - r'^\perp|, y) + G(|r^\perp - r'^\perp|, y; y_1) S(r', y) - G(r, y; y_1) \right].$$ (3.5)

The initial condition is set when the total rapidity coincides with the branching rapidity of the last common ancestor: $y = y_1$. In this case, the only possibility is that the onium has to branch at this very rapidity $y_1$ (or at the very beginning of its evolution), and each of its offspring must scatter with the nucleus. This condition can be translated into the following relation:

$$G(r, y_1; y_1) = \int dp_{1\to2}(r^\perp, r'^\perp) \left[ 1 - S(|r^\perp - r'^\perp|, y_1) \right] \left[ 1 - S(r', y_1) \right].$$ (3.6)

The evolution equations (3.2) and (3.5) for $T_2$ and $G$ can be solved numerically, but no analytical solutions are known. However, as we will shortly see, we can obtain exact asymptotic expressions for the ratios $T_2/T_1$ and $G/T_1$ in a picture expected to capture the main features of the QCD color dipole model and of more general branching random walks. For this purpose, we shall introduce a formulation in which $T_1$, $T_2$ and $G$ can be represented in terms of the dipole density and the nuclear scattering amplitude of a bare dipole as the starting point.

### 3.2.2 Frame-dependent formulation

We are going to formulate the solutions to the above evolution equations in such a way that it is useful to set up approximation schemes, from which we are able to find asymptotic expressions.

Consider a reference frame in which the nucleus is boosted to a rapidity $y_0$, and the onium is at rapidity $\tilde{y}_0 = y - y_0$. From now on, we use a notation with tilde $\tilde{y}_i$ for rapidities counted from the onium, and without tilde $y_i$ for rapidities counted from the nucleus ($\tilde{y}_i + y_i = y$). For the sake of convenience, instead of using as variables the transverse sizes $r$ of dipoles and the saturation momentum at $y$, $Q_s(y)$, we shall express all expressions in terms of the logarithms of these quantities, which are defined as

$$x \equiv \ln \frac{1}{r^2 Q_A^2}, \quad \text{and} \quad X_y \equiv \ln \frac{Q_s^2(y)}{Q_A^2}.$$ (3.7)

Let us start with the S-matrix element $S(x, y)$. Since all dipoles in the Fock state of the onium interact with the nucleus independently, its exact representation reads

$$S(x, y) = \left\langle \prod_{\{x_i\}} S(x_i, y_0) \right\rangle_{x, \tilde{y}_0}, $$ (3.8)

where the averaging $\langle \cdots \rangle_{x, \tilde{y}_0}$ is over all dipole configurations of the onium $x$ at rapidity $\tilde{y}_0$, represented by the set of logarithmic (log) sizes $\{x_i\}$. While on the left-hand side, $S(x, y)$ is the S-matrix element for the scattering of the evolved onium $x$ off the nucleus at the total rapidity $y$, the S-matrix element on the right-hand side $S(x_i, y_0)$ is for the scattering of an elementary dipole in the Fock state of the onium off the nucleus boosted at the rapidity $y_0$. Due to boost invariance, the former should be independent of the chosen frame, or $y_0$, used in the right-hand side.

One could check that the S-matrix element defined in Eq. (3.8) obeys the S-type BK equation (2.2). Indeed, increasing $y$ by $dy$ is tantamount to increase $\tilde{y}_0$ by the same amount $dy$ (while keeping $y_0$ unchanged). After the infinitesimal boost $dy$, the initial onium can branch into two dipoles $\overline{x} \equiv \ln 1/(r'^2 Q_A^2)$ and $\overline{\overline{x}} \equiv \ln 1/((r^\perp - r'^\perp)^2 Q_A^2)$, both of which develop into two sets of dipoles $\{\overline{x}_i\}$ and $\{\overline{\overline{x}}_i\}$, respectively, at the rapidity $\tilde{y}_0$. Otherwise, it remains unchanged. The decomposition into these two possibilities can be written as

$$\left\langle \prod_{\{x_i\}} S(x_i, y_0) \right\rangle_{x, \tilde{y}_0 + dy} = \left(1 - dy \int dp_{1 \to 2}(r^\perp, r'^\perp)\right) \left\langle \prod_{\{x_i\}} S(x_i, y_0) \right\rangle_{x, \tilde{y}_0}$$
$$+ dy \int dp_{1 \to 2}(r^\perp, r'^\perp) \left\langle \prod_{\{\overline{x}_i\}} S(\overline{x}_i, y_0) \right\rangle_{\overline{x}, \tilde{y}_0} \left\langle \prod_{\{\overline{\overline{x}}_i\}} S(\overline{\overline{x}}_i, y_0) \right\rangle_{\overline{\overline{x}}, \tilde{y}_0} . \quad (3.9)$$

After simple manipulations and the continuous limit $dy \to 0$, we recover the BK equation (2.2).

Let us introduce the number density $n(x)$ of dipoles of log size $x$ in the wave function of the onium. We can rewrite Eq. (3.8) as

$$S(x, y) = \left\langle \prod_{\{x'\}} [S(x', y_0)]^{n(x')dx'} \right\rangle_{x, \tilde{y}_0}$$
$$\underset{dx' \to 0}{=} \left\langle \exp\left[\int dx' n(x') \ln S(x', y_0)\right] \right\rangle_{x, \tilde{y}_0} \quad (3.10)$$
$$\equiv \left\langle e^{-I(y_0)} \right\rangle_{x, \tilde{y}_0} ,$$

where the product is now over all the bins in dipole size of with $dx'$, and we have defined

$$I(y_0) \equiv \int dx' n(x') \ln \frac{1}{S(x', y_0)} = \int dx' n(x') \ln \frac{1}{1 - T(x', y_0)}. \quad (3.11)$$

Since dipole evolution is a random process, $n(x')$ is a random density. The distribution of this random density would depend on the size of the initial onium and on the evolution rapidity $\tilde{y}_0$. From Eq. (3.10), the expression for the forward elastic amplitude $T_1 = 1 - S$ reads

$$T_1(x, y) \simeq \left\langle 1 - e^{-I(y_0)} \right\rangle_{x, \tilde{y}_0} . \quad (3.12)$$

Let us take the initial onium to be small such that its size is much smaller than the inverse saturation scale at the total rapidity $y$. We assume that the dipoles that effectively contribute to the integral all have log size $x'$ such that $T_1(x', y_0) \ll 1$. This is verified if the configurations of the onium Fock state which contain individual dipoles larger than the inverse nuclear saturation scale at $y_0$, i.e. $x' < X_{y_0}$, only bring a negligible contribution to the overall amplitude. We shall check *a posteriori* that it is a consistent assumption. With this approximation, we can expand the logarithm in Eq. (3.11) and obtain:

$$I(y_0) = \int dx' n(x') T_1(x', y_0), \tag{3.13}$$

which is the overlap of the dipole-nucleus scattering amplitude and the dipole density in the onium Fock state.

Let us now turn into the case of multiple scatterings. The complement to unity of $T_2$ includes the probabilities of no scattering and of having only one dipole in the Fock state of the onium involved in the scattering. Therefore, its exact formula reads

$$T_2(x, y; y_0) = \left\langle 1 - \left( 1 + \sum_{\{x_i\}} \frac{T_1(x_i, y_0)}{S(x_i, y_0)} \right) \prod_{\{x_i\}} S(x_i, y_0) \right\rangle_{x, \tilde{y}_0}. \tag{3.14}$$

Again, we can show that this expression obeys Eq. (3.2) in the same way as for the S-matrix element above.

Recall that relevant configurations contain only dipoles of log sizes $x'$ such that $T_1(x', y_0) \ll 1$, or equivalently $S(x', y_0) \simeq 1$. Going to the continuous limit, we get

$$T_2(x, y; y_0) \simeq \left\langle 1 - [1 + I(y_0)] e^{-I(y_0)} \right\rangle_{x, \tilde{y}_0}. \tag{3.15}$$

The equation for $G$ can be obtained from a derivative w.r.t $y_0$ of $T_2$, as shown in Eq. (3.4).

Our main task in this chapter is to evaluate the right-hand sides of Eqs. (3.12) and (3.15). Since the density $n(x')$ is a random quantity, and since its distribution is unknown, these evaluations cannot be done through a straightforward calculation. Instead, we will develop a simple model for the realizations of branching random walks and dipole evolution, which can quantify the above randomness and, hence, enables us to perform such task.

## 3.3  Model for dipole distribution

### Typical evolution

Since it is impossible to calculate the dipole distribution in an exact way, an approximation scheme is needed. We start with an onium of log-size $x$. At low rapidities $\tilde{y}_0 \sim 1$, since there are few dipoles, the density is very noisy. When $\tilde{y}_0$ becomes large, the density becomes smooth around the log-size $x$, since the typical number of dipole increases exponentially with $\tilde{y}_0$. However, the number density in the tail $|x - x'| \sim \chi'(\gamma_0)\tilde{y}_0$ is still low, and hence, the distribution remains noisy. We can take into account the effect of this statistical noise in the first approximation by putting a moving absorptive boundary, which is also known as the Brunet-Derrida cutoff [96], on the solution to the BFKL equation (the latter is the mean dipole density). This boundary is actually the largest-dipole tail of the dipole distribution in a typical evolution. This approximation gives the typical dipole density $\bar{n}(x'-x, \tilde{y}_0)$ of log-size $x'$ near the largest dipole, which reads [20]

$$\bar{n}(x'-x, \tilde{y}_0) = C_1(x'-x-\widetilde{X}_{\tilde{y}_0})e^{\gamma_0(x'-x-\widetilde{X}_{\tilde{y}_0})} \exp\left[-\frac{(x'-x-\widetilde{X}_{\tilde{y}_0})^2}{2\chi''(\gamma_0)\tilde{y}_0}\right]\Theta(x'-x-\widetilde{X}_{\tilde{y}_0}), \quad (3.16)$$

where the displacement $\widetilde{X}_{\tilde{y}_0}$, at large $\tilde{y}_0$, reads

$$\widetilde{X}_{\tilde{y}_0} = -\chi'(\gamma_0)\tilde{y}_0 + \frac{3}{2\gamma_0}\ln\tilde{y}_0 + \mathcal{O}. \quad (3.17)$$

When extrapolating to the nonasymptotic regime of $\tilde{y}_0$, the logarithmic singularity when $\tilde{y}_0 \to 0$ is regularized in such a way that $\widetilde{X}_{\tilde{y}_0} \xrightarrow{\tilde{y}_0 \to 0} 0$.

The formula (3.16) represents the dipole density in a typical realization of the dipole evolution, in the absence of a large fluctuation, in a region of size of order $\sqrt{\tilde{y}_0}$ from the typical log-size $x + \widetilde{X}_{\tilde{y}_0}$ of the largest dipole. The latter is the moving absorptive boundary mentioned above.

## Single fluctuation

As mentioned previously, we are interested in the nuclear scattering of a small onium whose size $x$ is in the window $1 < x - X_y \lesssim \sqrt{\chi''(\gamma_0)y}$, where $y$ is the total rapidity, which defines the scaling region up to strong inequalities. For such a small onium, all the dipoles in its typical configuration will be much smaller than the nuclear saturation scale, in every reference frame. The overlap between the mean density of a typical configuration with the dipole-nucleus scattering amplitude would then be negligible.

Therefore, on top of the above deterministic evolution, we assume that one single fluctuation occurs after some random evolution rapidity $0 < \tilde{y}_1 < \tilde{y}_0$ from the beginning of the dipole evolution. We assume that this fluctuation creates a dipole of size larger than the largest dipole in typical configurations by a factor $e^{\delta/2}$. In other words, the absolute difference between the log size of that large dipole and the typical log size of the largest dipole in typical configurations is $\delta$. Therefore, we will hereafter call $\delta$ the "size" of the fluctuation.

We need the distribution for the fluctuation size $\delta$ in our calculations. Since we are interested in large fluctuations, the particle distribution near the fluctuation is very dilute. We conjecture that the distribution for the fluctuation size can be approximated by the probability of observing the largest dipole with a log size shifted by $(-\delta)$ with respect to the mean-field tip. From the discussion on the dual interpretation of the BK equation in the previous chapter (see Section 2.1.3), the latter solves the BK equation. The rate for a fluctuation size $\delta$ at a large rapidity $\tilde{y}_1$ reads

$$p(\delta, \tilde{y}_1) = C\delta e^{-\gamma_0\delta} \exp\left(-\frac{\delta^2}{2\chi''(\gamma_0)\tilde{y}_1}\right)\Theta(\delta), \quad (3.18)$$

where $C$ is a constant. When the fluctuation occurs, it will develop into a smaller front. And the small onium always scatters exclusively with the nucleus throught this secondary front. Each elementary dipole $x'$ in the state of the onium interacts independently with an amplitude $\bar{T}_1(x', y_0)$ that solves the T-type BK equation and has the form given by Eq. (2.24). In the notation (3.7), we can write the latter as follows

$$\bar{T}_1(x', y_0) = C_2(x' - X_{y_0})e^{-\gamma_0(x' - X_{y_0})} \exp\left[-\frac{(x' - X_{y_0})^2}{2\chi''(\gamma_0)y_0}\right]\Theta(x' - X_{y_0}), \tag{3.19}$$

where, again, the saturation log-scale $X_{y_0}$ is regularized in such a way that $X_0 = 0$.

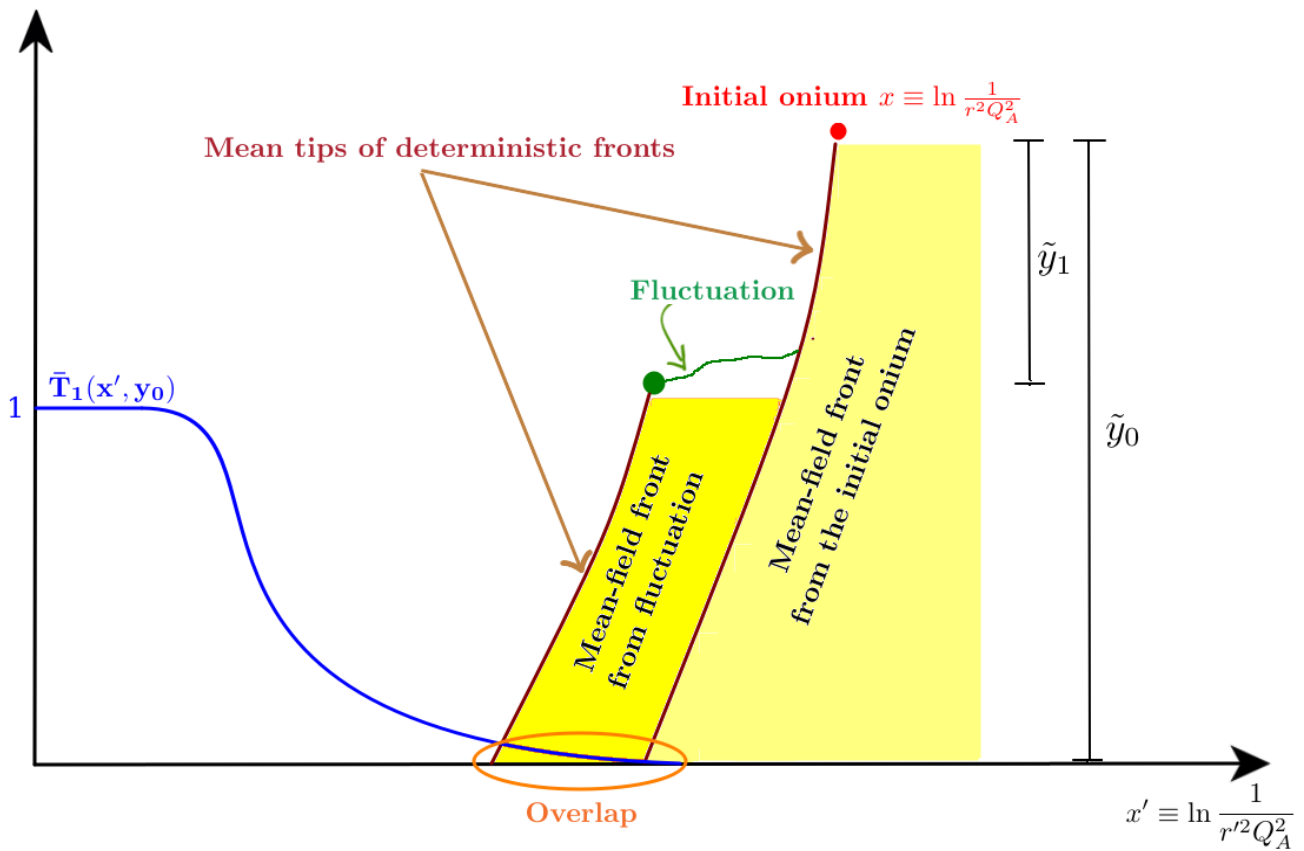

Figure 3.1: An illustration of the phenomenological model for dipole distribution in the nuclear scattering of a small onium. The initial onium $x$ develops into a deterministic front whose contribution to the scattering with the nucleus, represented by the dipole-nucleus amplitude $\bar{T}_1(x', y_0)$ (the blue curve), is negligible. The essence of the model is the occurrence of a fluctuation containing one unusual large dipole at some rapidity $\tilde{y}_1$, which builds up, by further dipole branchings, another deterministic front. The overlap with the nucleus (circled by the orange oval) is then dominated by the dipole density of this small front.

The model of dipole evolution presented here is a slightly modified version of the model for the evolution of general branching random walks, which was initially developed in Ref. [24] and applied to QCD in Ref. [20]. In that original model, the mean number density is deformed by large fluctuations which may occur in two different situations. First, they are likely to arise in the early stages of the evolution when the system is stochastic since the overall number of dipoles is small. Due to further rapidity evolution, the effect of the early fluctuations is to shift the mean-field value of the largest dipole at a rapidity $\tilde{y}_i$ by an random amount $\Delta_f$ whose probability distribution is denoted by $p_f(\Delta_f, \tilde{y}_i)$. Therefore, this type of fluctuations is referred to as "front fluctuations". At a large rapidity $\tilde{y}_j \gg 1$, when the total number of dipoles becomes large, fluctuations can still occur near the tip of the mean-field distribution, and hence, are referred to as "tip fluctuations". These tip fluctuations send randomly a small number (typically 1) of dipoles ahead of the mean-field tip by some distance $\Delta_t$ with the rate given by $p_t(\Delta_t)$. The expressions of $p_f(\Delta_f, \tilde{y}_i)$ and $p_t(\Delta_t)$ read [20, 24]

$$p_f(\Delta_f, \tilde{y}_i) \sim e^{-\gamma_0 \Delta_f} \Theta(\Delta_f), \quad \text{and} \quad p_t(\Delta_t) \sim e^{-\gamma_0 \Delta_t} \Theta(\Delta_t), \tag{3.20}$$

where for the former, it is expected to have a cut-off at $\Delta_f \sim \sqrt{\tilde{y}_i}$ given by a gaussian term (due to diffusion effect), which can be neglected for $\Delta_f \ll \sqrt{\tilde{y}_i}$. In fact, we can roughly recover Eq. (3.18) from Eq. (3.20). In particular, at the rapidity $\tilde{y}_1$, the mean-field front is deformed by a fluctuation $\delta$, which can be decomposed into a front fluctuation of width $\Delta_f$ and a tip fluctuation of size $\Delta_t = \delta - \Delta_f$. Given the probability distributions (3.20), the probability distribution of the whole fluctuation $\delta$ can be computed as

$$p(\delta, \tilde{y}_1) \sim \int_0^\delta d\Delta_f e^{-\gamma_0 \Delta_f} e^{-\gamma_0(\delta - \Delta_f)} = \delta e^{-\gamma_0 \delta}. \tag{3.21}$$

Up to a gaussian cutoff (which is the effect of diffusion), this coincides with Eq. (3.18). Therefore, we could think of the net fluctuation whose probability density is given by Eq. (3.18) as a combination of a front fluctuation and a tip fluctuation in general.

## 3.4 Heuristic calculations of scattering amplitudes

At this moment, we have the necessary ingredients to evaluate Eqs. (3.12) and (3.15). Before doing so, let us briefly summarize the picture. We are considering the scattering of a small dipole $x$ in the scaling windows off a large nucleus at the total rapidity $y$, in a frame where the nucleus is boosted to $y_0$ and the onium evolves to the remaining rapidity $\tilde{y}_0$. In the first place, the initial onium will develop into a deterministic front until a random evolution rapidity $\tilde{y}_1$. At this rapidity, a large fluctuation $\delta$ whose probability density is given by Eq. (3.18) occurs, which sends a dipole ahead of the mean-field boundary. A small deterministic front in the rapidity interval $\tilde{y}_0 - \tilde{y}_1$ which stems from this dipole and is characterized by a dipole density then overlaps with the dipole-nucleus scattering amplitude given by Eq. (3.19) to produce the dominant contribution to the scattering.

In this picture, the overlap (3.13) reads

$$I(y_0; \delta, y_1) \cong \int dx' \bar{n}(x' - \Xi_{\delta, \tilde{y}_1}, \tilde{y}_0 - \tilde{y}_1) \bar{T}_1(x', y_0), \tag{3.22}$$

where $\Xi_{\delta, \tilde{y}_1} \equiv x + \widetilde{X}_{\tilde{y}_1} - \delta$ is the log size of the lead dipole created by the large fluctuation at the rapidity $\tilde{y}_1$. The average over dipole configurations in Eqs. (3.12) and (3.15) in the current approximation is tantamount to the average over all possible fluctuations. Consequently, the expressions for $T_1$ and $T_2$ are given by

$$T_1(x, y) = \int_{y_0}^{y} dy_1 \int_0^{\infty} d\delta\, p(\delta, \tilde{y}_1) \left( 1 - e^{-I(y_0; \delta, y_1)} \right), \tag{3.23}$$

and

$$T_2(x, y; y_0) = \int_{y_0}^{y} dy_1 \int_0^{\infty} d\delta\, p(\delta, \tilde{y}_1) \left\{ 1 - [1 + I(y_0; \delta, y_1)]\, e^{-I(y_0; \delta, y_1)} \right\}. \tag{3.24}$$

We can also obtain a formula for $G$ itself in the framework of the phenomenological model. From Eq. (3.3), the only difference in the expressions for $G$ and $T_2$ is the integration over the fluctuation rapidity $y_1$. Indeed, the essence of the model is to single out one dipole in the state of the onium evolved to the rapidity $\tilde{y}_1$ that will play the role of the last common ancestor of all dipoles which scatter after evolution to the rapidity $\tilde{y}_0$. Therefore, the distribution $G$ reads

$$G(x, y; y_1) = \int_0^{\infty} d\delta\, p(\delta, \tilde{y}_1) \left\{ 1 - [1 + I(y_0; \delta, y_1)]\, e^{-I(y_0; \delta, y_1)} \right\}. \tag{3.25}$$

We introduce the logarithmic distance between the tip of the dipole ditribution (i.e. of the small front) and the top of the nuclear scattering amplitude (i.e. the saturation log-scale),

$$\Delta(y_0; \delta, y_1) \equiv \tilde{X}_{\tilde{y}_0 - \tilde{y}_1} + \Xi_{\delta, \tilde{y}_1} - X_{y_0}. \tag{3.26}$$

In other words, $\Delta(y_0; \delta, y_1)$ is the logarithm of the squared ratio of the size of the smallest dipole which would scatter with probability of order unity with the nucleus at rapidity $y_0$, and of the size of the largest dipole in the actual state of the onium at rapidity $\tilde{y}_0$. Substituting the expressions for the terms on the right-hand side of Eq. (3.26), the distance $\Delta(y_0; \delta, y_1)$ can be rewritten as

$$\Delta(y_0; \delta, y_1) = x - X_y - \delta + \frac{3}{2\gamma_0} \ln \frac{(\tilde{y}_0 - \tilde{y}_1) y_0 \tilde{y}_1}{y}. \tag{3.27}$$

As commented after Eqs. (3.17) and (3.19), the logarithmic term has to be regularized in the limits $y_1 \to y$ and $y_1 \to y_0$. Furthermore, with the considered choice of frame and parameters, it is always small compared to $x - X_y$.

We shall first show that it is safe to disregard fluctuations such that $\Delta(y_0; \delta, y_1) < 0$, or $\delta > \delta_0 \equiv x - X_y + \frac{3}{2\gamma_0} \ln \frac{(\tilde{y}_0 - \tilde{y}_1) y_0 \tilde{y}_1}{y}$. To this aim, we estimate Eq. (3.23) in that region by noticing that it is bound from above as

$$T_1(x, y)|_{\Delta < 0} \leq \int_{y_0}^{y} dy_1 \int_{\delta_0}^{\infty} d\delta\, p(\delta, \tilde{y}_1). \tag{3.28}$$

The integration over $\delta$ can be performed simply by a notice that the Gaussian factor in $p(\delta, \tilde{y}_1)$ can be replaced by an effective upper cut-off, set at $\delta_1 = \sqrt{2\chi''(\gamma_0)\tilde{y}_1}$. For the integration not to be null, the condition $\delta_1 > \delta_0$ should be satisfied, which implies $y_1 < y - (x - X_y)^2 / [2\chi''(\gamma_0)]$. After these manipulations, we obtain

$$T_1(x,y)|_{\Delta<0} \leq \frac{C}{\gamma_0} \frac{1}{y_0^{3/2}} (x - X_y) e^{-\gamma_0(x-X_y)} \times \int_{y_0}^{y - \frac{(x-X_y)^2}{[2\chi''(\gamma_0)]}} dy_1 \left( \frac{y}{(y-y_1)(y_1-y_0)} \right)^{3/2} \Bigg|_{\text{regularized}} , \quad (3.29)$$

where we have emphasized that the singularity at the lower bound of the integral needs to be regularized, so that the integral is finite of order 1. To see that, one could replace the lower bound by $y_0 + a$, where $a$ is a positive constant of order unity, and notice that the integration over the small domain $[y_0, y_0 + a]$ should be at most of order of $a$ after the regularization. The $y_1$-integral can then be computed, and the result is of order $(y/\tilde{y}_0)^{3/2} \sim 1$. From Eq. (3.29), we see that $T_1$ is suppressed by at least a factor $y_0^{3/2} \gg 1$ with respect to the expected result; see Eq. (3.19) with $y_0$ replaced $y$ and $x'$ replace by $x$. This proves that we can restrict ourselves to the region $\delta \leq \delta_0$.

Actually, a closer look would show that only the region with $\Delta \gtrsim \frac{3}{2\gamma_0} \ln y_0$ contributes significantly. This means that the relevant configurations contain dipoles such that their corresponding scattering amplitude, at rapidity $\tilde{y}_0$, is very small. This confirms our assumption that led to Eqs. (3.12) and (3.15) (also Eqs. (3.23) and (3.24)).

We see that the functions $T_1$, $T_2$ and $G$ are all written in terms of the probability density $p$ and the overlap $I$. For further calculations, let us write down the explicit expression for the latter, using Eq. (3.22) and the expression for $\bar{n}$ and $\bar{T}_1$,

$$I(y_0; \delta, y_1) = C_1 C_2 e^{\gamma_0(X_{y_0} - \widetilde{X}_{\tilde{y}_0 - \tilde{y}_1} - \Xi_{\delta, \tilde{y}_1})} \int dx' (x' - \widetilde{X}_{\tilde{y}_0 - \tilde{y}_1} - \Xi_{\delta, \tilde{y}_1})(x' - X_{y_0})$$

$$\times \exp\left[ -\frac{(x' - \widetilde{X}_{\tilde{y}_0 - \tilde{y}_1} - \Xi_{\delta, \tilde{y}_1})^2}{2\chi''(\gamma_0)(\tilde{y}_0 - \tilde{y}_1)} - \frac{(x' - X_{y_0})^2}{2\chi''(\gamma_0)y_0} \right] \Theta(x' - \widetilde{X}_{\tilde{y}_0 - \tilde{y}_1} - \Xi_{\delta, \tilde{y}_1})\Theta(x' - X_{y_0}).$$

$$(3.30)$$

We shall now compute the scattering amplitudes $T_1$ and $T_2$, and the distribution $G$ in the framework of the phenomenological model for dipole distribution, in two different types of reference frames, which are different to each other by the ordering of two variables $y_0$ and $(x - X_y)^2$. In particular, we are interested in

- a frame in which the nucleus is highly boosted: $1 \ll (x - X_y)^2 \ll y_0$, and

- a frame in which the nucleus is slightly boosted: $1 \ll y_0 \ll (x - X_y)^2$.

(See Fig. 3.2 for an illustration). For completeness, we also consider the frame that the nucleus is at rest to verify the boost invariance of the amplitude $T_1$, though this frame is not useful to compute the quantities $T_2$ and $G$ for multiple scatterings.

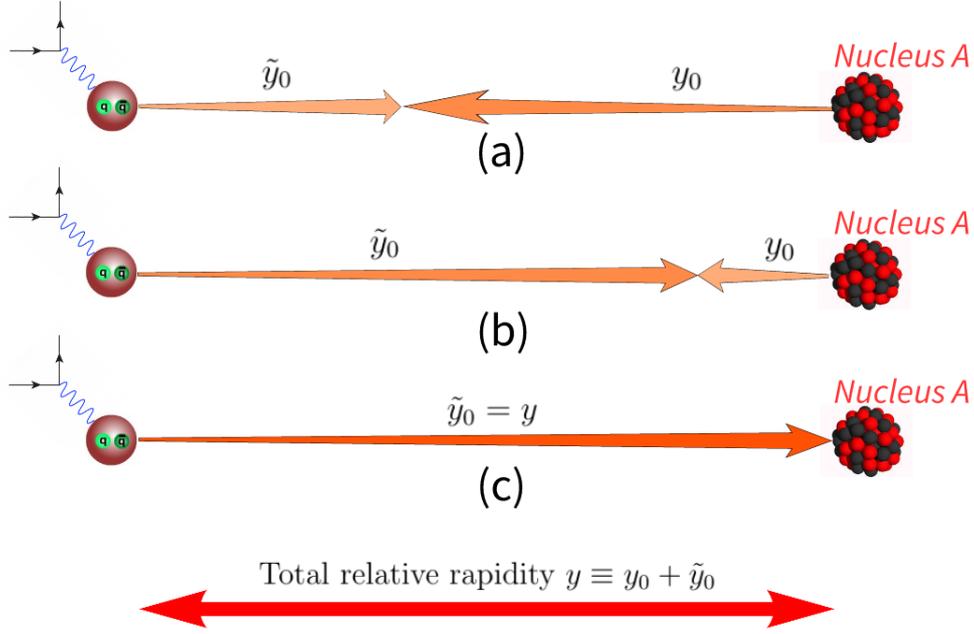

Figure 3.2: Three different frames for the onium-nucleus scattering with the total rapidity $y$ shared differently between the nucleus ($y_0$) and the onium ($\tilde{y}_0$): (a) the nucleus is strongly boosted, (b) the nucleus is slightly boosted, and (c) the nucleus is at rest.

### 3.4.1 Amplitudes in a frame in which the nucleus is highly boosted

Let us consider the frame in which the nucleus is boosted to a rapidity $y_0$ such that

$$1 \ll (x - X_y)^2 \ll y_0. \tag{3.31}$$

In the expression (3.30) for the overlap, we shift the integration variable $x'$ by defining $\bar{x} = x' - \widetilde{X}_{\tilde{y}_0 - \tilde{y}_1} - \Xi_{\delta, \tilde{y}_1}$, which represents the relative "distance" between the log size of the interacting dipoles and the log size of the largest dipole in the mean-field front that stems from the fluctuation. The overlap then reads

$$
\begin{aligned}
I(y_0; \delta, y_1) = {}& C_1 C_2 e^{-\gamma_0 \Delta(y_0; \delta, y_1)} \exp\left(-\frac{\Delta^2(y_0; \delta, y_1)}{2\chi''(\gamma_0) y_1}\right) \\
& \times \int_0^\infty d\bar{x}\, \bar{x}\, [\bar{x} + \Delta(y_0; \delta, y_1)] \exp\left[-\frac{y_1}{2\chi''(\gamma_0) y_0 (\tilde{y}_0 - \tilde{y}_1)} \left(\bar{x} + \frac{y_1 - y_0}{y_1} \Delta(y_0; \delta, y_1)\right)^2\right].
\end{aligned}
\tag{3.32}
$$

We observe that the integral is determined by a large integration region, up to $\bar{x} \sim \sqrt{y_0}$. Since $(x - X_y)^2 \ll y_0$, we can neglect $\Delta(y_0; \delta, y_1)$ compared to $\bar{x}$ in the second Gaussian term, and the integral can be evaluated exactly. Moreover, we will check *a posteriori* that typically, $\tilde{y}_1 \ll y$, or $y_1 \sim y$; hence, the first Gaussian factor involving $\Delta(y_0; \delta, y_1)$ can be set to unity. The evaluation yields the following result:

$$I(y_0; \delta, y_1) = \frac{C_1 C_2 \sqrt{\pi}}{4} e^{-\gamma_0 \Delta(y_0; \delta, y_1)} \left[ \frac{2\chi''(\gamma_0) y_0 (\tilde{y}_0 - \tilde{y}_1)}{y_1} \right]^{3/2}. \tag{3.33}$$

Substituting the expression (3.26) of $\Delta(y_0; \delta, y_1)$ into the above formula, we arrive at

$$I(y_0; \delta, y_1) = C_1 C_2 \sqrt{\frac{\pi}{2}} [\chi''(\gamma_0)]^{3/2} e^{-\gamma_0(x - X_y)} \left[ \frac{y}{y_1 \tilde{y}_1} \right]^{3/2} e^{\gamma_0 \delta}. \tag{3.34}$$

We eventually see that, $I$ appears to be independent of $y_0$. Let us isolate the $\delta-$ and $y_1-$independent factor by introducing the following notation:

$$p_1 \equiv C_1 C_2 \sqrt{\frac{\pi}{2}} [\chi''(\gamma_0)]^{3/2} e^{-\gamma_0(x - X_y)}. \tag{3.35}$$

The product $p_1 e^{\gamma \delta}$ is then just the overlap of the front of the nucleus with that of an onium if the latter were involved in a purely deterministic way starting at a log size $x - \delta$. Having $I$ at hand, we are now able to compute the functions $T_1$, $T_2$ and $G$.

### a. Forward elastic scattering amplitude $T_1$

Substituting the expressions of $I$ in Eq. (3.34) and of $p(\delta, \tilde{y}_1)$ in Eq. (3.18) into Eq. (3.23) and bearing in mind that $\Delta \geq 0$, we obtain

$$T_1(x, y) = C \int_{y_0}^{y} dy_1 \int_0^{\delta_0} d\delta \delta e^{-\gamma_0 \delta} \exp\left( -\frac{\delta^2}{2\chi''(\gamma_0)\tilde{y}_1} \right) \left\{ 1 - \exp\left[ -p_1 \left( \frac{y}{y_1 \tilde{y}_1} \right)^{3/2} e^{\gamma_0 \delta} \right] \right\}. \tag{3.36}$$

We shift the variable $\delta$ by defining a new variable $\delta' \equiv \delta + \frac{3}{2\gamma_0} \ln \frac{y}{y_1 \tilde{y}_1}$. Since the integration domain in $\delta'$ extends up to $\delta' \lesssim (x - X_y)$, a region much larger than any logarithm of rapidities appearing in this problem, the lower bound of the $\delta'-$integral can be kept to 0, and the logarithmic term in the Gaussian factor can be omitted. The upper bound can be set to $\infty$ since the region $\delta > \delta_0$ always gives a sub-dominant contribution. Therefore, Eq. (3.36) can be rewritten in term of the new variable $\delta'$ as

$$T_1(x, y) = C \int_0^{\infty} d\delta' \delta' e^{-\gamma_0 \delta'} \left[ 1 - \exp(-p_1 e^{\gamma_0 \delta'}) \right] \int_0^{\tilde{y}_0} d\tilde{y}_1 \left( \frac{y}{y_1 \tilde{y}_1} \right)^{3/2} \exp\left( -\frac{\delta'^2}{2\chi''(\gamma_0)\tilde{y}_1} \right), \tag{3.37}$$

where we have used the fact that $\tilde{y}_1 = y - y_1$. The integration over $\tilde{y}_1$ is dominated by the region $\tilde{y}_1 \ll y$, or $y_1 \simeq y$. Therefore we can replace the upper bound by $\infty$ and get

$$\int_0^{\tilde{y}_0} d\tilde{y}_1 \left( \frac{y}{y_1 \tilde{y}_1} \right)^{3/2} \exp\left( -\frac{\delta'^2}{2\chi''(\gamma_0)\tilde{y}_1} \right) \simeq \int_0^{\infty} \frac{d\tilde{y}_1}{\tilde{y}_1^{3/2}} \exp\left( -\frac{\delta'^2}{2\chi''(\gamma_0)\tilde{y}_1} \right) = \frac{\sqrt{2\pi \chi''(\gamma_0)}}{\delta'}. \tag{3.38}$$

Eq. (3.37) then becomes

$$T_1(x,y) = C\sqrt{2\pi\chi''(\gamma_0)} \int_0^\infty d\delta' e^{-\gamma_0\delta'} \left[1 - \exp(-p_1 e^{\gamma_0\delta'})\right].$$ (3.39)

From the aforementioned interpretation of $p_1 e^{\gamma_0\delta'}$, this expression effectively represents the nuclear scattering of an onium whose size, or equivalently whose evolution front, is shifted by a log size $\delta'$. This is due to a fluctuation happening at the very beginning of the evolution, $\tilde{y}_1 \ll \tilde{y}_0$ (or $y_1 \simeq y$), with the weight given by $e^{-\gamma_0\delta'}$. This is exactly the *front fluctuation*, which was already discussed in Section 3.3.

Equation (3.39) can be rewritten with the help of the integral $I_1$ defined in Appendix B as

$$T_1(x,y) = \frac{C}{\gamma_0}\sqrt{2\pi\chi''(\gamma_0)} \times I_1(p_1).$$ (3.40)

Using the evaluation of the integral $I_1$ in Eq. (B.2) and replacing $p_1$ by its definition in Eq. (3.35), the final result for the amplitude $T_1$ reads

$$T_1(x,y) \simeq CC_1C_2\pi[\chi''(\gamma_0)]^2(x - X_y)e^{-\gamma_0(x-X_y)}.$$ (3.41)

We have just recovered the scaling limit of the known solution to the BK equation; see Eq. (3.19) (with the substitutions $y_0 \to y$ and $x' \to x$). Due to the strong assumption $(x - X_y)^2 \ll y$, we cannot get consistently the finite-$y$ correction that appears in the form of a Gaussian factor $\exp[-(x-X_y)^2/(2\chi''(\gamma_0)y)]$. The calculation in this section indicates that, in the current considered frame, the realizations of the Fock state which trigger scattering events look like typical realizations, as far as their shape is concerned, but are overall shifted towards larger dipole sizes by a multiplicative factor (or additive term in the log scale), through a front fluctuation.

**b. Multiple scatterings: $T_2$ and $G$**

From Eq. (3.24) and the expression of $I(y_0; \delta, y_1)$ in Eq. (3.34), the full expression of the amplitude $T_2$ for scattering at least twice reads

$$T_2(x,y;y_0) = C\int_0^{\tilde{y}_0} d\tilde{y}_1 \int_0^{\delta_0} d\delta\delta e^{-\gamma_0\delta} \exp\left(-\frac{\delta^2}{2\chi''(\gamma_0)\tilde{y}_1}\right)$$
$$\times \left\{1 - \left[1 + p_1\left(\frac{y}{y_1\tilde{y}_1}\right)^{3/2}e^{\gamma_0\delta}\right]\exp\left[-p_1\left(\frac{y}{y_1\tilde{y}_1}\right)^{3/2}e^{\gamma_0\delta}\right]\right\}.$$ (3.42)

In the above formula, the dominant contribution still comes from the fluctuations of size $\delta \leq \delta_0$, as we will check *a posteriori*. The evaluation of $T_2$ goes along the very same lines as that of $T_1$ above. After performing the $y_1$ integration, we are left with an integral over the shifted variable $\delta'$,

$$T_2(x,y;y_0) = C\sqrt{2\pi\chi''(\gamma_0)} \int_0^\infty d\delta' e^{-\gamma_0\delta'} \left[1 - \left(1 + p_1 e^{\gamma_0\delta}\right)\exp\left(-p_1 e^{\gamma_0\delta}\right)\right].$$ (3.43)

Again, we make use of the integral $I_2$ defined in the Appendix to rewrite Eq. (3.43) as

$$T_2(x, y; y_0) = \frac{C}{\gamma_0} \sqrt{2\pi\chi''(\gamma_0)} \times I_2(p_1). \tag{3.44}$$

Substituting the formula for $I_2$ in Eq. (B.2) and the expression (3.35) of $p_1$ into Eq. (3.44), then dividing the latter by $T_1$ in Eq. (3.41) we eventually get

$$\frac{T_2(x, y; y_0)}{T_1(x, y)} \underset{y_0 \gg (x - X_y)^2}{=} \frac{1}{\gamma_0(x - X_y)}. \tag{3.45}$$

For fluctuations such that $\delta > \delta_0$, we have a very similar estimation as that for $T_1$,

$$T_2(x, y; y_0)|_{\delta > \delta_0} \leq C \int_0^{\tilde{y}_0} d\tilde{y}_1 \int_{\delta_0}^{\infty} d\delta \delta e^{-\gamma_0\delta} \exp\left(-\frac{\delta^2}{2\chi''(\gamma_0)\tilde{y}_1}\right) \lesssim \text{const} \times \frac{T_1}{y_0^{3/2}} \ll \frac{T_1}{\gamma_0(x - X_y)}. \tag{3.46}$$

This again proves that no single dipole has a significantly probability to scatter with the nucleus in relevant configurations within this picture.

Now we are going to evaluate the genealogy distribution $G$. Its explicit expression only differs from the expression for $T_2$ in Eq. (3.42) by the absence of the integration over $y_1$,

$$G(x, y; y_1) = C \int_0^{\delta_0} d\delta \delta e^{-\gamma_0\delta} \exp\left(-\frac{\delta^2}{2\chi''(\gamma_0)\tilde{y}_1}\right)$$
$$\times \left\{1 - \left[1 + p_1 \left(\frac{y}{y_1\tilde{y}_1}\right)^{3/2} e^{\gamma_0\delta}\right] \exp\left[-p_1 \left(\frac{y}{y_1\tilde{y}_1}\right)^{3/2} e^{\gamma_0\delta}\right]\right\}. \tag{3.47}$$

Again, since the integration region $\delta > \delta_0$ only gives an unimportant contribution, we can replace the upper bound $\delta_0$ by $\infty$. By a change of variable $t = Ae^{\gamma_0\delta}$, where $A \equiv p_1 \left(\frac{y}{y_1\tilde{y}_1}\right)^{3/2}$, Eq. (3.47) becomes

$$G(x, y; y_1) = \frac{C}{\gamma_0^2} A \int_A^{\infty} \frac{dt}{t^2} \ln\left(\frac{t}{A}\right) \exp\left[-\frac{\ln^2(t/A)}{2\gamma_0^2\chi''(\gamma_0)\tilde{y}_1}\right] \left\{1 - [1 + t] e^{-t}\right\} \tag{3.48}$$

Using the integrals $R$ and $S_k$ defined and evaluated in Appendix B, with $S_0$ up to next-to-leading-log order (Eq. (B.19)), we can rewrite $G$ as:

$$G(x, y; y_1) = \frac{C}{\gamma_0^2} A \left[R(A) - S_0(A) - S_1(A)\right] \simeq \frac{C}{\gamma_0^2} A \ln\frac{1}{A} \exp\left[-\frac{\ln^2\frac{1}{A}}{2\gamma_0^2\chi''(\gamma_0)\tilde{y}_1}\right], \tag{3.49}$$

with $\beta_0 \equiv 2\gamma_0^2\chi''(\gamma_0)$. Replacing $A$ and $p_1$ by their explicit expressions, we obtain

$$G(x, y; y_1) = \frac{CC_1C_2}{\gamma_0} \sqrt{\frac{\pi}{2}} [\chi''(\gamma_0)]^{3/2} \left(\frac{y}{y_1\tilde{y}_1}\right)^{3/2} (x - X_y) e^{-\gamma_0(x - X_y)} \exp\left[-\frac{(x - X_y)^2}{2\chi''(\gamma_0)\tilde{y}_1}\right]. \tag{3.50}$$

Divided by $T_1$, the distribution of the splitting rapidity $y_1$ of the slowest parent dipole of the set of dipoles which scatter reads

$$\frac{G(x, y; y_1)}{T_1(x, y)} = \frac{1}{\gamma_0} \frac{1}{\sqrt{2\pi\chi''(\gamma_0)}} \left[\frac{y}{y_1(y - y_1)}\right]^{3/2} \exp\left[-\frac{(x - X_y)^2}{2\chi''(\gamma_0)\bar{y}_1}\right]. \tag{3.51}$$

**Analogy to a genealogical problem of a branching-diffusion process**

Consider a branching-diffusion process (e.g., a BRW) on a real line $x$ evolving in time $t$ with a diffusion coefficient $D$ and a branching rate $r$. The mean density $\bar{n}$ of particles solves the following equation:

$$\partial_t \bar{n}(x, t) = \chi_F(-\partial_x; D, r)\bar{n}(x, t), \tag{3.52}$$

where $\chi_F(-\partial_x; D, r) \equiv D\partial_x^2 + r$ is the branching-diffusion linear kernel (this is the generalization of the FKPP kernel discussed in the previous chapter by including general constants $D$ and $r$). This kernel admits the eigenfunction $e^{-\beta x}$, corresponding to the eigenvalue $\chi_F(\beta; D, r) = D\beta^2 + r$. After some predefined time $T$, let us pick up two leftmost (or rightmost) particles, or any pair of particles in the tip (e.g., the $2^{nd}$ and the $5^{th}$ leftmost particles), and ask for the branching time $t_1$ of their last common ancestor (see Fig. 3.3). This is not exactly the same problem as that of the dipole evolution discussed above, but they are similar. According to Derrida and Mottishaw in Ref. [85], the asymptotic distribution of $t_1$ reads

$$p(t_1; T) = \frac{1}{\beta_0} \frac{1}{\sqrt{2\pi\chi_F'(\beta_0; D, r)}} \left[\frac{T}{t_1(T - t_1)}\right]^{3/2}. \tag{3.53}$$

where $\beta_0$ solves the equation $\beta_0\chi_F'(\beta_0; D, r) = \chi_F(\beta_0; D, r)$. One sees that, except for the Gaussian factor, there is a perfect analogy between this result (Eq. (3.53)) and our result presented in Eq. (3.51), up to the following substitutions:

$$y \leftrightarrow t, \quad y_1 \leftrightarrow t_1, \quad \chi(\gamma) \leftrightarrow \chi_F(\beta; D, r), \text{ and } \gamma_0 \leftrightarrow \beta_0. \tag{3.54}$$

### 3.4.2 Amplitudes in a frame in which the nucleus is slightly boosted

We now investigate the case in which the nucleus is slightly boosted, in such a way that

$$1 \ll y_0 \ll (x - X_y)^2. \tag{3.55}$$

Due to the reversed ordering between $y_0$ and $(x - X_y)^2$, some approximations at the very beginning of the calculations in the previous case are no longer valid in this type of frames. Therefore, this case is much more troublesome to deal with. We shall demonstrate how the estimations of $T_1$ and $T_2$ go, which will enable us to figure out how dominant configurations look like in this frame.

As usual, we will start by the calculation of the forward elastic scattering amplitude $T_1$. Bearing in mind that the configurations with a non-zero overlap with the saturation region of the nucleus contribute sub-dominantly, the overlap $I(y_0; \delta, y_1)$ is then small. So we can expand the exponential

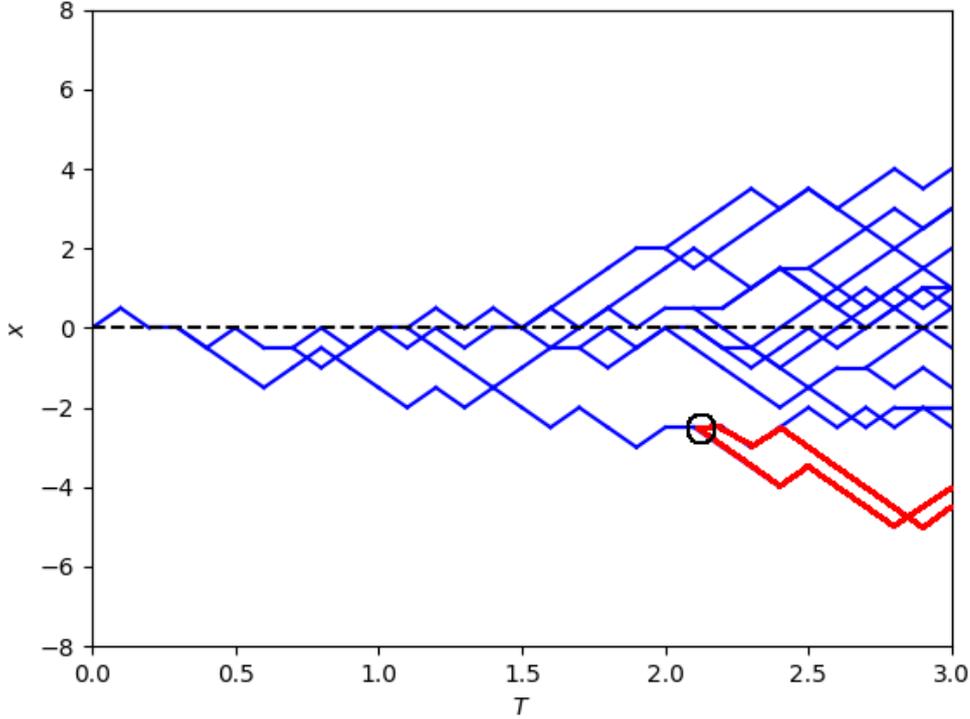

Figure 3.3: A realization of a branching random walk starting from a particle at $(T, x) = (0, 0)$ evolves up to $T = 3$. The branching of the last common ancestor of two leftmost particles is circled. The red lines represent the genealogical trajectories of the latter particles up to the last common ancestor.

in Eq. (3.23) to get

$$T_1(x, y) = \int_{y_0}^{y} dy_1 \int_0^{\delta_0} d\delta \, p(\delta, \tilde{y}_1) I(y_0; \delta, y_1).$$ (3.56)

Replacing the overlap $I$ by its expression, we obtain

$$T_1(x, y) = \frac{CC_1C_2}{y_0^{3/2}} e^{-\gamma_0(x - X_y)} \int_{y_0}^{y} dy_1 \int_0^{\delta_0} \delta d\delta \int_{\tilde{X}_{\tilde{y}_0 - \tilde{y}_1} + \Xi_{\delta, \tilde{y}_1}}^{\infty} dx'(x' - X_{y_0}) \underbrace{\exp\left[-\frac{(x' - X_{y_0})^2}{2\chi''(\gamma_0)y_0}\right]}_{\text{(I)}}$$

$$\times \left[\frac{y}{\tilde{y}_1(\tilde{y}_0 - \tilde{y}_1)}\right]^{3/2} (x' - \tilde{X}_{\tilde{y}_0 - \tilde{y}_1} - \Xi_{\delta, \tilde{y}_1}) \underbrace{\exp\left[-\frac{(x' - \tilde{X}_{\tilde{y}_0 - \tilde{y}_1} - \Xi_{\delta, \tilde{y}_1})^2}{2\chi''(\gamma_0)(\tilde{y}_0 - \tilde{y}_1)}\right]}_{\text{(II)}} \underbrace{\exp\left[-\frac{\delta^2}{2\chi''(\gamma_0)\tilde{y}_1}\right]}_{\text{(III)}}.$$ (3.57)

It is unfeasible to compute these nested integrals in a relatively straightforward way. Instead, we will exploit the Gaussian factors (I), (II) and (III) whose presence sets effective cutoffs, and hence, helps define the dominant contribution in the asymptotic limit. The first Gaussian factor (I) implies that the subdomain such that $(x' - X_{y_0}) \lesssim \sqrt{y_0}$ will contribute dominantly. Since $x'$ is larger than the leftmost tip (or the largest dipole) in the dipole distribution, the dominant contribution to $T_1$ comes from fluctuations such that the "distance" between the largest tip of the distribution and the top of the nucleus amplitude $\Delta(y_0; \delta, y_1)$ is also at most of order $\sqrt{y_0}$. Since $\Delta(y_0; \delta, y_1) \approx x - X_y - \delta$, the size of relevant fluctuations should be $\delta \sim (x - X_y)$, up to $\mathcal{O}(\sqrt{y_0})$. Combining with the factor (II), it is required that $\tilde{y}_0 - \tilde{y}_1 \lesssim y_0$, i.e., $\tilde{y}_1 \simeq \tilde{y}_0$ up to corrections of order $y_0$. Now, since $\delta \sim (x - X_y)$ and the initial onium $x$ is chosen to stay in the scaling region $(x - X_y) \ll \sqrt{y} \sim \sqrt{y_0}$, the Gaussian factor (III) can be safely set to unity.

Since the $y_1$ dependences in $\Delta$, or $\delta_0$, and $\widetilde{X}_{\tilde{y}_0 - \tilde{y}_1} + \Xi_{\delta, \tilde{y}_1}$ appear in the form of logarithmic terms, which are assumed to much smaller than any rapidity scales in the calculation, and since there is no exponential which can enhance those log contributions, we can replace $\delta_0$ by $(x - X_y)$ and $\widetilde{X}_{\tilde{y}_0 - \tilde{y}_1} + \Xi_{\delta, \tilde{y}_1}$ by $(x + \widetilde{X}_{\tilde{y}_0} - \delta)$. Then we can first perform the integral over $y_1$, which reads

$$\int_{y_0}^{\infty} \frac{dy_1}{(y_1 - y_0)^{3/2}} \left[ x' - (x + \widetilde{X}_{\tilde{y}_0} - \delta) \right] \exp \left[ - \frac{[x' - (x + \widetilde{X}_{\tilde{y}_0} - \delta)]^2}{2\chi''(\gamma_0)(y_1 - y_0)} \right] = \sqrt{2\pi \chi''(\gamma_0)}, \qquad (3.58)$$

where we have used the fact that $(\tilde{y}_1/y) \sim (\tilde{y}_1/\tilde{y}_0) \sim 1$, and hence, the integral is dominated by the region $y_1 \sim y_0$. Consequently, we can replace the upper bound $y$ by $\infty$.

In the next step, the integration over $x'$ is simple to perform,

$$\int_{x + \widetilde{X}_{\tilde{y}_0} - \delta}^{\infty} dx' (x' - X_{y_0}) \exp \left[ - \frac{(x' - X_{y_0})^2}{2\chi''(\gamma_0)y_0} \right] \simeq \int_{x - X_y - \delta}^{\infty} d\bar{x} \, \bar{x} \exp \left[ - \frac{\bar{x}^2}{2\chi''(\gamma_0)y_0} \right]$$
$$= \chi''(\gamma_0)y_0 \times \exp \left[ - \frac{(x - X_y - \delta)^2}{2\chi''(y_0)y_0} \right]. \qquad (3.59)$$

The only remaining intergration to perform is the integral over the fluctuation size $\delta$, which reads

$$\int_0^{x - X_y} d\delta \, \delta \exp \left[ - \frac{(x - X_y - \delta)^2}{2\chi''(\gamma_0)y_0} \right] \approx \sqrt{\frac{\pi}{2}} \sqrt{\chi''(\gamma_0)y_0} \times (x - X_y). \qquad (3.60)$$

We see that the integral over $\delta$ is dominated by a window of size $\sim \sqrt{y_0}$ near the upper bound $(x - X_y)$, which brings about a factor $\sqrt{y_0}$. Gathering all factors together, we get

$$T_1(x, y)|_{1 \ll y_0 \ll (x - X_y)^2} \simeq C C_1 C_2 \pi [\chi''(\gamma_0)]^2 (x - X_y) e^{-\gamma_0(x - X_y)}, \qquad (3.61)$$

which is perfectly identical to the expression (3.41) for $T_1$ in the case of the nucleus being highly boosted,

$$T_1(x, y)|_{1 \ll y_0 \ll (x - X_y)^2} = T_1(x, y)|_{1 \ll (x - X_y)^2 \ll y_0}. \qquad (3.62)$$

Therefore, we have verified explicitly that boost invariance actually holds in our phenomenological model.

It would be interesting to note that, while the form of $T_1$ is fully preserved, the physical pictures in the two frames are very dissimilar. In the current case with the slightly-boosted nucleus, the above analysis of the dominant integration domain implies that the fluctuation typically occurs late in the evolution of the onium, at a rapidity $\tilde{y}_1$ close to $\tilde{y}_0$, in a window of order $y_0$. The small front from the fluctuation is, hence, developed in a window of that size, in such a way that the overlap with the nucleus amplitude, whose scaling area has size of order $\sqrt{y_0}$, should be significant. This is necessary since the fluctuation needs to extend far out of the "mean field" region and thus, requires a large rapidity range to develop. The size of the front which results from the fluctuation is then of order $\sqrt{y_0}$, which is just the right size to have an optimal overlap with the front of the nucleus.

Returning to the previous case of the highly-boosted nucleus, we already pointed out that the fluctuation is likely to occur in the early state of the evolution ($\tilde{y}_1 \ll \tilde{y}_0$), and hence the fluctuation has a window of order $\tilde{y}_0 - \tilde{y}_1 \simeq \tilde{y}_0$ to develop its front. This is also the most favorable shape, since in this case, $y_0 \sim \tilde{y}_0$. Therefore, in both cases, the dominant configurations are selected by the universal requirement: the overlap between the dipole density and the nucleus amplitude should be optimized.

Now we move on to the calculation of the amplitude $T_2$. In this case, the expansion of the exponential in Eq. (3.24) is not usable, since we will loose the control of the overall constant factor multiplying the leading term. It turns out that the direct treatment for the integrals in Eq. (3.24) does not look possible.

Physically, to have at least two scatterings, the fluctuation should be typically pushed as far as the largest dipole in the distribution gets close to the top of the nucleus front, such that the overlap $I(\delta, y_1) \sim 1$. The typical values for the fluctuation is then limited in a narrow range of order unity. Consequently, the integral over the fluctuation's size $\delta$ would not generate a factor $\sqrt{y_0}$ as in the calculation for $T_1$ presented above. Therefore, one could guess the following parametric form of $T_2$:

$$T_2(x, y; y_0)|_{1 \ll y_0 \ll (x - X_y)^2} \sim \frac{T_1(x, y)}{\sqrt{y_0}}. \tag{3.63}$$

For a better estimation, we may use the fact that the distribution $G/T_1$ is boost invariant, and the integration of this ratio over the rapidity $y_1$ will give the ratio $T_2/T_1$. With Eq. (3.51), the latter reads

$$\frac{T_2(x, y; y_0)}{T_1(x, y)} = \frac{1}{\gamma_0} \frac{1}{\sqrt{2\pi \chi''(\gamma_0)}} \int_{y_0}^{y} dy_1 \left[ \frac{y}{y_1(y - y_1)} \right]^{3/2} \exp\left[ -\frac{(x - X_y)^2}{2\chi''(\gamma_0)(y - y_1)} \right]. \tag{3.64}$$

Since $\tilde{y}_1 \sim \tilde{y}_0 \sim y$, we can replace $(y - y_1)$ by $y$ in the above integral. The integral over $y_1$ is then trivial, and we obtain

$$\frac{T_2(x, y; y_0)}{T(x, y)} \bigg|_{1 \ll y_0 \ll (x - X_y)^2} \simeq \frac{1}{\gamma_0} \sqrt{\frac{2}{\pi \chi''(\gamma_0)}} \frac{1}{\sqrt{y_0}}. \tag{3.65}$$

The result has the same parametric form as that of the guess Eq. (3.63), but the overall constant is completely determined.

More generally, we can do the integration Eq. (3.64) in an exact way. By the change of variable

$u = y/(y - y_1)$, we have

$$
\begin{aligned}
\frac{T_2(x,y;y_0)}{T_1(x,y)} &= \frac{1}{\gamma_0} \frac{1}{\sqrt{2\pi\chi''(\gamma_0)}} \frac{1}{\sqrt{y}} \int_{y/\tilde{y}_0}^{\infty} \frac{u\,du}{(u-1)^{3/2}} \exp\left[-\frac{(x-X_y)^2}{2\chi''(\gamma_0)y}u\right] \\
&= \frac{1}{\gamma_0} \frac{1}{\sqrt{2\pi\chi''(\gamma_0)}} \frac{\exp\left[-\frac{(x-X_y)^2}{2\chi''(\gamma_0)y}\right]}{\sqrt{y}} \int_{y_0/\tilde{y}_0}^{\infty} dt \left(\frac{1}{t^{1/2}} + \frac{1}{t^{3/2}}\right) \exp\left[-\frac{(x-X_y)^2}{2\chi''(\gamma_0)y}t\right] \\
&= \frac{1}{\gamma_0} \frac{1}{\sqrt{2\pi\chi''(\gamma_0)}} \frac{\exp\left[-\frac{(x-X_y)^2}{2\chi''(\gamma_0)y}\right]}{\sqrt{y}} \left\{ \frac{\sqrt{2\pi\chi''(\gamma_0)y}}{x-X_y} \mathrm{erfc}\left(\frac{x-X_y}{\sqrt{\chi''(\gamma_0)y}}\sqrt{\frac{y_0}{\tilde{y}_0}}\right) \right. \\
&\quad \left. + 2\exp\left[-\frac{(x-X_y)^2}{2\chi''(\gamma_0)y}\frac{y_0}{\tilde{y}_0}\right] \sqrt{\frac{\tilde{y}_0}{y_0}} - \frac{\sqrt{2\pi}(x-X_y)}{\sqrt{2\chi''(\gamma_0)y}} \mathrm{erfc}\left(\frac{x-X_y}{\sqrt{\chi''(\gamma_0)y}}\sqrt{\frac{y_0}{\tilde{y}_0}}\right) \right\}.
\end{aligned}
\tag{3.66}
$$

As we are in the scaling region, $(x-X_y)^2 \ll y$, we can replace the complementary error functions $\mathrm{erfc}(u)$ and the exponential functions appearing on the above expression by 1 (actually, the exponential outside the curly bracket should disappear if we include the same factor in the expression of $T_1$). Also, due to the same reason, the last term is negligible compared to the first two terms in either cases. Finally, for the leading contribution, we get

$$
\frac{T_2(x,y;y_0)}{T_1(x,y)} \simeq \frac{1}{\gamma_0(x-X_y)} + \frac{1}{\gamma_0}\sqrt{\frac{2}{\pi\chi''(\gamma_0)}}\frac{1}{\sqrt{y_0}}\sqrt{\frac{\tilde{y}_0}{y}}.
\tag{3.67}
$$

In the first case, $1 \ll (x-X_y)^2 \ll y_0$, we recover Eq. (3.45) (the first term in Eq. (3.67)); while for the opposite ordering, we get Eq. (3.65) (the second term in Eq. (3.67) with $\tilde{y}_0 \approx y$).

### 3.4.3 Nucleus at rest

The forward elastic scattering amplitude $T_1$ was already analyzed in the rest frame of the nucleus ($y_0 = 0$) in Ref. [20]. A crucial point is that, unlike the above two frames, the nucleus in this case has not developed a universal front, and is characterized by a steep amplitude which can be approximated by a step function (see the initial conditions for the BK equation in the previous chapter). The scattering amplitude of a dipole becomes very small as soon as the size of this dipole gets smaller than the inverse saturation scale $1/Q_A$. Consequently, the overlap between the small front developed from the fluctuation at a rapidity $\tilde{y}_1$ ($\tilde{y}_0 - \tilde{y}_1 \ll 1$) and the tail with the nuclear amplitude is negligible. Instead, the dominant contribution should come from a fluctuation occuring at the very end of the evolution of the onium, which is referred to as the tip fluctuation. Therefore, the formulation for $T_1$ reads

$$
T_1(x,y)|_{y_0=0} = \int dx'\, p(x+\widetilde{X}_A - x', y) T_1(x', 0),
\tag{3.68}
$$

where $x + \widetilde{X}_y - x'$ is the probability of having dipoles of log size smaller than $x'$ in the distribution, i.e., the size of the fluctuation. We can replace $T(x', 0)$ by a Heaviside distribution supporting values of $x'$ such that $x' \leq \ln(1/Q_A)$,

$$T_1(x', 0) \simeq \Theta\left(-\ln\frac{1}{Q_A}\right).$$ (3.69)

Eq. (3.68) can be rewritten as

$$
\begin{aligned}
T_1(x, y)|_{y=0} &\propto \int_{-\infty}^{\ln\frac{1}{Q_A}} dx' (x + \widetilde{X}_y - x') e^{-\gamma_0(x + \widetilde{X}_y - x')} \exp\left[-\frac{(x + \widetilde{X}_y - x')^2}{2\chi''(\gamma_0)y}\right] \\
&\underset{\sim}{\propto} c\ (x - X_y) e^{-\gamma_0(x - X_y)},
\end{aligned}
$$ (3.70)

which can be obtained easily by noticing that the integral is dominated by a narrow range of $x'$ near the upper bound; hence, we can set the Gaussian factor to unity. The above result is what is expected at the parametric level. However it is not possible to relate the constant $c$ to the constants $C$, $C_1$ and $C_2$ defined in the phenomenological model before. This is because the latter are unambiguosly defined for evolved universal fronts, once a convention for the definition of the front position or saturation scale has been chosen. However, the transition from the initial condition to the well-defined front is not controlled analytically in the very early stage of the evolution.

Unlike $T_1$, which can be estimated at the parametric level, $T_2$ and $G$ cannot be calculated in this frame. Indeed, their evaluation requires a precise understanding of the particle distribution near the tip. However, the fluctuation in the current case is typically at the end of the evolution of the onium, and hence, does not have enough rapidity to develop into a well-defined front. Its particle content is still a puzzle, which in turn prevents our effort to calculate $T_2$ and $G$.

We have derived the expressions for the ratios $T_2/T_1$ and $G/T_1$ with well-defined overall constant factors, based on the assumptions of our phenomenological model. Whether these constants are the correct ones for branching random walks and for the QCD color dipole model depends on the ability of the phenomenological picture to capture accurately enough the features of the latter models. Therefore, we shall perform numerical calculations to check the validity of the model.

## 3.5 Numerical evaluation of the ancestry distribution

In this section, we are going to check the result we have obtained from the phenomenological model for the distribution of the branching rapidity of the last common ancestor by solving numerically the exact evolution equations governing it. We will consider both the QCD evolution equation (see Eq. (3.5)) and its corresponding version for branching random walk models. While the former can give us a visualization of the numerical solutions in comparison to the analytical result, the latter can help us assess quantitatively how the solutions approach the asymptotics, since we can technically go to much higher rapidities in this case. The employment of the BRW model in this check is motivated by the property that the asymptotics is the same for all models in the universality class of branching-diffusion processes, as discussed in the previous chapter. The only difference when switching between models is to change the numerical values of the kernel parameters $\gamma_0$, $\chi''(\gamma_0)$ and $\chi'(\gamma_0)$, which depend on the detailed elementary processes.

In order to compare more easily different values of $y$, it is useful to introduce the rapidity overlap $q \equiv \tilde{y}_1/y$. Its distribution is simply given by Eq. (3.51) up to a change of variable,

$$\pi_\infty(q; x, y) = \frac{1}{\sqrt{y}} \frac{1}{\gamma_0} \frac{1}{\sqrt{\pi\chi''(\gamma_0)}} \frac{1}{q^{3/2}(1-q)^{3/2}} \exp\left[-\frac{(x-X_y)^2}{2\chi''(\gamma_0)y}\frac{1}{q}\right], \qquad (3.71)$$

where the $\infty$ subscript is to remind that this distribution is valid for asymptotic values of $y$. In most cases, we will restrict ourselves to values of $q$ such that $q \gg (x - X_y)^2/[2\chi''(\gamma_0)y]$ (the right-hand side is small in the scaling region we are interested in). Therefore, we can neglect the Gaussian factor in Eq. (3.71), and the $y$ dependence simply appears as a prefactor.

### 3.5.1 QCD evolution equations

As being discussed previously, the probability $G$ solves the evolution equation (3.5), which depends on the solutions to the BK equation for either the S-matrix element $S$ (Eq. (2.2)) or the forward amplitude $T_1$ (2.3), in the framework of the QCD dipole model. While it is impossible to find exact solutions to such equations, we can solve them numerically. The strategy is simple: we evolve the S-matrix element $S$ from the initial condition at zero rapidity to the rapidity $y_1$, then use this solution to construct the initial condition for $G$ according to Eq. (3.6). We then further advance $G$ to the total rapidity $y$ by the virtue of Eq. (3.5), and eventually divide it by $T_1$ to get the desired distribution. For the initial condition for $T_1$, we choose the MV model (2.4), with parameters set as follows:

$$Q_A = 1 \text{ GeV}, \quad \Lambda_{QCD} = 0.2 \text{ GeV}. \qquad (3.72)$$

In order to satisfy the condition $1 \ll x - X_y \ll \sqrt{\chi''(\gamma_0)y}$ (here we recover the factor $\chi''(\gamma_0)$), or $1 \ll \ln[1/(r^2 Q_s^2(y))] \ll \sqrt{\chi''(\gamma_0)y}$, we pick the onium size $r$ such that

$$\ln\frac{1}{r^2 Q_s^2(y)} = \sqrt{\kappa}\, y^{1/4}, \qquad (3.73)$$

with a constant $\kappa$ which enables us to move more or less deep into the scaling region by varying its value. This choice of $r$ is proportional to the geometric average of the two boundaries of the scaling region.

We plot the overlap distribution $\pi(q)$ rescaled by the factor $\sqrt{y}$ for different values of the total rapidity $y$ in Fig. 3.4, with $\kappa = 20$. The numerical solutions are shown to have the similar shape of the expected asymptotics, and get closer to it when increasing the total rapidity $y$. However, the convergence seems to be very slow. Unfortunately, we cannot go to a very high rapidity in this case due to technical reasons. Therefore, it is a big challenge to test the convergence to the analytical asymptotic result just by using the numerical solutions to the QCD evolution equations. Instead, we will employ the asymptotic universality of branching-diffusion processes, see the next section.

In addition to check the convergence to the asymptotics, we also plot the distribution for dif-ferent values of $\kappa$, i.e. for different onium sizes, for a fixed value of the total rapidity to see its behavior when approaching the right boundary of the scaling region (see Fig. 3.5). We see that the distribution in the region of small values of the overlap $q$, corresponding to small $\tilde{y}_1$, is suppressed. In addition, this suppression seems to be similar to a Gaussian suppression, which appears in Eq. (3.71). These observations indicate qualitatively that the analytical asymptotic expression found from the phenomenological model for the ancestry distribution is reasonable.

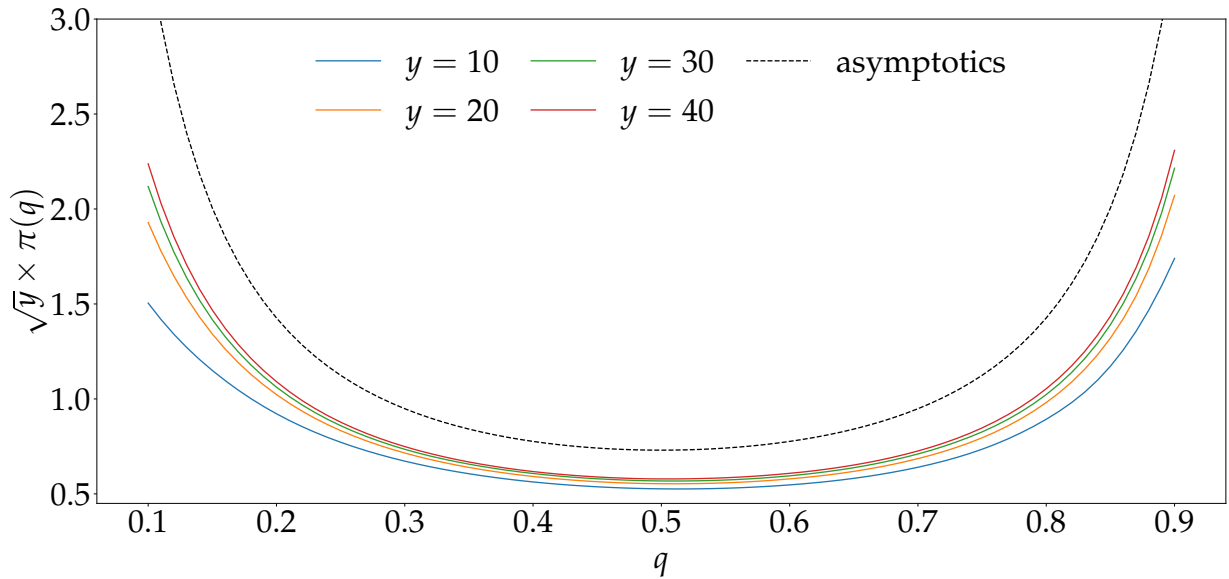

Figure 3.4: Distribution of the rapidity overlap at different values of the total rapidity from the solutions to the QCD evolution equations (full lines). The onium size $r$ is picked according to the condition $\ln[1/(r^2 Q_s^2(y))] = \sqrt{20}y^{1/4}$. The expected asymptotics is also superimposed for comparison (dashed line).

### 3.5.2 A branching random walk model

To see better how the distribution approaches its asymptotics, we consider a branching random walk in discrete space and time, give by a lattice with parameters $(\delta x, \delta y)$, which is defined as follows. After an evolution step $\delta y$, a particle on site $x$ can jump to the site on the left $(x - \delta x)$, or on the right $(x + \delta x)$ with respective probabilities $\frac{1}{2}(1 - \delta y)$. Otherwise, it may branch into two particles on the same site $x$ with probability $\delta y$.

This is exactly the BRW model defined to construct and to implement the Monte Carlo algorithm in the previous chapter. This model differs from the QCD dipole evolution by the facts that in the latter, the diffusion and the branching occur at the same time through a single process, and that QCD is a theory in the continuum. However, these two points should not alter the asymptotic behavior of the distribution we are considering.

By the previous discusssions, the equivalence to the forward elastic scattering amplitude $T_1$ in QCD is the probability $\mathcal{T}_1(x, y)$ to find at least one particle to the right of the site $x$ at the rapidity $y$.

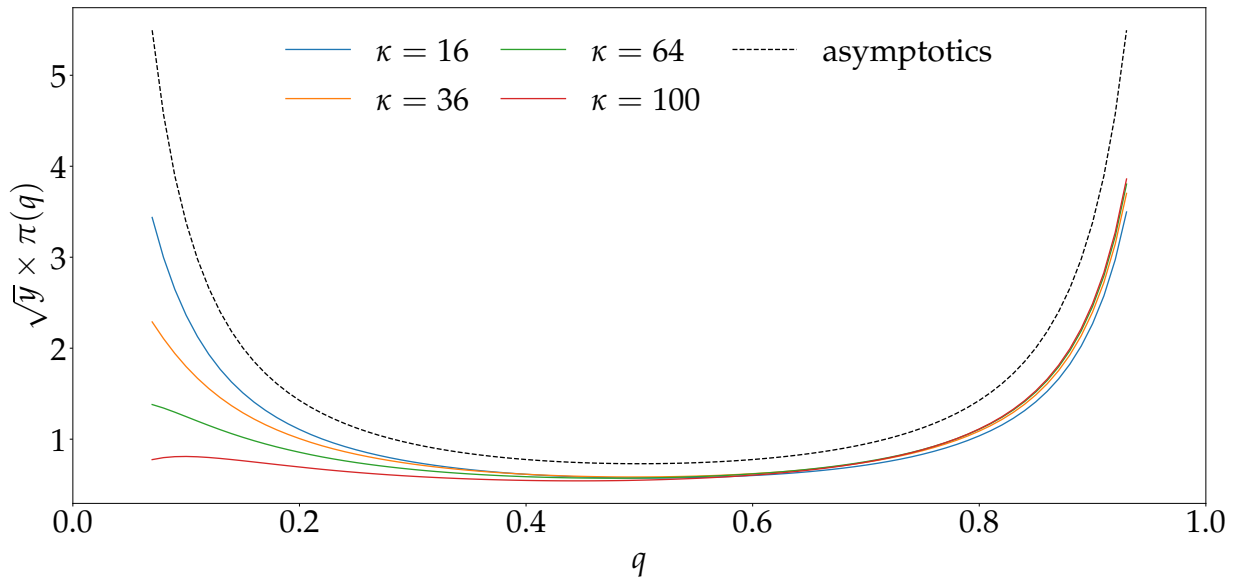

Figure 3.5: Distribution of the rapidity overlap at different sizes of the onium (full lines), when the total rapidity is fixed at $y = 40$. The expected asymptotics is also superimposed for comparison (dashed line).

The latter solves the finite difference equation (2.51), which can be rewritten in our new notations as

$$\mathcal{T}_1(x, y + \delta y) = \frac{1}{2}(1 - \delta y)\left[\mathcal{T}_1(x - \delta x, y) + \mathcal{T}_1(x + \delta x, y)\right] + \delta y \mathcal{T}_1(x, y)\left[2 - \mathcal{T}_1(x, y)\right], \quad (3.74)$$

which is analogous to the T-type BK equation in QCD and the FKPP equation for branching diffusion. The initial condition is given by the Heaviside distribution $\mathcal{T}(x, 0) = \Theta(-x)$. The branching diffusion kernel of Eq. (3.74) linearized near the unstable fixed point $\mathcal{T} \sim 0$ accepts $e^{\gamma x}$ as the eigenfunction, with the corresponding eigenvalue given by

$$\chi(\gamma) = \frac{1}{\delta y} \ln\left[\frac{1}{2}(1 - \delta y)(e^{\gamma \delta x} + e^{-\gamma \delta x}) + 2\delta y\right]. \quad (3.75)$$

Now we turn into the genealogical problem. The equivalence to the QCD evolution equation for the probability $G$ in this BRW model reads

$$G(x, y + \delta y; y_1) = \frac{1}{2}(1 - \delta y)\left[G(x - \delta x, y; y_1) + G(x + \delta x, y; y_1)\right] + 2\delta y G(x, y; y_1)S(x, y), \quad (3.76)$$

with the initial condition given by

$$G(x, y_1; y_1) = T^2(x, y_1). \quad (3.77)$$

$G(x, y; y_1)$ in this model is precisely the probability for the last common ancestor of all particles to

the right of $x$ at the rapidity $y$ to branch at the rapidity $y_1$ (see Fig. 3.6).

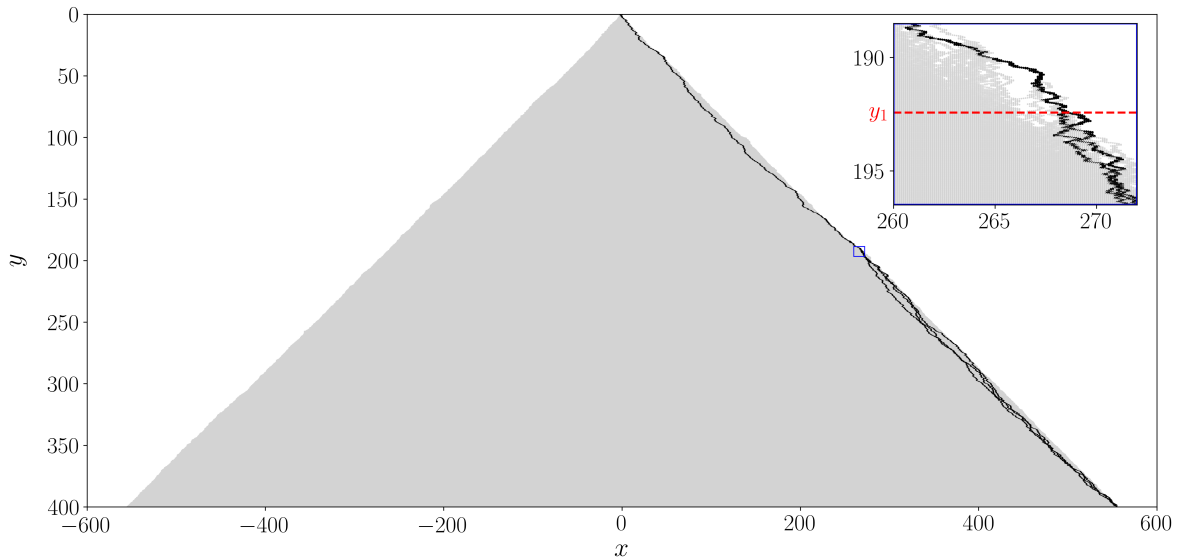

Figure 3.6: A realization of the BRW model described in this section evolving up to $y = 400$ in which there are particles to the right of the position $X \simeq X_{y=400} + \sqrt{2} \times 400^{1/4}$. The grey zone is the set of occupied lattice sites for all values of the rapidity. The black lines represent the worldlines of all the particles that end up with a position to the right of $X$ (including $X$) at the final rapidity $y = 400$. The last common ancestor of these particle branches at the rapidity $y_1 \approx 192.43$. The inset is a zoom on the region around the branching rapidity $y_1$, which illustrates that this common ancestor is from a large fluctuation, as assumed in the phenomenological model. This realization is generated using the algorithm presented in the previous chapter. Figure is adapted from Ref. [26].

For the implementation, we set the lattice parameters to the following values:

$$\delta x = 0.1, \quad \delta y = 0.01. \tag{3.78}$$

With this choice of the lattice parameters, the values of $\gamma_0$, $\chi'(\gamma_0)$ and $\chi''(\gamma_0)$, where $\gamma_0$ solves $\gamma_0 \chi'(\gamma_0) = \chi(\gamma_0)$, are given by

$$\gamma_0 = 1.43195 \cdots, \quad \chi'(\gamma_0) = 1.39436 \cdots, \quad \chi''(\gamma_0) = 0.96095 \cdots. \tag{3.79}$$

As in the previous section, in order to optimally satisfy the condition that $x$ is well located in the scaling region, $1 \ll x - X_y \ll \sqrt{y}$, we set $x$ to a value $X$ such that

$$X \simeq X_y + \sqrt{\kappa} y^{1/4}, \tag{3.80}$$

(see Eq. (3.73)). The constant $\kappa$ is now picked in the set $\{1, 2, 4\}$. Since we are working on a lattice and the right hand side is generally off the lattice, it is not possible to set $X$ exactly to the value on the right hand side. Instead, we pick the closest lattice site to the left of the latter.

The current model allows us to boost $y$ up to $\mathcal{O}(10^6)$. In Fig. 3.7, we plot the distribution

$\pi(q)$ of the rapidity overlap $q$ at a finite rapidity, rescaled by $\sqrt{y}$, for different rapidities and for $\kappa = 2$. The asymptotic distribution $\pi_\infty(q)$ in Eq. (3.71), with the Gaussian factor suppressed, is also superimposed.

We see that, when increasing the rapidity, the distribution obtained from the numerical solutions get closer to its expected asymptotics. To perceive this convergence quantitatively, we pick a value of $q$ and plot the quantity $1 - \pi_y(q)/\pi_\infty(q)$, which is expected to tend to 0 when $y \to \infty$, as a function of $\ln y/\sqrt{y}$. For $q = 0.5$, the result is shown in Fig. 3.8. (We also checked for some other values of $q$ which are not close to 0 and 1, and got the same results.)

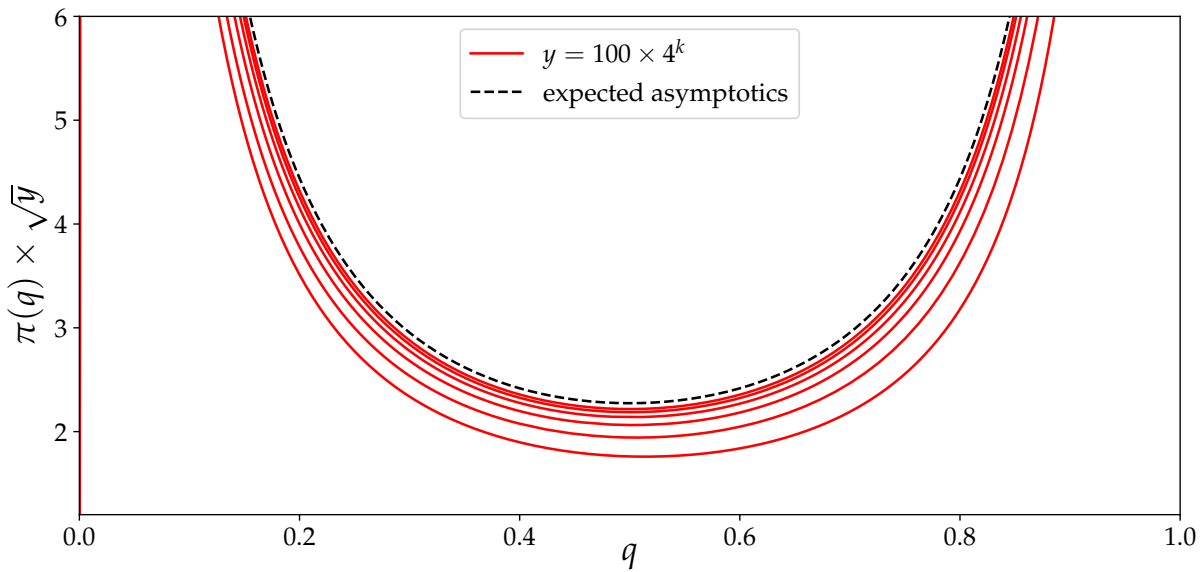

Figure 3.7: Distribution of the rapidity overlap for different finite rapidities $y = 100 \times 10^k$ (full lines), with $k$ running from 1 to 6 from bottom to top. $X$ is set to the value $X_y + \sqrt{2}y^{1/4}$ in each case. The asymptotic distribution (without Gaussian factor) is also plotted for comparison (dashed line). Figure is adapted from Ref. [26].

For a better estimation, we fit the following function to the numerical data points:

$$1 - \frac{\pi_y(q = 0.5)}{\pi_\infty(q = 0.5)} \underset{\text{fit}}{\simeq} a + \frac{b_{11} \ln y + b_{12}}{\sqrt{y}}. \tag{3.81}$$

The values of the parameters $a$, $b_{11}$ and $b_{12}$ obtained from the fit is shown in Table 3.1. They appear to be reasonable: $a$ is close to zero by the order of one percent, while $b_{11}$ and $b_{12}$ are of order 1. Therefore, we conclude that our analytical formula in Eq. (3.71) is well-supported by the numerical calculation.

From these numerical results, it is also important to note that the finite-$y$ corrections are significantly sizable and the convergence to the asymptotics is slow. Figure 3.8 and the fit seem to indicate that the leading finite-rapidity correction to the asymptotic distribution $\pi_\infty(q)$ may take the form of a multiplicative factor $(1 + \text{const} \times \ln y/\sqrt{y})$. However, there is no theory up-to-date that may enable us to understand the form of the distribution beyond the asymptotic level.

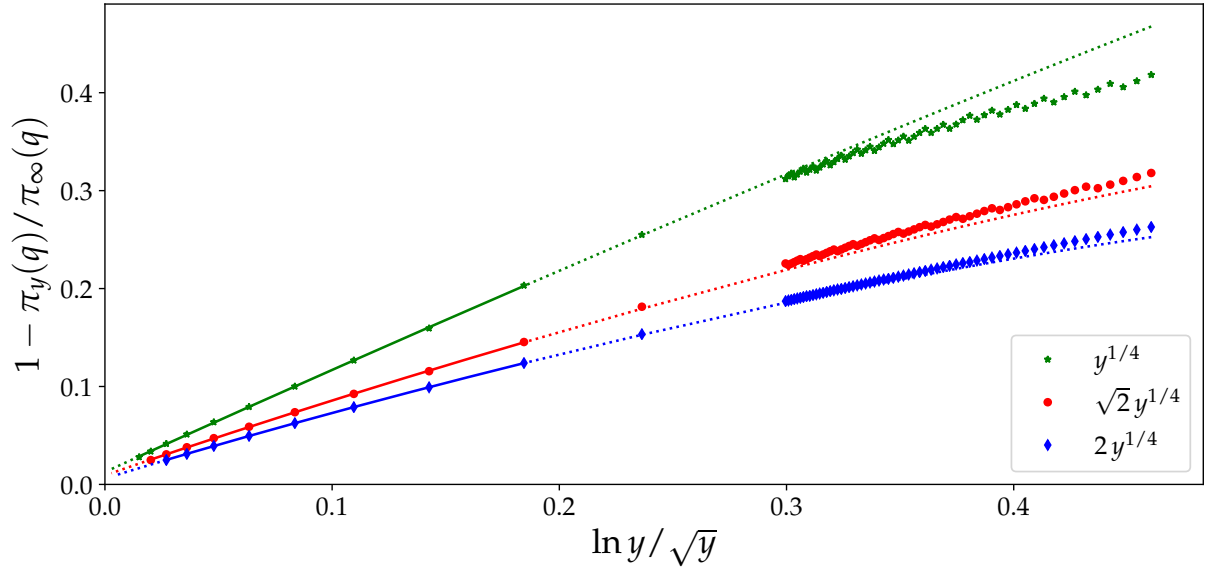

Figure 3.8: The unity complement of the ratio between the finite-rapidity and the asymptotic distributions at the overlap $q = 0.5$ for values of $\kappa \in \{1, 2, 4\}$ and for different values of the total rapidity $y$ as a function of $\ln y/\sqrt{y}$. The points come from the numerical solutions of Eqs. (3.74) and (3.76). The continuous lines represent the fit according to the function (3.81), while the dotted lines are extrapolations outside the domain in which the fit is performed. The oscillations observed in the small-$y$ tail is due to the fact that $X$ is not exactly at $X_y + \sqrt{\kappa}y^{1/4}$, but at the nearest site compared to the latter. Figure is adapted from Ref. [26].

Table 3.1: Values of the parameters in Eq. (3.81) obtained from the fit to the numerical points shown in Fig. 3.8.

| $X - X_y$ | $a$ | $b_{11}$ | $b_{12}$ |
|---|---|---|---|
| $y^{1/4}$ | $1.26 \times 10^{-2}$ | $1.10$ | $-0.52$ |
| $\sqrt{2}y^{1/4}$ | $0.91 \times 10^{-2}$ | $0.90$ | $-1.17$ |
| $2y^{1/4}$ | $0.59 \times 10^{-2}$ | $0.81$ | $-1.28$ |

## 3.6 Summary

In the scattering of an onium off a large nucleus at a large rapidity $Y$, in a frame in which the onium is boosted to a large rapidity $\tilde{Y}_0$, the latter interacts via its highly-evolved quantum state which can be represented by a set of color dipoles of various sizes, which are randomly distributed. As the interaction between each elementary dipole with the nucleus is encoded in a scattering amplitude whose values monotonically decrease from 1 to 0 as the dipole size gets smaller, it turns out that only a subset of those dipoles actually scatter off the nucleus in a particular realization. Requiring to have at least one dipole in such subsets defines the forward elastic scattering amplitude $T_1$. Another interesting quantity is the amplitude $T_2$ to have at least two dipoles in the set that scatter, which provides a measure of the correlations of the dipoles involved in the interaction. The detailed scattering configuration would depend upon the kinematical regime of interest.

In the current chapter, we have developed a phenomenological picture for the dipole distribution of a small onium evolved to a high rapidity in the scattering off a large nucleus. The picture involves of the mean-field evolution and a rare fluctuation occuring in the course of the evolution of the onium in such a way that it creates a large dipole of size much deviated from the typical value of the tip of the dipole distribution at the rapidity of the fluctuation. We have shown that the dominant pattern of the partonic configurations qualitatively contigents on the chosen reference frame, which is to guarantee that the overlap between the dipole distribution and the dipole-nucleus amplitude is optimal. In particular, if the nucleus is boosted to a rapidity $Y_0$ much larger than $\ln^2[r^2 Q_s^2(Y)]$, the fluctuation substantially takes place at the top of the evolution, and effectively shifts the whole distribution toward larger sizes. This type of fluctuations is therefore called the "front fluctuation". On the other hand, if the nucleus is less boosted, $1 \ll \bar{\alpha} Y_0 \ll \ln^2[r^2 Q_s^2(Y)]$, the fluctuation occurs preferably near the tip of the well-developed front of the onium at a rapidity $\tilde{y}_1$ close to the bottom of the evolution. Finally, in the extreme case when the nucleus stays at rest, the scattering is dominated by a large fluctuation at the very end of the evolution, which would create a dipole of size at least touching the inverse saturation scale of the nucleus.

Apart from the forward elastic scattering amplitude $T_1$, which is boost invariant, switching to a frame such that the fluctuation is sufficiently developed enables us to calculate the asymptotic amplitude of scatttering twice $T_2$ whose ratio to $T_1$ is determined with no free parameters,

$$\frac{T_2(r, Y; Y_0)}{T_1(r, Y)} \simeq \frac{1}{\gamma_0 \ln[1/(r^2 Q_s^2(Y))]} + \frac{1}{\gamma_0} \sqrt{\frac{2}{\pi \chi''(\gamma_0)}} \frac{1}{\sqrt{\bar{\alpha} Y_0}} \sqrt{\frac{Y - Y_0}{Y}}, \tag{3.82}$$

in which we have relaxed to the QCD variables.

More interestingly, we have deduced the full analytical expression for the distribution of the branching rapidity $Y_1$ of the last common ancestor of the set of dipoles that effectively interact with the nucleus at an asymptotic total rapidity $Y$. In the QCD variables, it reads

$$\frac{G(r, Y; Y_1)}{T_1(r, Y)} = \frac{1}{\sqrt{\bar{\alpha}}} \frac{1}{\gamma_0} \frac{1}{\sqrt{2\pi \chi''(\gamma_0)}} \left[ \frac{Y}{Y_1(Y - Y_1)} \right]^{3/2} \exp\left[ -\frac{\ln^2[r^2 Q_s^2(Y)]}{2\chi''(\gamma_0)\bar{\alpha}(Y - Y_1)} \right]. \tag{3.83}$$

This equation makes the main qualitative result for the chapter. The expressions in Eqs. (3.82) and (3.83) are valid for the onium size located in the scaling region, which is defined as

$$1 \ll \ln \frac{1}{r^2 Q_s^2(Y)} \ll \sqrt{\bar{\alpha} Y}. \tag{3.84}$$

In another aspect, we expect Eq. (3.83) to represent the distribution of the branching time of the last common ancestor of all particles that end up to the right of some predefined position $x$ in the scaling region for general one-dimensional branching random walk models, after the following identifications of time and space variables:

$$t \leftrightarrow \bar{\alpha}Y, \quad t_1 \leftrightarrow \bar{\alpha}Y_1,$$
$$x \leftrightarrow \ln[1/(r^2 Q_A^2)], \quad X_t \leftrightarrow \ln[Q_s^2(Y)/Q_A^2], \tag{3.85}$$

and with the kernel parameters depending on the specific elementary processes. Furthermore, we observed that our result coincides with a conjecture by Derrida and Mottishaw [85] for the distribution of the branching rapidity of the last common ancestor of the two particles picked in the tip. While they derived that distribution based on the generalized random energy model [97, 98], our result is from the phenomenological picture of branching random walks. Indeed our approach cannot be applied to the type of genealogical problems addressed by Derrida and Mottishaw to find the complete distribution, since we do not have an enough understanding on the distributions of particles near the tip and on their correlations. Finding a good description for the latter still needs more efforts.

In addition to the asymptotic understanding of the ancestry distribution, it is also important to develop a formalism which enables the calculation of finite-rapidity (time) corrections, as the convergence to the asymptotics appears to be very slow. It is of interest not only for particle physics but also for the investigation of branching diffusion processes in statistical physics, and therefore, an exciting challenge for further developments.

Last but not least, the probabilities $T_2$ and $G$ discussed in this chapter are related to the diffractive scattering cross sections for the diffractive dissociation of a small onium off a large nucleus. Understanding the latter is our initial motivation for the current project. We will discuss diffractive disscociation in the next chapter, with a reminiscence to those probabilities.

# Diffractive dissociation

## Contents

In the previous chapter, we introduced a phenomenological model of dipole distribution which enabled us to study the structure of the dipole evolution of a small onium in the scattering off a large nucleus. For the current chapter, we shall keep working with the nuclear scattering of small onia, focusing particularly on an analytical study of the diffractive dissociation process at asymptotics. In addition, the diffractive dissociation of a virtual photon will also be analyzed numerically based on QCD evolution equations aiming at producing predictions for future colliders. These are our recent results [27, 28] on diffraction.

## 4.1 Diffractive phenomena in high-energy scattering

In high-energy particle scattering, diffraction is defined to be a process in which there is an angular region with no particle produced in the final state, which is referred to as the rapidity gap [99, 100] (see also Ref. [101]). It has been observed in the scattering of hadrons and nuclei [102–107]. There were also direct evidences for diffraction in deep-inelastic electron-proton scattering at DESY HERA [108–110]: it was reported that about 10% of the inclusive DIS cross section is diffractive.

The history of diffraction in the physics of high-energy nuclear scattering can be traced back to the 1950s, when that term was first introduced in the literature [111–114]. Good and Walker [115] then provided a modern description for hadronic diffraction and established an elegant theoretical framework in which inelastic diffraction is related to the dispersion of the forward elastic scattering amplitude. In the mean time, it was also formulated in the framework of the Regge theory [116–119] in which diffraction is due to the exchanges of color singlet objects called "pomerons". The discussions in those studies all focused on soft diffraction. It was not until 1985 that the hard diffraction was first discussed by Ingleman and Schlein [120]. The striking experimental observations at DESY HERA [108–110] and Tevatron [106, 107], as mentioned above, then made an impressive boost to the physics of diffraction.

We are interested in hard diffraction in DIS, which is traditionally divided into two classes: the quasi-elastic scattering in which the diffractive system is typically a vector meson or a hadronized open quark-antiquark pair, and the diffractive dissociation in which the virtual photon is dissociated into an inclusive set of particles in the final state. While there has been a great advance in the study of the former recently (see e.g. Ref. [121] for a review), the diffractive dissocation has drawn less attention. To gain insights into the latter phenomenon is our spotlight in this chapter. As a major motivation, the detail investigation of diffraction, including diffractive dissociation, is a main goal at future electron-ion colliders [1, 3, 122].

Diffractive DIS can be described on the basis of the color dipole formulation, in which the virtual photon interacts via its quark-antiquark dipole state (onium). Nikolaev and Zakharov [31, 123, 124] were pioneers in using the color dipole approach for the inclusive diffractive dissociation of a virtual photon off a proton. Employing this approach together with a description of saturation effect, Golec-Biernat and Wusthoff [51, 52] were able to succesfully describe the HERA diffractive data. On a further development of the dipole formulation including the high-energy evolution, Mueller [32] showed that, at high energy and in the limit of large number of colors, the S-matrix of the hadronic scattering of the dipole state of the virtual photon is diagonal in the transverse size (see Chapter 1), which consequently enables to link the Good and Walker mechanism [115] to QCD to describe the hard diffractive dissociation in DIS [19, 23, 125, 126]. Based on this QCD color dipole model, Kovchegov and Levin [127–129] established an elegant formulation, which provide detailed predictions for diffractive cross sections in the scattering of an onium with the nucleus in the form of nonlinear evolution equations whose leading-order version was then studied in detail numerically [130–132].

Interestingly enough, the Kovchegov-Levin formulation enables us to address the distribution of rapidity gaps (see Eqs. (4.7) and (4.8) below), which shows up as an important observable. On the experimental side, the existence of rapidity gap is a typical signature to detect diffractive events. Also, since the partonic content of the hadrons has signature in the final state, such as the rapidity gap, such observable could provide indications on the microscopic mechanism of diffraction. As a matter of fact, a recent study [22, 23] has suggested a picture at the partonic-level for diffraction in the nuclear scattering of a small onium, from which the authors enabled to derive the asymptotic rapidity gap distribution.

In this chapter, we shall investigate the diffractive dissociation of a virtual photon off a large nucleus, focusing on the diffractive scattering cross sections and the rapidity gap distributions. Our aims are twofold. First, we would like to develop a theoretical formulation from which one can derive diffractive observables in an onium-nucleus scattering in a particular kinematic regime of interest. Second, we aim to produce predictions for the distribution of rapidity gaps in realistic kinematics of future electron ion colliders, based on the numerical solutions of the original Balitsky-Kovchegov and the Kovchegov-Levin equations at their next-to-leading versions. Both analyses are all based on the QCD color dipole picture, which will be recalled in the next section before coming to the main results.

## 4.2  Dipole formulation for diffractive dissociation in DIS

As discussed in Chapter 1, deep-inelastic virtual photon-nucleus scattering at high-energy can be conveniently described in a frame, e.g. the target rest frame, in which the virtual photon ($\gamma^*$) interacts with the nucleus ($A$) via its color-singlet quark-antiquark dipole state, i.e. onium. Such dipole picture allows for the following dipole factorization for the total cross section:

$$\sigma_{tot}^{\gamma^* A}(Q^2, Y) = \int d^2 r^\perp \int_0^1 dz \sum_{p=L,T;f} \left| \Psi_p^{\gamma^* \to q_f \bar{q}_f}(r, z, Q^2) \right|^2 \sigma_{tot}^{q\bar{q}A}(r, Y), \tag{4.1}$$

which is a virtuality ($Q^2$)-dependent and rapidity ($Y$)-dependent weighted average of the total cross section of the onium-nucleus scattering $\sigma_{tot}^{q\bar{q}A}$ over onium transverse sizes $r^\perp$ and over longitudinal momentum fractions $z$ of the virtual photon carried by the quark (or the antiquark) (see Eq. (1.70)). The rapidity $Y$ encodes the squared center-of-mass energy $\hat{s}$ of the process as $Y = \ln[(\hat{s} + Q^2)/Q^2]$. The weight is given by the probability density functions $\left| \Psi_{L,T}^{\gamma^* \to q_f \bar{q}_f}(r, z, Q^2) \right|^2$ of the quantum pair creation $\gamma^* \to q_f \bar{q}_f$ in the longitudinal (L) and the transverse (T) polarizations for a quark flavor $f$, whose expressions are given by [38]

$$\left| \Psi_L^{\gamma^* \to q_f \bar{q}_f}(r, z, Q^2) \right|^2 = \frac{\alpha_{EM} N_c}{2\pi^2} 4Q^2 z^2 (1-z)^2 e_f^2 K_0^2(r\alpha_f), \tag{4.2a}$$

$$\left| \Psi_T^{\gamma^* \to q_f \bar{q}_f}(r, z, Q^2) \right|^2 = \frac{\alpha_{EM} N_c}{2\pi^2} e_f^2 \left\{ \alpha_f^2 K_1^2(r\alpha_f) \left[ z^2 + (1-z)^2 \right] + m_f^2 K_0^2(r\alpha_f) \right\}, \tag{4.2b}$$

where $\alpha_{EM}$ is the electromagnetic coupling and $\alpha_f^2 \equiv Q^2 z(1-z) + m_f^2$, with $m_f$ and $e_f$ being the mass and the electric charge of a quark of flavor $f$, respectively. $K_{0,1}(x)$ are the modified Bessel functions of the second kind.

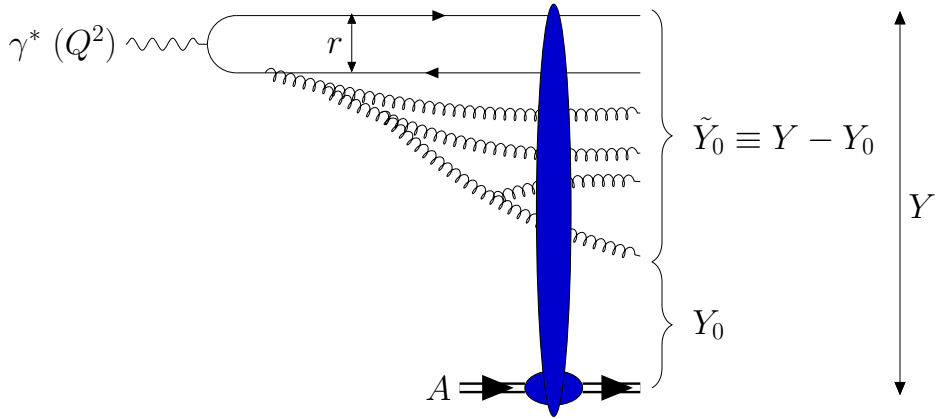

Figure 4.1: A diagrammatic visualization of the diffractive dissociation of a virtual photon in the dipole picture. The onium state dressed by quantum radiations of the virtual photon is, after the interaction with the nucleus, dissociated into a set of final-state particles, while the nucleus is kept intact. There is a gap in rapidity between the slowest particle in that set and the nucleus, which defines diffractive event. Figure is adapted from Ref. [28].

The diffractive dissociation can also be formulated within the dipole picture. The difference between diffraction and the generic DIS process is the fact that, while produced particles can distribute at any rapidity in the latter, there is a rapidity gap with no particle observed for the former case (see Fig. 4.1). This gap is deserved to the color-singlet nature of the overall gluonic exchange state in diffraction. Similar to Eq. (4.1), the diffractive cross section for the diffractive dissociation of a virtual photon with a minimal gap $Y_0$ ($0 < Y_0 \leq Y$) can be factorized as

$$\sigma_{diff}^{\gamma^* A}(Q^2, Y, Y_0) = \int d^2 r^\perp \int_0^1 dz \sum_{p=L,T;f} \left| \psi_p^f(r, z, Q^2) \right|^2 \sigma_{diff}^{q\bar{q}A}(r, Y, Y_0). \tag{4.3}$$

The factorizations in Eqs. (4.1) and (4.3) indicate that it is natural to analyse the nuclear scattering, including the diffractive dissociation, of an onium. Indeed, the latter is better controlled theoretically. And, due to the weight average, the behaviors of the nuclear scattering of the onium would be present in that of the virtual photon, which helps gain some insights into the latter process, at least at a qualitative level.

We thus now focus on the diffractive dissociation of an onium. As a reminder, the total cross section $\sigma_{tot}^{q\bar{q}A}(r, Y)$ is related to the forward nuclear elastic scattering amplitude $T_1(r, Y)$ (assuming the impact parameter independence) of an onium of size $r$ at the total rapidity $Y$ at a fixed impact parameter by $\sigma_{tot}^{q\bar{q}A}(r, Y) = \sigma_0 2 T_1(r, Y)$, where $\sigma_0 \sim \pi R_A^2$ ($R_A$ is nuclear radius) is a surface that stems from the integration over impact parameter. At high energy and in the limit of large $N_c$, $T_1$ solves the T-type BK equation given in Eq. (2.3), assuming large nucleus. We also hereafter denote by $\sigma_T(r, Y)$ the total cross section per unit impact parameter, $\sigma_T(r, Y) = 2T_1(r, Y)$.

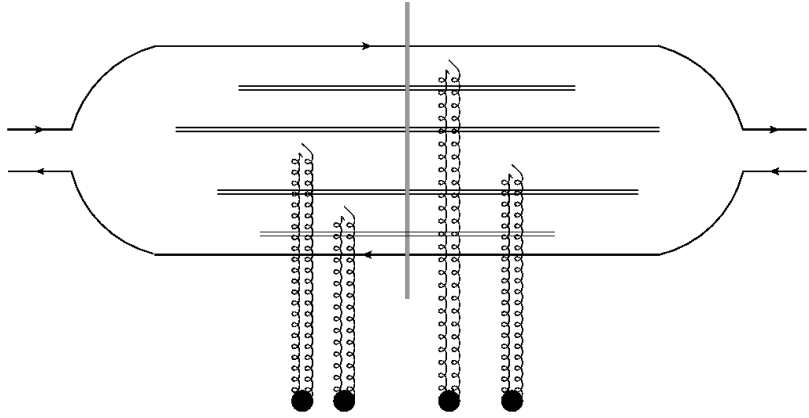

Figure 4.2: A large-$N_c$ diagram contributing to the cross section for diffractive dissociation of an onium. The grey vertical line represents the final state at the lightcone time $x^+ = +\infty$. The amplitude is to the left of this line, while the conjugate amplitude is to the right. Interacting dipoles in the Fock state of the onium exchange gluons with the nucleons (black dots) constituting the nucleus.

As for diffractive dissociation, we define $\sigma_D(r, Y; Y_0)$ to be the diffractive cross section per unit impact parameter with rapidity gap not less than $Y_0$. The cross section $\sigma_{diff}^{q\bar{q}A}(r, Y; Y_0)$ is then related to $\sigma_D(r, Y; Y_0)$ as

$$\sigma_{diff}^{q\bar{q}A}(r, Y; Y_0) = \sigma_0 \sigma_D(r, Y; Y_0), \tag{4.4}$$

Figure 4.2 represents a diagram contributing to the cross section $\sigma_D(r, Y; Y_0)$. At large $N_c$, Kovchegov and Levin [127] (see also Ref. [37]) established an evolution equation governing the high-energy evolution of $\sigma_D(r, Y; Y_0)$. This equation reads

$$
\begin{aligned}
\partial_Y \sigma_D(r, Y; Y_0) = \bar{\alpha} \int dp_{1\to2}(r^\perp, r'^\perp) \big[ & \sigma_D(r', Y; Y_0) + \sigma_D(|r^\perp - r'^\perp|, Y; Y_0) - \sigma_D(r, Y; Y_0) \\
& -2\sigma_D(r', Y; Y_0)T_1(|r^\perp - r'^\perp|, Y) - 2\sigma_D(|r^\perp - r'^\perp|, Y; Y_0)T_1(r', Y) \\
& +\sigma_D(r', Y; Y_0)\sigma_D(|r^\perp - r'^\perp|, Y; Y_0) + 2T_1(r', Y)T_1(|r^\perp - r'^\perp|, Y) \big].
\end{aligned} \tag{4.5}
$$

The initial condition for Eq. (4.5) can be set at $Y = Y_0$: in this case, the scattering is elastic,

$$\sigma_D(r, Y_0; Y_0) = T_1^2(r, Y_0). \tag{4.6}$$

In this chapter, we are interested in the following quantities:

i. the ratios between the diffractive cross section and the total cross section: $\sigma_D(r, Y; Y_0)/\sigma_T(r, Y)$ (for onium) and $\sigma_{diff}^{\gamma^*A}(Q^2, Y; Y_0)/\sigma_{tot}^{\gamma^*A}(Q^2, Y)$ (for virtual photon); and

ii. the distributions of rapidity gaps, which are defined as the ratios between the diffractive cross

section at a *fixed* rapidity gap $Y_{gap}$ and the total cross section:

$$\pi(r, Y; Y_{gap}) \equiv -\frac{1}{\sigma_T} \frac{\partial \sigma_D}{\partial Y_0}\bigg|_{Y_0 = Y_{gap}} \tag{4.7}$$

for an onium in the initial state, and

$$\Pi(Q^2, Y; Y_{gap}) \equiv -\frac{1}{\sigma_{tot}^{\gamma^* A}} \frac{\partial \sigma_{diff}^{\gamma^* A}}{\partial Y_0}\bigg|_{Y_0 = Y_{gap}} \tag{4.8}$$

for a virtual photon.

The diffractive-to-total cross section ratios estimate the possibility of observing diffractive events in the scattering. Meanwhile, the distributions of rapidity gaps tell us how likely to observe a diffractive event of gap size $Y_{gap}$, which may help in the selection of gap events at colliders.

The analysis of the above quantities requires the solutions to the Kovchegov-Levin (KL) equation (4.5). Due to its complex structure, to solve it analytically, in a direct way, is still not possible. However, it can be solved numerically, and we shall use its numerical solutions to investigate the quantities of interest at kinematics accessible at future electron-ion colliders. Prior to that, in the next section, we are going to introduce a formulation for diffraction based on which we can derive analytical solutions to the Kovchegov-Levin equation at an asymptotically large rapidity for a small onium size $r$, based on the phenomenological model for dipole distribution presented in the previous chapter and on a probabilistic picture of scattering cross sections.

## 4.3 Analytical asymptotics for diffractive dissociation of an onium

We consider the scattering of an onium of size $r$ off a large nucleus $A$, at the total relative rapidity $Y$. The onium is picked such that its size is well located in the scaling region, $0 \ll \ln[1/(r^2 Q_s^2(Y))] \ll \sqrt{\bar{\alpha} Y}$. Inherited from the previous chapter, as we are dealing with the leading-order evolution, we shall rescale rapidity variables by multiplying them by $\bar{\alpha}$, and the new rescaled rapidity variables are denoted by lowercase letter, for example, $y \equiv \bar{\alpha} Y$. In addition, it is convenient to use the logarithms of transverse sizes and saturation scales defined in Eq. (3.7) instead of the original variables.

We are going to use the aforementioned phenomenological model of dipole distribution for our next calculations. Therefore, it is necessary to consider a reference frame in which the rapidity is shared between the projectile and the target: the nucleus is boosted to the rapidity $y_0$ and the onium evolves to the rapidity $\tilde{y}_0 \equiv y - y_0$. All the relevant rapidities are assumed to be large, $y, y_0, \tilde{y}_0 \gg 1$, so that the variables of well-developed mean-field fronts can be used properly.

### 4.3.1 Formulation of scattering cross sections

#### a. Frame-dependent representation of cross sections

In the reference frame we are considering, by the dipole branching process, the Fock state of the evolved onium at the rapidity $\tilde{y}_0$ consists of elementary dipoles of various log sizes $x'$. The scattering cross sections can be expressed using the S-matrix element $S(x', y_0)$ for the scattering of an elementary dipole $x'$ off a nucleus boosted to $y_0$, as in Eqs. (3.8) and (3.10). To this aim, let us denote $\mathcal{S}(y_0)$ being the S-matrix element for the scattering of a particular realization of the Fock state of the onium at $\tilde{y}_0$, represented by the number density $n(x')$, off a nucleus at $y_0$. In term of the density $n$, $\mathcal{S}(y_0)$ reads

$$
\begin{aligned}
\mathcal{S}(y_0) &= \prod_{\{x_i\}} S(x_i, y_0) = \prod_{x'} [S(x', y_0)]^{n(x')dx'} \\
&= \exp\left(-\int dx' n(x') \ln \frac{1}{S(y_0, x')}\right) \equiv e^{-I(y_0)};
\end{aligned}
\tag{4.9}
$$

see Eqs. (3.10) and (3.11). Again, note that, since $n$ is a random distribution, $I(y_0)$ is also a random quantity.

With the help of $\mathcal{S}(y_0)$, we can rewrite the S-matrix element for the nuclear scattering of the initial onium $x$ at the total rapidity $y$ as

$$
S(x, y) = \langle \mathcal{S}(y_0) \rangle_{x, \tilde{y}_0}.
\tag{4.10}
$$

By the optical theorem, the total cross section reads

$$
\sigma_T(x, y) = 2 \langle 1 - \mathcal{S}(y_0) \rangle_{x, \tilde{y}_0}.
\tag{4.11}
$$

$\sigma_T(x, y)$ should be boost-invariant: while $y_0$ is present in the right-hand side of the representation (4.11), the explicit expression of $\sigma_T(x, y)$ does not depend on it. We showed in the previous chapter that, in the framework of our phenomenological model, this is in fact verified.

Now we move on to the diffractive cross section. According to the Good-Walker mechanism [115], diffractive dissociation cross section ($\sigma_{DD}$) is related to the dispersion of the forward elastic scattering amplitudes when evaluated in the different Fock states,

$$
\sigma_{DD}(x, y; y_0) = \langle x, \tilde{y}_0 | \mathbb{T}^\dagger(y_0) \mathbb{T}(y_0) | x, \tilde{y}_0 \rangle - |\langle x, \tilde{y}_0 | \mathbb{T}(y_0) | x, \tilde{y}_0 \rangle|^2,
\tag{4.12}
$$

where $|x, \tilde{y}_0\rangle$ is the quantum state of the initial onium of log size $x$ evolved to the rapidity $\tilde{y}_0$, and $\mathbb{T}(y_0)$ is the interacting (transition) matrix (T-matrix) representing the interaction of a dipole state when the latter traverses the nucleus at rapidity $y_0$. $\sigma_{DD}(x, y; y_0)$ is different from $\sigma_D(x, y; y_0)$ by the elastic term. In other words, $\sigma_{DD}(x, y; y_0)$ is the cross section for inelastic diffraction. Introducing

$\{|m\rangle\}$ as a complete set of dipole states, we can rewrite Eq. (4.12) as

$$
\sigma_{DD}(x, y; y_0) = \sum_{m,m',m''} \langle x, \tilde{y}_0|m\rangle \langle m|\mathbb{T}^\dagger(y_0)|m'\rangle \langle m'|\mathbb{T}(y_0)|m''\rangle \langle m''|x, \tilde{y}_0\rangle
$$
$$
- \left| \sum_{m,m'} \langle x, \tilde{y}_0|m\rangle \langle m|\mathbb{T}(y_0)|m'\rangle \langle m'|x, \tilde{y}_0\rangle \right|^2 . \tag{4.13}
$$

In the color dipole model for the onium-nucleus scattering at high energy, the dipoles are eigenstates of the T-matrix. Therefore, we can rewrite Eq. (4.13) as

$$
\sigma_{DD}(x, y; y_0) = \sum_m \langle x, \tilde{y}_0|m\rangle \left| \langle m|\mathbb{T}(y_0)|m\rangle \right|^2 \langle m|x, \tilde{y}_0\rangle - \left| \sum_m \langle x, \tilde{y}_0|m\rangle \langle m|\mathbb{T}(y_0)|m\rangle \langle m|x, \tilde{y}_0\rangle \right|^2
$$
$$
= \sum_m |\psi_m(x, \tilde{y}_0)|^2 \left| \langle m|\mathbb{T}(y_0)|m\rangle \right|^2 - \left| \sum_m |\psi_m(x, \tilde{y}_0)|^2 \langle m|\mathbb{T}(y_0)|m\rangle \right|^2 , \tag{4.14}
$$

where $\psi_m(x, \tilde{y}_0) \equiv \langle m|x, \tilde{y}_0\rangle$ is the probability amplitude for the Fock state of the onium $x$ at the rapidity $\tilde{y}_0$ to be the dipole state $m$. We denote by $\mathcal{T}(y_0)$ the T-matrix element for a particular scattering dipole state, which is the forward elastic amplitude for the scattering of a particular realization of the onium Fock state at $\tilde{y}_0$ with the nucleus at $y_0$. From Eq. (4.14), the diffractive dissociation cross section reads

$$
\sigma_{DD}(x, y; y_0) = \left\langle \mathcal{T}^2(y_0) \right\rangle_{x,\tilde{y}_0} - \left\langle \mathcal{T}(y_0) \right\rangle_{x,\tilde{y}_0}^2 , \tag{4.15}
$$

where the average over possible dipole realizations $\langle \cdots \rangle_{x,\tilde{y}_0}$ is corresponding to the sum over dipole eigenstates $|m\rangle$ with weights $|\psi_m(x, \tilde{y}_0)|^2$ in Eq. (4.14). Since $\mathcal{T}(y_0) = 1 - \mathcal{S}(y_0)$, we can also write the cross section $\sigma_{DD}$ as:

$$
\sigma_{DD}(x, y; y_0) = \left\langle \mathcal{S}^2(y_0) \right\rangle_{x,\tilde{y}_0} - \left\langle \mathcal{S}(y_0) \right\rangle_{x,\tilde{y}_0}^2 . \tag{4.16}
$$

To obtain a representation for $\sigma_D$, we just need to add the elastic contribution $\langle 1 - \mathcal{S}(y_0) \rangle_{x,\tilde{y}_0}^2$. The diffractive scattering cross section with a minimal gap $y_0$ eventually reads

$$
\sigma_D(x, y; y_0) = \left\langle [1 - \mathcal{S}(y_0)]^2 \right\rangle_{x,\tilde{y}_0} \tag{4.17}
$$

Finally, the inelastic cross section is the difference between the total cross section and the diffractive cross section,

$$
\sigma_{in}(x, y; y_0) = \left\langle 1 - \mathcal{S}^2(y_0) \right\rangle_{x,\tilde{y}_0} . \tag{4.18}
$$

In comparison to the total cross section $\sigma_T$, the cross sections $\sigma_D$ and $\sigma_{in}$ depend on $y_0$. The rapidity gap distribution can be expressed in term of either $\sigma_D$, as in Eq. (4.7), or $\sigma_{in}$,

$$\pi(x, y; y_{gap}) \equiv -\frac{1}{\sigma_T} \frac{\partial \sigma_D}{\partial y_0}\bigg|_{y_0=y_{gap}} = \frac{1}{\sigma_T} \frac{\partial \sigma_{in}}{\partial y_0}\bigg|_{y_0=y_{gap}}. \tag{4.19}$$

So far, the above representation of the S-matrix element and cross sections with $\mathcal{S}(y_0)$ given by Eq. (4.9) is valid in the dipole model for the QCD evolution of the onium Fock state and with the assumption that the nucleus is large.

### b. Probabilistic picture for cross sections of small onia

Let us recall that we are interested in the nuclear scattering of small onia with size in the scaling regime. In this case, the scattering is effectively dominated by the Fock state configurations in which the probability for each individual dipole $x'$ to scatter with the nucleus is very small (i.e., $S(x', y_0) \sim 1$), which was already verified in the framework of our phenomenological model of dipole distribution. This implies that the probability for the same dipole to scatter more than once is negligible. Then, we can approximate $I(y_0)$ by Eq. (3.13), which is now denoted by $I^{(1)}(y_0)$,

$$I^{(1)}(y_0) = \int dx' n(x') \bar{T}_1(x', y_0). \tag{4.20}$$

In the last chapter, this integral was called "the overlap". It corresponds to the sum of diagrams in which one single dipole in one given realization of the onium Fock state interacts by exchanging a single color-singlet two-gluon state with the nucleus. The approximation leading to Eq. (4.20) shall be referred to as the "single-exchange approximation". This is the meaning of the superscript (1) in the notation $I^{(1)}(y_0)$ of the overlap.

For our purpose, let us define

$$F_N(\mathcal{I}) \equiv \frac{\mathcal{I}^N}{N!}, \tag{4.21}$$

and

$$G_k(\mathcal{I}) \equiv F_k(\mathcal{I}) e^{-\mathcal{I}}. \tag{4.22}$$

$F_N\left[I^{(1)}(y_0)\right]$ $(N \geq 0)$ is the quantum-mechanical amplitude corresponding to the sum of all diagrams in which $N$ dipoles in the onium Fock state exchange color-singlet two-gluon states with the nucleus at the rapidity $y_0$. Meanwhile, $G_k(\mathcal{I})$ is unitarized,

$$\sum_{k \geq 0} G_k(\mathcal{I}) = 1. \tag{4.23}$$

With this property, $G_k\left[I^{(1)}(y_0)\right]$ accepts a probabilistic interpretation: when choosing scattering configurations with a weight given by their amplitude $F_N\left[I^{(1)}(y_0)\right]$, it represents the probability to pick those in which $k$ dipoles interact. Note that, both $F_N$ and $G_k$ are evaluated for a given realization of the onium Fock state.

We can rewrite the cross sections in Eqs. (4.11), (4.17) and (4.18) in terms of $F_N$ and $G_k$ in the single-exchange approximation. In particular, the total cross section reads

$$\sigma_T(x, y) = 2 \left( 1 - \left\langle G_0 \left[ I^{(I)}(y_0) \right] \right\rangle_{x, \tilde{y}_0} \right) \equiv 2 \sum_{k=1}^{\infty} w_k(x, y; y_0), \tag{4.24}$$

where $w_k$ are the average weights (over all possible realizations) to select scattering configurations with $k$ interacting dipoles,

$$w_k(x, y; y_0) \equiv \left\langle G_k \left[ I^{(I)}(y_0) \right] \right\rangle_{x, \tilde{y}_0}. \tag{4.25}$$

Note that, each weight, except for $w_0$, may *a priori* depend on $y_0$. The weight $w_0$ should be boost invariant since it is precisely the S-matrix element.

In the single-exchange approximation, the diffractive cross section can be rewritten as

$$\begin{aligned}
\sigma_D(x, y; y_0) &= \left\langle 1 - 2e^{-I^{(1)}(y_0)} + \left[ e^{-I^{(1)}(y_0)} \right]^2 \right\rangle_{x, \tilde{y}_0} = \left\langle e^{-I^{(1)}(y_0)} \left( e^{-I^{(1)}(y_0)} + e^{I^{(1)}(y_0)} - 2 \right) \right\rangle_{x, \tilde{y}_0} \\
&= \left\langle 2 \sum_{\substack{k \geq 2 \\ k \text{ even}}} \frac{\left[ I^{(1)}(y_0) \right]^k}{k!} e^{-I^{(1)}(y_0)} \right\rangle_{x, \tilde{y}_0} = 2 \sum_{\substack{k \geq 2 \\ k \text{ even}}} w_k(x, y; y_0),
\end{aligned} \tag{4.26}$$

which is two times the weight of the graphs with even number of participating dipoles. This relation suggests that the diffraction of a small dipole is dominated by an even number of exchanges at the interaction time.

Lastly, the inelastic scattering can be expressed as twice the weight of having an odd number of exchanges,

$$\sigma_{in}(x, y; y_0) = 2 \sum_{\substack{k \geq 1 \\ k \text{ odd}}} w_k(x, y; y_0). \tag{4.27}$$

We realize from this probablistic formulation that the calculation of the scattering cross sections requires the average weights of selecting a particular number of interacting dipoles. We shall present our estimation of the latter based on our aforementioned phenomenological picture of dipole distribution, together with establishing an exact evolution equation of their generating function. Before continuing with further calculations, we are going to show that, the representations of the diffractive cross section $\sigma_D$ are egligible, i.e., they indeed obey the KL equation (4.5).

### c. Connection to the Kovchegov-Levin equation

To the aim to recover the KL equation, let us introduce the probability $S_{in}(r, y; y_0)$ that there is no inelastic scattering between the state of the onium at $\tilde{y}_0$ and the nucleus at $y_0$,

$$S_{in}(r, y; y_0) = \left\langle \left[ \mathcal{S}(y_0) \right]^2 \right\rangle_{r, \tilde{y}_0}, \tag{4.28}$$

where we have reused the original transverse size variable instead of the log transverse size notation.

Now we boost the scattering by an infinitesimal rapidity $dy$ by assuming that the onium is boosted while keeping the rapidity of the nucleus unchanged. We further put this infinitesimal boost at the begining of the QCD evolution of the onium. This is the technique we already employed in the previous chapters to construct various evolution equations. By doing so, we are left with two possibilities. In the first place, there is no dipole branching after $dy$; then $S_{in}(r, y + dy; y_0) = S_{in}(r, y; y_0)$. Otherwise, the initial onium may branch into two daughter dipoles $r'$ and $|r^\perp - r'^\perp|$ after $dy$. In this case, for having no inelastic scattering between the state of the onium at $\tilde{y}_0 + dy$ with the nucleus, the state of each daughter dipole after the evolution over $\tilde{y}_0$ should not scatter inelastically. Gathering those two cases in one equation, we have

$$\underbrace{\left\langle [\mathcal{S}(y_0)]^2 \right\rangle_{r, \tilde{y}_0 + dy}}_{S_{in}(r, y + dy; y_0)} = \left( 1 - dy \int dp_{1 \to 2}(r^\perp, r'^\perp) \right) \underbrace{\left\langle [\mathcal{S}(y_0)]^2 \right\rangle_{r, \tilde{y}_0}}_{S_{in}(r, y; y_0)}$$
$$+ dy \int dp_{1 \to 2}(r^\perp, r'^\perp) \underbrace{\left\langle [\mathcal{S}(y_0)]^2 \right\rangle_{r', \tilde{y}_0}}_{S_{in}(r', y; y_0)} \underbrace{\left\langle [\mathcal{S}(y_0)]^2 \right\rangle_{|r^\perp - r'^\perp|, \tilde{y}_0}}_{S_{in}(|r^\perp - r'^\perp|, y; y_0)}. \tag{4.29}$$

Letting $dy \to 0$, we realize that $S_{in}$ obeys the S-type BK equation. Since $S_{in} = 1 - \sigma_{in}$, it is followed that the inelastic cross section solves the T-type BK equation.

Now we turn into diffraction. From Eqs. (4.17) and (4.28), $S_{in}$ and $\sigma_D$ are related to each other as

$$\sigma_D(r, y; y_0) = 1 - 2T_1(r, y) + S_{in}(r, y; y_0). \tag{4.30}$$

By taking the derivative with respect to $y$ both sides of Eq. (4.30), using the facts that $T_1$ and $S_{in}$ obey the T-type and the S-type BK equations, respectively, it is straightforward to see that $\sigma_D$ solves the KL equation (4.5).

The remaining thing to deal with is the initial condition. At $y = y_0$, or $\tilde{y}_0 = 0$, the Fock state of the onium is just itself. Therefore,

$$S_{in}(r, y_0; y_0) = S^2(r, y_0). \tag{4.31}$$

Substituting this into Eq. (4.30), with $y = y_0$, we recover the initial condition for the diffractive cross section in Eq. (4.6). Therefore, we have shown that, the representation (4.17) of the diffractive cross section is valid.

In the above paragraph, we have considered an onium of generic size. Now we retrieve our key assumption that the onium is picked in the scaling region, and hence, recall the probabilistic representation (4.26) of the diffractive cross section. Let us define $W_E(r, y; y_0)$ to be the weight of the graphs with an even number of participating dipoles,

$$W_E(r, y; y_0) = \sum_{\substack{k \geq 2 \\ k \text{ even}}} w_k \left( x \equiv \ln \frac{1}{r^2 Q_A^2}, y; y_0 \right). \tag{4.32}$$

Then for an odd number of exchanges,

$$W_O(r, y; y_0) = T_1(r, y) - W_E(r, y; y_0). \tag{4.33}$$

The diffractive and the inelastic cross sections are just twice $W_E$ and $W_O$, respectively.

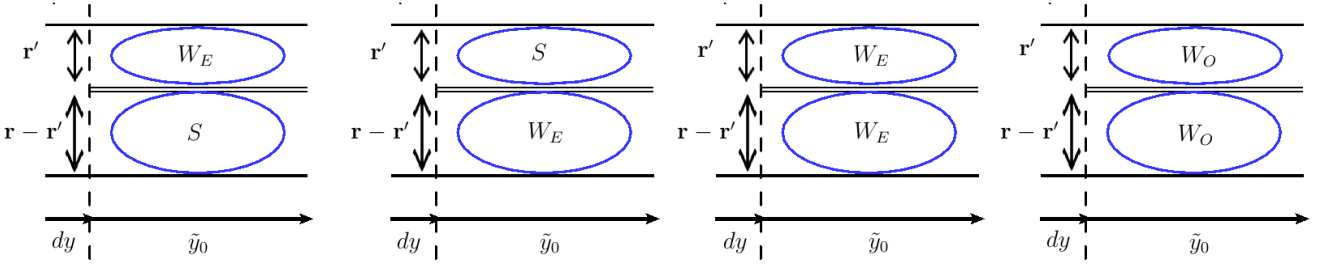

Figure 4.3: Contributions of dipole branching events to the evolution of $W_E$.

Using the well-known technique presented above, we can easily establish an evolution equation for $W_E$. In particular, we advance the system by a rapidity step $dy$ by boosting the onium from its rest frame. If the onium does not branch, $W_E(r, y + dy; y_0)$ is just $W_E(r, y; y_0)$. In case it splits into two offspring $r'$ and $|\mathbf{r} - \mathbf{r}'|$, one has following possibilities (see Fig. 4.3). If only one offspring interacts, it should scatter an even number of times. Otherwise, if both offspring interact, to have an even number of exchanges in the initial onium $r$ boosted to $\tilde{y}_0 + dy$ is equivalent to have in the states of both offspring either an even or an odd number of interacting dipoles. Therefore, we have

$$
\begin{aligned}
W_E(r, y + dy; y_0) = {} & \left(1 - dy \int dp_{1\to 2}(r^\perp, r'^\perp)\right) W_E(r, y; y_0) \\
& + dy \int dp_{1\to 2}(r^\perp, r'^\perp) \big\{ W_E(r', y; y_0) \left[1 - T_1(|r^\perp - r'^\perp|, y)\right] + W_E(|r^\perp - r'^\perp|, y; y_0) \left[1 - T_1(r', y)\right] \\
& \quad + W_E(r', y; y_0) W_E(|r^\perp - r'^\perp|, y; y_0) \\
& \quad + \left[T_1(r', y) - W_E(r', y; y_0)\right] \left[T_1(|r^\perp - r'^\perp|, y) - W_E(|r^\perp - r'^\perp|, y; y_0)\right] \big\}.
\end{aligned}
\tag{4.34}
$$

Passing to the limit $dy \to 0$, we obtain the following evolution equation for $W_E$:

$$
\begin{aligned}
\partial_y W_E(r, y; y_0) = \int dp_{1\to 2}(r^\perp, r'^\perp) \big[ & W_E(r', y; y_0) + W_E(|r^\perp - r'^\perp|, y; y_0) - W_E(r, y; y_0) \\
& - 2 W_E(r', y; y_0) T_1(|r^\perp - r'^\perp|, y) - 2 W_E(|r^\perp - r'^\perp|, y; y_0) T_1(r', y) \\
& + 2 W_E(r', y; y_0) W_E(|r^\perp - r'^\perp|, y; y_0) + T_1(r', y) T_1(|r^\perp - r'^\perp|, y) \big].
\end{aligned}
\tag{4.35}
$$

Multiplying both sides of Eq. (4.35) by 2, we get the KL equation for $\sigma_D = 2W_E$.

For the initial condition at $y = y_0$, since the Fock state of the onium at $\tilde{y}_0 = 0$ contains just one dipole (the initial one), there is no way for this state to scatter more than once in the single-exchange approximation we are considering. Consequently,

$$
W_E(r, y_0; y_0) = 0.
\tag{4.36}
$$

The term $T_1^2(r, y_0)$ is consistently suppressed in $2W_E(r, y_0; y_0)$, due to the single-exchange approximation. Therefore, we can conclude that we have recovered the KL evolution equation for the diffractive cross section in the limit of interest.

Using the same technique, we can also establish the equation for $W_O$: it obeys the T-type BK equation, with the initial condition $W_O(r, y_0; y_0) = T_1(r, y_0)$; see Ref. [27].

The probabilistic representation in the single-exchange limit we are considering for diffraction has just been shown to be consistent with the KL formulation. The basic building-blocks in this representation are the average weights $w_k$. In the next part, we are going to compute these weights, which then enable us to deduce the desired quantities.

### 4.3.2 Calculation of the weights of the number of participating dipoles

#### a. Heuristic calculations using the phenomenological picture

We now compute the average weights $w_k$ based on their definition (4.25) as the probability to have exactly $k$ dipoles in the Fock state of the onium evolved to rapidity $\tilde{y}_0$ effectively interacting with the nucleus at the rapidity $y_0$, and on the phenomenological model for dipole distribution presented in the previous chapter. It is always understood that the rapidities $y_0$ and $\tilde{y}_0$ are large, so that the variables of asymptotic fronts can be properly used.

Let us first recap the main features of the phenomenological model, for which we shall also reiterate some important formulae. The key assumption of the model is that the onium evolves deterministically except for one single fluctuation consisting in one unusually large dipole produced at some random rapidity $\tilde{y}_1$, of log transverse size $x + \widetilde{X}_{\tilde{y}_1} - \delta$, which subsequently evolves to the rapidity $\tilde{y}_0$. In general, the deterministic evolution of an initial dipole of log size $\mathfrak{X}$ in the rapidity $\Delta \tilde{y}$ results in a Fock state characterized by a mean-field dipole density $\bar{n}(x' - \mathfrak{X}, \Delta \tilde{y})$ for a dipole of size $x'$, which is the solution to the BFKL equation supplemented by a cut-off on the large-dipole tail. This density reads

$$\bar{n}(x' - \mathfrak{X}, \Delta \tilde{y}) = C_1 \left( x' - \mathfrak{X} - \widetilde{X}_{\Delta \tilde{y}} \right) e^{\gamma_0 \left( x' - \mathfrak{X} - \widetilde{X}_{\Delta \tilde{y}} \right)} \exp \left[ -\frac{\left( x' - \mathfrak{X} - \widetilde{X}_{\Delta \tilde{y}} \right)^2}{2 \chi''(\gamma_0) \Delta \tilde{y}} \right] \Theta \left( x' - \mathfrak{X} - \widetilde{X}_{\Delta \tilde{y}} \right),$$

(4.37)

where $\widetilde{X}_{\Delta \tilde{y}} = -\chi'(\gamma_0) \Delta \tilde{y} + \frac{3}{2\gamma_0} \ln \Delta \tilde{y}$. In the current context, the dipole density is generated by both the evolution of the initial onium $\{\mathfrak{X} = x, \Delta \tilde{y} = \tilde{y}_0\}$ and the evolution of the large dipole created by the rare fluctuation $\{\mathfrak{X} = x + \widetilde{X}_{\tilde{y}_1} - \delta, \Delta \tilde{y} = \tilde{y}_0 - \tilde{y}_1\}$. At the scattering rapidity $\tilde{y}_0$, the number density of dipoles of size $x'$ is given by the sum of these two contributions,

$$n(x') = \bar{n}(x' - x, \tilde{y}_0) + \bar{n}(x' - x - \widetilde{X}_{\tilde{y}_1} + \delta, \tilde{y}_0 - \tilde{y}_1).$$

(4.38)

The size $\delta$, which is the difference between the log sizes of the large dipole by the fluctuation and of the largest dipole in the typical mean-field configuration at $\tilde{y}_1$, $x + \widetilde{X}_{\tilde{y}_1}$, is a random variable distributed according to the following probability distribution:

$$p(\delta, \tilde{y}_1) = C\delta e^{-\gamma_0 \delta} \exp\left(-\frac{\delta^2}{2\chi''(\gamma_0)\tilde{y}_1}\right)\Theta(\delta). \tag{4.39}$$

Now we turn into the overlap $I^{(1)}(y_0)$ in Eq. (4.20) between the dipole density $n(x')$ and the nuclear scattering amplitude $\bar{T}_1(x', y_0)$. Since the onium is small (in the scaling region), the nuclear overlap of the mean density generated by the deterministic evolution of the initial onium (the first term in Eq. (4.38)) is negligible in comparison to the contribution from the fluctuation. Therefore, in our model, the overlap reads

$$I^{(1)}_{\delta, \tilde{y}_1}(x, y; y_0) = \int dx' \bar{n}(x' - x - \tilde{X}_{\tilde{y}_1} + \delta, \tilde{y}_0 - \tilde{y}_1)\bar{T}_1(x', y_0), \tag{4.40}$$

where $\bar{T}_1(x', y_0)$ is given by Eq. (3.19).

$I^{(1)}_{\delta, \tilde{y}_1}(x, y; y_0)$ is rigorously not boost invariant. The detailed calculation of the overlap depends upon the choice of the reference frame, as presented in the previous chapter. For the frame in which the nucleus is highly boosted, $y_0 \ll (x - X_y)^2$, and in the regime of interest, a calculation of the overlap was presented in the previous chapter, and led to the following expression:

$$I^{(1)}_{\delta, \tilde{y}_1}(x, y; y_0) = C_1 C_2 \sqrt{\frac{\pi}{2}}[\chi''(\gamma_0)]^{3/2} e^{-\gamma_0(x-X_y)}\left(\frac{y}{y_1 \tilde{y}_1}\right)^{3/2} e^{\gamma_0 \delta}; \tag{4.41}$$

see Eq. (3.34). It turns out that, in the frame with the slightly-boosted nucleus, the optimal overlap is also of the form (Eq. (4.41)). Again, this does not mean that the overlap is boost invariant, but reflect the requirement that the dominant configurations should optimize the overlap. In any case, the optimal overlap (Eq. (4.41)) is effectively independent of $y_0$. Therefore, we shall suppress the $y_0$ dependence of $I^{(1)}_{\delta, \tilde{y}_1}$ in the following calculations.

In the phenomenological model, the weights $w_k$ of numbers of interacting dipoles, which is defined in Eq. (4.25), is formulated as

$$w_k(y, x; y_0) = \int_{y_0}^{y} dy_1 \int_0^{+\infty} d\delta p(\delta, \tilde{y}_1) \frac{1}{k!}[I^{(1)}_{\delta, \tilde{y}_1}(x, y)]^k e^{-I^{(1)}_{\delta, \tilde{y}_1}(x, y)}, \tag{4.42}$$

where, again, the average over all possible configurations is replaced by the integrations over the fluctuation size $\delta$ and over the rapidity $y_1$ (measured from the nucleus) at which the fluctuation occurs, weighted by the distribution $p(\delta, \tilde{y}_1)$.

We start by computing the integral over $\delta$,

$$\pi_k(x, y; y_1) \equiv \int_0^{+\infty} d\delta p(\delta, \tilde{y}_1) \frac{1}{k!}[I^{(1)}_{\delta, \tilde{y}_1}(x, y)]^k e^{-I^{(1)}_{\delta, \tilde{y}_1}(x, y)}. \tag{4.43}$$

Replacing $p(\delta, \tilde{y}_1)$ by its expression in Eq. (4.39) and using $I^{(1)}_{\delta, \tilde{y}_1}$, which is now denoted $I$, as a new integration variable instead of $\delta$, the integral $\pi_k(x, y; y_1)$ becomes

$$\pi_k(x, y; y_1) = \frac{C}{\gamma_0} \frac{1}{k!} I_{0,\tilde{y}_1}^{(1)}(x, y) \int_{I_{0,\tilde{y}_1}^{(1)}(x,y)}^{+\infty} dI \ln \frac{I}{I_{0,\tilde{y}_1}^{(1)}(x, y)} I^{k-2} e^{-I} \exp\left(-\frac{\ln^2\left[I/I_{0,\tilde{y}_1}^{(1)}\right]}{2\gamma_0^2 \chi''(\gamma_0)\tilde{y}_1}\right). \quad (4.44)$$

The integral in Eq. (4.44) is the integral $S_k$ ($k \geq 1$) defined and evaluated in Appendix B, with $A \equiv I_{0,\tilde{y}_1}^{(1)}(x, y)$ and $\beta_0 \equiv 2\gamma_0^2 \chi''(\gamma_0)$. Using Eqs. (B.10) and (B.13), we find

$$\pi_{k=1}(x, y; y_1) = \gamma_0 \chi''(\gamma_0) C \frac{1}{k!} \tilde{y}_1 I_{0,\tilde{y}_1}^{(1)}(x, y) \left[1 - \exp\left(-\frac{\ln^2(I_{0,\tilde{y}_1}^{(1)}(x, y))}{2\gamma_0 \chi''(\gamma_0)\tilde{y}_1}\right)\right], \quad (4.45)$$

and

$$\pi_{k\geq 2}(x, y; y_1) = \frac{C}{\gamma_0} I_{0,\tilde{y}_1}^{(1)}(x, y) \frac{1}{k(k-1)} \ln \frac{1}{I_{0,\tilde{y}_1}^{(1)}(x, y)} \exp\left(-\frac{\ln^2(I_{0,\tilde{y}_1}^{(1)}(x, y))}{2\gamma_0 \chi''(\gamma_0)\tilde{y}_1}\right). \quad (4.46)$$

Replacing $I_{0,\tilde{y}_1}^{(1)}(x, y)$ by its expression in Eq. (4.41), we obtain the following final expressions for the density $\pi_k(x, y; y_1)$ of the rapidity $y_1$:

(i) Case $k = 1$:

$$\pi_{k=1}(x, y; y_1) = c\sqrt{\frac{\chi''(\gamma_0)}{2\pi}} e^{-\gamma_0(x-X_y)} \frac{y^{3/2}}{y_1^{3/2} \tilde{y}_1^{1/2}} \left[1 - \exp\left(-\frac{(x-X_y)^2}{2\chi''(\gamma_0)\tilde{y}_1}\right)\right]. \quad (4.47)$$

(ii) Case $k \geq 2$:

$$\pi_{k\geq 2}(x, y; y_1) = \frac{c}{\gamma_0} \frac{1}{\sqrt{2\pi \chi''(\gamma_0)}} \frac{(x-X_y)e^{-\gamma_0(x-X_y)}}{k(k-1)} \left(\frac{y}{y_1 \tilde{y}_1}\right)^{3/2} \exp\left(-\frac{(x-X_y)^2}{2\chi''(\gamma_0)\tilde{y}_1}\right), \quad (4.48)$$

where

$$c \equiv CC_1 C_2 \pi [\chi''(\gamma_0)]^2. \quad (4.49)$$

Now we move on to the weights $w_k$. For $k = 1$, the integration over $y_1$ reads

$$\int_{y_0}^{y} dy_1 \frac{y^{3/2}}{y_1^{3/2} \tilde{y}_1^{1/2}} \left[1 - \exp\left(-\frac{(x-X_y)^2}{2\chi''(\gamma_0)\tilde{y}_1}\right)\right] = 2\sqrt{\tilde{y}_0} \left[1 - \exp\left(-\frac{(x-X_y)^2}{2\chi''(\gamma_0)\tilde{y}_0}\right)\right]$$
$$+ \sqrt{\frac{2\pi}{\chi''(\gamma_0)}}(x-X_y)\text{erfc}\left(\frac{x-X_y}{\sqrt{2\chi''(\gamma_0)\tilde{y}_0}}\right). \quad (4.50)$$

In the scaling region, $(x-X_y) \ll \sqrt{y} \sim \sqrt{y_0}$, we get the following expression for the weight $w_1$:

$$w_1(x, y; y_0) \simeq c(x-X_y)e^{-(x-X_y)}. \quad (4.51)$$

The integration over $y_1$ for the weights $w_{k\geq 2}$ can be written in terms of the error function and of the elementary functions as

$$\int_{y_0}^{y} dy_1 \left(\frac{y}{y_1 \tilde{y}_1}\right)^{3/2} \exp\left(-\frac{(x-X_y)^2}{2\chi''(\gamma_0)\tilde{y}_1}\right) = 2\sqrt{\frac{\tilde{y}_0}{yy_0}} \exp\left(-\frac{(x-X_y)^2}{2\chi''(\gamma_0)\tilde{y}_0}\right)$$
$$+ \frac{\sqrt{2\pi\chi''(\gamma_0)}}{x-X_y}\left(1 - \frac{(x-X_y)^2}{\chi''(\gamma_0)y}\right) \text{erfc}\left(\frac{x-X_y}{\sqrt{2\chi''(\gamma_0)}}\sqrt{\frac{y_0}{y\tilde{y}_0}}\right) \exp\left(-\frac{(x-X_y)^2}{2\chi''(\gamma_0)y}\right). \tag{4.52}$$

In the scaling regime of interest, it boils down to two simple terms. Therefore, the weights $w_k$ with $k \geq 2$ eventually read

$$w_{k\geq 2}(x,y;y_0) = \frac{c}{\gamma_0}\frac{1}{k(k-1)}\left(1 + \sqrt{\frac{2}{\pi\chi''(\gamma_0)}}\frac{x-X_y}{\sqrt{y_0}}\right)e^{-\gamma_0(x-X_y)}. \tag{4.53}$$

Interestingly enough, the $k$ dependence comes as an overall factor, from which we can deduce the following simple expression for the ratio $w_{k\geq 2}/w_2$:

$$\frac{w_{k\geq 2}}{w_2} = \frac{2}{k(k-1)}. \tag{4.54}$$

This ratio shows that the distribution of the number of partipating dipoles decays slowly at large $k$. As a matter of fact, the mean participant number is formally infinite. Therefore, once the multiple scatterings are relevant, the events which involve a large number of interacting dipoles are not rare at all.

### b. Generating function

We first observe that the set of weights $\{w_k; k \geq 0\}$ obey a hierarchy of evolution equations,

$$\partial_y w_k(r,y;y_0) = \int dp_{1\to 2}(r^\perp, r'^\perp)\left[\sum_{j=0}^{k} w_j(r',y;y_0)w_{k-j}(|r^\perp - r'^\perp|,y;y_0) - w_k(r,y;y_0)\right], \tag{4.55}$$

with the initial condition $w_k(r,y_0;y_0) = \delta_{k,0}S(r,y_0) + \delta_{k,1}T_1(r,y_0)$. This hierarchy can be proved straightforwardly with the technique used to derive different evolution equations above, by noticing that when the initial onium branches after an infinitesimal boost, one should take into account all the cases in which the numbers of participating dipoles of two offspring add up to $k$. When $k = 0$, this hierarchy degenerates into a closed equation: This is precisely the BK equation for the $S$- matrix element $S(r,y) \equiv w_0(r,y;y_0)$.

One can construct the ordinary generating function for the weights $w_k$,

$$\tilde{w}_\lambda(r,y;y_0) = \sum_{k=0}^{\infty} \lambda^k w_k(r,y;y_0). \tag{4.56}$$

This generating function satisfies the following properties. First, the condition that the sum of all possible values of participating dipoles should be unitary can be expressed in terms of the generating function that $\tilde{w}_{\lambda=1}(r, y; y_0) = 1$. Second, the generating function at $\lambda = 0$ coincides with the S-matrix element, $\tilde{w}_{\lambda=0}(r, y; y_0) = S(r, y)$. Third, the difference between the diffractive cross section and the inelastic cross section is related to the difference between the values of the generating function evaluated at two different values of $\lambda$:

$$\sigma_D(r, y; y_0) - \sigma_{in}(r, y; y_0) = 2(\tilde{w}_{\lambda=-1}(r, y; y_0) - \tilde{w}_{\lambda=0}(r, y; y_0)).  \quad (4.57)$$

Finally, and most interestingly, it turns out to solve the S-type BK equation,

$$\partial_y \tilde{w}_\lambda(r, y; y_0) = \int dp_{1\to 2}(r^\perp, r'^\perp) \left[ \tilde{w}_\lambda(r', y; y_0)\tilde{w}_\lambda(|r^\perp - r'^\perp|, y; y_0) - \tilde{w}_\lambda(r, y; y_0) \right].  \quad (4.58)$$

which can be proved using Eq. (4.55). The initial condition at $y = y_0$ is given by

$$\tilde{w}_\lambda(r, y_0; y_0) = 1 - (1 - \lambda)T(r, y_0).  \quad (4.59)$$

Therefore, if we know the solution of the evolution equation (4.58), with the initial condition (4.59), we can deduce the expressions of the weights, and hence, of the scattering cross sections. In the following, we shall try to conjecture the solution to this equation based on the traveling wave property of the asymptotic solution of the BK equation and on the above heuristic calculation of the weights within the phenomenological model.

**Traveling wave solution in the asymptotic limit**

In the infinite-rapidity limit, the solution to the BK equation Eq. (4.58) converges to a traveling wave. In particular, the generating function $\tilde{w}_\lambda(x, y; y_0)$ at an asymptotic large rapidity $y$ tends to a function of $x - X_y + f_{y_0}(\lambda)$ only, where $f_{y_0}(\lambda)$ is a "delay function" that vanishes for $\lambda = 0$. The term "delay" comes from the fact that, the position of the front $X_y$ is pulled back by the distance $f_{y_0}(\lambda)$ with an initial condition of the form (4.59) when $0 < \lambda < 1$.

When furthermore $x - X_y + f_{y_0}(\lambda)$ is taken finite but large, by choosing an appropriate value of $x$ well-located in the scaling regime, the analytic form for the shape of the traveling wave is given by

$$1 - \tilde{w}_\lambda(x, y; y_0) = c_w \left[ x - X_y + f_{y_0}(\lambda) \right] e^{-\gamma_0 \left[ x - X_y + f_{y_0}(\lambda) \right]}.  \quad (4.60)$$

where $c_w$ is an undetermined constant of order unity.

For the delay function, we can guess its form from the above heuristics of the weights $w_k$ from the phenomenological model. Substituting the expressions of the weights $w_{k\geq 1}(x, y; y_0)$ in Eqs. (4.51) and (4.53), bearing in mind that $w_0(x, y; y_0) = S(x, y) = 1 - c(x - X_y)e^{-\gamma_0(x - X_y)}$, into the definition of the generating function, we get

$$
\begin{aligned}
1 - \tilde{w}_\lambda &\simeq c \left\{ (1-\lambda)(x - X_y) - \frac{1}{\gamma_0} \left( 1 + \sqrt{\frac{2}{\pi \chi''(\gamma_0)}} \frac{x - X_y}{\sqrt{y_0}} \right) \left[ \sum_{k=2}^{\infty} \frac{\lambda^k}{k(k-1)} \right] \right\} e^{-\gamma_0(x - X_y)} \\
&= c(1-\lambda) \left\{ (x - X_y) + \frac{1}{\gamma_0} \left( 1 + \sqrt{\frac{2}{\pi \chi''(\gamma_0)}} \frac{x - X_y}{\sqrt{y_0}} \right) \left[ \ln \frac{1}{1-\lambda} - \frac{\lambda}{1-\lambda} \right] \right\} e^{-\gamma_0(x - X_y)} \\
&\simeq c(1-\lambda) \left\{ (x - X_y) \left( 1 + \frac{1}{\gamma_0} \ln \frac{1}{1-\lambda} \sqrt{\frac{2}{\pi \chi''(\gamma_0)}} \frac{1}{\sqrt{y_0}} \right) \right. \\
&\quad \left. + \frac{1}{\gamma_0} \ln \frac{1}{1-\lambda} \left( 1 - \frac{1}{\gamma_0} \sqrt{\frac{2}{\pi \chi''(\gamma_0)}} \frac{1}{\sqrt{y_0}} \right) \left( 1 + \frac{1}{\gamma_0} \ln \frac{1}{1-\lambda} \sqrt{\frac{2}{\pi \chi''(\gamma_0)}} \frac{1}{\sqrt{y_0}} \right) \right\} e^{-\gamma_0(x - X_y)} \\
&\simeq c(1-\lambda)^{1 - \frac{1}{\gamma_0} \sqrt{\frac{2}{\pi \chi''(\gamma_0)}} \frac{1}{\sqrt{y_0}}} \left[ x - X_y + \frac{1}{\gamma_0} \ln \frac{1}{1-\lambda} \left( 1 - \frac{1}{\gamma_0} \sqrt{\frac{2}{\pi \chi''(\gamma_0)}} \frac{1}{\sqrt{y_0}} \right) \right] e^{-\gamma_0(x - X_y)},
\end{aligned}
\tag{4.61}
$$

or

$$
\begin{aligned}
1 - \tilde{w}_\lambda(x, y; y_0) \simeq c &\left[ x - X_y + \frac{1}{\gamma_0} \ln \frac{1}{1-\lambda} \left( 1 - \frac{1}{\gamma_0} \sqrt{\frac{2}{\pi \chi''(\gamma_0)}} \frac{1}{\sqrt{y_0}} \right) \right] \\
&\times \exp \left[ -\gamma_0 \left( x - X_y + \frac{1}{\gamma_0} \ln \frac{1}{1-\lambda} \left( 1 - \frac{1}{\gamma_0} \sqrt{\frac{2}{\pi \chi''(\gamma_0)}} \frac{1}{\sqrt{y_0}} \right) \right) \right].
\end{aligned}
\tag{4.62}
$$

In Eq. (4.61), we have added and removed terms of order $\mathcal{O}(1/\sqrt{y_0})$ compared to the leading term $(x - X_y)$ in the curly bracket, which is acceptable at this level of approximation. We have also used the following relation:

$$
(1-\lambda) \ln \frac{1}{1-\lambda} = \lambda - \sum_{k \geq 2} \frac{\lambda^k}{k(k-1)},
\tag{4.63}
$$

and the expansion

$$
(1-\lambda)^{-\frac{1}{\gamma_0} \sqrt{\frac{2}{\pi \chi''(\gamma_0)}} \frac{1}{\sqrt{y_0}}} \underset{y_0 \gg 1}{=} 1 + \frac{1}{\gamma_0} \ln \frac{1}{1-\lambda} \sqrt{\frac{2}{\pi \chi''(\gamma_0)}} \frac{1}{\sqrt{y_0}} + \mathcal{O}\left( \frac{1}{y_0} \right).
\tag{4.64}
$$

Comparing Eqs. (4.60) and (4.62) in parallel, we conjecture that the delay function has the following form

$$
f_{y_0}(\lambda) = \frac{1}{\gamma_0} \ln \frac{1}{1-\lambda} \left( 1 - \frac{1}{\gamma_0} \sqrt{\frac{2}{\pi \chi''(\gamma_0)}} \frac{1}{\sqrt{y_0}} \right).
\tag{4.65}
$$

The approximations leading to this conjecture implies that this solution is valid for large $y_0$ such that $|\ln(1-\lambda)| \ll \sqrt{y_0}$, which looks somehow very limiting for $y_0$. However, as we are interested in the expansion of the generating function around $\lambda = 0$, this condition is not so restrictive. We

shall present a check for the conjecture (4.65) of the delay function later, based on the numerical solutions to the evolution equation of the generating function.

### 4.3.3 Diffractive cross sections

With the asymptotic expressions for the weights $w_k$ in hand, we are now able to derive the physical observables of interest. In particular, we are now going derive the expressions for the diffractive cross section with a minimal gap $y_0$ and the rapidity gap distribution. We shall then connect our result to a recent study also on the diffractive gap distribution.

#### a. Analytical asymptotics

We first see that, the asymptotic formula of the total cross section for the onium log size $x$ chosen in the scaling region is given by

$$\sigma_T(x,y) = 2\sum_{k\geq 1} w_k \approx 2w_1 = 2c(x-X_y)e^{-\gamma_0(x-X_y)}, \tag{4.66}$$

which completely agrees with the result obtained in Chapter 3. In Eq. (4.66), the sum is dominated by $w_1$: the weights $w_{k\geq 2}$ are suppressed as they are of order $\mathcal{O}\left\{\min\left[1/(x-X_y), 1/(\sqrt{y_0})\right]\right\}$ compared to the leading contribution. This shows that, for the nuclear scattering of a small onium, the total cross section is mainly due to one single exchange between the onium Fock state and the nucleus.

The diffractive cross section can be obtained by doubling the weight of having an even exchange, with the weight for no exchange being excluded. Using Eq. (4.53), the above expression for the total cross section $\sigma_T$ and the identity

$$\sum_{\text{even } k\geq 2} \frac{1}{k(k-1)} = \ln 2, \tag{4.67}$$

we arrive at the following simple asymptotic expression for the diffractive-to-total cross section ratio:

$$\frac{\sigma_D(x,y;y_0)}{\sigma_T(x,y)} = \frac{\ln 2}{\gamma_0}\left(\frac{1}{x-X_y} + \sqrt{\frac{2}{\pi\chi''(\gamma_0)}}\frac{1}{\sqrt{y_0}}\right), \tag{4.68}$$

which is valid for $x$ picked in the scaling region and $y_0 \gg 1$.

Let us now interpret two terms in Eq. (4.68). The fluctuation creating a large dipole occurs most likely either in the beginning of the evolution (leading to a dissociative but small mass event), or close to the scattering rapidity $\tilde{y}_0$ (leading to a gap size close to $y_0$). The first configuration is dominant when $y_0 \gg (x-X_y)^2$, leading to the first term in Eq. (4.68). The second term would dominate the diffractive cross section for the opposite ordering of $y_0$ and $(x-X_y)^2$, the case in which the second configuration is most probable.

In the same manner, the distribution of rapidity gaps is obtained by doubling the sum of the densities $\pi_{k\geq 2}$ in Eq. (4.48) (with $k$ even), then dividing the result by the total cross section $\sigma_T$. It eventually reads

$$\pi(x,y;y_{gap}) = \frac{\ln 2}{\gamma_0\sqrt{2\pi\chi''(\gamma_0)}} \left[\frac{y}{y_{gap}(y-y_{gap})}\right]^{3/2} \exp\left[-\frac{(x-X_y)^2}{2\chi''(\gamma_0)(y-y_{gap})}\right] \qquad (4.69)$$

As in the above cross sections, the distribution (4.69) is expected to be valid for a large total rapidity $y$ and for $x$ chosen in the scaling region, $1 \ll x - X_y \ll \sqrt{y}$. Additionally, the gap $y_{gap}$ should satisfy the condition $y_{gap}, y - y_{gap} \gg 1$.

The rapidity gap distribution (4.69) is very similar to the distribution of the branching rapidity of the last common ancestor of the set of dipoles which scatter in Eq. (3.51). The only difference between these two distributions is an extra factor $\ln 2$ in the former, which comes from the fact that two distributions are related to two different sets of the weight $w_k$. While the gap distribution is related to even numbers of participating dipoles, the distribution of the last common ancestor sums up the contributions of all possible numbers $k$ of dipoles which interact, starting from $k=2$,

$$\frac{G(x,y;y_1)}{T(x,y)} = \frac{\sum_{k=2}^{\infty} \pi_k(x,y;y_1)}{\sum_{k=1}^{\infty} w_k(x,y;y_1)}. \qquad (4.70)$$

Consequently, the identity (4.67) is replaced, in this case, by $\sum_{k=2}^{\infty} 1/[k(k-1)]$, which is unity.

### b. Connection to a recent picture of rapidity gap events

As a matter of fact, for the rapidity gap distribution (4.69), our new result is the determination of the overall constant. The functional form of the distribution was first found in Ref. [22], based on a prototype of the phenomenological model presented in Chapter 3. We shall now relate the reasoning leading to the (incomplete) asymptotic rapidity gap distribution used in the mentioned reference to the probabilistic description of diffraction of a small onium presented in this chapter.

Let us first briefly revisit the picture of diffraction in Ref. [22]. We start by defining $\mathcal{P}(x,y;\mathfrak{X})$ the probability of having at least one dipole whose log size is smaller than $\mathfrak{X}$ in the Fock state of an onium $x$ at rapidity $y$. As discussed in Chapter 2, it solves the T-type BK equation, with the initial condition given by the step function,

$$\mathcal{P}(x,y=0;\mathfrak{X}) = \Theta(\mathfrak{X}-x). \qquad (4.71)$$

We know that, when $y \to \infty$, the solution to the BK equation with the initial condition (4.71) tends to a traveling wave. For $x$ in the scaling region, $1 \ll (x-X_y)^2 \ll y$, $\mathcal{P}$ reads

$$\mathcal{P}(x,y;\mathfrak{X}) = c_{\mathcal{P}}(x+\tilde{X}_y-\mathfrak{X})e^{-\gamma_0(x+\tilde{X}_y-\mathfrak{X})} \exp\left[-\frac{(x+\tilde{X}_y-\mathfrak{X})^2}{2\chi''(\gamma_0)y}\right], \qquad (4.72)$$

where $c_{\mathcal{P}}$ is an unknown constant. We recall that, $x+\tilde{X}_y$ is the log size of the largest dipole in a typical configuration of the onium Fock state at the rapidity $y$.

Now let us eyeball the onium-nucleus system at a rapidity $0 < y_g < y$ counting from the nucleus. The initial onium then evolves to the rapidity $y - y_g$. We assume that both $y_g$ and $y - y_g$ are large parameters. The Fock state of the onium at $y - y_g$, which is a stochastic ensemble of dipoles, may contain a few unusually large dipoles of log sizes smaller than the logarithm of the nuclear saturation momentum at $y_g$, i.e. in the saturation region, which is generated by a rare fluctuation. These dipoles will be probed by the nucleus with a probability of order unity ($T \sim 1$). Consequently, the ratio of the elastic cross section to the total cross section reaches its maximal value,

$$\frac{\sigma_{el}}{\sigma_T} \simeq \frac{1}{2}, \tag{4.73}$$

which characterizes the scattering of quantum particles off a black disk (the black disk limit). Furthermore, the elastic scattering corresponds to the diffraction of particles in the shadow of the disk. Therefore, this configuration will result in a diffractive dissociation event with rapidity gap $y_g$.

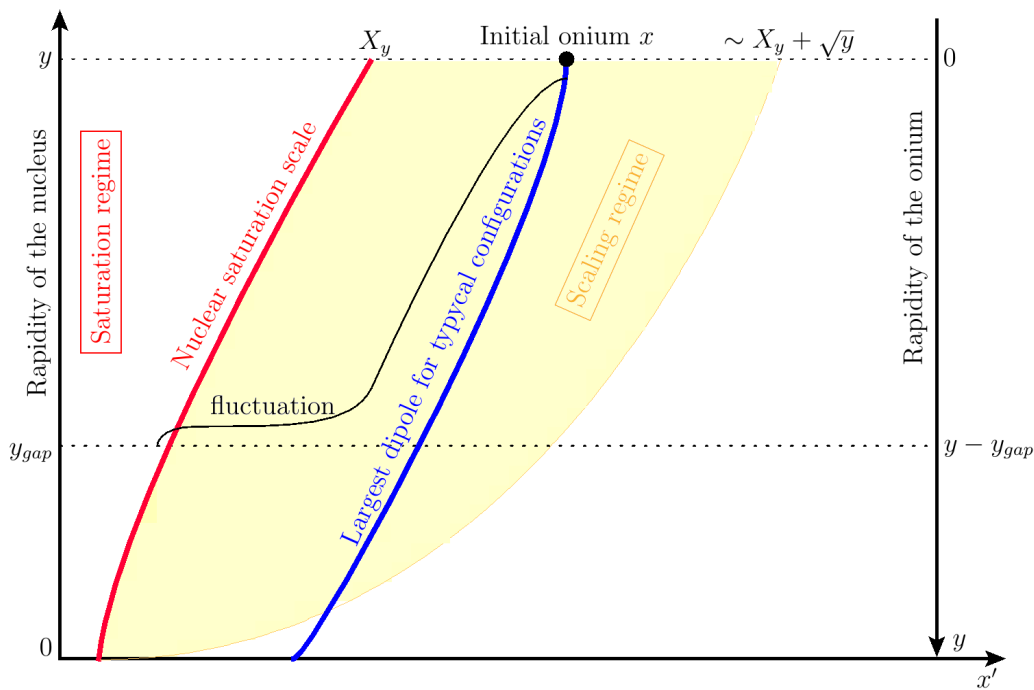

Figure 4.4: An illustration of the picture of diffractive dissociation event in Ref. [22]. The vertical axis shows the rapidity of the nucleus (upward) and of the onium (downward) in such a way that at each slice, the sum of their rapidities is the total relative rapidity $y$. The nucleus is represented by the nuclear saturation scale (red line), which is the onset of the saturation regime (to the left of that scale). The scaling region is close to the saturation line to the right (orange domain). The blue line represents the largest dipole size in a typical realization of the Fock state of the onium $x$ at each rapidity during the evolution. Diffraction with a gap $y_{gap}$ is due to a rare fluctuation at $y_{gap}$ creating a large dipole inside the saturation regime.

Within this picture, the diffractive cross section at a fixed rapidity gap $y_{gap}$ is proportional to the probability of having a dipole whose log size is smaller than $X_{y_{gap}}$ in the Fock state of the onium at the rapidity $y - y_{gap}$. From Eq. (4.72), after dividing the diffractive cross section to the total cross section, the rapidity gap distribution reads

$$\pi(x, y; y_{gap}) = c_D \left[ \frac{y}{y_{gap}(y - y_{gap})} \right]^{3/2} \exp \left[ -\frac{(x - X_y)^2}{2\chi''(\gamma_0)(y - y_{gap})} \right], \qquad (4.74)$$

where $c_D$ is a constant which cannot be determined from this picture. An schematic illustration of the picture is presented in Fig. 4.4.

It should be noticed that, this picture of diffraction suggests a possible connection between the rapidity gap distribution in diffraction and the genealogical distribution of the splitting time of the last common ancestor of the dipoles which scatter [22, 23, 133–135]. This is a motivation for us to study the genealogical problem in the dipole evolution.

To link to the probabilistic description of diffraction, we notice that the above picture is in the rest frame of the nucleus. In this frame, the total cross section is dominated by the tip fluctuation occuring at the very end of the evolution of the onium, sending exclusively a dipole into the saturation regime; see Chapter 3. Since such fluctuation does not have enough rapidity to develop further, it is unlikely to find more than one dipole to effectively interact with the nucleus. Consequently, it is unlikely to have a diffractive dissociation event, since the latter is due to an even number of participants. To have at least two interacting dipoles, we need a fluctuation at an intermediate rapidity $0 < y_g < y$ consisting a dipole inside the saturation region. This configuration of the dipole evolution then favors the diffractive events.

### 4.3.4 Numerical check for the delay function

In the previous chapter, we already argued that it is technically not possible to use the numerical solution to the original QCD evolution equations to check the asymptotics. Instead, one uses their equivalent version for a BRW model introduced in the last two chapters. Using that model, we are going to check that the conjecture of the delay function in Eq. (4.65) is consistent with numerical calculations.

The function $v_\lambda \equiv 1 - \tilde{w}_\lambda$, in the discrete model of BRW of interest, obeys the equivalence of the T-type BK equation,

$$\begin{aligned} v_\lambda(x, y + \delta y; y_0) = \frac{1}{2}(1 - \delta y) \left[ v_\lambda(x - \delta x, y; y_0) + v_\lambda(x + \delta x, y; y_0) \right] \\ + \delta y \ v_\lambda(x, y; y_0)[2 - v_\lambda(x, y; y_0)]. \end{aligned} \qquad (4.75)$$

with the initial condition given by $v_\lambda(x, y_0; y_0) = (1 - \lambda)T_1(x, y_0)$. The amplitude $T_1(x, y)$ evolves according to the same evolution equation (4.75), from the step initial condition $T(x, 0) = \Theta(-x)$, which is tantamount to the MV or the GBW amplitudes.

In order to measure the delay function with a parameter $\lambda$, the numerical strategy is to have numerics for $v_0$ and $v_\lambda$, and then, to compute the difference in the position between these two fronts. We first advance the amplitude $T_1$ from the step function at zero rapidity to $y_0$ according to Eq. (4.75), with $v_\lambda$ replaced by $T_1$. For $v_0$, we further evolve $T(x, y_0)$ to the final rapidity $y$. Meanwhile, for $v_\lambda$, we multiply $T(x, y_0)$ by $(1 - \lambda)$ and, then, advance the result to $y$. For each front, the front position $x_p$ can be determined from the condition $v_\lambda(x_p, y; y_0) = 1/2$. We repeat the calculation of the delay function for different values of $\lambda$, $y$ and $y_0 < y$.

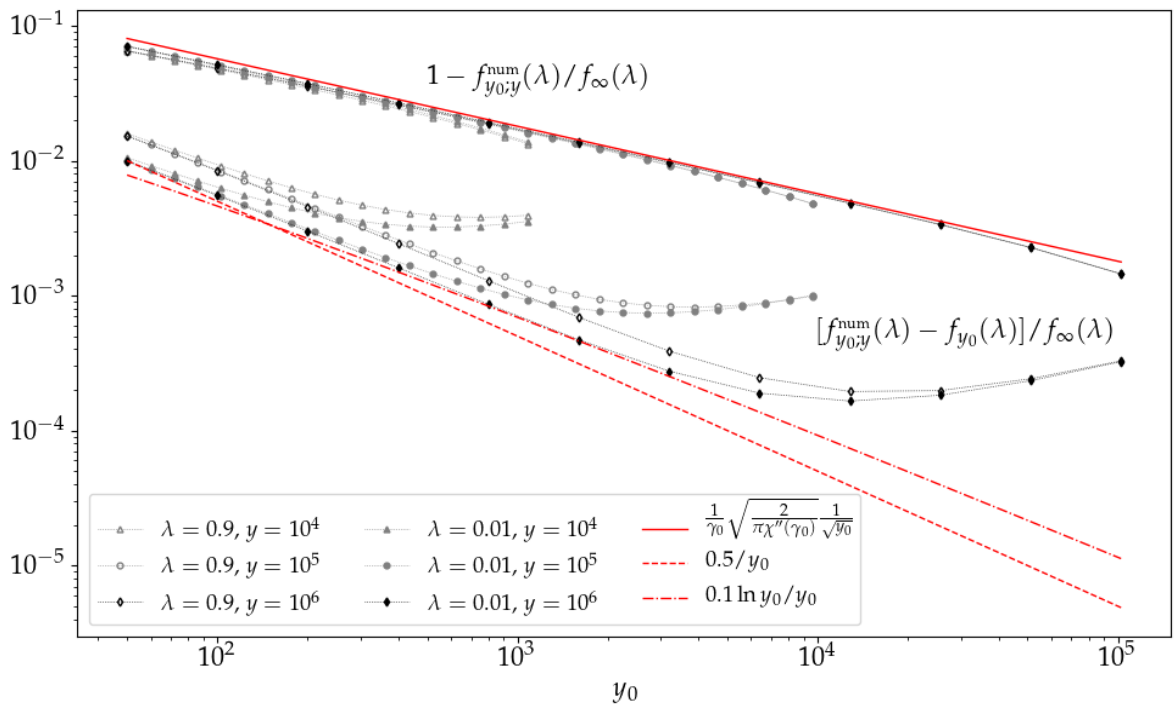

Figure 4.5: Comparison of the delay function $f_{y_0}(\lambda)$ extracted from the numerical solutions to the exact evolution equation and its conjectured formula (4.65), as a function of $y_0$. The points represent the data for $1 - f_{y_0;y}^{num}(\lambda)/f_\infty(\lambda)$, which should tend to $\frac{1}{\gamma_0}\sqrt{\frac{2}{\pi \chi''(\gamma_0)}} \frac{1}{\sqrt{y_0}}$ (full red line) at asymptotically large $y$, and for the difference $[f_{y_0;y}^{num}(\lambda) - f_{y_0}(\lambda)]/f_\infty(\lambda)$. Two different values of $\lambda$ are considered (0.9 and 0.01); and for each $\lambda$, we pick three values of rapidity $y$ ($10^3$, $10^4$ and $10^5$). We plot also functions proportional to $1/y_0$ (dashed line) and $\ln y_0/y_0$ (dashed-dotted line) for comparison. Figure is adapted from Ref. [27]

The results are shown in Fig. 4.5. We consider the two following quantities:

$$1 - \frac{f_{y_0;y}^{num}(\lambda)}{f_\infty(\lambda)}, \quad \text{and} \quad \frac{f_{y_0;y}^{num}(\lambda) - f_{y_0}(\lambda)}{f_\infty(\lambda)}, \tag{4.76}$$

where we have added the $y$ dependence to the numerical solutions, since they are evaluated numerically at a finite rapidity. According to Eq. (4.65), the former quantity should tend to $\frac{1}{\gamma_0}\sqrt{\frac{2}{\pi \chi''(\gamma_0)}} \frac{1}{\sqrt{y_0}}$ at asymptotics. We see that, in the relevant parametric domain, $1 \ll y_0 \ll y$, all numerical points almost superimpose and approach the asymptotic conjecture (see the upper set of points). Futhermore, at larger $y$, the agreement gets better.

The lower set of points represent the second quantity in Eq. (4.76), which is the difference between the numerical data at finite $y$ and the asymptotic conjecture. For a fixed value of $\lambda$, the points in the domain $y_0 \ll y$ for different $y$ overlap. We see that, the mismatch is accordant with a function that decreases with $y_0$ as $1/y_0$ or $\ln y_0/y_0$.

## 4.4 Numerical evaluation of diffractive cross sections of a virtual photon

In this section, we are going to present a numerical study of the diffractive dissociation of a virtual photon off a large nucleus, focusing on the rapidity gap distribution, at the rapidities which are accessible at future electron-ion colliders. We will come back to the original variables for transverse sizes and transverse momenta instead of log variables, and the rapidity (uppercase notation) instead of the rescaled rapidity (lowercase notation). We shall start with a brief recall of the theoretical framework and the choice of kinematics for the numerical calculation. A detailed study of the diffractive onium-nucleus scattering for different onium sizes will be provided prior to presenting predictions for the virtual photon-nucleus scattering.

### 4.4.1 Theoretical framework

The numerical study relies on the QCD dipole model of the nuclear scattering of a virtual photon of virtuality $Q^2$ at high energy. In this model, the total cross section $\sigma_{tot}^{\gamma^*A}(Q^2, Y)$ and the diffractive cross section $\sigma_{diff}^{\gamma^*A}(Q^2, Y; Y_0)$ with a minimal gap $Y_0$ can be factorized according to the dipole factorization in Eqs. (4.1) and (4.3), respectively.

At leading order, the onium forward elastic scattering amplitude $T_1(r, Y)$ obeys the T-type BK equation (2.3), while the onium diffractive scattering cross section $\sigma_D(r, Y; Y_0)$ with a minimal gap $Y_0$ solves the KL equation (4.5). For the sake of convenience, instead of solving the KL equation for $\sigma_D$, we will solve the evolution equation for the dipole inelastic scattering cross section $\sigma_{in}(r, Y; Y_0) = \sigma_T(r, Y) - \sigma_D(r, Y; Y_0)$, which can be shown straightforwardly to be the T-type BK equation. We also refer this equation to as the KL equation for the inelastic cross section $\sigma_{in}$. The initial condition for $\sigma_{in}$ at $Y = Y_0$ is given by

$$\sigma_{in}(r, Y_0; Y_0) = 2T_1(r, Y_0) - T_1^2(r, Y_0). \tag{4.77}$$

In the meantime, the MV amplitude (Eq. (2.4)) is chosen as the initial condition for the forward amplitude $T_1$.

The BK equation for the amplitude $T_1$ is known not only at leading order but also at next-to-leading order [58–65]. However, the KL equation beyond the leading order has not been established. The only known subleading correction to the KL equation comes from the running of the strong coupling [128, 129]. To include such corrections, we simply replace the dipole splitting kernel $\bar{\alpha} dp_{1\to2}(r^\perp, r'^\perp)$ at leading order by its theoretical-motivated running-coupling version. In the following, we will denote the former by $dp^{LO}(r, r')$,

$$dp^{LO}(r^\perp - r'^\perp) \equiv \bar{\alpha} dp_{1\to 2}(r^\perp, r'^\perp). \qquad (4.78)$$

Different prescriptions for the running-coupling kernel were proposed [58, 59, 136, 137]. In the current analysis, we work with the following ones:

(i) the Balitsky prescription [58]:

$$dp^{Bal}(r^\perp, r'^\perp) = \bar{\alpha}(r^2)\frac{d^2r'^\perp}{2\pi}\left[\frac{r^2}{r'^2|r^\perp - r'^\perp|^2} + \frac{1}{r'^2}\left(\frac{\bar{\alpha}(r'^2)}{\bar{\alpha}(|r^\perp - r'^\perp|^2)} - 1\right) \right.$$
$$\left. + \frac{1}{|r^\perp - r'^\perp|^2}\left(\frac{\bar{\alpha}(|r^\perp - r'^\perp|^2)}{\bar{\alpha}(r'^2)} - 1\right)\right], \qquad (4.79)$$

(ii) the so-called "parent dipole" prescription [136]:

$$dp^{Bal}(r^\perp, r'^\perp) = \bar{\alpha}(r^2)\frac{d^2r'^\perp}{2\pi}\frac{r^2}{r'^2|r^\perp - r'^\perp|^2}. \qquad (4.80)$$

Notice again that $\bar{\alpha}$ is kept fixed at a predefined value in Eq. (4.78), while runs with transverse scales in Eqs. (4.79) and (4.80)

We follow Refs. [137, 138] to regularize the running-coupling constant $\bar{\alpha}(r^2)$ to avoid the issue of the Landau pole. In particular, for the dipole sizes under some threshold $r \leq r_c$, the coupling is given by

$$\bar{\alpha}(r^2) = \frac{12N_c}{(11N_c - 2N_f)\ln\left(\frac{4C^2}{r^2\Lambda_{QCD}^2}\right)}, \qquad (4.81)$$

where the number of quark flavors $N_f$ and the number of colors $N_c$ are fixed at the values $N_c = N_f = 3$. The constant $C$ reflects the uncertainty in the Fourier transform from momentum space to coordinate space. In the meantime, for larger dipole sizes, $r > r_c$, the coupling is frozen to a fixed value $\bar{\alpha}_c \equiv \bar{\alpha}(r_c)$. This regularization is motivated by theoretical studies of the Schwinger-Dyson equations for the gluon propagator in the infra-red regime (IR) and lattice QCD [139–141] results which suggest that the strong coupling freezes to a constant value between 0.5 and 0.7 in the IR.

We also notice that, in case of the onium-nucleus scattering, the analytical asymptotic expression for the rapidity gap distribution is now available with fixed coupling, as presented in the previous section. However, there are still no analytical calculations for such quantity in the running-coupling case. One motivation of this numerical analysis is to check whether the prediction (4.69) for the asymptotic shape of the rapidity gap distribution already manifests at a finite rapidity, and whether the running-coupling effects could significantly modify the shape of the distribution.

Since our aim is to produce predictions for future electron-ion colliders, we select kinematics accessible at those machines. Therefore, we pick two values for the total relative rapidity: $Y = 6$ and $Y = 10$. The former value is accessible at BNL-EIC for low to moderate center-of-mass energies, such as $\sqrt{s_{eA}} = 90\ GeV$ or $\sqrt{s_{eA}} = 45\ GeV$ (with $A \geq 56$) [3], and at CERN-LHeC for $\sqrt{s_{ePb}} = 877\ GeV$ [1]. Meanwhile, the latter is reachable for the electron-ion collisions with the center-of-mass energy $\sqrt{s_{ePb}} = 877\ GeV$ at CERN-LHeC [1]. As for the photon virtuality $Q^2$, we choose pertubative values in the range $1 - 10\ GeV^2$. A detailed numerical set up is given in the Appendix C.

### 4.4.2 Diffractive onium-nucleus scattering

The behavior of the nuclear scattering of an onium depends on its relative size compared to the inverse saturation scale $2/Q_s(Y)$ of the nucleus. The latter separates two regimes of interest: the dilute regime ($r < 2/Q_s(Y)$) in which the scattering probability is small, and the saturation regime ($r \geq 2/Q_s(Y)$) with the scattering amplitude of order unity. For this reason, it is convenient to introduce the following scaling variable

$$\tau \equiv \ln \frac{2}{rQ_s(Y)}. \tag{4.82}$$

Its name is from the fact that, at an asymptotically high rapidity and in the region $1 \ll \tau \ll \sqrt{Y}$, the forward elastic amplitude $T_1$ is effectively a function only of this variable; see the last two chapters. The saturation momentum $Q_s(Y)$ can be extracted from the numerical solutions of the BK equation for $T_1$ by using the condition $T_1(r = 2/Q_s(Y), Y) = 0.5$. Positive values of $\tau$ parametrize the dilute regime, while its negative values encode the saturation region.

Let us start with onia of sizes larger than the inverse saturation momentum. Figure 4.6 displays the rapidity gap distributions $\pi(r, Y; Y_{gap})$ for the diffractive dissociation of onia in the saturation region ($\tau \leq 0$). As the onium size goes more deeply inside the saturation regime, the nucleus appears more likely to be a black disk. At this limit, there should be an equal probability of $1/2$ for the scattering to be purely elastic or inelastic. Such two contributions are excluded in the definition of the diffractive dissociation distribution of gaps $\pi(r, Y; Y_{gap})$ ($0 < Y_{gap} < Y$). Consequently, the contribution from the diffractive dissociation is suppressed as $\tau$ becomes more negative, or the onium becomes larger in size, as shown in Fig. 4.6. The suppression is apparently stronger if one takes into account the running of the strong coupling.

We now move on to the distributions for onia picked in the dilute regime, which are plotted in Fig. 4.7. The shapes of the distributions between the fixed and the running coupling are not similar. However, the distributions for both fixed and running coupling schemes are shown to share some common properties. First, large-gap events are more probable for the sizes close to the inverse saturation scale, while small-gap events are dominant for the sizes much smaller than $2/Q_s(Y)$. Viewing from the rest frame of the nucleus, this property can be explained qualitatively using the phenomenological model, in which the diffractive dissociation of a small onium with a fixed gap size $Y_{gap}$ is triggered by a large-dipole fluctuation in the onium Fock state at $Y - Y_{gap}$ creating a dipole larger than the inverse saturation scale $2/Q_s(Y_{gap})$. For onium sizes close to the saturation line, the favored fluctuations are those of small width, which are easy to happen at the early stage of the evolution. On the other hand, if the onium is far from the saturation boundary, the size of the fluctuation should be large, and hence, it needs an enough rapidity span to develop. Therefore, it is more likely for the fluctuation to occur at the downstream of the evolution. Another similarity between two scenarios is the behavior of the distributions when approaching the color transparency limit [31], $r \to 0$. At such limit, the large fluctuation is less probable: a big price should be paid to have a fluctuation with a very large size. Consequently, the contribution of the diffractive dissociation should be suppressed. In this case, the inelastic contribution dominates the total cross section, and even the total cross section rapidly approaches zero.

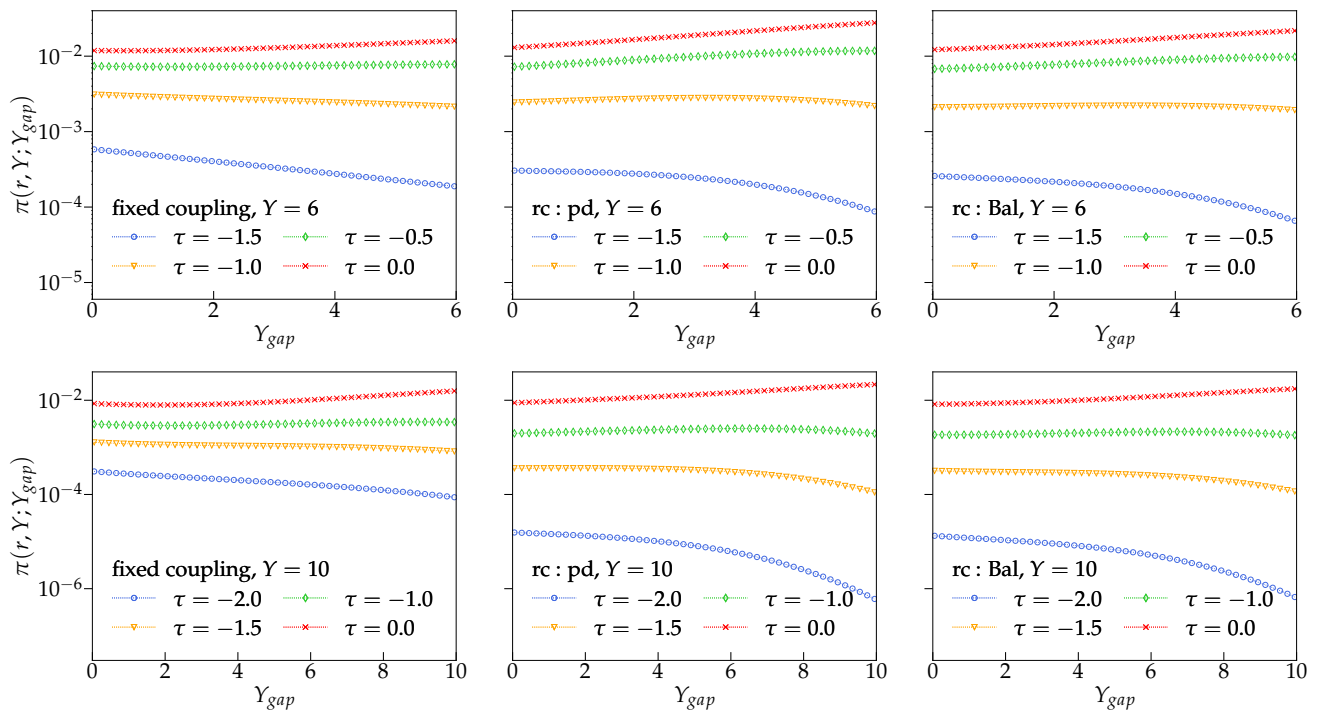

Figure 4.6: Rapidity gap distributions for the diffractive scattering of onia of different sizes in the saturation region ($\tau \leq 0$) at the total relative rapidities $Y = 6$ (first row) and $Y = 10$ (second row) with both fixed and running coupling scenarios. Figure is adapted from Ref. [28].

Interestingly, when the coupling is fixed, the shape predicted by the asymptotic distribution (4.69) is already exhibited at realistic rapidities ($Y = 6$ and $Y = 10$). In order to check that this peculiar shape corresponding indeed to the onset of the asymptotics in Eq. (4.69), we push the calculation to a higher value of the total relative rapidity, in particular $Y = 30$ (see Fig. 4.8). Note that this value of rapidity cannot be accessible at planned electron-ion colliders. Focusing on the fixed-coupling panel, the convex shape of the distribution in the dilute region (see $\tau = 3.4$) looks more similar to the predicted asymptotics. However, finite-rapidity corrections are still sizeable at this rapidity, which would screen the asymptotic appearance. Furthermore, that convex shape also is also seen in the distributions for onium sizes in the saturation region.

In summary, the rapidity gap distribution for the diffractive dissociation of an onium off a large nucleus depends upon the regime (either dilute or saturation) where the onium resides, and is suppressed when the onium size become very different from the inverse saturation momentum of the nucleus. The running of the strong coupling modifies the shape of the distribution in comparison to the fixed coupling scenarios, however, it is not very susceptible to the selection of the running coupling prescription. We also checked that, in the case of fixed coupling, the peculiar convex shape from the asymptotic prediction already shows in the distribution at finite, realistic, rapidities for onium sizes in the dilute regime and not very distant from the inverse saturation line.

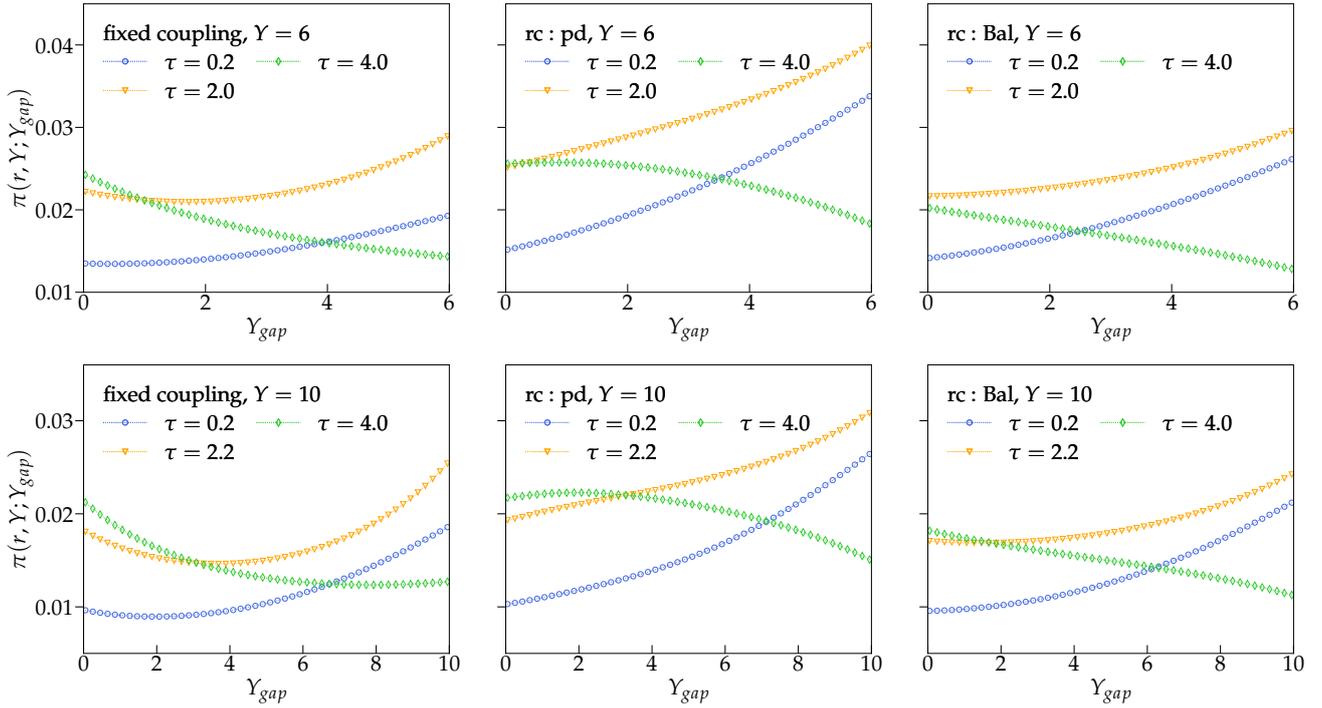

Figure 4.7: Rapidity gap distributions for the diffractive scattering of onia of different sizes in the dilute region ($\tau > 0$) at the total relative rapidities $Y = 6$ (first row) and $Y = 10$ (second row) with both fixed and running coupling scenarios. Figure is adapted from Ref. [28].

### 4.4.3 Predictions for the diffractive dissociation of a virtual photon

We start by plotting the diffractive cross section $\sigma_{diff}^{\gamma^*A}(Q^2, Y; Y_0)$ with a minimal gap $Y_0$ normalized to the total cross section $\sigma_{tot}^{\gamma^*A}(Q^2, Y)$ for the diffractive scattering of a virtual photon; see Fig. 4.9. This quantity estimates the rate of the diffractive events, including the (quasi-)elastic contributions, and how close to the black-disk limit we are. As shown in Fig. 4.9, this ratio decreases slowly with the virtuality $Q^2$. It is closer to the black-disk limit when the scale ratio $Q^2/Q_s^2(Y)$ gets smaller, as the onium states of larger sizes, in the saturation region, are more probable to be probed. The predictions with the running-coupling inclusion are a bit higher than those with the fixed-coupling kernel, by a few percent; and the rates with the Balitsky prescription are closer to the latter. For example, taking $Q^2 = 2GeV^2$, the fixed-coupling equations predict about $20\% - 28\%$ of total events are diffractive at $Y = 6$, and about $25\% - 34\%$ at $Y = 10$, depending on which threshold $Y_0$ is considered. Replacing the fixed-coupling kernel by the Balitsky kernel, such percentages rise to about $22\% - 30\%$ at $Y = 6$, and about $28\% - 36\%$ at $Y = 10$.

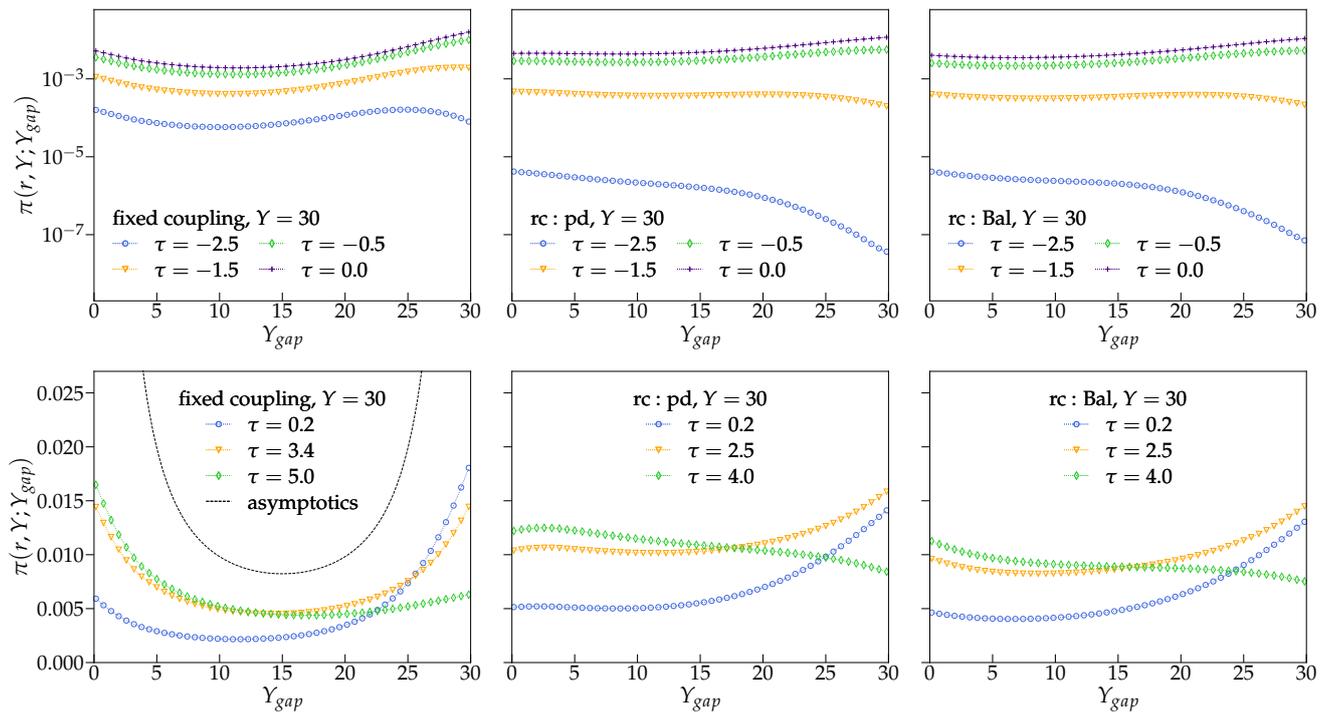

Figure 4.8: Rapidity gap distributions for the diffractive scattering of onia of different sizes in the saturation region ($\tau \leq 0$; first row) and the dilute region ($\tau > 0$; second row) for the total relative rapidity $Y = 30$ with both fixed and running coupling scenarios. In the case of fixed coupling and $\tau > 0$, the analytical asymptotic prediction (4.69) is superimposed for comparison. Figure is adapted from Ref. [28].

Figure 4.10 shows the numerical predictions of the rapidity gap distribution for different kinematics and scenarios. With the chosen set of the virtuality $Q^2$, the quantity $\ln[Q/Q_s(Y)]$, which gives the typical value of the scaling variable $\tau$ in Eq. (4.82), is not far from 0, which suggests that the dominant contribution should come from the onium sizes close to the saturation line. We see that the gap distribution also depends on the relative ratio between the virtuality and the nuclear saturation momentum, $Q^2/Q_s(Y)^2$. Unlike the diffractive-to-total cross section ratio, it decreases when that momentum ratio becomes smaller, i.e. when getting closer to the scattering off a black disk. As pointed out before, this is because only diffractive dissociation is included in the definition of the distribution. We note however that, by the above discussion on the color transparency limit, the distribution should also be suppressed at large $Q^2$ such that $Q \gg Q_s(Y)$. The distribution in such regime is not considered in this analysis.

With the current choices of kinematics, both fixed and running coupling scenarios predict a inclination to have diffractive events with large rapidity gap $Y_{gap}$ (close to the total relative $Y$). However, there is a difference between the two cases: there is an enhancement for the distribution of the gaps close to 0 for the fixed coupling case, which becomes more obvious at a higher rapidity. This is the manifestation of the peculiar convex shape discussed previously, which reflects the analytical prediction from the asymptotic calculations.

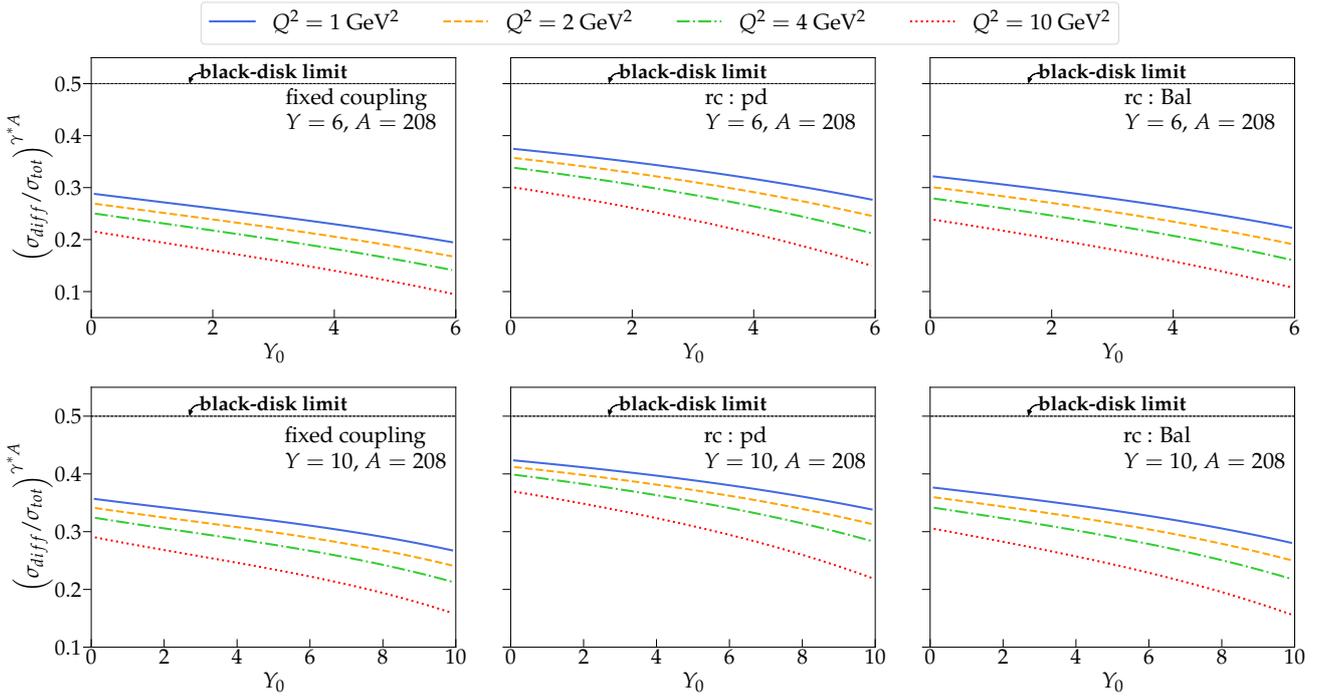

Figure 4.9: The diffractive-to-total cross section ratio $(\sigma_{diff}/\sigma_{tot})^{\gamma^*A}$ as a function of the minimal rapidity gap $Y_0$ at different $Q^2$ considering two values of the total rapidity $Y = 6$ (first row) and $Y = 10$ (second row). Figure is adapted from Ref. [28].

Since the gap distribution is shown to be sensitive to the scale ratio $Q/Q_s(Y)$, and since $Q_s(Y)$ is set to grow with the nuclear mass number $A$ as $Q_s^2(Y) \sim A^{1/3}$ [30, 49, 142, 143] (see the appendix), the distribution should depend on the nuclear mass number $A$, as reported in Fig. 4.11. In particular, it is suppressed as the virtual photon of a fixed virtuality scatters off a larger nucleus. Owing to the fact that the nuclear dependence of the saturation scale is mild, this suppression appears fairly weak.

We can transform the distribution of the rapidity gap $Y_{gap}$ into the distribution of the (squared) invariant mass $M_X^2$ of the inclusive set of final state particles $X$ from the diffractive dissociation of a virtual photon. This distribution reads

$$\mathcal{M}_{diff}(Q^2, Y; M_X^2) \equiv \frac{1}{\sigma_{tot}^{\gamma^*A}} \frac{d\sigma_{diff}^{\gamma^*A}}{dM_X^2} = \frac{\Pi(Q^2, Y; Y_{gap})}{M_X^2 + Q^2}, \qquad (4.83)$$

with

$$Y_{gap} = Y - \ln \frac{M_X^2 + Q^2}{Q^2}. \qquad (4.84)$$

We shall refer $\mathcal{M}_{diff}$ to as the diffractive mass spectrum. Figure 4.12 shows its behavior when varying either $Q^2$ or $A$ and keeping the remainder fixed. The spectra from both fixed and running coupling equations have the same property: the low mass regime dominates over the high mass regime. One can see that, as the photon becomes more virtual, the diffractive events with low dissociated mass get suppressed significantly, while the high mass domain is slightly enhanced. And if the nucleus becomes heavier, the mass spectra also go down, as in the case of the gap distributions. However, the nuclear dependence of the mass spectum appears to be much milder compared to that of the rapidity gap distribution.

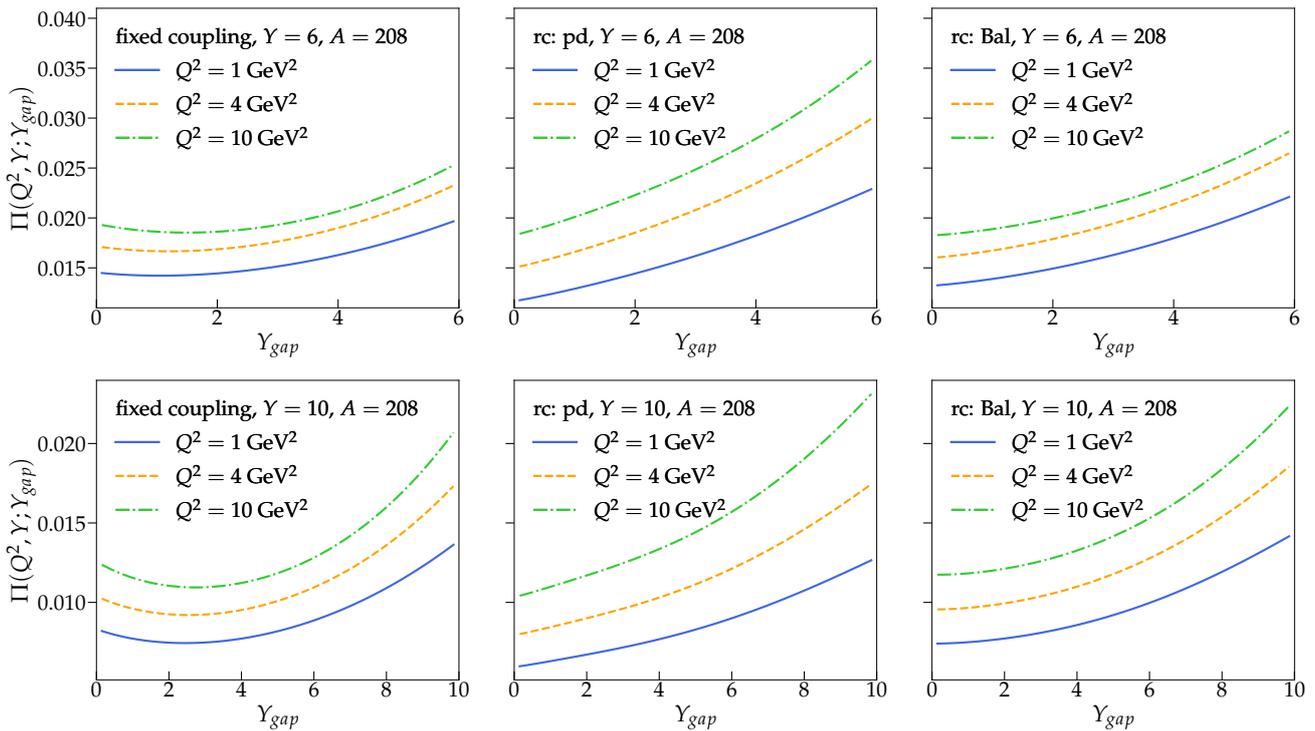

Figure 4.10: The rapidity gap distribution as a function of the rapidity gap $Y_{gap}$ for the diffractive dissociation of a virtual photon at different values of the kinematic variables $Q^2$ and $Y$. Figure is adapted from Ref. [28].

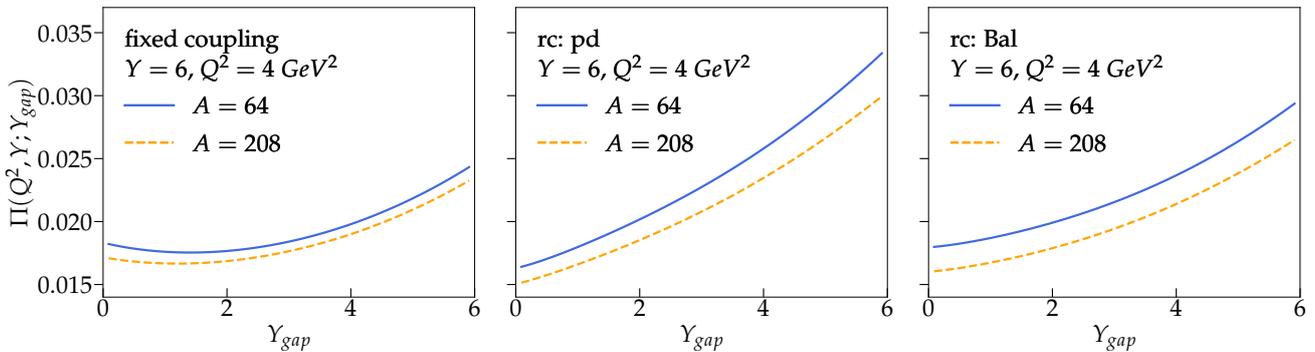

Figure 4.11: The rapidity gap distribution for two different nuclei $A = 64$ and $A = 208$. The kinematic variables are fixed at $Y = 6$ and $Q^2 = 4 GeV^2$. Figure is adapted from Ref. [28].

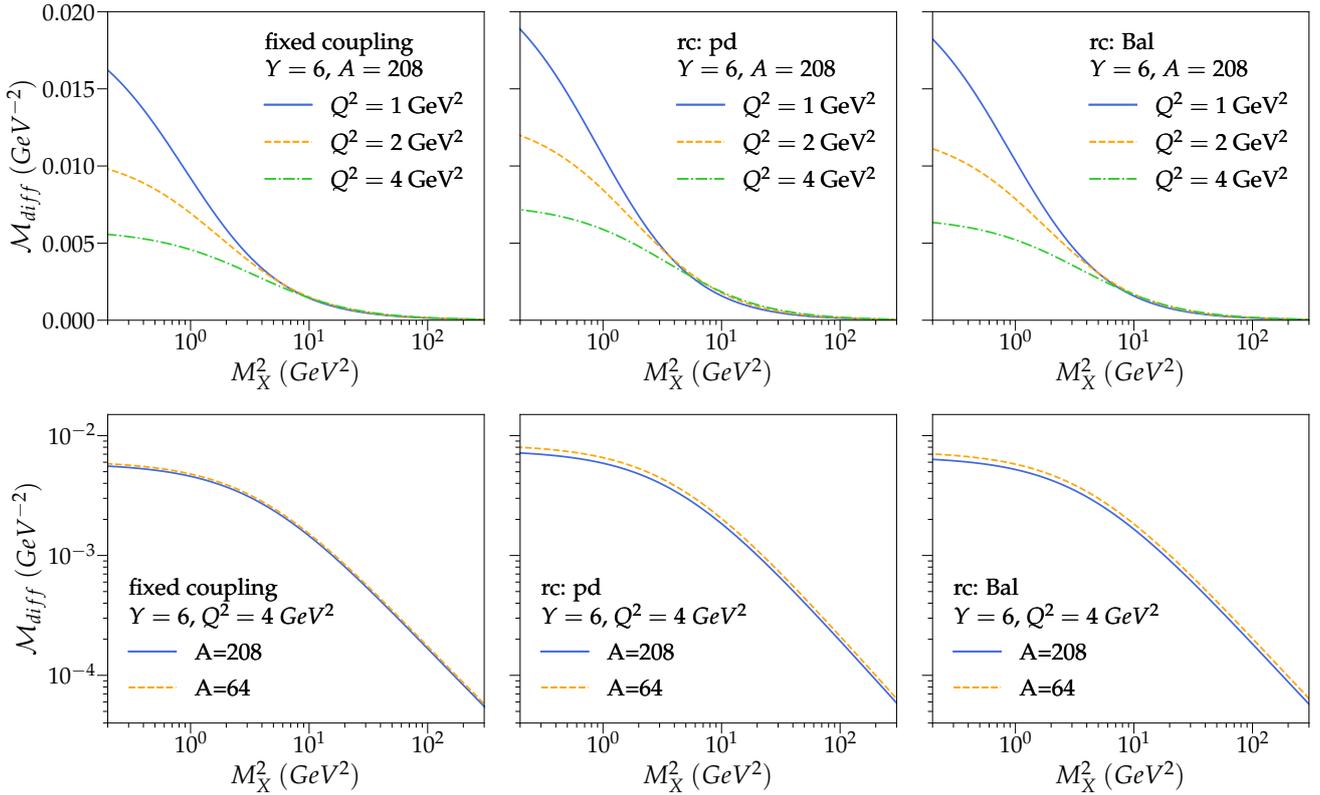

Figure 4.12: The diffractive mass spectra at different virtualities $Q^2$ when $A$ is fixed (first row) and at two different nuclear mass numbers $A$ when $Q^2$ is fixed (second row) for the total rapidity $Y = 6$. The former is adapted from Ref. [28].

### 4.4.4 Running of the strong coupling for diffractive dissociation

The inclusion of the running coupling correction amounts to slow down the dipole evolution by suppressing the emission of small dipoles in the quantum state of the onium [136, 137]. Consequently, in the wave function of an initial onium of size larger than the saturation line ($\tau < 0$), the emissions of large dipoles inside the saturation region are favoured. The scattering is then more elastic, leading to the stronger suppression of the diffractive dissociation when moving deeply into the saturation regime. In addition, at large onium sizes, the running-coupling kernels tend to a universal form. Therefore, the rapidity gap distributions for the two chosen running-coupling prescriptions look very similar deeply inside the saturation region.

In addition, with our choices of kinematics, the dominant domain for the size of the onium state in the virtual photon-nucleus is in the vicinity of the saturation scale. With the suppression of small-dipole emissions, the nuclear scattering of the onium states of the virtual photon is more elastic. As a result, the diffractive-to-total cross section ratio gets closer to the black-disk limit when taking into account the running-coupling correction. This could also explain the observation that, the running-coupling equations lead to a more significant dominance of the large-gap domain over the small-gap one.

Let us now apply, in a very naive way, the aforementioned phenomenological model used in Refs. [22, 23] for the running coupling case. We still base on the twofold representation of the BK equation in which the latter controls the evolution of the probability $P(r, \tilde{Y}; R)$ of having at least one dipole of size larger than some scale $R$ in the Fock state of the initial onium $r$ at the rapidity $\tilde{Y}$. In the scaling region, it reads

$$P(r, \tilde{Y}; R) \simeq C_{rc} \tilde{Y}^{1/6} \left[ \frac{R_{rc}^2(\tilde{Y})}{R^2} \right]^{\gamma_0} \mathrm{Ai} \left[ \xi_1 + \frac{3\xi_1}{4\beta_c} \frac{\ln\left( \frac{R^2}{R_{rc}^2(\tilde{Y})} \right)}{\tilde{Y}^{1/6}} \right], \tag{4.85}$$

where $\xi_1 = -2.338\cdots$ is the rightmost zero of the Airy function $\mathrm{Ai}(\xi)$, and

$$\beta_c \equiv \frac{3}{4}\xi_1 \left( \frac{\chi''(\gamma_0)}{\sqrt{1.5\gamma_0\chi(\gamma_0)}} \right)^{1/3} = -5.36\cdots \tag{4.86}$$

The function $R_{rc}(\tilde{Y})$ is the typical largest dipole size of the Fock state (mean tip) in the running-coupling scenario, which is expected to grow with $\tilde{Y}$ in a similar way to the saturation momentum. It reads

$$R_{rc}^2(\tilde{Y}) \sim r^2 \exp\left[ \alpha_c(\tilde{Y} + \delta_1)^{1/2} + \beta_c(\tilde{Y} + \delta_2)^{1/6} \right] \tag{4.87}$$

where $\alpha_c = \sqrt{(8\chi'(\gamma_0))/3} \simeq 3.61$, and $\delta_{1,2}$ encode finite-rapidity subleading corrections. According to the phenomenological model for diffraction, to have a diffractive event with rapidity gap $Y_{gap}$, there should be a large fluctuation in the wave function of the onium at $Y - Y_{gap}$ sending particle to the nuclear saturation regime at $Y_{gap}$. For such argument to be valid, the mean-field front of the onium at $Y - Y_{gap}$ and the saturation regime of the nucleus at $Y_{gap}$ should necessarily not overlap each other. This condition can be simply expressed as

$$R_{sc}^2(Y - Y_{gap}) < \frac{1}{Q_{s,rc}^2(Y_{gap})}, \tag{4.88}$$

where $Q_{s,rc}(Y)$ is the nuclear saturation scale at $Y$ with the running-coupling correction. The latter reads

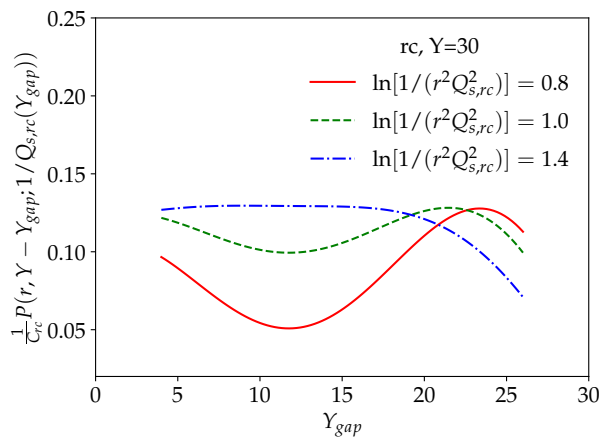

Figure 4.13: The probability $P(r, Y - Y_{gap}; 1/Q_{s,rc}(Y_{gap}))$ with the overall constant excluded as a function of $Y_{gap}$. Three different onium sizes are selected. The total rapidity is $Y = 30$.

$$Q_{s,rc}^2(Y_{gap}) \simeq \Lambda_{QCD}^2 \exp\left[ \alpha_c(Y_{gap} + \delta_2)^{1/2} + \beta_c(Y_{gap} + \delta_3)^{1/6} \right], \tag{4.89}$$

where, as in Eq. (4.87), $\delta_{2,3}$ are subleading corrections. In the spirit of the phenomenological model, the gap distribution is given by

$$\pi_{rc}(r, Y; Y_{gap}) \propto P(r, Y - Y_{gap}; 1/Q_{s,rc}(Y_{gap})). \qquad (4.90)$$

Figure 4.13 shows the probability $P(r, Y - Y_{gap}; 1/Q_{s,rc}(Y_{gap}))$ of the form (4.85) as a function of the rapidity gap $Y_{gap}$ for $Y = 30$, with the overall unknown constant being ignored. We see that the curves share some features with the numerics of the dilute regime in the running-coupling case, as shown in Fig. 4.8. In particular, for the onium sizes close to the saturation boundary, the large-gap domain are more favored. Meanwhile, the small-gap domain dominates the distribution for larger onium sizes. Although this is just a naive estimation, it may suggest that, to a certain extent, one could adapt the asymptotic calculation for the fixed-coupling case to that for the running-coupling scenario.

### 4.4.5 Comparisons to other studies

A recent study [144] on diffractive scattering in electron-ion collisions also made predictions on the diffractive-to-total cross section ratio. In particular, at $Q^2 = 2\ GeV^2$ diffraction is predicted to account for about 20% of the events in the ePb collisions, which does not vary much at different momentum fractions $x$ (or correspondingly at different rapidity $Y$), based on several models. In fact, that prediction is rather close to our above prediction when the minimal gap $Y_0$ is large, with the fixed-coupling or the running-coupling Balitsky kernels. For other cases, our predictions is fairly higher than those of the cited study.

The shape of the mass spectra from our calculation is quite similar to that of the same quantity shown in Ref. [3] based on the models of saturation [145, 146], and of leading-twist shadowing [147, 148]. In comparison to the results of the former model, our results expose two differences. First, the mass spectra from a model of saturation have a local maximum in the low-mass domain [3], which does not appear in our predictions. Second, while our results predict a slight enhancement in the high-mass regime at higher $Q^2$, the mass spectra from that saturation model appear to be suppressed at all possible values of the invariant mass when increasing the virtuality.

Finally, we can comment on the nuclear dependence of the difractive distributions. The suppression of the mass spectra when scattering off a larger nucleus seems to qualitatively agree with the results from the leading-twisted shadowing model [3, 147, 148]. Consequently, our results may reflect the nuclear shadowing effect on the diffractive gap (mass) distributions.

## 4.5 Summary

The dipole factorization of the high-energy nuclear scattering of a virtual photon allows to formulate the diffractive scattering process in term of the more fundamental object, the onium, and hence, promotes the study of the diffractive dissociation of the latter.

We have presented that the diffractive scattering of an onium off a large nucleus at a large rapidity, in such a kinematic regime that the total cross section is small, can be described by a purely probabilistic formulation. In particular, it is twice the probability of having in the Fock state of the onium at the scattering rapidity an even number of participating dipoles. This classical-like formulation is unforeseen, since diffraction is typically a quantum mechanical phenomenon with no classical counterpart.

Such probabilistic formulation and the phenomenological model of dipole distribution have enabled us to derive the complete expressions of the diffractive-to-total cross section ratio requiring a minimal rapidity gap $Y_0$, and of the distribution of rapidity gaps $Y_{gap}$ for the scattering of an onium of size $r$ in the geometric scaling region at an asymptotic total relative rapidity $Y$. The former reads

$$\frac{\sigma_D(r, Y; Y_0)}{\sigma_T(r, Y)} = \frac{\ln 2}{\gamma_0} \left[ \frac{1}{\ln[1/(r^2 Q_s^2(Y))]} + \sqrt{\frac{2}{\pi \chi''(\gamma_0)}} \frac{1}{\sqrt{\bar{\alpha} Y_0}} \right], \tag{4.91}$$

where the nuclear saturation scale grows with $Y$ as $Q_s^2(Y) = Q_A^2 e^{\chi'(\gamma_0)\bar{\alpha} Y_0}/(\bar{\alpha} Y)^{3/(2\gamma_0)}$. Meanwhile, the asymptotic rapidity gap distribution is given by

$$\pi(r, Y; Y_{gap}) = \frac{\ln 2}{\gamma_0 \sqrt{2\pi \chi''(\gamma_0)}} \frac{1}{\sqrt{\bar{\alpha}}} \left[ \frac{Y}{Y_{gap}(Y - Y_{gap})} \right]^{3/2} \exp\left( -\frac{\ln^2[r^2 Q_s^2(Y)]}{2\chi''(\gamma_0)\bar{\alpha}(Y - Y_{gap})} \right). \tag{4.92}$$

We have also found that the weight $w_k$ of having $k$ participating dipoles (for $k \geq 2$) decays gradually like $1/[k(k-1)]$. This implies that, events with a large number of color singlet exchanges between the onium and the nucleons constituting the nucleus are typical for diffraction, which is consistent with the general expectation that diffraction is sensitive to the onset of saturation.

Employing the dipole model for diffractive dissociation, we have also performed a numerical evaluation of the diffractive cross sections and of the rapidity gap distributions using both the fixed-coupling and the running-coupling evolution equations for kinematics accessible at future electron-ion machines. The results predict a significant ratio of diffractive events at the chosen kinematical variables. Interestingly, at realistic rapidities, the numerics for the rapidity gap distributions in the fixed-coupling case already exhibit the shape predicted by Eq. (4.92) at asymptotics. In addition, while there is a siginificant difference between the distributions deduced within the fixed-coupling scenario and those from the running-coupling equations, they are not very sensitive to the choice of the prescription to taking into account the running of the strong coupling in the dipole kernel.

The predictions presented in Eqs. (4.91) and (4.92) may be viewed as a good starting point for the construction of a model for diffractive dissociation in real electron-ion collisions at future colliders. However, for the model to be realistic, it would be extremely useful to find a systematic way to compute the next-to-leading order corrections, presumably of relative order $\ln Y/\sqrt{Y}$ or $1/\sqrt{Y}$, which makes a potential future development. Another possible development is to extend the current analytical calculation for the case of including the running-coupling correction.

Let us close this chapter by refering back to general branching random walk processes. In the context of the latter, the weights $w_k$ could be interpreted as the rate to select exactly $k$ particles in the tip according to a particular distribution which is taken to be the initial condition for the F-KPP equation. Therefore, a more rigorous derivation of the weights $w_k$ beyond the heuristics is of great interest to understanding the tip region of a general branching random walk, which, as already mentioned previously, has many applications in different fields of science. The generating function method presented in Section 4.3.2 could pave a promising way for this potential development.

# Conclusions and outlooks

Each chapter has its own summary at the end. Here we would like to draw some main points from what we have discussed through the whole of the thesis.

This thesis focused on the deep-inelastic virtual photon-nucleus scattering at high energy, which is related to the nuclear scattering of an onium, a color-singlet quark-antiquark dipole, by the dipole factorization. The latter process can be described, at large number of colors, by using the QCD color dipole model in which soft-gluon emissions in the wave function of the onium is replaced by a dipole branching process. Within this formulation, the dipole evolution is a peculiar one-dimensional branching random walk. As a matter of fact, the Balitsky-Kovchegov (BK) equation describing the rapidity evolution of QCD amplitudes is in the same universality class of the F-KPP equation, which controls the time evolution of branching-diffusion processes on a line.

The nuclear scattering of an onium, in a frame in which the latter is highly evolved, is due to the interaction of a particular subset of dipoles in the onium Fock state, which is generated by dipole branching process, with the nucleus. In the scaling region, in which the probability for the same dipole in the Fock state of the onium to scatter more than once is negligible, the scattering is triggered by a large fluctuation which creates at least one dipole of large transverse size beyond the typical configuration at a certain rapidity during the evolution. The dominant realization of the fluctuation is selected in such a way that the overlap between the dipole density and the dipole-nucleus amplitude, equiped with a probability density of the fluctuation size, is optimal, which eventually guarantees the boost invariance of the forward elastic scattering amplitude. Consequently, the fluctuation looks very different in different frames, from the rest frame of the nucleus to a frame in which the nucleus is significantly boosted. Eventually, the dipole density at the scattering rapidity is generated by the combination of the mean-field evolution and a rare fluctuation, which is the essence of the phenomenological picture for dipole distribution in the onium-nucleus scattering.

The phenomenological model for dipole distribution allows the freedom to select a frame in which one can derive the asymptotic distribution for the branching rapidity of the last common ancestor of the set of dipoles which interact, when this set consists of at least two dipoles. This genealogical problem for the QCD dipole evolution is in analogy to another one for more general one-dimensional branching random walks: the probability distributions in the two problems are very similar in their analytical forms.

In addition to the scattering configurations, we discussed also diffractive dissociation of onia for which an equation was written down 20 years ago by Kovchegov and Levin (KL), but no analytical solution had been found. We found that the diffractive cross section for a small onium is twice the probability to have an even, non-zero, number of interacting dipoles in the onium Fock state, and that, while the total cross section is dominated by one single exchange, the events with a large number of participants are typical for diffraction. Interestingly, using the phenomenological model for dipole distribution, we are able to, from that probabilistic description of diffraction, derive the parameter-free asymptotics expressions of the diffractive cross section with a minimum gap, and of the rapidity gap distribution. Furthermore, within the dipole formulation, using the numerical solutions to the original QCD evolution equations (BK and KL) and their extension taking into account the running of the strong coupling, we investigated numerically diffractive dissociation of a virtual photon for the kinematics accessible at future electron-ion colliders. Predictions on the shape of the rapidity gap distribution and on the diffractive-to-total cross section ratio were presented. As an interesting point, the analysis of the rapidity gap distribution showed a connection between the distribution shape at realistic rapidities and the predictions of the phenomenological model at asymptotics.

Since the investigation of the nuclear scattering of small onia indicated the importance to characterize the dipole distribution in the region close to the largest dipole, and since the dipole evolution belongs to the class of one-dimensional branching random walk, we established a Monte Carlo algorithm to generate a tip region of an one-dimensional branching random walk evolved to large time, which provides a numerical tool to study the particle distribution near a tip and genealogical structure of the evolution in both typical and rare realizations. The algorithm could also be adapted to the continuous limit of the branching random walk - the branching Brownian motion, which offers a starting point for analytical studies of the tip region.

Not only the works presented in the thesis answered some of our questions, they also opened potential questions for further studies. As mentioned previously, in QCD, possible developments include the extension of the analytical study to the sub-asymptotic regime, and to the running-coupling case, which would be important for further phenomenological applications in future electron-ion machines. One could also question on the possibility of extending the current analytical studies to other dilute-dense systems, such as the proton-ion collisions. On the statistical side, the construction of a theoretical formulation to calculate tip observables is a promising outlook.

# Spinor matrix elements

The derivation of the color dipole model in Chapter 1 requires the following two matrix elements in the eikonal limit:

$$
\begin{aligned}
\Gamma_1^\mu(p', p) &= \bar{u}_r(p')\gamma^\mu u_s(p), \\
\Gamma_2^\mu(p', p) &= \bar{v}_r(p')\gamma^\mu v_s(p).
\end{aligned}
\tag{A.1}
$$

Here we will compute the first current, based on Eqs. (1.30) and (1.35). The expression for the second term can be deduce directly from

$$
\bar{u}_r(p)\gamma^\mu u_s(p') = \bar{v}_r(p')\gamma^\mu v_s(p).
\tag{A.2}
$$

We can choose two particular basis vectors $w_r$ ($r = \pm 1/2$) which satisfy the condition (1.30) as follows:

$$
\begin{aligned}
w_{1/2}(p) &= 2^{-1/4}\sqrt{p^+}(1, 0, 1, 0)^T, \\
w_{-1/2}(p) &= 2^{-1/4}\sqrt{p^+}(0, 1, 0, -1)^T,
\end{aligned}
\tag{A.3}
$$

where $T$ stands for the matrix transpose operator. The novel normalization reads

$$
w_r^\dagger(p')w_s(p) = \sqrt{2p'^+p^+}\delta_{rs}.
\tag{A.4}
$$

We also use the Dirac representation for the gamma matrices,

$$
\gamma^0 = \begin{pmatrix} \mathbb{1}_{2\times 2} & 0 \\ 0 & -\mathbb{1}_{2\times 2} \end{pmatrix}, \quad \gamma^i = \begin{pmatrix} 0 & \sigma^i \\ -\sigma^i & 0 \end{pmatrix},
\tag{A.5}
$$

where $\sigma^i$ ($i = 1, 2, 3$) are the Pauli matrices,

$$
\sigma^1 = \begin{pmatrix} 0 & 1 \\ 1 & 0 \end{pmatrix}, \sigma^2 = \begin{pmatrix} 0 & -i \\ i & 0 \end{pmatrix}, \sigma^3 = \begin{pmatrix} 1 & 0 \\ 0 & -1 \end{pmatrix}.
\tag{A.6}
$$

## "Plus" component

Substituting the expression of the spinor $u$ in Eq. (1.35) into the expression of the $\Gamma_1^\mu$ and keeping only the plus component, we have

$$\Gamma_1^+(p',p) = \left[\bar{w}_r(p') + w_r^\dagger(p')(-\gamma^\perp \cdot p'^\perp + m)\frac{\gamma^0\gamma^+}{2p'^+}\right]\gamma^+\left[w_s(p) + \frac{\gamma^+}{2p^+}(\gamma^\perp \cdot p^\perp + m)w_s(p)\right]. \quad (A.7)$$

Since $(\gamma^+)^2 = 0$, the only surviving term is

$$\Gamma_1^+(p',p) = \bar{w}_r(p')\gamma^+ w_r(p) = \sqrt{2}w^\dagger(p')\Lambda_+ w_s(p). \quad (A.8)$$

Since $w_s$ is a vector of the subspace image of the projector $\Lambda_+$, then $\Lambda_+ w_s = w_s$. Using Eq. (A.4), we obtain the following final expression for the plus component:

$$\boxed{\Gamma_1^+(p',p) = 2\sqrt{p'^+p^+}\delta_{rs}.} \quad (A.9)$$

## "Minus" component

The minus component $\Gamma_1^-$ reads

$$\begin{aligned}\Gamma_1^-(p',p) &= \left[\bar{w}_r(p') + w_r^\dagger(p')(-\gamma^\perp \cdot p'^\perp + m)\frac{\gamma^0\gamma^+}{2p'^+}\right]\gamma^-\left[w_s(p) + \frac{\gamma^+}{2p^+}(\gamma^\perp \cdot p^\perp + m)w_s(p)\right] \\ &= w_r^\dagger(p')(-\gamma^\perp \cdot p'^\perp + m)\frac{\gamma^0\gamma^+\gamma^-\gamma^+}{4p'^+p^+}(\gamma^\perp \cdot p^\perp + m)w_s(p),\end{aligned} \quad (A.10)$$

where other terms after the expansion vanish due to the fact that $\Lambda_- w_r = 0$, as $w_s$ is on the subspace image of the projector $\Lambda_+$. Using the properties $\gamma_-\gamma_+ = \sqrt{2}\gamma_0\gamma_+$ and $(\Lambda_+)^2 = \left[(\gamma_0\gamma_+)/\sqrt{2}\right]^2 = (\gamma_0\gamma_+)/\sqrt{2}$, and the anticommutation relation of the gamma matrices, we get

$$\Gamma_1^-(p',p) = w_r^\dagger(p')\frac{(-\gamma^\perp \cdot p'^\perp + m)(\gamma^\perp \cdot p^\perp + m)}{\sqrt{2}p'^+p^+}w_s(p). \quad (A.11)$$

Let us evaluate the numerator of the second factor by expanding it,

$$\begin{aligned}(-\gamma^\perp \cdot p'^\perp + m)(\gamma^\perp \cdot p^\perp + m) &= -\gamma^i\gamma^j p'_i p_j + m^2\mathbb{1}_{4\times 4} + m\gamma^j(p^j - p'^j) \\ &= \begin{pmatrix}\sigma^i\sigma^j & 0 \\ 0 & \sigma^i\sigma^j\end{pmatrix}p'_i p_j + m^2\mathbb{1}_{4\times 4} + m\gamma^j(p^j - p'^j).\end{aligned} \quad (A.12)$$

Using $\sigma^i\sigma^j = \delta^{ij}\mathbb{1}_{2\times 2} + i\varepsilon^{ij}{}_k\sigma^k$ ($\varepsilon^{ijk}$ is the Levi-Civita symbol), Eq. (A.12) becomes

$$(p'^\perp \cdot p^\perp)\mathbb{1}_{4\times 4} + i(p'^\perp \times p^\perp)\begin{pmatrix}\sigma^3 & 0 \\ 0 & \sigma^3\end{pmatrix} + m^2\mathbb{1}_{4\times 4} + m\gamma^j(p^j - p'^j), \quad (A.13)$$

where the cross product of two transverse vectors $p'^\perp$ and $p^\perp$ is a scalar, $p'^\perp \times p^\perp = \det(p'^\perp, p^\perp)$. Now, using Eqs. (A.3) to (A.5) we have

$$w_r^\dagger(p') \begin{pmatrix} \sigma^3 & 0 \\ 0 & \sigma^3 \end{pmatrix} w_s(p) = \sqrt{2p'^+ p^+}\, 2r\, \delta_{rs},$$
$$w_r^\dagger(p') \gamma^j p^j w_s(p) = -\sqrt{2p'^+ p^+}\, 2r(p^1 - 2irp^2)\delta_{r,-s}. \tag{A.14}$$

All in all, we obtain the following expression for the minus component:

$$\boxed{\Gamma_1^-(p',p) = \frac{\delta_{rs}}{\sqrt{p'^+ p^+}} \left[ p'^\perp \cdot p^\perp + m^2 + 2ir(p'^\perp \times p^\perp) \right] - \frac{2r\delta_{r,-s}}{\sqrt{p'^+ p^+}} \left[ (p^1 - 2irp^2) - (p'^1 - 2irp'^2) \right].} \tag{A.15}$$

## Transverse components

For the transverse components, we have

$$\Gamma_1^i(p',p) = \left[ \bar{w}_r(p') + w_r^\dagger(p')(-\gamma^\perp \cdot p'^\perp + m)\frac{\gamma^0 \gamma^+}{2p'^+} \right] \gamma^i \left[ w_s(p) + \frac{\gamma^+}{2p^+}(\gamma^\perp \cdot p^\perp + m)w_s(p) \right]$$
$$= w_r^\dagger(p')(-\gamma^\perp \cdot p'^\perp + m)\frac{\gamma^i}{\sqrt{2p'^+}} w_s(p) - w_r^\dagger(p')\frac{\gamma^i}{\sqrt{2p^+}}(\gamma^\perp \cdot p^\perp + m)w_s(p) \tag{A.16}$$
$$= w_r^\dagger(p')\left( -\frac{\gamma^\perp \cdot p'^\perp}{p'^+}\frac{\gamma_i}{\sqrt{2}} - \frac{\gamma_i}{\sqrt{2}}\frac{\gamma^\perp \cdot p^\perp}{p^+} \right)w_s(p) + \frac{m}{\sqrt{2}}\left( \frac{1}{p'^+} - \frac{1}{p^+} \right)w_r^\dagger(p')\gamma_i w_s(p),$$

where we have used the facts that $w_{r,s} = \Lambda_+ w_{r,s}$ and $(\gamma^+)^2 = 0$. Using Eqs. (A.3) to (A.5), we obtain

$$w_r^\dagger(p') \gamma^i w_s(p) = -\sqrt{2p'^+ p^+}\, 2r(\delta^{i1} - 2ir\delta^{i2})\delta_{r,-s},$$
$$w_r^\dagger(p') \gamma^i \gamma^j p^j w_s(p) = -\sqrt{2p'^+ p^+}(p^i + 2ir\varepsilon^{ij}p^j)\delta_{rs},$$
$$w_r^\dagger(p') \gamma^j p'^j \gamma^i w_s(p) = -\sqrt{2p'^+ p^+}(p'^i - 2ir\varepsilon^{ij}p'^j)\delta_{rs}, \tag{A.17}$$

where $\varepsilon^{ij}$ is the Levi-Civita symbol. Substituting Eq. (A.17) into Eq. (A.16), one gets

$$\boxed{\Gamma_1^i(p',p) = \delta_{rs}\sqrt{p'^+ p^+}\left[ \frac{p'^i - 2ir\varepsilon^{ij}p'^j}{p'^+} + \frac{p^i + 2ir\varepsilon^{ij}p^j}{p^+} \right] - \delta_{r,-s}2rm\sqrt{p'^+ p^+}\left( \frac{1}{p'^+} - \frac{1}{p^+} \right)(\delta^{i1} - 2ir\delta^{i2})} \tag{A.18}$$

We see that the spinor matrix element $\Gamma_1^\mu(p',p)$ contain both helicity-flip and helicity-non-flip terms (except for the plus component). In the eikonal limit, its becomes

$$\Gamma_1^\mu(p,p) = 2p^\mu \delta_{rs}. \tag{A.19}$$

That is, only the helicity-non-flip term survives.

# Some useful integrals

## B.1 Integrals used in Chapters 3 and 4

The calculations presented in Chapters 3 and 4 involve some integrals which are defined as follows:

$$
\begin{aligned}
I_1(A) &\equiv \int_1^\infty \frac{dt}{t^2} \left[ 1 - e^{-At} \right], \\
I_2(A) &\equiv \int_1^\infty \frac{dt}{t^2} \left[ 1 - (1+At)e^{-At} \right], \\
R(I_0) &\equiv \int_A^\infty \frac{dI}{I^2} \ln\left(\frac{I}{A}\right) \exp\left[ -\frac{\ln^2(I/A)}{\beta_0 \tilde{y}_1} \right], \\
S_{k\geq 0}(A) &\equiv \int_A^\infty \frac{dI}{I} \ln\left(\frac{I}{A}\right) I^{k-1} e^{-I} \exp\left[ -\frac{\ln^2(I/A)}{\beta_0 \tilde{y}_1} \right].
\end{aligned}
\tag{B.1}
$$

We are going to evaluate them in the limit $A \ll 1$, keeping only the leading term. By a change of variable $u = At$, the first two integral can be rewritten as

$$
\boxed{
\begin{aligned}
I_1(A) &= A \int_A^\infty \frac{du}{u^2} \left[ 1 - e^{-u} \right] = 1 - e^{-A} + A\Gamma(0,A) \underset{A \ll 1}{=} A \ln\frac{1}{A} + \mathcal{O}(A), \\
I_2(A) &= A \int_A^\infty \frac{du}{u^2} \left[ 1 - (1+u)e^{-u} \right] = I_1(A) - A\Gamma(0,A) = 1 - e^{-A} \underset{A \ll 1}{=} A + \mathcal{O}(A^2).
\end{aligned}
}
\tag{B.2}
$$

For the integral $R$, the integration by parts leads to

$$
\begin{aligned}
R(I_0) &= \frac{\beta_0 \tilde{y}_1}{2A} - \frac{\beta_0 \tilde{y}_1}{2} \int_A^\infty \frac{dI}{I^2} \exp\left[ -\frac{\ln^2(I/A)}{\beta_0 \tilde{y}_1} \right] = \frac{\beta_0 \tilde{y}_1}{2A} - \frac{\beta_0 \tilde{y}_1}{2A} \int_0^\infty dt e^{-t} \exp\left[ -\frac{t^2}{\beta_0 \tilde{y}_1} \right] \\
&= \frac{\beta_0 \tilde{y}_1}{2A} \left\{ 1 - \frac{\sqrt{\pi \beta_0 \tilde{y}_1}}{2} \exp\left(\frac{\beta_0 \tilde{y}_1}{4}\right) \left[ 1 - \mathrm{erf}\left(\frac{\sqrt{\beta_0 \tilde{y}_1}}{2}\right) \right] \right\}.
\end{aligned}
\tag{B.3}
$$

The relevant limit is $\tilde{y}_1 \gg 1$, in which case we can use the following expansion of the error function:

$$
\mathrm{erf}\left(\frac{\sqrt{\beta_0 \tilde{y}_1}}{2}\right) = 1 - \frac{e^{-\frac{\beta_0 \tilde{y}_1}{4}}}{\sqrt{\pi}} \left[ \frac{2}{\sqrt{\beta_0 \tilde{y}_1}} - \frac{4}{\beta_0 \tilde{y}_1} \frac{1}{\sqrt{\beta_0 \tilde{y}_1}} + \mathcal{O}\left(\frac{1}{\tilde{y}_1^2 \sqrt{\tilde{y}_1}}\right) \right].
\tag{B.4}
$$

$R$ then becomes

$$R(I_0) \underset{\tilde{y}_1 \gg 1}{\simeq} \frac{1}{A} \left[ 1 + \mathcal{O}\left( \frac{1}{\tilde{y}_1} \right) \right].$$ (B.5)

Now we deal with the integral $S_k$ (with an interger $k \geq 0$). They can be rewritten in the form of a series as

$$\begin{aligned} S_k(A) &= \sum_{j=0}^{\infty} \frac{(-1)^j}{j! \, (\beta_0 \tilde{y}_1)^j} \int_A^{\infty} \frac{dI}{I} \ln^{2j+1}\left( \frac{I}{A} \right) I^{k-1} e^{-I} \\ &\equiv \sum_{j=0}^{\infty} \frac{(-1)^j}{j! \, (\beta_0 \tilde{y}_1)^j} H_{2j+1}^{(k)}, \end{aligned}$$ (B.6)

where

$$H_{2j+1}^{(k)} \equiv \int_A^{\infty} \frac{dI}{I} \ln^{2j+1}\left( \frac{I}{A} \right) I^{k-1} e^{-I}.$$ (B.7)

We will estimate the integral $H_{2j+1}^{(k)}$ integral in three separate cases: $k \geq 2$ and $k = 1$, $k = 0$.

**Case $k \geq 2$.** With the help of the incomplete gamma function $\Gamma(a, x)$, the integral $H_{2j+1}^{(k)}$ reads

$$H_{2j+1}^{(k)} = \frac{\partial^{2j+1}}{\partial \alpha^{2j+1}} \bigg|_{\alpha=0} \left[ A^{-\alpha} \times \Gamma(\alpha + k - 1, A) \right].$$ (B.8)

The value of $\alpha$ (before being eventually set to zero) can be restricted in a small interval $(-\epsilon, \epsilon)$ around 0. So, for $k \geq 2$, as $A \ll 1$, we can approximate $\Gamma(\alpha + k - 1, A)$ by $\Gamma(\alpha + k - 1, 0)$. The derivative in Eq. (B.8) then appears as a sum of terms containing powers of $\ln(1/A)$. Keeping only the leading log term, we get

$$H_{2j+1}^{(k)} \simeq \Gamma(k-1) \ln^{2j+1} \frac{1}{A} = (k-2)! \ln^{2j+1} \frac{1}{A}.$$ (B.9)

Substituting this into Eq. (B.6), and resumming the leading log series, the integral $S_k$ for $k \geq 2$ reads

$$S_{k \geq 2}(A) \simeq (k-2)! \ln \frac{1}{A} \exp \left[ -\frac{\ln^2 \frac{1}{A}}{\beta_0 \tilde{y}_1} \right].$$ (B.10)

**Case $k = 1$.** By integration by parts, we have

$$\begin{aligned} H_{2j+1}^{(1)} &= \int_A^{\infty} \frac{dI}{I} \ln^{2j+1}\left( \frac{I}{A} \right) e^{-I} = \frac{1}{2j+2} \int_A^{\infty} dI \ln^{2j+2}\left( \frac{I}{A} \right) e^{-I} \\ &= \frac{1}{2j+2} \frac{\partial^{2j+2}}{\partial \alpha^{2j+2}} \bigg|_{\alpha=0} \left[ A^{-\alpha} \times \Gamma(\alpha + 1, A) \right]. \end{aligned}$$ (B.11)

With the same argument to the case $k \geq 2$, we can approximate $\Gamma(\alpha + 1, A)$ by $\Gamma(\alpha + 1, 0)$. The integral $H^{(1)}{}_{2j+1}$ in the leading log approximation reads

$$H^{(1)}_{2j+1} \simeq \frac{1}{2j+2} \ln^{2j+2} \frac{1}{A}. \tag{B.12}$$

Resumming the leading log series, we obtain the following expression for $S_1$:

$$\boxed{S_1(A) = \frac{\beta_0 \tilde{y}_1}{2} \left[ 1 - \exp\left( -\frac{\ln^2 \frac{1}{A}}{\beta_0 \tilde{y}_1} \right) \right].} \tag{B.13}$$

**Case** $k = 0$. By integration by parts, we get

$$
\begin{aligned}
H^{(0)}_{2j+1} &= \int_A^\infty \frac{dI}{I^2} \ln^{2j+1}\left(\frac{I}{A}\right) e^{-I} \\
&= \frac{1}{2j+2} \left[ \int_A^\infty \frac{dI}{I^2} \ln^{2j+2}\left(\frac{I}{A}\right) e^{-I} + \int_A^\infty \frac{dI}{I} \ln^{2j+2}\left(\frac{I}{A}\right) e^{-I} \right] \\
&= \frac{1}{2j+2} \left[ H^{(0)}_{2j+2} + H^{(1)}_{2j+2} \right] = \frac{1}{2j+2} \left[ H^{(0)}_{2j+2} + \frac{1}{2j+3} \ln^{2j+3} \frac{1}{A} \right].
\end{aligned}
\tag{B.14}
$$

From this, we have the following recurrence relation:

$$H^{(0)}_m = m H^{(0)}_{m-1} - \frac{1}{m+1} \ln^{m+1} \frac{1}{A}, \quad (m \geq 1). \tag{B.15}$$

By induction, we can prove the following general formula for $H^{(0)}_m$:

$$
\begin{aligned}
H^{(0)}_m &= H^{(0)}_0 m! - \sum_{i=-1}^{m-2} \frac{m!}{(m-i)!} \ln^{m-i} \frac{1}{A} \\
&= m! \left( \frac{e^{-A}}{A} - \Gamma(0, A) \right) - \sum_{i=-1}^{m-2} \frac{m!}{(m-i)!} \ln^{m-i} \frac{1}{A} \\
&= \frac{m!}{A} \left[ 1 + \mathcal{O}(A) \right] - \frac{1}{m+1} \ln^{m+1} \frac{1}{A} \left[ 1 + \mathcal{O}\left( \frac{1}{\ln \frac{1}{A}} \right) \right]
\end{aligned}
\tag{B.16}
$$

or,

$$H^{(0)}_{2j+1} = \frac{(2j+1)!}{A} \left[ 1 + \mathcal{O}(A) \right] - \frac{1}{2j+2} \ln^{2j+2} \frac{1}{A} \left[ 1 + \mathcal{O}\left( \frac{1}{\ln \frac{1}{A}} \right) \right] \tag{B.17}$$

Resumming only leading log terms, while keeping only the term $j = 0$ for the first part (containing $1/A$) of Eq. (B.17), $S_0$ finally reads

$$\boxed{S_0(A) \underset{\text{LL}}{\simeq} \frac{1}{A} - \frac{\beta_0 \tilde{y}_1}{2} \left[ 1 - \exp\left( -\frac{\ln^2 \frac{1}{A}}{\beta_0 \tilde{y}_1} \right) \right].} \tag{B.18}$$

When keeping up to next-to-leading-log order, $S_0$ reads

$$\boxed{S_0(A) \underset{\text{NLL}}{\simeq} \frac{1}{A} - \frac{\beta_0 \tilde{y}_1}{2}\left[1 - \exp\left(-\frac{\ln^2 \frac{1}{A}}{\beta_0 \tilde{y}_1}\right)\right] - \ln\frac{1}{A}\exp\left(-\frac{\ln^2\frac{1}{A}}{\beta_0\tilde{y}_1}\right).} \tag{B.19}$$

## B.2 Other integrals

In the thesis, we made use of the digamma function,

$$\psi(x) \equiv \frac{1}{\Gamma(x)}\frac{d\Gamma(x)}{dx}, \tag{B.20}$$

and its following integral representation:

$$\psi(x) = -\gamma_E + \int_0^1 \frac{1-t^x}{1-t}dt. \tag{B.21}$$

($\gamma_E = 0.57721\cdots$ is the Euler constant.)

The following integral was also needed:

$$\int \frac{d^2l^\perp}{2\pi}\, \frac{l^\perp \cdot n^\perp}{l^2}e^{il^\perp \cdot x^\perp} = i\frac{x^\perp \cdot n^\perp}{x^2}, \tag{B.22}$$

where $l^\perp, n^\perp$ and $x^\perp$ are two-dimensional transverse vectors. Let us now prove the latter. Denote by $\phi, \alpha$ and $\beta$ the angle between those vectors measured counter-clockwise in reference to $n^\perp$ as $\phi = (x^\perp, l^\perp), \alpha = (n^\perp, x^\perp)$, and $\beta = (n^\perp, l^\perp)$ $(\cos\beta = \cos(\phi+\alpha))$. We can rewrite the leff-hand side of Eq. (B.22) as

$$\begin{aligned}\int \frac{d^2l^\perp}{2\pi}\,\frac{l^\perp \cdot n^\perp}{l^2}e^{il^\perp \cdot x^\perp} &= \int_0^\infty dl \int_{-\pi}^\pi \frac{d\phi}{2\pi}n\cos\beta\, e^{ilx\cos\phi}\\ &= \int_0^\infty dl\int_{-\pi}^\pi \frac{d\phi}{2\pi}n\left(\cos\phi\cos\alpha - \sin\phi\sin\alpha\right)e^{ilx\cos\phi}\\ &= in\cos\alpha\int_0^\infty J_1(lx)dl = i\frac{x^\perp \cdot n^\perp}{x^2},\end{aligned} \tag{B.23}$$

where $J_1(x)$ is the first-order Bessel function of the first kind (qed).

# Numerical setup for solving the evolution equations

In this appendix, we present the numerical scheme to solve the BK equations for the $T_1$ and $\sigma_{in}$ both the fixed-coupling and the running-coupling scenarios, which is used for the numerical analyses in Chapter 3. The equation for the latter is equivalent to the KL equation.

To solve such integro-differential equations, we use the fourth-order Runge-Kutta method with rapidity step $dY = 10^{-2}$. The solutions are stored in a grid of the dipole size variable $r$ in which 1000 points are spaced equally in the logarithmic scale in the range $10^{-14} \leq r\Lambda_{QCD} \leq 10^2$. Integrals are computes using the mid-point quadrature scheme. For $r\Lambda_{QCD} < 10^{-14}$ (the color transparency limit), we use the power-law extrapolation, while for $r\Lambda_{QCD} > 10^2$ (the saturation limit), we set the solutions to 1.

Different parameters for the calculation are set as follows:

(i) The QCD parameter $\Lambda_{QCD} = 0.217$ . This value is obtained by requiring that the value of the running strong coupling at the mass of the $Z^0$ boson is $\bar{\alpha}(r^2 = 4C^2/M_{Z^0}^2) = 0.1104$ [149], with $M_{Z^0} \simeq 91.18 \ GeV$.

(ii) Fixed coupling $\bar{\alpha} = 0.14$.

(iii) The frozen value of the running coupling $\bar{\alpha}_c = 0.5$.

(iv) The constant $C$ in the expression of the running coupling is set to the value $C^2 = 6.5$ [138].

(v) Nuclear saturation scale at zero rapidity $Q_A^2 = 0.26A^{1/3}Q_{p0}^2$, where $A$ is the nuclear mass number and $Q_{p0}$ is the saturation scale of the proton, which is assumed to be $\Lambda_{QCD}$ . The factor 0.26 leads to the smallness of the ratio $Q_A^2/Q_{p0}^2$, which was interpreted as the weak nuclear enhancement [146].

(vi) Quark masses $m_u = m_d = m_s = 140 \ MeV$, $m_c = 1.5 \ GeV$. Active quark flavors in the sums appearing in Eq. (4.2) are determined from the condition $Q^2 > 4m_f^2$.

To check the validation of the numerical calculation, we extract the saturation momenta from the numerical solutions for the forward elastic amplitude $T_1$ and plot them as functions of the rapidity (for $A = 208$) (see Fig. C.1). When the coupling is fixed, the following function is fitted to the numerical data:

$$Q_s^{(fc)} = a_f \exp \left( b_f Y - c_f \ln Y \right). \tag{C.1}$$

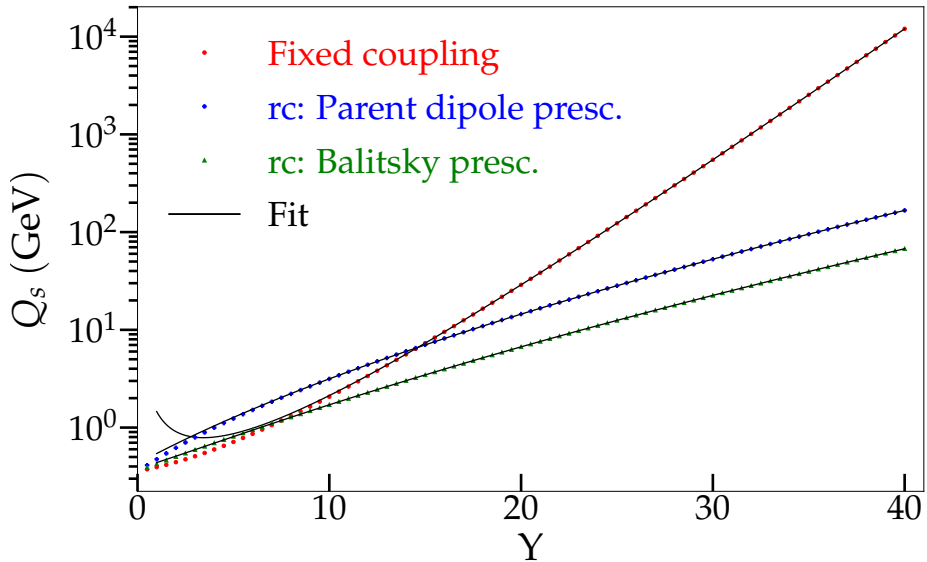

Figure C.1: Saturation momenta from the solutions for the amplitude $T_1$ in three different schemes. The black lines are fitting results using Eqs. (C.1) and (C.2).

In the case of running coupling, the fitting function reads

$$Q_s^{(rc)} = a_r \exp \left( b_f (Y + d_{r1})^{1/2} - c_f (Y + d_{r2})^{1/6} \right). \tag{C.2}$$

Fitting parameters are shown in Table C.1. The fitting values of the parameters $b_f, c_f$ and $b_r$ are close to their established theoretical values [55, 56, 142], which reads $b_f = \bar{\alpha} \chi'(\gamma_0)/2 \simeq 0.342$, $c_f = 3/(4\gamma_0) \simeq 1.195$ and $b_r = \alpha_c/2 \simeq 1.804$, respectively.

Table C.1: Values of the fitting parameters in Eqs. (C.1) and (C.2) obtained from the fits to the corresponding numerical data for the saturation momenta.

| **Kernel** | $a_f$ | $\mathbf{b_f}$ | $c_f$ | | |
|---|---|---|---|---|---|
| Fixed coupling | 1.035 | 0.342 | 1.174 | | |
| **Kernel** | $a_r$ | $\mathbf{b_r}$ | $c_r$ | $d_{r1}$ | $d_{r2}$ |
| rc: parent dipole | 0.419 | 1.805 | 3.374 | 7.862 | 11.131 |
| rc: Balitsky | 0.111 | 1.810 | 3.352 | 9.432 | 4.635 |

**Titre:** Propriétés statistiques des configurations partoniques et dissociation diffractive dans la diffusion électron-noyau à haute énergie.

**Mots clés:** chromodynamique quantique, diffusion électron-noyau, dissociation diffractive, modèle des dipôles de couleur, marche aléatoire avec branchements.

**Résumé:** Dans cette thèse, nous étudions les propriétés statistiques des états quantiques d'un quarkonium, et nous en déduisons des prédictions pour les sections efficaces de dissociation diffractive qui seront mesurées aux futurs collisionneurs électrons-ions.

Dans le cadre du modèle des dipôles de couleur de la chromodynamique quantique (QCD), on montre que de tels états peuvent être représentés par un ensemble de dipôles généré par un processus stochastique défini par un branchement binaire particulier. En premier lieu, les événements d'interaction d'un dipôle de couleur avec un noyau lourd, dans le régime dans lequel les paramètres de la réaction sont définis de sorte que la section efficace totale soit petite, sont induits par des fluctuations partoniques rares, dont la distribution dépend du référentiel choisi. Il s'avère que la liberté de sélectionner un référentiel permet de déduire une expression analytique asymptotique de la distribution de la rapidité du premier branchement du dipôle parent le plus lent dans l'ensemble des dipôles qui interagissent. Notre étude montre l'importance de bien comprendre la distribution des dipôles et leurs corrélations dans ces fluctuations particulières, dont les propriétés sont communes à une vaste classe de modèles de marches aléatoires branchantes. Dans ce but, nous développons un nouvel algorithme de Monte Carlo pour générer la région frontalière d'une marche aléatoire branchante unidimensionnelle.

De plus, notre approche nous permet de calculer la section efficace diffractive conditionnée à un "gap" de rapidité minimal $Y_0$ ou la distribution des "gaps" de rapidité $Y_{gap}$ dans la dissociation diffractive d'un petit dipôle sur un noyau lourd, dans une région paramétrique bien définie. Nous obtenons ainsi des solutions asymptotiques à l'équation de Kovchegov-Levin pour la section efficace de dissociation diffractive nucléaire d'un dipôle à haute énergie. Enfin, nous présentons des prédictions quantitatives pour la distribution des "gaps" de rapidité dans le domaine cinématique des futurs collisionneurs électron-ion, sur la base de solutions numériques de l'équation originale de Kovchegov-Levin et de son extension à une constante de couplage forte courante.

**Title:** Statistical properties of partonic configurations and diffractive dissociation in high-energy electron-nucleus scattering.

**Keywords:** quantum chromodynamics, electron-nucleus collision, diffractive dissociation, color dipole model, branching random walk.

**Abstract:** In this thesis, we study the detailed partonic content of the quantum states of a quark-antiquark color dipole subject to high-energy evolution, which are represented by a set of dipoles generated by a stochastic binary branching process, in the scattering off a large nucleus. We also produce predictions for diffractive dissociation in electron-ion collisions, based on the dipole picture of quantum chromodynamics (QCD). Our main results can be captured as follows.

First, the scattering events of a color dipole, when parameters are set in such a way that the total cross section is small, are triggered by rare partonic fluctuations, which look different as seen from different reference frames. It turns out that the freedom to select a frame allows to deduce an asymptotic expression for the rapidity distribution of the first branching of the slowest parent dipole of the set of those which scatter. In another aspect, such study implies the importance of the characterization of particle distribution in the frontier region in the states generated by the QCD dipole branching, and more generally, by any one-dimensional branching random walk model. To this aim, we develop a Monte Carlo algorithm to generate the frontier region of a binary branching random walk on a real line.

Furthermore, with the above statistical description, we are able to calculate the diffractive cross section demanding a minimal rapidity gap $Y_0$ and the distribution of rapidity gaps $Y_{gap}$ in the diffractive dissociation of a small dipole off a large nucleus, in a well-defined parametric region. They are the asymptotic solutions to the so-called Kovchegov-Levin equation, which describes diffractive dipole dissociation at high energy. Additionally, we present predictions for the distribution of rapidity gaps in realistic kinematics of future electron-ion machines, based on the numerical solutions of the original Kovchegov-Levin equation and of its next-to-leading extension taking into account the running of the strong coupling.